\documentclass[11pt]{article}
\usepackage{graphicx}
\usepackage{amssymb}
\usepackage{amsmath}
\usepackage{amsfonts}
\usepackage{bm}
\usepackage[square,numbers,sort&compress]{natbib}

\usepackage{epsfig}

\usepackage{amssymb}
\usepackage{amsmath}

\usepackage{lineno}

\allowdisplaybreaks[3]
\makeatletter
\@addtoreset{equation}{section}
\makeatother

\newcommand\cev[1]{\overleftarrow{#1}}
\newcommand{\clebz}[3]{(\, #1\,\,0\,\,#2\,\,0|#3\,\,0\,)} 

\newcommand{\bra}[1]{\langle \, #1 \, |}
\newcommand{\ket}[1]{| \, #1 \, \rangle}
\newcommand{\Tr}{\text{Tr}}
\usepackage{bm}
\newcommand{\Slash}[1]{\ooalign{\hfil/\hfil\crcr$#1$}}

\newcommand{\be}{\begin{eqnarray}}
\newcommand{\ee}{\end{eqnarray}}
\newcommand{\calO}{{\cal O}}

\usepackage{sectsty} 

\usepackage{authblk} 

\usepackage{hyperref}

\begin{document}

\title{Heavy Hadrons in Nuclear Matter}

\date{}

\author[1,2]{Atsushi~Hosaka}
\author[3]{Tetsuo~Hyodo}
\author[4]{Kazutaka~Sudoh}
\author[3,5]{Yasuhiro~Yamaguchi}
\author[6]{Shigehiro~Yasui\thanks{yasuis@th.phys.titech.ac.jp}}
\affil[1]{Research Center for Nuclear Physics (RCNP), Osaka University, Ibaraki, Osaka, 567-0047, Japan}
\affil[2]{J-PARC Branch, KEK Theory Center, Institute of Particle and Nuclear Studies, KEK, Tokai, Ibaraki, 319-1106, Japan}
\affil[3]{Yukawa Institute for Theoretical Physics, Kyoto University, Kyoto, 606-8317, Japan}
\affil[4]{Nishogakusha University, 6-16, Sanbancho, Chiyoda, Tokyo, 102-8336, Japan}
\affil[5]{Istituto Nazionale di Fisica Nucleare (INFN), Sezione di Genova, via Dodecaneso 33, 16146 Genova, Italy}
\affil[6]{Department of Physics, Tokyo Institute of Technology, Tokyo 152-8551, Japan}

\maketitle

\begin{abstract}%
Current studies on heavy hadrons in nuclear medium are reviewed with a summary of the basic theoretical concepts of QCD, namely chiral symmetry, heavy quark spin symmetry, and the effective Lagrangian approach.
The nuclear matter is an interesting place to study the properties of heavy hadrons from many different points of view.
We emphasize the importance of the following topics: (i) charm/bottom hadron-nucleon interaction, (ii) structure of charm/bottom nuclei, and (iii) QCD vacuum properties and hadron modifications in nuclear medium.
We pick up three different groups of heavy hadrons, quarkonia ($J/\psi$, $\Upsilon$), heavy-light mesons ($D$/$\bar{D}$, $\bar{B}$/$B$) and heavy baryons ($\Lambda_{c}$, $\Lambda_{b}$).
The modifications of those hadrons in nuclear matter provide us with important information to investigate the essential properties of heavy hadrons.
We also give the discussions about the heavy hadrons, not only in infinite nuclear matter, 
 but also in finite-size atomic nuclei with finite baryon numbers, to serve future experiments.

\end{abstract}


\tableofcontents


\section{Introduction}
\label{sec:introduction}

It is an important problem to understand hadron properties based on the fundamental theory of the strong interaction, Quantum Chromodynamics (QCD).  
Due to the non-trivial features of the QCD dynamics at low energies, 
the hadron physics shows us many interesting and even unexpected non-trivial phenomena.  
The fact that hadronic phenomena are so rich implies that various studies from many different views are useful and indispensable to reveal the nature of the hadron dynamics. 
Not only isolated hadrons but also hadronic matter under extreme conditions of high temperature, of high baryon density, and of many different flavors provide important hints to understand the hadron dynamics.

One of familiar forms of hadronic matter is the atomic nucleus, the composite system of protons and neutrons.   
The nuclear physics has been developed so far, based on various phenomenological approaches (shell models, collective models, and so on).  
Recently, ab-initio calculations are being realized such that many-body nuclear problems are solved starting from the bare nucleon-nucleon interaction determined phenomenologically with high precision~\cite{Bertsch:2007,Bogner:2009bt,Lacroix:2010qn}.
Yet a large step forward has been made; the lattice QCD  has derived the nucleon-nucleon interaction~\cite{Ishii:2006ec,Aoki:2009ji}.
Thus the so far missing path from QCD to nucleus is now being exploited.

Nevertheless, if we look at the problem, for instance, of neutron stars, we confront with a difficulty in explaining the so-called twice the solar mass problem.  
Because of the high density environment in the inside of the neutron star, the strangeness, the third flavor of quarks,  appears as an explicit degree of freedom.  
This occurs primarily in the form of hyperons forming hypernuclei, composite systems of hyperons, such as $\Lambda$ and $\Sigma$, together with protons and neutrons.  
Hypernuclear physics is then an active field, where one of the current goals is to determine the two-body and even three-body forces for hyperons and nucleons to explain the stability of the massive neutron stars~\cite{Lacroix:2010qn,Hashimoto:2006aw,Botta:2012xi}.

The dynamics of strange hadrons can provide the new energy scale, several hundred MeV, which is much larger 
than nuclear physics scale of order a few or ten MeV at most (single-particle motion, surface vibration, rotation of deformed nuclei, nucleon pairings and so on).
As a famous example actively studied,
 we expect that (anti-)kaons appear in nuclear matter as an active 
degrees of freedom~\cite{Akaishi:2002bg}.  
This is another possible form of dense strangeness flavored matter.
Because of that large energy scale, we need to consider the properties of hadrons explicitly.\footnote{This may be contrasted with the traditional view of nuclear physics, namely the nucleon dynamics is not seen directly, but it is ``renormalized" to the low energy degrees of freedom such as the collective modes (e.g. surface vibration, rotation, nucleon pairings)~\cite{ring2004nuclear}.}
Anti-kaons in the nuclear matter have long been discussed in various respects.  
They interact with nucleons attractively in particular in S-wave. 
Because of the suppressed kinetic energy due to their rather large mass, even small attraction is enough to trap them in a nucleus.  
A well-known and the simplest system of such is $\Lambda(1405)$ as a quasi-bound state of $\bar KN$, the first negative parity excited state of $\Lambda$~\cite{Hyodo:2011ur}.  
Partly due to the difficulty of the quark model in explaining the state despite its general success, such an idea was proposed many years ago~\cite{Dalitz:1967fp}.  
After some time, it has been revived due to the developments of  chiral theories of QCD.
It has now become one of active subjects to confirm its nature of the $\bar KN$ quasi-bound state.
Once this will turn out to be the case, the impact on the hadron and nuclear physics is very large, 
where we expect to see many rich and unexpected phenomena.  

The strangeness does not only bring new phenomena but also plays a role of impurity to analyze aspects of the strong interaction  that we cannot see easily without it.  
First of all, obviously, the strange quark brings another energy scale such that we may be able to see the QCD dynamics at various energy scales.  
One example is already seen above; the attractive force between the anti-kaon and the nucleon as a consequence of the low energy theorems of the spontaneously broken chiral symmetry has a different energy scale from that between the pion and the nucleon.\footnote{We notice that, for the light $u$ and $d$ flavors, the first excited state of the nucleon, $N(1535)$ with spin-parity $J^{P}=1/2^{-}$, is well explained as an orbital excitation of valence quarks (P-wave excitation), but with an $s$ quark the corresponding state, $\Lambda(1405)$ with $J^{P}=1/2^{-}$, shows up with very different structure governed by the $\bar{K}N$-$\pi \Sigma$ dynamics.}
Another example is the mass inversion of the $\Lambda(1820)$-$\Sigma(1775)$ of spin and parity 
$J^P = 5/2^-$ in comparison with the ground states $\Lambda(1116)$-$\Sigma(1190)$.  
These examples show that by using an impurity the properties of the system changes, and it brings us useful information to understand the underlying mechanism of hadron dynamics.  

Turning to the nuclear matter, study of hadrons in nuclear medium provides also a unique tool for investigating the vacuum structure of QCD.
In the QCD vacuum, it is known that there appear several different types of the quark and gluon condensates as a result of non-perturbative effects from QCD at low energy.
Those condensates are decisive for the hadron properties (mass, interaction and so on).
One of the most important condensates is the chiral condensate induced by the dynamical breaking of chiral symmetry.
In fact, the light mass of the pion can be explained as the nature of the Nambu-Goldstone boson, which appears as the lowest energy states in the symmetry-broken vacuum.
The small but finite mass of the pion is understood by the Gell-Mann--Oaks--Renner relation~\cite{GellMann:1968rz}, where the pion mass is related to the small explicit breaking of chiral symmetry for $u$, $d$ quarks and to the chiral condensate in vacuum.
As a consequence of the spontaneous symmetry breaking, the interaction of pions with a matter field is constrained by chiral symmetry, such as by the Weinberg-Tomozawa interaction~\cite{Weinberg:1966kf,Tomozawa:1966jm},\footnote{This is the driving interaction to provide the strong $\bar{K}N$ attraction for $\Lambda(1405)$, when the $\bar{K}$ meson is regarded as the Nambu-Goldstone boson, as stated above.} as well as by the Goldberger-Treiman relation for the axial-vector coupling.
As another example, the gluon condensate is related to the hadron mass generation as a consequence of the scale anomaly (the trace anomaly).
However, it is still a non-trivial problem how those quark and gluon condensates affect the hadron properties.
This problem can be accessed by observing the modifications of hadrons when those condensates change in nuclear matter (see Refs.~\cite{Hayano:2008vn,Leupold:2009kz} for recent reviews).
For example, it is known both in experiments and in theories that the spectra of vector mesons change in atomic nuclei.
In this respect, $\phi$ meson with $s$ quarks in nuclear matter, which is related to the problem of the $s\bar{s}$ contents in a nucleon, is also interesting~\cite{Hatsuda:1991ez,Gubler:2014pta,Gubler:2016itj}. 
Therefore, the nuclei can be used as a good stage to investigate the hadron properties from QCD.

Having said so much, now we attempt to make further extensions of flavors to the {\it heavy flavor} region,
namely the {\it charm} and {\it bottom} flavors, in the hadronic matter.
The introduction of heavy quarks changes the nuclear system to be free from the constraint governed by chiral symmetry.
The problem is very challenging because so far we have little experimental data.  
Because the charm (bottom) quark is very heavy as compared to the $u$, $d$ and $s$ quarks, it is natural that they do not appear in ordinary hadronic matter, nuclei.  
But they must show up under  extreme conditions which we do not know much about.  
Thus we expect yet unexperienced phenomena in the presence of the charm (bottom) quark.
Not only that, it should provide  useful information for the QCD dynamics of hadrons.  
In fact, in the past decade we have experienced
exciting time for discoveries of many exotic charm (and bottom as well) hadrons, called $X$, $Y$, $Z$ states~\cite{Swanson:2006st,Brambilla:2010cs,Hosaka:2016pey} as well as $P_{c}(4380)$ and $P_{c}(4450)$ states~\cite{Aaij:2015tga} and $X(5586)$ state~\cite{D0:2016mwd}, which were observed in accelerator facilities.
Their existence strongly suggests that our naive picture of hadrons, namely three quarks for a baryon and quark-antiquark for a meson, is not sufficient, as already suggested by $\Lambda(1405)$.  
The understanding of exotic hadrons
 is the latest
  hot topic in hadron physics~\cite{Swanson:2006st,Brambilla:2010cs,Hosaka:2016pey}.  

There are novel features for charm and bottom quarks, which are qualitatively different from the light quarks~\cite{Neubert:1993mb,Manohar:2000dt}.

First, charm and bottom quarks have heavy masses: $m_{c}=1.275\pm0.025$ GeV and $m_{b}=4.66\pm0.03$ GeV~\cite{Agashe:2014kda}.
Those masses are much larger than the typical scale of the low energy QCD, 
$\Lambda_{\rm QCD}=214\pm 7$ MeV ($\overline{MS}$ scheme with $N_{f}=5$~\cite{Agashe:2014kda}.), which gives the energy scales of the hadron dynamics.
Therefore, naively to say, we may expect that a charm (bottom) quark can play the role of an ``impurity particle".
This is a new degree of freedom in the low energy QCD, which would not be much affected by the change of vacuum.
This property shows up when we study the change of heavy hadrons in nuclear medium; we can separate the change of the light degrees of freedom from the heavy quark (the Born-Oppenheimer approximation).

Second, the heavy quark interaction has a special property in coupling to a gluon field;
the spin-flip process of the heavy quark is suppressed by the $1/m_{Q}$ factor with the heavy quark mass $m_{Q}$.
Especially, it becomes completely zero when the heavy quark limit ($m_{Q}\rightarrow \infty$) is adopted.
The suppression of the spin-flip interaction is helpful to separate the light degrees of freedom from the heavy quark.
In fact, this property leads the heavy quark spin symmetry as a novel symmetry in the heavy quark sector, which plays the significant role in the heavy hadron dynamics.

Those two properties of a heavy quark, namely the separation of degree of freedom and the suppression of the spin-flip process, are crucial to explain many properties (mass splittings, branching ratios of decays and so on) of the charm/bottom hadrons.

\begin{figure}[tbp]
\begin{center}
\includegraphics[width=11cm,bb=0 0 842 595]{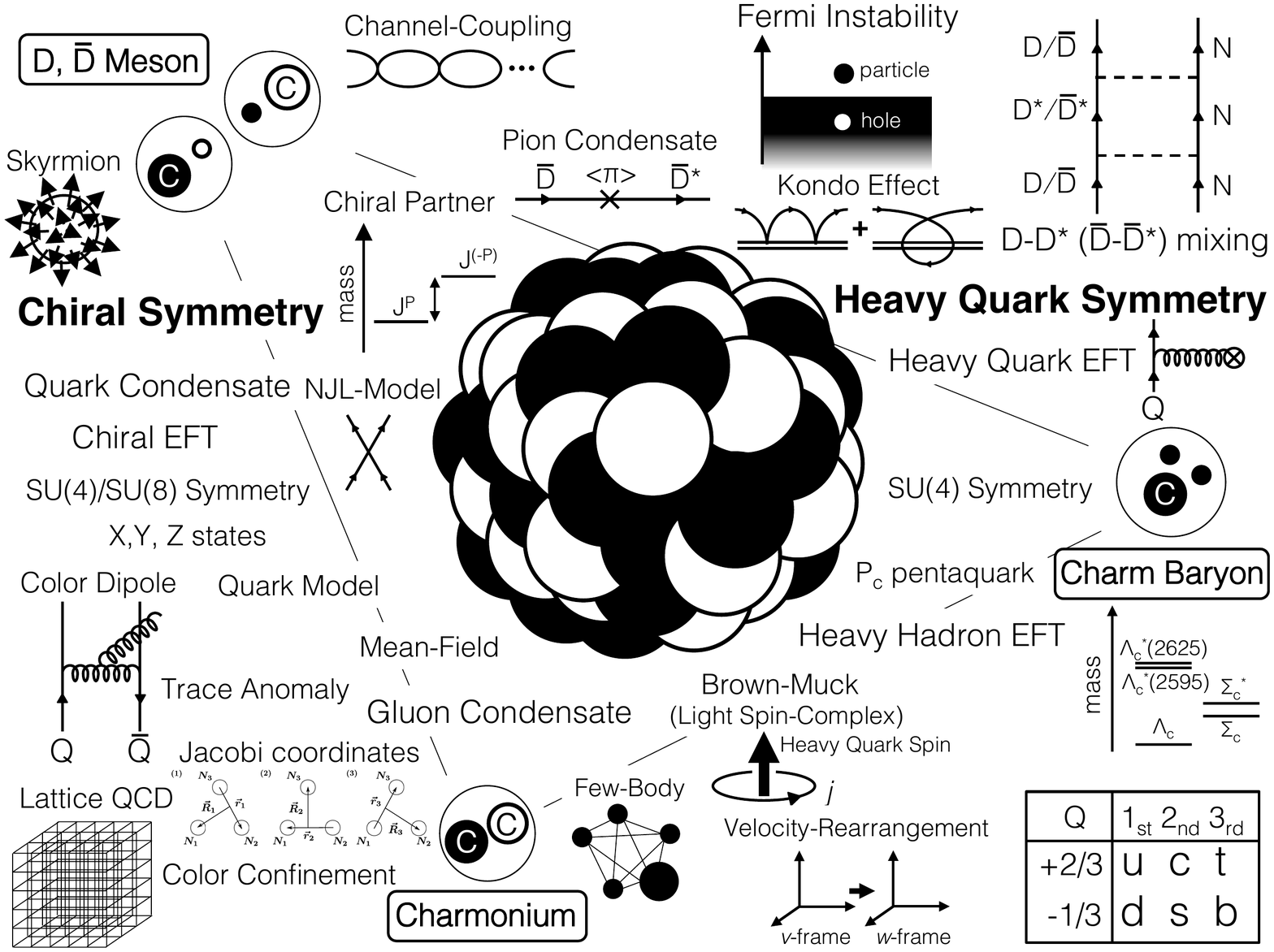}
\caption{A schematic figure for topics covered by studies of heavy hadrons in nuclear medium.}
\label{fig:160518}
\end{center}
\end{figure}

Given those unique properties of the heavy quarks, we expect it quite interesting to study the ``{\it heavy-flavor nuclei}" which are nuclei 
 containing a charm quark or a bottom quark as an impurity particle, 
with
 the extension of flavors from up, down and strangeness to charm and bottom.
There are many open problems, which should be addressed in the study of the heavy-flavor nuclei: how the nuclear structure changes by a heavy impurity hadron, how the hadron-hadron interactions as well as the hadron masses are affected by the change of the QCD vacuum, what kind of low energy mode heavy-flavor nuclei can have, and so on (Fig.~\ref{fig:160518}).

Our purpose in this review is to survey the preceding and current studies of the charm/bottom hadrons in heavy-flavor nuclei.
The main issues are summarized 
as
the following three items.

\begin{enumerate}
\item Charm/bottom hadron-nucleon interaction
\item Structure of charm/bottom nuclei 
\item QCD vacuum and hadron properties in nuclear medium 
\end{enumerate}

1.~{\it Charm/bottom hadron-nucleon interaction.---}
Charm/bottom hadron-nucleon interaction is one of the most basic 
ingredients
to study the charm/bottom nuclei.
However, the biggest problem is that there is only poor information from experiments, especially at low-energy scattering.
In literature, instead, there have been theoretical studies about various types of the charm/bottom hadron-nucleon interaction.
As a naive extension from SU(3) symmetry, which is valid in light flavors up to strangeness,  we may consider the SU(4) flavor symmetry including a charm flavor.
We may also consider SU(5) symmetry up to the bottom flavor.
Although the SU(4) and/or SU(5) symmetries would be useful for classifying the hadron states, we have to keep in mind that those symmetries 
cannot be applied
for the hadron spectroscopy.\footnote{See for example Ref.~\cite{georgi1999lie}.}
When we regard the mass of the charm/bottom quark $m_{Q}$ much heavier than the typical low-energy scale $\Lambda_{\mathrm{QCD}}$, it is a natural way to consider the heavy quark limit ($m_{Q} \rightarrow \infty$, $\Lambda_{\mathrm{QCD}}/m_{Q} \rightarrow 0$) as the leading approximation.

Whatever symmetries are adopted for heavy quarks,
it is an important question to ask what kind of hadron interactions is working.
At long distances, there are various types of meson exchange forces such as vector meson exchanges providing strong repulsion or attraction (depending on baryon charges)~\cite{Sakurai:1960ju}, and a pion exchange leading to tensor force which is crucial for the binding of the deuteron~\cite{Tornqvist:1991ks,Manohar:1992nd}.
At short distances, it is also possible to have direct quark exchanges~\cite{Oka:1980ax,Oka:1981ri,Oka:1981rj} and multi-gluon (Pomeron) exchanges~\cite{Brodsky:1989jd}, and so on.
Recently, it has become possible to study the hadron interactions from the first principle due to the rapid development of the lattice QCD computations~\cite{Ishii:2006ec,Yamazaki:2011nd,Beane:2011iw}.
We will overview the current status of the understanding of the heavy-hadron interaction in various approaches.

\vspace{1em}

2.~{\it Structure of charm/bottom nuclei.---}
Based on the charm/bottom hadron-nucleon interaction, we investigate the properties of the charm/bottom nuclei as many-body systems in several theoretical approaches.
In our approach, we regard the nuclear matter as an almost free Fermi gas, and investigate the medium effects for the charm/bottom hadrons by considering the Pauli exclusion effect from the occupied Fermi sea of the nucleons (Sect.~\ref{sec:nuclear_matter}).
By this method, we obtain the effective masses, the effective coupling constants, and the decay widths of the charm/bottom hadrons in nuclear medium (cf.~G-matrix formalism~\cite{Bethe:1971xm}).
We note that the nuclear matter as a free Fermi gas is quite unstable for any small attraction between nucleons, leading to another state as the most stable state (cf. the BCS instability in superconductivity~\cite{abrikosov1975methods}).\footnote{This is an example of the quantum fluctuations, which are important in the state with finite baryon number density. Such fluctuation effect is suppressed at finite temperature.}
This is called the Fermi instability.
It is an important subject to investigate the effect of the Fermi instability when the heavy hadron exists as an impurity particle in nuclear matter.\footnote{The Kondo effect is a known phenomena caused by the Fermi instability when the heavy impurity particle with non-Abelian interaction exists (see~Sect.~\ref{sec:D_mesons}).}
As a feedback effect, the behavior of nucleons in nuclear matter is also affected by the existence of heavy hadrons, and it can change the nuclear matter.
Eventually, we have to analyze the dynamics both for heavy hadrons and nucleons in a self-consistent way.

\vspace{1em}

3.~{\it QCD vacuum and hadron properties in nuclear medium.---}
The quark condensates and gluon condensates are directly related to the properties of the QCD ground state.
When those quantities are modified
in nuclear matter, 
the modification affects
the properties of hadrons in nuclear matter (Sect.~\ref{sec:QCDSR}).
Here we consider several basic topics of the QCD vacuum properties.
As is known, chiral symmetry plays an
important role for the generation of the hadron masses and interactions at low energy as the result of the dynamical breaking in vacuum.
In the light flavor sector, it has been studied that, as a precursory signal, the partial restoration of chiral symmetry inside nuclei can be observed through the change of hadron properties (e.g.~mass modification)~\cite{Hatsuda:1994pi}.
We may ask what kind of condensate is responsible for the properties of heavy hadrons in nuclear medium.
For example, the gluon condensate is an interesting quantity for heavy hadrons, because it will be expected that the gluon dynamics dominates over the light quark dynamics in heavy flavor sector~\cite{Luke:1992tm,Klingl:1998sr}.
The light quark condensate $\langle \bar{q}q \rangle$ is also important for the mass generation of the heavy-light mesons as well as of heavy-light-light and heavy-heavy-light baryons.

\vspace{1em}

From the above considerations, we will discuss how the charm/bottom hadrons behave inside nuclei.
However, we have to say that this field is still in progress, and the systematic knowledge 
has not yet been obtained so far.
The purpose of this review is, therefore, to summarize the current results in the theoretical schemes studied so far, to point out the important views and unsolved problems, and to motivate the readers to study further in coming future.

It is worthwhile to remind the readers of the important points of view through this review.
First, we emphasize the role of ``symmetry", such as chiral symmetry and heavy quark symmetry, in the heavy hadron systems.
Symmetries enable us to understand general features of the physical systems in a model-independent manner, although we have to rely on model-dependent calculations to obtain numerical results to be connected with experimental data in many cases.
Second, we emphasize the importance of the ``finite size" of the charm/bottom nuclei.
Of course, it is very useful in theoretical analysis to consider the infinite volume of nuclear matter, because the theoretical treatment is much easier than that in finite systems.
However, we should not ignore the properties of finite systems, such as surface effects and discrete energy levels, which are characteristic aspects of nuclei.
The few-body calculation is also important to understand the charm/bottom nuclei to compare the theoretical results with experimental data.
In literature, there have been some few-body calculations performed so far, but the applied systems are still limited.
As one of the techniques of the few-body calculation, we will pick up the Gaussian expansion method and explain some details which would provide us with a useful tool to investigate the charm/bottom nuclei.

This review is organized as follows.
In Sect.~\ref{sec:theory}, we will overview the basics of the theoretical approaches.
We summarize the basic properties of QCD (Sect.~\ref{sec:Properties_of_QCD}),
give explanation about chiral symmetry and heavy quark symmetry (Sect.~\ref{sec:symmetries}), and introduce hadron effective theories.
 (Sect.~\ref{sec:hadronic_effective_theories}).
We discuss some technical descriptions for nuclei, such as few-body calculation, propagators of nucleons in nuclear matter, and general properties of chiral symmetry in nuclear medium (Sect.~\ref{sec:finite_density}).
In the following sections, we survey the current status of theoretical studies of the charm/bottom nuclei.
Here we separate the discussions about the heavy hadrons according to the combinations of light quark $q$ and heavy quark $Q$:
 (i) $\bar{Q}Q$ mesons (e.g.~$\eta_{c}$ and $J/\psi$) in Sect.~\ref{sec:charmonia},
 (ii) $\bar{q}Q$ and $q\bar{Q}$ mesons (e.g.~$D$ and $D^{\ast}$, $\bar{D}$ and $\bar{D}^{\ast}$) in Sect.~\ref{sec:D_mesons}
 and (iii) $Qqq$ baryons (e.g.~$\Lambda_{c}$, $\Sigma_{c}$ and $\Sigma_{c}^{\ast}$) in Sect.~\ref{sec:charm_baryons}, respectively.
For each type of heavy hadron, we focus on the three different properties: (a) heavy hadron-nucleon interaction, (b) few-body systems and (c) heavy hadrons in nuclear matter.
The last section is devoted to summary and future perspectives.

\section{Theoretical basics}
\label{sec:theory}

The basic theory of hadrons is, needless to say, given by the quantum chromodynamics (QCD).
In contrast, the effective theories are often used in the actual studies of heavy hadrons rather than QCD.
In this section, we will see theoretical tools for heavy hadrons in nuclear systems, by considering how the heavy hadron effective theory is connected to QCD through the symmetries, such as chiral symmetry and heavy quark symmetry.

\subsection{Properties of QCD}
\label{sec:Properties_of_QCD}

Let us start from the QCD Lagrangian
\begin{align}
    {\cal L}_{\mathrm{QCD}} 
    &= - \frac{1}{4} G_{\mu\nu}^{a} G^{a \mu\nu} 
    + \bar{q} 
    (i\Slash{D}-m_{\mathrm{q}})q .
    \label{eq:QCD}
\end{align}
The first term is the pure gluonic part made of the gluon field $A_{\mu}^{a}$ $(a=1,\dots, 8)$ which belongs to the adjoint representation of color SU(3) symmetry. The field strength tensor is defined by
\begin{align}
    G_{\mu\nu}^{a}
    &= 
    \partial_{\mu}A_{\nu}^{a}
    -\partial_{\nu}A_{\mu}^{a}
    -gf^{abc}A_{\mu}^{b}A_{\nu}^{c} ,
\end{align}
where $g$ is the gauge coupling and $f^{abc}$ are the structure constants defined by $[T^{a},T^{b}]=if^{abc}T^{c}$ with $T^{a}$ being the generator of the color SU(3) symmetry. The quark field $q$ belongs to the fundamental representation of color SU(3), and the coupling to the gauge field is given by the covariant derivative 
\begin{align}
    D_{\mu}
    &= 
    \partial_{\mu}
    +igA_{\mu}^{a}T^{a} .
\end{align}
The generators in the fundamental representation are expressed by the three-by-three Gell-Mann matrices as $T^{a}=\lambda^{a}/2$. The quark field has six flavor indices
\begin{align}
    q
    &=\begin{pmatrix}
    u & d & s & c & b & t
    \end{pmatrix}^{t} .
\end{align}
Each component carries the flavor quantum number (isospin $I$, third component of the isospin $I_{3}$, strangeness $S$, charm $C$, bottomness $B$ and topness $T$) as follows:
\begin{align}
    u &: I=1/2, I_{3}=+1/2, \\
    d &: I=1/2, I_{3}=-1/2, \\
    s &: S=-1, \\
    c &: C=+1, \\
    b &: B=-1, \\
    t &: T=+1. 
\end{align}
In Eq.~\eqref{eq:QCD}, $m_{\rm q}$ is the mass of the quark with the flavor q generated by the Higgs mechanism.

QCD is a renormalizable quantum field theory, and the running coupling constant decreases at high energy due to the asymptotic freedom.
At low energy, the coupling constant blows up and the perturbative calculation breaks down. This occurs at the energy scale $\Lambda_{\rm QCD}$ which is specified below.
As a result, in contrast to the simplicity of the QCD Lagrangian, the content which is induced from the Lagrangian is much complicated and provides rich structure of vacuum.
According to the renormalization group, the fine structure ``constant" $\alpha_{s}=g^{2}/4\pi$ is not actually the constant, but runs as a function of the energy scale parameter $\mu$ by
\begin{align}
 \mu^{2} \frac{\mathrm{d}\alpha_{s}}{\mathrm{d}\mu^{2}} = \beta(\alpha_{s}) \equiv - \left( b_{1} \alpha_{s}^{2}+b_{2}\alpha_{s}^{3}+ \cdots \right),
 \label{eq:renormalization_QCD}
\end{align}
where $\beta(\alpha_{s})$ is the Gell-Mann--Low beta function.
The explicit form of the beta function is given in powers of $\alpha_{s}$ in perturbative calculation at high energy scale, where the coefficients $b_{1}$, $b_{2}$, $\cdots$, at each order can be calculated.
For example, it is known that we obtain $b_{1}=(33-2N_{f})/12\pi$ with the  number of flavors $N_{f}$ at one-loop level.
At this order, the solution of Eq.~(\ref{eq:renormalization_QCD}) is given by
\begin{align}
 \alpha_{s}(\mu^{2}) = \cfrac{4\pi}{\left(11-\frac{2}{3}N_{f} \right) \ln \left( \frac{\mu^{2}}{\Lambda_{\mathrm{QCD}}^{2}} \right)},
  \label{eq:renormalization_QCD_sol}
\end{align}
with the parameter $\Lambda_{\mathrm{QCD}}$ for $\Lambda_{\mathrm{QCD}} \ll \mu$.
The value of $\Lambda_{\mathrm{QCD}}$ can be estimated by fitting the coupling strength $\alpha_{s}(\mu_{0}^{2})$ to reproduce experimental observables at some high energy scale $\mu_{0}$. The explicit value depends on the number of flavors and the renormalization scheme adopted in the calculation of the renormalization group equation. From the recent analysis~\cite{Agashe:2014kda}, it is determined as $\Lambda_{\rm QCD}=214\pm 7$ MeV in the $\overline{MS}$ scheme with $N_{f}=5$.
Note that the solution (\ref{eq:renormalization_QCD_sol}) is correct only at high energy scale more than a few GeV, because the perturbative expansion in terms of $\alpha_{s}$ in the right hand side of Eq.~(\ref{eq:renormalization_QCD}) should be valid at perturbative level.
Nevertheless, when we regard $\mu$ as small quantity and set $\mu \simeq \Lambda_{\mathrm{QCD}}$,
we find that $\alpha_{s}$ becomes a large coupling constant.
Such a strong coupling indicates the break-down of the perturbative calculation.
Therefore, we need essentially non-perturbative approaches to analyze the low energy phenomena of QCD.
It is very difficult to accomplish this by analytical calculations, and we need the numerical calculation by the lattice QCD simulations as first principle approach.
However, in some cases, we can use a {\it phenomenological} framework by focusing on the proper symmetries in QCD and assuming that the hadronic degrees of freedom are fundamental.
In the following, we will explain two symmetries, chiral symmetry for light quarks and spin symmetry for heavy quarks, and introduce the hadron effective theories based on those symmetries.

\subsection{Symmetries}
\label{sec:symmetries}

\subsubsection{Chiral symmetry}
\label{sec:chiral_symmetry}

\paragraph{Light quarks and chiral symmetry}
Chiral symmetry is a guiding principle to study the low-energy phenomena of the strong interaction, which is approximately realized in the light quark sector of QCD~\cite{Donoghue:1992dd,Hosaka:2001ux,Scherer:2012xha}. The light quarks satisfy
\begin{align}
    m_{\rm q} \ll \Lambda_{\rm QCD}
    \label{eq:lightquark} ,
\end{align}
namely, the quarks whose masses are much smaller than the QCD scale. For up and down quarks, Eq.~\eqref{eq:lightquark} is well satisfied. For these light quarks, it is reasonable to start from the massless Lagrangian
\begin{align}
    {\cal L}_{\mathrm{massless}} 
    &=  \bar{q} 
    i\Slash{D}q , \label{eq:massless}
\end{align}
with $q=(u,d)^{\mathrm{t}}$
and to treat the effect of the quark mass $m_{\rm q}$ perturbatively. For the discussion of chiral symmetry, we introduce the projection operators 
\begin{align}
    P_{L}
    &= \frac{1-\gamma_{5}}{2} , \quad
    P_{R}
    = \frac{1+\gamma_{5}}{2} ,
\end{align}
and define the left- and right-handed quarks as
\begin{align}
    q_{L} 
    &=  P_{L}q,
    \quad 
    q_{R}
    =  P_{R}q .
\end{align}
In the massless Lagrangian~\eqref{eq:massless}, the left-handed quarks are separated from the right-handed quarks by this decomposition:
\begin{align}
    {\cal L}_{\mathrm{massless}} 
    &=  \bar{q}_{L} i\Slash{D}q_{L}
    +\bar{q}_{R} i\Slash{D}q_{R} .
\end{align}
This Lagrangian is invariant under global $\textrm{U}(2)_{R}\otimes \textrm{U}(2)_{L}$ transformation. The vector $\textrm{U}(1)$ part is realized as the conservation of the quark number, while the axial $\textrm{U}(1)_{A}$ symmetry is explicitly broken by quantum anomaly~\cite{tHooft:1986nc}. The invariance under the $\textrm{SU}(2)_{R}\otimes \textrm{SU}(2)_{L}$ transformation is called chiral symmetry:
\begin{align}
    q_{R} 
    &\to Rq_{R},\quad
    R=e^{i\theta_{R}^{i}T^{i}} \in \textrm{SU}(2)_{R} , \label{eq:right} \\
    q_{L} 
    &\to Lq_{L},\quad
    L=e^{i\theta_{L}^{i}T^{i}} \in \textrm{SU}(2)_{L},\label{eq:left}
\end{align}
with $i=1,\dots ,3$ and $T^{i}$ being the generator of SU(2). The scalar quark bilinear form mixes the left- and right-handed components:
\begin{align}
    \bar{q}q
    &=\bar{q}_{L}q_{R}+\bar{q}_{R}q_{L} .
    \label{eq:qbarq}
\end{align}
The quark mass term in Eq.~\eqref{eq:QCD} therefore breaks chiral symmetry explicitly. In this way, chiral symmetry is realized in the massless limit of the QCD Lagrangian. This ideal $m_{\rm q}\to 0$ limit in QCD is called chiral limit. 

The transformation laws~\eqref{eq:right} and \eqref{eq:left} are generated by $(T^{i},0)$ and $(0,T^{i})$ of the Lie algebra $\mathcal{G}$. We define the vector and axial transformations as
\begin{align}
    q_{R} 
    &\to e^{i\theta_{V}^{i}T^{i}} q_{R},\quad
    q_{L} 
    \to e^{i\theta_{V}^{i}T^{i}} q_{L}, 
    \label{eq:vector}\\
    q_{R} 
    &\to e^{i\theta_{A}^{i}T^{i}} q_{R},\quad
    q_{L} 
    \to e^{-i\theta_{A}^{i}T^{i}} q_{L} 
    \label{eq:axial},
\end{align}
which are generated by $(T^{i},T^{i})$ and $(T^{i},-T^{i})$, respectively. If the mass of the up quark equals the mass of the down quark, the quark mass term~\eqref{eq:qbarq} is invariant under the vector transformation $(e^{i\theta_{V}^{i}T^{i}},e^{i\theta_{V}^{i}T^{i}})\in \textrm{SU}(2)_{V}$ which is called isospin symmetry.

\paragraph{Spontaneous and explicit breaking of chiral symmetry}
In general, symmetries of the Lagrangian may be broken by the ground state of the theory (vacuum $\ket{0}$). This is the phenomena called spontaneous symmetry breaking~\cite{Nambu:1961tp,Nambu:1961fr,Goldstone:1961eq}. In QCD, the vacuum expectation value of the $\bar{q}q$ operator is known to be nonzero:
\begin{align}
    \bra{0}\bar{q}q\ket{0}
    &\neq 0 .
\end{align}
This quark condensate breaks chiral symmetry spontaneously as expected
in Eq.~\eqref{eq:qbarq}. In the chiral limit, vectorial $\textrm{SU}(2)_{V}$ subgroup $(V,V)$ of chiral $\textrm{SU}(2)_{R}\otimes \textrm{SU}(2)_{L}$ symmetry remains unbroken, thanks to the Vafa-Witten theorem~\cite{Vafa:1983tf}. This indicates that $\bra{0}\bar{u}u\ket{0}=\bra{0}\bar{d}d\ket{0}$ in the chiral limit. Thus, the spontaneous breaking patten of chiral symmetry is
\begin{align}
    \textrm{SU}(2)_{R}\otimes \textrm{SU}(2)_{L}
    &\to \textrm{SU}(2)_{V} .
    \label{eq:SSB}
\end{align}
As a consequence of the spontaneous symmetry breaking, a massless Nambu-Goldstone (NG) boson emerges for each broken generator in the Lorentz invariant system. In the case of two-flavor chiral symmetry~\eqref{eq:SSB}, there appear three massless pions, $\pi^{0}$ and $\pi^{\pm}$, corresponding to three broken generators $(T^{i},-T^{i})$. 

The explicit symmetry breaking is caused by the small but nonzero masses of the up and down quarks. This effect can be introduced as an external current. The couplings to the scalar current $s$ and the pseudoscalar current $p$ are given by   
\begin{equation}
    \mathcal{L}_{\rm ext}
    =\bar{q}(s-i\gamma_{5}p)q
    =
    \bar{q}_{L}\mathcal{M}^{\dag}q_{R}
    +\bar{q}_{R}\mathcal{M}q_{L} ,
\end{equation}
where we denote
\begin{equation}
    s+ip= \mathcal{M} .
\end{equation}
The Lagrangian is invariant if the external fields transform as
\begin{align}
    \mathcal{M}^{\dag}
    &\to L\mathcal{M}^{\dag}R^{\dag},\quad
    \mathcal{M}
    \to R\mathcal{M}L^{\dag} .
    \label{eq:Mtrans}
\end{align}
In the isospin symmetric limit $m_{u}=m_{d}=\hat{m}$, the quark masses are introduced by identifying
\begin{equation}
    \mathcal{M}
    = \mathcal{M}^{\dag}
    =
    \begin{pmatrix}
    \hat{m} & 0 \\
    0 & \hat{m}
    \end{pmatrix} ,
    \label{eq:quarkmass}
\end{equation}
which is not invariant under Eq.~\eqref{eq:Mtrans} and breaks chiral symmetry. Equations~\eqref{eq:Mtrans} and \eqref{eq:quarkmass} will be used in Section~\ref{sec:chiral_effective_theory} to construct the effective Lagrangian.

The explicit breaking due to the quark masses causes the perturbative corrections to the chiral limit. For instance, the pions become massive due to the explicit breaking. When the explicit breaking is small, the pion mass $m_{\pi}$ is given by the Gell-Mann--Oakes--Renner (GMOR) relation~\cite{GellMann:1968rz}
\begin{align}
    F^{2} m_{\pi}^{2}
    &
    =-\hat{m}\bra{0}\bar{q}q\ket{0}
    +\mathcal{O}(\hat{m}^{2}),
    \label{eq:GMOR}
\end{align}
where $F \simeq 93$ MeV
 is the pion decay constant.

For later convenience, here we briefly mention the trace anomaly in QCD in the chiral limit. Consider the global scale transformation of the space-time $x_{\mu}\to \sigma x_{\mu}$ with a parameter $\sigma$. The quark and gluon fields are transformed according to their canonical dimensions as 
\begin{align}
    q(x)
    &
    \to \sigma^{3/2}q(\sigma x),\quad
    A_{\mu}^{a}(x)
    \to \sigma A_{\mu}^{a}(\sigma x) .
\end{align}
The QCD action in the chiral limit ($m_{\rm q}=0$) is invariant under this symmetry. The conserved current associated with this transformation $\Delta^{\mu}$ is called dilatational current, and can be related with the energy-momentum tensor as $\Delta^{\mu}=x_{\nu}\Theta^{\nu\mu}$. The conservation of the dilatational current is, however, broken explicitly by the quantum effect, as shown in Refs.~\cite{Collins:1976yq,Nielsen:1977sy}:
\begin{align}
    \partial_{\mu}\Delta^{\mu}
    &
    =\Theta^{\mu}_{\mu}
    =\frac{\beta(\alpha_{s})}{2g}G_{\mu\nu}^{a}G^{a\mu\nu} ,
\end{align}
where $\beta(\alpha_{s})$ is the beta function in Eq.~\eqref{eq:renormalization_QCD}. 
The scale invariance of the classical Lagrangian is thus broken with a nonzero gauge coupling constant $g$. Because this anomaly relates the trace of the energy-momentum tensor with the gluon fields, it is called the trace anomaly.

\subsubsection{Heavy quark symmetry}
\label{sec:heavy_quark_symmetry}

\paragraph{Heavy quark effective theory}
In contrast to the limit of the massless fermion,
we now consider the limit of infinitely massive fermion.
Around this limit, we can obtain the heavy quark effective theory as an expansion by $1/m_{Q}$ with heavy quark mass $m_{Q}$.
This limit corresponds to regarding the charm and bottom quark masses infinitely large.
Let us start by separating the QCD Lagrangian (\ref{eq:QCD}) as 
\begin{align}
 {\cal L}_{\mathrm{QCD}} = {\cal L}_{\mathrm{heavy}} + {\cal L}_{\mathrm{light}},
\end{align}
with the heavy quark part
\begin{align}
 {\cal L}_{\mathrm{heavy}} = \sum_{Q} \bar{Q} (iD\hspace{-0.6em}/-m_{Q})Q,
\end{align}
and the light quark and gluon part
\begin{align}
 {\cal L}_{\mathrm{light}} = - \frac{1}{4} G_{\mu\nu}^{a} G^{a \mu\nu} + \sum_{q} \bar{q} (iD\hspace{-0.6em}/-m_{q})q.
 \label{eq:QCD_light}
\end{align}
In the former, $Q$ is the heavy quark field, and in the latter, $q$ is the light quark field.

Let us consider large $m_{Q}$ for a heavy quark. Here we focus on the system with a single heavy flavor.
It is convenient to introduce $v$-frame with four-velocity $v^{\mu}$, in which frame the heavy quark is at rest.
We regard that the most of energy-momentum of the heavy quark $p$ is given by its on-mass-shell component.
Namely, we separate
\begin{align}
 p^{\mu} = m_{Q} v^{\mu} + k^{\mu},
\end{align}
where we suppose that the off-mass-shell component, $k^{\mu}$, which is a residual part, is much smaller than $m_{Q}$ ($k^{\mu} \ll m_{Q}$).
The four-velocity $v^{\mu}$ satisfies $v^{\mu}v_{\mu}=1$ and $v^{0}>0$ from the on-mass-shell condition and the propagation into the positive direction of time.\footnote{We should not confuse the spatial component of $v^{\mu}$, $\vec{v}$, with the three-dimensional velocity $\vec{u}$. They are related by $\vec{v}=\vec{u}/\!\sqrt{1-|\vec{u}\,|^{2}}$.}
Thus, the heavy quark momentum is separated to the on-mass-shell part and the off-mass-shell part.

Let us define the effective field for the positive energy state in the heavy quark limit:
\begin{align}
 Q_{v}(x) = \frac{1+v\hspace{-0.5em}/}{2} e^{im_{Q} v\cdot x} Q(x),
\end{align}
with the original heavy quark field $Q(x)$.
We notice that $Q_{v}(x)$ satisfies the condition
\begin{align}
 v\hspace{-0.5em}/ \, Q_{v}(x) = Q_{v}(x).
 \label{eq:positive_projection}
\end{align}
In the rest frame $v^{\mu}=(1,\vec{0}\,)$, the projection operator $(1+v\hspace{-0.5em}/)/2$ picks up the upper two components in four-component Dirac spinor in the standard representation.
The factor $e^{im_{Q} v\cdot x}$ leaves only the residual momentum scale, $k^{\mu}$, as a dynamical variable in the effective field $Q_{v}$.
Hence $Q_{v}$ has no explicit $m_{Q}$ dependence.
Similarly, we define the effective field for the negative energy state as
\begin{align}
 {\cal Q}_{v} = \frac{1-v\hspace{-0.5em}/}{2} e^{im_{Q} v\cdot x} Q(x),
\end{align}
which satisfies the condition $v\hspace{-0.5em}/ \, {\cal Q}_{v} = - {\cal Q}_{v}$.
Note the sign in the projection operator $(1-v\hspace{-0.5em}/)/2$ because it picks up the lower two components in the Dirac spinor in the rest frame.

When we use two effective fields $Q_{v}(x)$ and ${\cal Q}_{v}(x)$,
we can rewrite ${\cal L}_{\mathrm{heavy}}$ as
\begin{align}
 {\cal L}_{\mathrm{heavy}}
 =
 \bar{Q}_{v} v\!\cdot\!iDQ_{v} - \bar{\cal Q}_{v} (v\!\cdot\!iD+2m_{Q}) {\cal Q}_{v}
 +\bar{\cal Q}_{v} i\hspace{0.2em}/\hspace{-0.6em}D_{\perp} Q_{v}
 +\bar{Q}_{v} i\hspace{0.2em}/\hspace{-0.6em}D_{\perp} {\cal Q}_{v},
\end{align}
with $D_{\perp}^{\mu} = D^{\mu} - v^{\mu} v \cdot D$.
It is important to note that the mass of $Q_{v}$ is ``zero" while the mass of ${\cal Q}_{v}$ is $2m_{Q}$.
This is because we measure the energy as the residual momentum for the positive energy state.
Hence we may eliminate the negative energy state by regarding $2m_{Q}$ as a large quantity.
Physically, this corresponds to neglect of the creation of $\bar{Q}Q$ pairs in the state.
By using the equation of motion for ${\cal Q}_{v}$,
\begin{align}
 \left( v \!\cdot\! iD + 2m_{Q} \right) {\cal Q}_{v} = i\hspace{0.2em}/\hspace{-0.6em}D_{\perp} Q_{v},
 \label{eq:equation_of_motion}
\end{align}
we eliminate ${\cal Q}_{v}$ in ${\cal L}_{\mathrm{heavy}}$, and obtain
\begin{align}
{\cal L}_{\mathrm{heavy}}
&=
\bar{Q}_{v}
\left(
 v \!\cdot\! iD + i\hspace{0.2em}/\hspace{-0.6em}D_{\perp} \frac{1}{v \!\cdot\! iD + 2m_{Q}} i\hspace{0.2em}/\hspace{-0.6em}D_{\perp}
\right) Q_{v} \nonumber \\
&=
 \bar{Q}_{v} v \!\cdot\! iD Q_{v}
 + \bar{Q}_{v} \frac{(iD_{\perp})^{2}}{2m_{Q}} Q_{v} - g_{s} \bar{Q}_{v} \frac{\sigma_{\mu\nu}G^{\mu\nu}}{4m_{Q}} Q_{v} + {\cal O}(1/m_{Q}^{2}),
\end{align}
up to and including ${\cal O}(1/m_{Q})$.
We note that this Lagrangian is valid at tree level, because we have used the equation of motion (\ref{eq:equation_of_motion}).
As we will see, the first and second terms are not affected by quantum fluctuations (loop corrections),
while the third term is.
To take this effect into account, we introduce the Wilson coefficient $c(\mu)$ at some energy scale $\mu$.
Finally, we obtain the result
\begin{align}
 {\cal L}_{\mathrm{heavy}} =
   \sum_{Q=1}^{N_{h}}
    \left(  \bar{Q}_{v} v \!\cdot\! iD Q_{v}
 + \bar{Q}_{v} \frac{(iD_{\perp})^{2}}{2m_{Q}} Q_{v} - c(\mu) g_{s} \bar{Q}_{v} \frac{\sigma_{\mu\nu}G^{\mu\nu}}{4m_{Q}} Q_{v} + {\cal O}(1/m_{Q}^{2}) \right),
 \label{eq:HQET_Lagrangian}
\end{align}
where the sum over various heavy flavors is taken ($N_{h}$ the number of heavy flavor).
This framework is called the heavy quark effective theory (HQET).
Equation (\ref{eq:HQET_Lagrangian}) is the effective Lagrangian in the HQET.
The similar procedure is applicable to higher orders.
In the higher orders, not only the heavy-quark--gluon couplings, but also light-quark--heavy-quark couplings enter into the effective Lagrangian.
The analysis up to ${\cal O}(1/m_{Q}^{n})$ with $n\ge 2$ can be found in e.g. Ref.~\cite{Bauer:1997gs} for ${\cal O}(1/m_{Q}^{2})$ and Ref.~\cite{Balzereit:1998jb} for ${\cal O}(1/m_{Q}^{3})$ (see also Ref.~\cite{Manohar:1997qy} and references therein).

\paragraph{Heavy quark symmetry}
In the heavy quark limit ($m_{Q} \rightarrow \infty$), the leading term $v\cdot iD$ in Eq.~(\ref{eq:HQET_Lagrangian}) is important to realize
 the heavy spin-flavor $\mathrm{SU}(2N_{h})$ symmetry.
This symmetry implies simultaneously (i) heavy-quark spin symmetry and (ii) heavy-quark flavor symmetry.
As for (i), the operator changing the heavy quark spin is absent in $v\cdot iD$, and hence the heavy quark spin is a conserved quantity for any non-perturbative dynamics acting on the heavy quark.
As for (ii), there is no heavy flavor dependence in $v\cdot iD$, and hence the interchange of two heavy flavors, such as charm and bottom quarks, leaves the interaction invariant.
Those symmetries are called the heavy-quark symmetry (HQS).
When we do not consider weak decay processes replacing a bottom quark by a charm quark,
we sometimes focus only on the heavy quark spin symmetry in the heavy-hadron dynamics.
In the following, we simply use the heavy quark symmetry to indicate the heavy quark spin symmetry. 

It is important to note that the heavy quark symmetry holds at any energy scale from low energy scale to high energy scale.
In this sense, the heavy quark symmetry may be contrasted to chiral symmetry, which is broken dynamically in the QCD vacuum.
We can observe the approximate realization of the heavy quark symmetry in the heavy hadron spectroscopy.
Let us consider the heavy hadrons composed of $\bar{q}Q$ with a light quark $q$ and a heavy quark $Q$.
We decompose the total spin of the heavy hadron $\vec{J}$ into the heavy quark spin $\vec{S}$ and the remaining part $\vec{j}$:
\begin{align}
 \vec{J} = \vec{S} + \vec{j},
 \label{eq:J_decomposition}
\end{align}
where $\vec{J}$ is conserved and $\vec{S}$ is conserved in the heavy quark limit.\footnote{We notice the conservation of $\vec{J}$ is valid only when the vacuum is rotationally invariant. If the rotational symmetry is broken, we cannot apply the following discussion. For example, such situation can happen when the external field (e.g. a magnetic field~\cite{Suzuki:2016kcs}) breaks the rotational symmetry.}
The conservation of each $\vec{J}$ and $\vec{S}$ leads to the conservation of $\vec{j}$ through Eq.~(\ref{eq:J_decomposition}).
Although we have denoted the light quark by $q$, this is, in field theoretical view,  an ensemble of many particle states, such as $q+\bar{q}qq+ \bar{q}qg+\cdots$ with a light antiquark $\bar{q}$ and a gluon $g$, namely light degrees of freedom except for the heavy quark.
To obtain the explicit form of the wave function is a highly non-perturbative problem. 
Nevertheless, $\vec{j}$ is a conserved quantity regardless of any complex structure of $q+\bar{q}qq+ \bar{q}qg+\cdots$.
This complex object conserving $\vec{j}$ is called ``brown muck"~\cite{Neubert:1993mb}.

The conservation of light degrees of freedom immediately leads to the unique properties of heavy hadrons in the spectroscopy.
As we have seen, the heavy quark spin $\vec{S}$ and the brown muck spin $\vec{j}$ are independently conserved quantities in the heavy quark limit.
Therefore, we conclude that the heavy hadrons are two degenerate states with spin
\begin{align}
 J_{\pm} = j\pm1/2,
 \label{eq:HQS_multiplet}
\end{align}
for $j \ge 1/2$.
Those two degenerate states are called the HQS doublet.
 In case of $j=0$, we have only one state with spin $J=1/2$.
 This state is called the HQS singlet.

As examples in the meson sector, we consider $D$ ($J^{P}=0^{-}$; 1870 MeV) and $D^{\ast}$ ($1^{-}$; 2010 MeV) mesons, whose mass difference is about 140 MeV, for charm, and $\bar{B}$ ($J^{P}=0^{-}$; 5280 MeV) and $\bar{B}^{\ast}$ ($1^{-}$; 5325 MeV) mesons, whose mass difference is about 45 MeV, for bottom.
By regarding those mass differences as small quantities, we can consider that $D$ and $D^{\ast}$ ($\bar{B}$ and $\bar{B}^{\ast}$) mesons are approximately degenerate states in mass.
The degenerate states correspond to the heavy hadrons with a common brown muck with spin and parity $j^{{\cal P}}=1/2^{-}$ for $\bar{q}$ in $\bar{q}Q$ in the heavy quark limit.
Therefore, $D$ and $D^{\ast}$ mesons as well as $\bar{B}$ and $\bar{B}^{\ast}$ mesons are classified as the HQS doublet.

In the baryon sector, we consider $\Sigma_{c}$ ($J^{P}=1/2^{+}$; 2455 MeV) and $\Sigma_{c}^{\ast}$ ($3/2^{+}$; 2520 MeV), whose mass difference is 65 MeV, for charm, and $\Sigma_{b}$ ($J^{P}=1/2^{+}$; 5811 MeV) and $\Sigma_{b}^{\ast}$ ($3/2^{+}$; 5832 MeV), whose mass difference is 21 MeV.
Again, we can regard that $\Sigma_{c}$ and $\Sigma_{c}^{\ast}$ ($\Sigma_{b}$ and $\Sigma_{b}^{\ast}$) baryons are approximately degenerate states,
which correspond to the heavy hadrons with brown muck of $I=1$ and $j^{{\cal P}}=1^{+}$ for $qq$ in $qqQ$ in the heavy quark limit.
Therefore, $\Sigma_{c}$ and $\Sigma_{c}^{\ast}$ baryons as well as $\Sigma_{b}$ and $\Sigma_{b}^{\ast}$ baryons are classified as the HQS doublet.
We note that the ground state baryon $\Lambda_{c}$ ($J^{P}=1/2^{+}$; 2287 MeV) and $\Lambda_{b}$ ($1/2^{+}$; 5620 MeV) have no state in near mass region, and hence they are classified to be the HQS singlet.

The structure of the brown muck in heavy hadrons is seen also in the branching ratios of the decay widths.
Let us suppose that the initial heavy hadron $\Psi_{J'}^{\prime(j')}$ with total spin $J'$ and brown muck spin $j'$ decays to the final heavy hadron $\Psi_{J}^{(j)}$ with total spin $J$ and brown muck spin $j$ by emitting a light hadron, e.g. a pion $\pi$ with relative angular momentum $L$.
Due to the independent conservations of the heavy quark spin $\vec{S}$ and the brown muck spin $\vec{j}$,
we see that the strength of the decay widths
is parametrized as
\begin{align}
 \Gamma \left[ \Psi_{J'}^{\prime(j')} \rightarrow \Psi_{J}^{(j)}+\pi \right]
\propto
(2j+1)(2J'+1)
\left|
\left\{
\begin{array}{ccc}
 L & j' & j \\
 1/2 & J & J'  \\  
\end{array}
\right\}
\right|^{2}
+{\cal O}(1/M),
\label{eq:Isgur_Wise_decay}
\end{align}
by neglecting the corrections with the heavy hadron mass $M$~\cite{Isgur:1991wq}.
Here $J^{(\prime)}=j^{(\prime)}\pm1/2$.
In realistic application to experimental data, we need to include the phase space factor due to the different mass thresholds.
Including this factor, we can reproduce the branching ratios of the known decay patterns of $D_{0}^{\ast}$ ($J^{P}=0^{+}$), $D_{1}$ ($1^{+}$) mesons to $D$, $D^{\ast}$ mesons with a pion emission~\cite{Manohar:2000dt}.
We note that the similar result is applicable to any light hadrons other than a pion.

As far as we follow the heavy quark symmetry, the brown muck structure in Eqs.~(\ref{eq:HQS_multiplet}) and (\ref{eq:Isgur_Wise_decay}) is quite general.
They should hold not only in normal hadrons (mesons and baryons), but also in exotic hadrons (multiquarks, hadronic molecules, quark-gluon hybrids, and so on) as well as in nuclei and nuclear matter with a heavy quark~\cite{Yasui:2013vca,Yamaguchi:2014era}.
For example, let us consider the case that a $\bar{D}$ meson or a $\bar{D}^{\ast}$ meson exists in nuclear matter.
If we consider the heavy quark limit, the masses of $\bar{D}$ and $\bar{D}^{\ast}$ mesons should be regarded as the same.
This is also the case in nuclear matter.
As far as the rotational symmetry is unbroken, the spin of the brown muck should be conserved in nuclear matter.
Hence the $\bar{D}$ and $\bar{D}^{\ast}$ mesons in nuclear matter should be degenerate as in vacuum, as long as the heavy quark symmetry is adopted.

Interestingly, the brown muck can exhibit a new structure when the heavy hadron combined with light hadrons forms an extended object.
As an example, let us consider a $\bar{D}^{(\ast)}N$ two-boy system
 which 
 is composed of a $\bar{D}^{(\ast)}$ meson and a nucleon.
In the heavy quark limit, the object carrying the conserved spin is given as the brown muck.
In the case of $\bar{D}^{(\ast)}N$, the brown muck should have the structure $qN$ where $q$ is the light quark inside the $\bar{D}^{(\ast)}$ meson.
The quark-hadron structure like $qN$, which includes both quarks and hadrons as effective degrees of freedom, can be called the ``light spin-complex (or spin-complex)", because the spin is the most important conserved quantity in this complex system~\cite{Yasui:2013vca}.
This is an interesting unit of matter which appears in the extended heavy hadron systems.
Furthermore, when $\bar{D}^{(\ast)}N$ exists in nuclear matter, it is surrounded by many pairs of a nucleon and a hole, $NN^{-1}$, as the low-energy degrees of freedom around the Fermi energy.
In the heavy quark limit, the spin-complex is given by $q+qNN^{-1}+q(NN^{-1})(NN^{-1})+\dots$, where $q$ is again the light quark inside the $\bar{D}^{(\ast)}$ meson.
More examples will be shown in Sect.~\ref{sec:D_mesons} and Sect.~\ref{sec:charm_baryons}.

We have to keep in mind that the mass splitting due to the HQS breaking, about 140 MeV between a $\bar{D}$ meson and a $\bar{D}^{\ast}$ meson, is still larger than the typical energy scales in nuclear dynamics, for example 40 MeV as the Fermi energy of a nucleon at normal nuclear matter density.
Hence, the approximate mass degeneracy as the HQS doublet may not be clearly identified within low energy scales.
To identify the HQS doublet structure precisely, we need to include the dynamics at large energy scales in nucleon-nucleon interaction, where the scattering energy reaches a few hundred MeV or more.
In bottom sector, the HQS structure will be found at lower energy scales, because the mass splitting, about 45 MeV between a $B$ meson and a $B^{\ast}$ meson, is comparable with the Fermi energy of the nuclear matter.
The HQS doublet in the baryon sector, $\Sigma_{c}$ and $\Sigma_{c}^{\ast}$ for charm and $\Sigma_{b}$ and $\Sigma_{b}^{\ast}$ for bottom, would provide a more appropriate situation to study the HQS structure in experiments, because their mass splittings, about 65 MeV and about 20 MeV, are smaller.

Regardless to the finite masses of heavy quarks, we emphasize that the HQS plays the essentially important role in the dynamics of heavy hadrons in nuclear matter to achieve the systematic understanding of their properties, for example, mass shifts, binding energies and decay widths in medium.
As we will see examples in later sections, the binding energies for each component in the HQS doublet exhibit an interesting behavior in a view of the HQS, though the mass thresholds are different due to the HQS breaking.

\paragraph{\mbox{\boldmath $1/m_{Q}$} corrections}

In realistic world, the heavy quark mass is not infinitely large, but is finite.
It is, therefore, important to consider the corrections to the heavy quark limit when we compare theoretical results with experimental data at quantitative level.
It is one of the biggest advantages of the heavy quark effective theory that the finite mass corrections are included systematically by a series expansion of $1/m_{Q}$.
For charm and bottom cases, we can regard the correction terms are small in comparison with the leading terms.
We may consider, for example, that the leading order correction is of the order of $\Lambda_{\mathrm{QCD}}/m_{Q}$.
We can estimate that $\Lambda_{\mathrm{QCD}}/m_{c}$ for charm and $\Lambda_{\mathrm{QCD}}/m_{b}$ for bottom are about 16\% and 4\%, respectively, with $m_{c}\simeq 1.3$ GeV, $m_{b}\simeq 4.7$ GeV, and $\Lambda_{\rm QCD}\simeq 0.21$ GeV ($\overline{MS}$ scheme with $N_{f}=5$)~\cite{Agashe:2014kda}.

To investigate the correction terms, the following guides are important~\cite{Luke:1992cs}.

\begin{enumerate}
\item The leading order terms and the next-to-leading order terms are related by the velocity rearrangement (or reparametrization).
\item There exists a term at the next-to-leading order, which breaks the heavy quark symmetry.
\end{enumerate}

Let us explain 1 and 2 in some details.

As for 1, we remind ourselves that the velocity-frame with four-velocity $v^{\mu}$ was chosen in defining the heavy quark effective theory.
However, the choice of the velocity-frame should be arbitrary, and hence another velocity-frame with four velocity $w^{\mu} \neq v^{\mu}$ could be chosen.
The two velocity-frames with $v^{\mu}$ and $w^{\mu}$ are related by the Lorentz boost.
This is called the velocity rearrangement (or reparametrization).
Let us consider the invariance of the effective Lagrangian (\ref{eq:HQET_Lagrangian}) in the velocity rearrangement.
The Lorentz boost is given by
\begin{align}
 v^{\mu} &\rightarrow v^{\mu} + \varepsilon^{\mu}/m_{Q}, \\
 k^{\mu} &\rightarrow k^{\mu} - \varepsilon^{\mu},
\end{align}
by regarding $\varepsilon^{\mu}$ is a quantity much smaller than $m_{Q}$.
To consider the $1/m_{Q}$ correction, we assume that this Lorentz boost is valid up to and including ${\cal O}(1/m_{Q})$.
For the condition $v_{\mu}v^{\mu}=1$ to hold after the Lorentz boost, we have to impose a constraint condition
\begin{align}
 v \!\cdot\! \varepsilon = 0,
\end{align}
provided that ${\cal O}(1/m_{Q}^2)$ is neglected.
We also impose that the projection condition $v\hspace{-0.5em}/ Q_{v}=Q_{v}$ for the effective field $Q_{v}$ is invariant under this Lorentz boost,
\begin{align}
 Q_{v} \rightarrow e^{i \varepsilon \cdot x} \left( 1+ \frac{\varepsilon\hspace{-0.5em}/}{2m_{Q}} \right) Q_{v}.
\end{align}
According to this Lorentz transformation, the leading part ${\cal L}_{0}$ and the next-to-leading part ${\cal L}_{1}$ in Eq.~(\ref{eq:HQET_Lagrangian}),
\begin{align}
 {\cal L}_{0} &= \bar{Q}_{v} v \!\cdot\! iD Q_{v}, \\
 {\cal L}_{1} &= \bar{Q}_{v} \frac{(iD_{\perp})^{2}}{2m_{Q}} Q_{v} - c(\mu) g_{s} \bar{Q}_{v} \frac{\sigma_{\mu\nu}G^{\mu\nu}}{4m_{Q}} Q_{v},
\end{align}
are transformed as
\begin{align}
{\cal L}_{0} & \rightarrow  {\cal L}_{0} + \frac{1}{m_{Q}} \bar{Q}_{v} i \varepsilon \!\cdot\! D Q_{v}, \\
{\cal L}_{1} & \rightarrow  {\cal L}_{1} - \frac{1}{m_{Q}} \bar{Q}_{v} i \varepsilon \!\cdot\! D Q_{v}, 
\end{align}
respectively.
Although each of ${\cal L}_{0}$ and ${\cal L}_{1}$ is not invariant, 
their sum ${\cal L}_{0}+{\cal L}_{1}$ is invariant under the Lorentz boost.
Therefore, we confirm that the effective Lagrangian in the HQET is invariant up to and including ${\cal O}(1/m_{Q})$.

Concerning Eq.~(\ref{eq:HQET_Lagrangian}), because the concrete expression of the effective Lagrangian at the next-to-leading order is already known, we may consider that the invariance under the velocity rearrangement may not bring new information.
However, this is important to analyze higher order terms because this scheme is quite general.
For example, higher order terms ${\cal O}(1/m_{Q}^{n})$ with $n \ge 2$ can be investigated by imposing several constraints among several terms~\cite{Manohar:1997qy}.
This scheme is useful also to construct the higher order terms in the heavy hadron effective theory, as shown in Sect.~\ref{sec:heavy_hadron_effective_theory}.

As for 2, there is a spin-flip coupling for a heavy quark and a gluon, namely the magnetic coupling, at ${\cal O}(1/m_{Q})$.
In the rest frame $v^{\mu}=(1,\vec{0})$, for example, $\sigma^{\mu\nu}$ becomes
\begin{align}
 \bar{Q}_{v} \sigma^{\mu\nu} Q_{v}
=
\bar{Q}_{v}
\epsilon^{ijk}
\left(
\begin{array}{cc}
 \sigma^{k} & 0  \\
 0 & 0
\end{array}
\right)
Q_{v}, \hspace{0.5em}\mathrm{with}\hspace{0.5em} \mu,\nu=i,j,k=1,2,3,
\end{align}
and, it flips the heavy quark spin in the positive energy component.
Therefore the heavy quark spin is not a conserved quantity at ${\cal O}(1/m_{Q})$.
This interaction is analogous to the fine structure of the electron states in atoms.

The properties 1 and 2 should hold, not only at the quark level, but also at the hadron level.
We will discuss how those two points play the role for the heavy hadron effective theory in Sect.~\ref{sec:heavy_hadron_effective_theory}.

According to the HQET, the physical quantities concerning a heavy quark can be expanded by $1/m_{Q}$.
The expansion should hold for any environment with finite temperature and baryon number density.
It should hold also in nuclei and nuclear matter with a heavy quark.
In the rest frame of the heavy hadron $v_{r}^{\mu}=(1,\vec{0}\,)$, the heavy hadron mass is parametrized as
\begin{align}
 M_{H}(T,\rho) = m_{Q} + \bar{\Lambda}(T,\rho) - \frac{\lambda_{1}(T,\rho)}{2m_{Q}} + 4\vec{S} \!\cdot\! \vec{j} \,  \frac{\lambda_{2}(T,\rho;m_{Q})}{2m_{Q}} + {\cal O}(1/m_{Q}^{2}),
\end{align}
with coefficients $\bar{\Lambda}(T,\rho)$, $\lambda_{1}(T,\rho)$ and $\lambda_{2}(T,\rho;m_{Q})$,
which are dependent on the temperature $T$ and the baryon number density $\rho$.
In order for this expansion to be valid, the temperature $T$ as well as the chemical potential $\mu$ corresponding to $\rho$ should be much smaller than the heavy quark mass $m_{Q}$.
We note that $\bar{\Lambda}(T,\rho)$, $\lambda_{1}(T,\rho)$ and $\lambda_{2}(T,\rho;m_{Q})$ are defined by
\begin{align}
 \bar{\Lambda}(T,\rho) &= \frac{1}{2} \langle H_{v_{r}} | {\cal H}_{0} | H_{v_{r}} \rangle, \label{eq:expansion_1} \\
 \lambda_{1}(T,\rho) &= \frac{1}{2} \langle H_{v_{r}} | \bar{Q}_{v_{r}} (iD_{\perp})^{2} Q_{v_{r}} | H_{v_{r}} \rangle, \label{eq:expansion_2} \\
 8\vec{S} \!\cdot\! \vec{j} \, \lambda_{2}(T,\rho;m_{Q}) &= \frac{1}{2} c(\mu) \langle H_{v_{r}} | \bar{Q}_{v_{r}} g_{s} \sigma_{\alpha\beta}G^{\alpha\beta} Q_{v_{r}} | H_{v_{r}} \rangle . \label{eq:expansion_3}
\end{align}
$\bar{\Lambda}(T,\rho)$ is related to the energy momentum tensor in the heavy quark limit.
${\cal H}_{0}$ is the Hamiltonian corresponding to the Lagrangian Eq.~(\ref{eq:QCD_light}) for the light degrees of freedom.
Note that $\lambda_{2}(T,\rho;m_{Q})$ receives the logarithmic correction by $1/m_{Q}$ due to the quantum correction, as indicated by the Wilson coefficient 
$c(\mu \simeq m_{Q})$
in Eq.~(\ref{eq:HQET_Lagrangian}).

Interestingly, Eqs.~(\ref{eq:expansion_2}) and (\ref{eq:expansion_3}) are rewritten as
\begin{align}
 \lambda_{1}(T,\rho) &= - m_{Q} \langle H_{v_{r}} | \bar{Q}_{v_{r}} g_{s} \vec{x} \!\cdot\!\vec{E} Q_{v_{r}} | H_{v_{r}} \rangle, \label{eq:lambda_1} \\
 8\vec{S} \!\cdot\! \vec{j} \, \lambda_{2}(T,\rho;m_{Q})
 &=
 \frac{1}{2} c(\mu) \langle H_{v_{r}} | \bar{Q}_{v_{r}} g_{s} \vec{\sigma} \!\cdot\! \vec{B} Q_{v_{r}} | H_{v_{r}} \rangle, \label{eq:lambda_2}
\end{align}
where $E^{i}=-G^{0i}$ and $B^{i}=\varepsilon^{ijk}G^{jk}$ are electric and magnetic components, respectively, of the gluon field in the rest frame.
The first equation was derived in Ref.~\cite{Neubert:1993zc} (see also Ref.~\cite{Bigi:1997fj}).
In Eq.~(\ref{eq:lambda_1}), $\vec{x}$ denotes the position of the center-of-mass of the system, which should coincide with the position of the heavy quark in the heavy-quark limit.
Therefore, $\lambda_{1}(T,\rho)$ and $\lambda_{2}(T,\rho;m_{Q})$ are related to the coupling strengths between the heavy quark and the electric and magnetic gluons, respectively.
As the environment changes, the values of $\lambda_{1}(T,\rho)$ and $\lambda_{2}(T,\rho;m_{Q})$ change also.
Because the heavy quark would not be affected by the environment so much,
we would consider that the change of $\lambda_{1}(T,\rho)$ and $\lambda_{2}(T,\rho;m_{Q})$ is reflected by the change of the coupling of the heavy quark to gluon fields in the environment (see Sect.~\ref{sec:D_mesons}).

\subsection{Hadronic effective theories}
\label{sec:hadronic_effective_theories}

In the low energy region, QCD is not directly solvable due to the non-perturbative dynamics.
Because the quarks and gluons are confined in the QCD vacuum,
it is essentially important to use hadrons as effective degrees of freedom instead of the quarks and gluons.
As effective theories for heavy hadrons, we introduce the chiral effective theory (Sect.~\ref{sec:chiral_effective_theory}) and the heavy hadron effective theory (Sect.~\ref{sec:heavy_hadron_effective_theory}).
Those effective theories are constructed according to chiral symmetry and heavy quark symmetry, as discussed in Sect.~\ref{sec:symmetries}.\footnote{We will use different conventions for various variables such as the pion decay constants and the field normalizations adopted in the chiral effective theory and in the heavy hadron effective theory.}

\subsubsection{Chiral effective theory}
\label{sec:chiral_effective_theory}

Because of the nonperturbative nature, QCD cannot be directly used to calculate the low-energy hadronic phenomena. Nevertheless, it is possible to establish an effective description, focusing on the symmetry principles and the low energy excitations. The effective field theory (EFT) is a useful technique to extract the relevant physics in the low energy domain. A traditional example is the effective Lagrangian for Quantum Electrodynamics (QED)~\cite{Heisenberg:1935qt}. In QED, the lowest energy excitation is the massless photon. When the energy of the system is much smaller than the electron mass, the dynamics of the photons can be described by the effective Lagrangian which is constrained by the symmetries of the underlying fundamental theory, QED. At present, EFT is vastly utilized in various fields of physics~\cite{Leutwyler:1993gf,Braaten:2007nq,Dubovsky:2011sj,Watanabe:2014fva}.

In the light quark sector of QCD, systematic EFT approach is developed as chiral perturbation theory (ChPT)~\cite{Weinberg:1979kz,Gasser:1983yg,Gasser:1984gg,Scherer:2012xha}. As discussed in section~\ref{sec:chiral_symmetry}, Nambu-Goldstone theorem ensures the appearance of almost massless pions as a consequence of the spontaneous breaking of chiral symmetry. The low-energy phenomena of the strong interaction is therefore determined by the effective Lagrangian of pions. A unique feature of ChPT is the spontaneous breakdown of chiral symmetry, which gives the constraints on the dynamics of NG bosons (the low energy theorems). General prescription to construct an effective Lagrangian with the nonlinear realization is given in Refs.~\cite{Coleman:1969sm,Callan:1969sn,Bando:1987br}, when the symmetry $G$ of the Lagrangian is spontaneously broken into a subgroup $H\subset G$ by the vacuum in the fundamental theory. 

In the two-flavor QCD in the chiral limit, chiral symmetry $G=\textrm{SU}(2)_{R}\otimes \textrm{SU}(2)_{L}$ breaks down to the isospin symmetry $ H=\textrm{SU}(2)_{V}$. To construct the effective Lagrangian, we define the chiral field 
\begin{equation}
    u(\phi)
    =\exp\left\{\frac{i\phi}{2F}\right\} ,
\end{equation}
where $F \simeq 93$ MeV is the pion decay constant in the two-flavor chiral limit and three NG boson fields $\pi^{0},\pi^{\pm}$ are collected in a $2\times 2$ matrix as 
\begin{equation}
    \phi=\begin{pmatrix}
    \pi^{0} & \sqrt{2}\pi^{+} \\
    \sqrt{2}\pi^{-} & -\pi^{0}
    \end{pmatrix}.
 \label{eq:phi_def}
\end{equation}
Because the generator of the broken symmetry is $(T^{i},-T^{i})$, [see Eq.~\eqref{eq:axial}], the representative of the coset space $G/H$ can be parametrized by the NG boson fields $\phi$ as
\begin{align}
    \xi(\phi)
    =(u(\phi),u^{\dag}(\phi)).
\end{align}
Using the unique decomposition of the group element $g\xi(\phi) =\xi(\phi^{\prime})h(\phi,g)$ with $h(\phi,g)\in H$, we define the nonlinear transformation law of the NG boson fields under $g=(R,L)$ through
\begin{align}
    (u(\phi),u^{\dag}(\phi))
    \stackrel{g}{\to} (R,L)(u(\phi),u^{\dag}(\phi)) (h^{\dag}(\phi,g),h^{\dag}(\phi,g)), 
\end{align}
which can be simplified as the transformation law of the chiral field
\begin{align}
    u(\phi)
    \stackrel{g}{\to} Ru(\phi)h^{\dag}(\phi,g)= h(\phi,g)u(\phi)L^{\dag} .
    \label{eq:utrans}
\end{align}
Because $h(\phi,g)$ depends on $\phi$, Eq.~\eqref{eq:utrans} defines the nonlinear transformation of the NG boson fields. Under the unbroken symmetry $g=h\in H$, this reduces to a linear transformation.

The fundamental quantity to construct the effective Lagrangian is the Maurer-Cartan 1-form $\alpha_{\mu}=i^{-1}(u(\phi),u^{\dag}(\phi))^{-1}\partial_{\mu}(u(\phi),u^{\dag}(\phi))$, which can be decomposed into
\begin{align}
    \alpha_{\mu\parallel}
    &=\frac{1}{2i}
    [u^{\dag}(\phi)\partial_{\mu}u(\phi)
    +u(\phi)\partial_{\mu}u^{\dag}(\phi)] , \\
    \alpha_{\mu\perp}
    &=\frac{1}{2i}
    [u^{\dag}(\phi)\partial_{\mu}u(\phi)
    -u(\phi)\partial_{\mu}u^{\dag}(\phi)] .
\end{align}
The transformation laws are given by
\begin{align}
    \alpha_{\mu\parallel}
    &\stackrel{g}{\to} 
    h(\phi,g)
    \alpha_{\mu\parallel}
    h^{\dag}(\phi,g)
    +\frac{1}{i} h(\phi,g)\partial_{\mu} h^{\dag}(\phi,g) , \label{eq:aparalleltrans}\\
    \alpha_{\mu\perp}
    &\stackrel{g}{\to}
    h(\phi,g)
    \alpha_{\mu\perp}
    h^{\dag}(\phi,g) \label{eq:aperptrans}.
\end{align}
It is illustrative to notice that these quantities correspond to the vector current and the axial current made by pions:
\begin{align}
    \alpha_{\mu\parallel}
    &
    = -\frac{i}{8F^{2}}(\phi\partial_{\mu}\phi-\partial_{\mu}\phi\phi)+\mathcal{O}(\phi^{4}), \label{eq:aparallelexpand} \\
    \alpha_{\mu\perp}
    &
    = \frac{\partial_{\mu}\phi}{2F}+\mathcal{O}(\phi^{3}).
    \label{eq:aperpexpand}
\end{align}
Correspondence to the notation in Ref.~\cite{Scherer:2012xha} is understood as the chiral connection $\Gamma_{\mu}=i\alpha_{\mu\parallel}$ and the chiral vielbein $u_{\mu}=-2\alpha_{\mu\perp}$. The effective field theory is based on the most general Lagrangian with $\alpha_{\mu\parallel}$ and $\alpha_{\mu\perp}$ which is invariant under Eqs.~\eqref{eq:aparalleltrans} and \eqref{eq:aperptrans}. There are, however, infinitely many invariant terms. We thus need to introduce a power counting scheme in order to establish the hierarchy of different terms. For the description of the low-energy phenomena, derivative expansion of the effective Lagrangian is useful. This corresponds to the expansion in powers of the four-momentum of pions $p$ over an ultraviolet momentum scale $\Lambda$. We estimate $\Lambda\sim 1$ GeV, either by the chiral symmetry breaking scale, $\Lambda\sim 4\pi F$ or by the lowest energy of the non-NG mode in QCD, the mass of the rho meson, $\Lambda\sim m_{\rho}$.

Now our task is to construct the most general Lagrangian out of $\alpha_{\mu\parallel}$ and $\alpha_{\mu\perp}$, with the smallest number of derivatives. There is only one term
\begin{equation}
    \mathcal{L}_{2}=-F^{2}\Tr(\alpha_{\mu\perp}\alpha^{\mu}_{\perp}) ,
    \label{eq:L2}
\end{equation}
where the coefficient $-F^{2}$ is chosen to obtain the correct normalization of the kinetic term of pions. In the purely pionic sector, it is convenient to define
\begin{equation}
    U(\phi)
    =u^{2}(\phi)=\exp\left\{\frac{i\phi}{F}\right\},
\end{equation}
because the transformation law does not contain $h$:
\begin{equation}
    U(\phi)
    \stackrel{g}{\to} RU(\phi)L^{\dag} .
    \label{eq:Utrans}
\end{equation}
With the $U$ field, the Lagrangian~\eqref{eq:L2} can be expressed as
\begin{equation}
    \mathcal{L}_{2}
    =
    \frac{F^{2}}{4}\Tr(\partial_{\mu}U\partial^{\mu}U^{\dag}) .
\end{equation}
The invariance under Eq.~\eqref{eq:Utrans} is also clear. Because there are two derivatives, the lowest order Lagrangian is counted as $\mathcal{O}(p^{2})$. This Lagrangian contains even powers of $\phi$ fields, corresponding to the kinetic terms, the four-point vertices with two derivatives, and so on. The coefficient $F^{2}/4$ is chosen such that the kinetic terms are properly normalized. Applying the power counting scheme for the Green's functions, we find that the one-loop diagrams with the $\mathcal{O}(p^{2})$ interactions are counted as $\mathcal{O}(p^{4})$. Thus, the ultraviolet divergences in the one-loop diagrams can be renormalized by the counter terms in the higher-order Lagrangian at $\mathcal{O}(p^{4})$. In this way, systematic order-by-order renormalization is guaranteed.

Now we introduce the quark mass term which breaks chiral symmetry explicitly. We note that there are many ways to break chiral symmetry. In the construction of the chiral effective Lagrangian, it is important to break the chiral symmetry as in the same manner with the symmetry breaking in the fundamental theory of QCD. For this purpose, we first construct chiral \textit{invariant} Lagrangian with the transformation law of $\mathcal{M}$ in Eq.~\eqref{eq:Mtrans}, and then substitute the quark mass matrix as in Eq.~\eqref{eq:quarkmass} to break symmetry. Because we consider the perturbative expansion around the chiral limit, the quark masses are regarded as small. The invariant structure with single $\mathcal{M}$ is given by
\begin{equation}
    \Tr(\mathcal{M} U^{\dag}+U\mathcal{M}^{\dag}) .
\end{equation}
Motivated by the GMOR relation~\eqref{eq:GMOR}, we assign the counting of  $\mathcal{O}(p^{2})$ for the quark mass matrix $\mathcal{M}$. The lowest order $\mathcal{O}(p^{2})$ chiral Lagrangian with explicit symmetry breaking is now given by
\begin{equation}
    \mathcal{L}_{2}
    =
    \frac{F^{2}}{4}\Tr(\partial_{\mu}U\partial^{\mu}U^{\dag})
    +\frac{F^{2}}{4}\Tr(\chi U^{\dag}+U\chi^{\dag}),
    \label{eq:L2ex}
\end{equation}
with 
\begin{equation}
    \chi
    =2B\mathcal{M}
    =2B
    \begin{pmatrix}
    \hat{m} & 0 \\
    0 & \hat{m}
    \end{pmatrix} ,
\end{equation}
where the low energy constant $B$ is determined by using the definition of the chiral condensate $\bra{0}\bar{q}q\ket{0}=\partial \bra{0}\mathcal{H}\ket{0}/\partial \hat{m}$ as
\begin{equation}
    B
    =-\frac{\bra{0}\bar{q}q\ket{0}}{2F^{2}} .
\end{equation}
By expanding Eq.~\eqref{eq:L2ex}, we can verify that the pion mass follows the GMOR relation~\eqref{eq:GMOR}. 

So far we have concentrated on the system with only pions. In QCD, the baryon number is conserved as a consequence of the vector $U(1)$ symmetry. Therefore, the system can be classified by the baryon number $B$. Let us first consider the baryon number $B=1$ sector. In the nonlinear realization~\cite{Coleman:1969sm,Callan:1969sn} the nucleons can be incorporated as matter fields, which belong to a representation of the unbroken symmetry $H$. In the two-flavor sector, the proton $p$ and the neutron $n$ belong to the fundamental representation of the isospin SU(2):
\begin{equation}
    N
    =
    \begin{pmatrix}
    p \\
    n
    \end{pmatrix} ,
\end{equation}
which transforms as
\begin{equation}
    N
    \stackrel{g}{\to} h(\phi,g) N,
    \quad 
    \bar{N}
    \stackrel{g}{\to} \bar{N}h^{\dag}(\phi,g)
    \label{eq:nucleon} .
\end{equation}
Now we construct the most general invariant Lagrangian with a nucleon bilinear form. Because $h(\phi,g)$ contains the pion field $\phi$, the kinetic term of baryon is not invariant by itself. Instead, we can define the covariant derivative $D_{\mu}$ as
\begin{equation}
    D_{\mu}
    =\partial_{\mu}+i\alpha_{\mu\parallel} ,
\end{equation}
which gives the homogeneous transformation:
\begin{equation}
    D_{\mu}N
    \stackrel{g}{\to} h(\phi,g) D_{\mu}N .
\end{equation}
The most general Lagrangian with only single derivative is 
\begin{equation}
    \mathcal{L}_{\pi N}^{(1)}
    =\bar{N}
    \left(i\Slash{D}-m-g_{A}\gamma^{\mu}\gamma_{5}\alpha_{\mu\parallel}\right)
    N ,
\end{equation}
where $m$ and $g_{A}$ are the mass and the axial charge of the nucleon in the chiral limit, respectively. This leading order term is counted as $\mathcal{O}(p^{1})$. We note that the mass term $\bar{N}N$ is allowed for the matter field. Because of the additional mass scale $m$ in the chiral limit, there arises the power counting violation problem in the loop calculation~\cite{Gasser:1987rb}. To recover the consistent power counting beyond the tree level, it is necessary either to employ the heavy baryon formalism~\cite{Jenkins:1990jv,Bernard:1992qa} or to employ the improved renormalization schemes, such as the infrared regularization~\cite{Becher:1999he} and the extended on mass shell scheme~\cite{Gegelia:1999gf,Fuchs:2003qc}.

The coupling to the pions are generated through $\alpha_{\mu\parallel}$ and $\alpha_{\mu\perp}$. From the expansions~\eqref{eq:aparallelexpand} and \eqref{eq:aperpexpand}, we obtain the two-pion coupling called the Weinberg-Tomozawa (WT) term and the one-pion coupling called Yukawa term:
\begin{align}
    \mathcal{L}_{\rm WT}
    &=\frac{i}{8F^{2}}\bar{N}
    \gamma^{\mu}(\phi\partial_{\mu}\phi-\partial_{\mu}\phi\phi)
    N , \\
    \mathcal{L}_{\rm Yukawa}
    &=-\frac{g_{A}}{2F}\bar{N}
    \gamma^{\mu}\gamma_{5}\partial_{\mu}\phi N .
    \label{eq:Yukawa}
\end{align}
We note that the coupling strength of the Yukawa term is determined by the axial charge of the nucleon $g_{A}$, while the strength of the WT term is determined only by the pion decay constant $F$, in accordance with the low-energy theorem~\cite{Weinberg:1966kf,Tomozawa:1966jm}. In this way, the constraints on the pion-nucleon coupling from chiral symmetry are properly encoded in the effective Lagrangian. Following the same procedure, it is also possible to introduce other hadrons to determine their coupling with pions based on chiral symmetry.

ChPT can be generalized to three flavors with the inclusion of the kaons and the eta meson. In this case, however, there can be non-NG excitations which have a comparable mass with the threshold energies of the NG bosons, because of the substantially large strange quark mass. For instance, the scalar mesons $f_{0}(980)$ and $a_{0}(980)$ exist around the $\bar{K}K$ threshold, and the $\Lambda(1405)$ resonance lies below the $\bar{K}N$ threshold. While naive perturbation theory cannot be applied to these sectors, it is shown that the unitarization of the chiral interaction is useful~\cite{Hyodo:2011ur,Oller:1997ti,Oller:1997ng,Kaiser:1995eg,Oset:1997it,Oller:2000fj}. By performing the nonperturbative resummation of the interactions derived in ChPT, the s-wave near-threshold resonances can be dynamically generated. The driving force is the Weinberg-Tomozawa interaction; in the flavor symmetric limit, the low-energy $s$-wave scattering of the NG boson with a target hadron in the representation $T$ in channel $\alpha$ is given by
\begin{align}
    V_{\rm WT}(\sqrt{s})
    &=-\frac{(\sqrt{s}-M)}{4F^{2}}[C_{2}(T)+N_{F}-C_{2}(\alpha)] ,
\end{align}
where $\sqrt{s}$ is the total energy, $M$ is the mass of the target hadron, and $C_{2}(R)$ is the quadratic Casimir of SU($N_{f}$) for the representation $R$~\cite{Hyodo:2006yk,Hyodo:2006kg}. It is remarkable that the sign and the strength of the driving interaction is solely determined by the group theoretical structure.

In the sector with baryon number $B=2$, we need to consider the nuclear forces~\cite{Weinberg:1990rz,Weinberg:1991um}. The nuclear force can be induced by the one-pion exchange from the Yukawa vertex~\eqref{eq:Yukawa}, while the power counting also allows the four-nucleon contact interaction, which is responsible for the short-range part of the nuclear force. The shallow bound state of deuteron and the large scattering length of the $^{1}S_{0}$ channel indicate the existence of an additional small momentum scale, which mandates the nonperturbative resummation of the interaction derived from ChPT Lagrangian. For more details, see the review articles~\cite{Epelbaum:2008ga,Machleidt:2011zz}. The applications of ChPT to nuclear matter and finite nuclei are discussed in Ref.~\cite{Holt:2013fwa}.

\subsubsection{Heavy hadron effective theory}
\label{sec:heavy_hadron_effective_theory}


We consider the heavy hadron effective theory as an effective theory of QCD at low energy.
As discussed in Sect.~\ref{sec:heavy_quark_symmetry},
one of the most important symmetries in heavy hadrons is the heavy quark (spin) symmetry in the heavy quark limit ($m_{Q} \rightarrow \infty$).
The correction terms deviated from the heavy quark limit are taken systematically by a series of the $1/m_{Q}$ expansion.
In addition, due to the presence of light quarks in the QCD vacuum, we need to consider also chiral symmetry for investigating the properties of heavy hadrons.
Therefore, we have to consider both chiral symmetry and heavy quark symmetry (see Refs.~\cite{Manohar:2000dt,Casalbuoni:1996pg}.).

\paragraph{Heavy meson effective theory}

We consider the heavy hadron effective theory for heavy-light mesons, $Q\bar{q}$ ($D$, $D^{\ast}$, $\bar{B}$, $\bar{B}^{\ast}$) mesons and $q\bar{Q}$ ($\bar{D}$, $\bar{D}^{\ast}$, $B$, $B^{\ast}$) mesons.
We mainly focus on the coupling to pions, because the possible terms in the effective Lagrangian are well controlled by chiral symmetry and  several unknown parameters for coupling constants are well limited.

Let us consider the effective field for $Q\bar{q}$ mesons.\footnote{The present formalism was given for small brown muck spin in Refs.~\cite{Georgi:1990cx,Mannel:1990vg} and was extended to general cases in Ref.~\cite{Falk:1991nq}. Here we follow the description in Ref.~\cite{Grinstein:1995uv}.}
We introduce the spinor $u_{Q\,\alpha}$ for the heavy quark $Q$ and the spinor $v_{q\,\beta}$ for the light antiquark $\bar{q}$
 with the Dirac indices $\alpha$ and $\beta$, which are defined in the four-velocity $v$-frame.
We express the $Q\bar{q}$ meson field by bispinor form
\begin{align}
u_{Q\,\alpha} \bar{v}_{q\,\beta},
\end{align}
with $v\hspace{-0.5em}/ u_{Q}=u_{Q}$ and $\bar{v}_{q} v\hspace{-0.5em}/=\bar{v}_{q}$.
We note that $\bar{q}$ stands for the brown muck whose quantum number is the same as the light antiquark, but contains non-perturbative components like $q\bar{q}\bar{q}$ and $g\bar{q}$, as discussed in Sect.~\ref{sec:heavy_quark_symmetry}.
The important property of $\bar{q}$ is chiral symmetry only. 
In the rest frame $v^{\mu}=(1,\vec{0}\,)$, we define the spin operator
\begin{align}
S^{i} = \frac{1}{2} \gamma_{5} \gamma^{0} \gamma^{i} =
\frac{1}{2}
\left(
\begin{array}{cc}
 \sigma^{i} & 0 \\
 0 & \sigma^{i}
\end{array}
\right),
\label{eq:spin_operator}
\end{align}
with $i=1,2,3$.
The matrix is expressed in the Dirac representation.
The spin up and down states of $u_{Q}$ are given by $u^{(1)}_{Q\alpha}=\delta_{1\alpha}$ and $u^{(2)}_{Q\alpha}=\delta_{2\alpha}$.
In contrast, the spin up and down states of $v_{q}$ are given by $v^{(2)}_{q\alpha}=-\delta_{4\alpha}$ and $v^{(1)}_{q\alpha}=-\delta_{3\alpha}$.
We note the relation $\vec{S}(u_{Q} \bar{v}_{q}) = (\vec{S}u_{Q}) \bar{v}_{q} + u_{Q} (\vec{S}\bar{v}_{q})$.
The spin 0 state is given by
\begin{align}
u_{Q}^{(1)} \bar{v}_{q}^{(1)} + u_{Q}^{(2)} \bar{v}_{q}^{(2)}
=
\frac{1+\gamma^{0}}{2} \gamma_{5},
\end{align}
and the spin 1 state is by
\begin{align}
u_{Q}^{(1)}\bar{v}_{q}^{(2)} &= \frac{1+\gamma^{0}}{2} /\hspace{-0.5em}\epsilon^{(+)}, \\
\frac{1}{\sqrt{2}} \left( u_{Q}^{(1)} \bar{v}_{q}^{(1)} - u_{Q}^{(2)} \bar{v}_{q}^{(2)} \right) &= \frac{1+\gamma^{0}}{2} /\hspace{-0.5em}\epsilon^{(0)}, \\
u_{Q}^{(2)}\bar{v}_{q}^{(1)} &= \frac{1+\gamma^{0}}{2} /\hspace{-0.5em}\epsilon^{(-)},
\end{align}
with the polarization vector ${\epsilon^{(\lambda)}}^{\mu}$ ($\lambda=\pm,0$) for spin 1, whose explicit forms are ${\epsilon^{(\pm)}}^{\mu}=\frac{1}{\sqrt{2}}(0,1,\pm i, 0)$ and ${\epsilon^{(0)}}^{\mu}=(0, 0, 0, 1)$ (cf.~Ref.~\cite{Grinstein:1995uv}).
After the Lorentz transformation from the rest frame to general frame with four-velocity $v^{\mu}$, we obtain the spin 0 state
\begin{align}
u_{Q}^{(1)} \bar{v}_{q}^{(1)} + u_{Q}^{(2)} \bar{v}_{q}^{(2)}
\rightarrow
\frac{1+v\hspace{-0.5em}/}{2} \gamma_{5},
\end{align}
and the spin 1 state
\begin{align}
u_{Q}^{(1)}\bar{v}_{q}^{(2)} &\rightarrow \frac{1+v\hspace{-0.5em}/}{2} /\hspace{-0.5em}\epsilon^{(+)}, \\
\frac{1}{\sqrt{2}} \left( u_{Q}^{(1)} \bar{v}_{q}^{(1)} - u_{Q}^{(2)} \bar{v}_{q}^{(2)} \right) &\rightarrow \frac{1+v\hspace{-0.5em}/}{2} /\hspace{-0.5em}\epsilon^{(0)}, \\
u_{Q}^{(2)}\bar{v}_{q}^{(1)} &\rightarrow \frac{1+v\hspace{-0.5em}/}{2} /\hspace{-0.5em}\epsilon^{(-)},
\end{align}
noting that $(1+\gamma^{0})/2$ is changed by the Lorentz transformation.
From this result, we find that the heavy-light meson fields, $P_{v}$ for spin 0 and $P_{v}^{\ast \mu}$ for spin 1, can be parametrized by the bispinor representations
\begin{align}
\frac{1+v\hspace{-0.5em}/}{2} \gamma_{5} P_{v}, \hspace{1em} \frac{1+v\hspace{-0.5em}/}{2} /\hspace{-0.6em}P_{v}^{\ast}.
\end{align}
According to the spin transformation $S_{Q}$ for the heavy quark $Q$ and the chiral transformation $U_{q}$ for the light antiquark $q$
\begin{align}
u_{Q}\bar{v}_{q} \rightarrow S_{Q} \left(u_{Q}\bar{v}_{q}\right) U^{\dag}_{q}.
\end{align}
$P_{v}$ and $P_{v}^{\ast \mu}$ are transformed as
\begin{align}
\frac{1+v\hspace{-0.5em}/}{2} \gamma_{5} P_{v} &\rightarrow S_{Q} \frac{1+v\hspace{-0.5em}/}{2} \gamma_{5} P_{v} U^{\dag}_{q}, \label{eq:transformation_0} \\
\frac{1+v\hspace{-0.5em}/}{2} /\hspace{-0.6em}P_{v}^{\ast} &\rightarrow S_{Q} \frac{1+v\hspace{-0.5em}/}{2} /\hspace{-0.6em}P_{v}^{\ast} U^{\dag}_{q}. \label{eq:transformation_1}
\end{align}
Here we use $S_{v}^{\mu} v\hspace{-0.5em}/ = v\hspace{-0.5em}/ S_{v}^{\mu}$.
Because the spin 0 and 1 states are degenerate in the heavy quark limit,
and they have the common transformation properties for the heavy quark spin symmetry as well as for chiral symmetry,
we introduce
\begin{align}
H_{v}(x) = \frac{1+v\hspace{-0.5em}/}{2} \left( P^{\ast}_{v\,\mu}(x) \gamma^{\mu} + i P_{v}(x) \gamma_{5} \right),
\label{eq:H_field}
\end{align}
as a superposed field including the spin 0 and 1 states.
The transformation of $H_{v}$ is given by
\begin{align}
H_{v} \rightarrow S_{Q} H_{v} U^{\dag}_{q},
\end{align}
from Eqs.~(\ref{eq:transformation_0}) and (\ref{eq:transformation_1}).

Let us consider the interaction of the $q\bar{Q}$ meson with pions.
The pion dynamics is given by chiral symmetry and by momentum expansion of the pion, as the Nambu-Goldstone modes in the QCD vacuum with the broken chiral symmetry (Sect.~\ref{sec:chiral_effective_theory}).
As for the latter, we consider the lowest order contribution, ${\cal O}(q)$, with respect to the pion momentum $q$.
We consider the chiral transformation, $V^{\mu}(x) \rightarrow U_{q} V^{\mu}(x) U_{q}^{\dag} +i U_{q} \partial ^{\mu} U_{q}^{\dag}$ and $A^{\mu} \rightarrow U_{q} A^{\mu} U_{q}^{\dag}$, for the vector current $V^{\mu}(x)=\frac{i}{2}(\xi^{\dag}\partial^{\mu}\xi+\xi\partial^{\mu}\xi^{\dag})$ and the axial-vector current $A^{\mu}(x)=\frac{i}{2}(\xi^{\dag}\partial^{\mu}\xi-\xi\partial^{\mu}\xi^{\dag})$ with the pion field $\xi=\exp(i\phi/\sqrt{2}f_{\pi})$ for $\phi$ defined in Eq.~(\ref{eq:phi_def}).\footnote{Note that the notations are different from those in Sect.~\ref{sec:chiral_effective_theory} as $f_{\pi}=\sqrt{2}F$, $\xi=u$, $V^{\mu}=-a^{\mu}_{\parallel}$, $A^{\mu}=-a^{\mu}_{\perp}$ and $U_{q}=h$.} 
Then, we find that the effective Lagrangian
\begin{align}
{\cal L}_{\mathrm{heavy-light}} = \mathrm{Tr}\bar{H}_{v} v \!\cdot\! iD H_{v} + g \mathrm{Tr} \bar{H}_{v} H_{v} \gamma_{\mu} \gamma_{5} A^{\mu} + {\cal O}(1/M),
\label{eq:L_heavy_hadron_pion}
\end{align}
is invariant under both heavy quark spin symmetry and chiral symmetry.
At the leading order, the correction by the heavy hadron mass $M$ is neglected.
In the first term, $D^{\mu}H_{v}=\partial^{\mu}H_{v}-iV^{\mu}H_{v}$ is the covariant derivative.\footnote{We note that the covariant derivative for $\bar{q}Q$ meson is defined by $D^{\mu}H_{v}=\partial^{\mu}H_{v}+iH_{v}V^{\mu}$~\cite{Manohar:2000dt}.}
The trace $\mathrm{Tr}$ is taken over the Dirac spinor and the isospin.
In the second term, $g$ is the coupling constant in the axial-vector interaction.
Note that the axial-vector current is given as $A^{\mu}\simeq-\partial^{\mu} {\cal \phi}/\sqrt{2}f_{\pi}$ for lowest number of pions, hence the single pion coupling to $H_{v}$ field, which is relevant to the one-pion-exchange potential, is supplied by this term.
It is also important to note that the single pion coupling vertex has the structure of $q^{i} P_{v}^{\ast i \dag}P_{v}$ (or $q^{i} P_{v}^{\dag} P_{v}^{\ast i}$) and $q^{i} \varepsilon^{ijk} P_{v}^{\ast j \dag} P_{v}^{\ast k}$ ($i,j,k=1,2,3$) with $q^{i}$ being the three-dimensional momentum of the pion. 
Hence the single pion coupling causes the spin-flip of the heavy-light mesons~\cite{Cohen:2005bx,Yasui:2009bz}.\footnote{This form is comparable with the non-relativistic form of the axial-vector coupling of $\pi NN$, $\vec{q}\!\cdot\!\vec{\sigma}$, with $\vec{\sigma}$ being the spin operator acting on the nucleon.}
More precisely to say, only the light component $q$ of $q\bar{Q}$ flips, while the heavy quark $Q$ is not due to the conservation of the heavy quark spin (cf.~Sect.~\ref{sec:heavy_quark_symmetry}).
The explicit forms of the potential between a heavy-light meson ($\bar{q}Q$, $q\bar{Q}$) and a nucleon will be given in Sect.~\ref{sec:D_mesons}.

\paragraph{\mbox{\boldmath $1/M$} corrections}
We consider the $1/M$ corrections for the effective Lagrangian (\ref{eq:L_heavy_hadron_pion}) for $q\bar{Q}$ meson.
To include the $1/M$ correction in the effective Lagrangian, the velocity rearrangement and the heavy quark spin breaking are important~\cite{Luke:1992cs,Kitazawa:1993bk} (see also Ref.~\cite{Yasui:2013xr}), as introduced in Sect.~\ref{sec:heavy_quark_symmetry}.

To consider the velocity rearrangement, we introduce the ``original" four velocity defined by
\begin{align}
 {\cal V}^{\mu} = \frac{v^{\mu}+iD^{\mu}/M}{|v^{\mu}+iD^{\mu}/M|},
\end{align}
containing the residual momentum.
This satisfies the normalization condition ${\cal V}_{\mu}{\cal V}^{\mu}=1$.
Up to and including ${\cal O}(1/M)$, ${\cal V}^{\mu}$ is expanded as
\begin{align}
 {\cal V}^{\mu} = v^{\mu} + \frac{1}{M} \left( iD^{\mu} - v^{\mu} v \!\cdot\! iD \right) + {\cal O}(1/M^{2}),
 \label{eq:original_four_velocity}
\end{align}
which satisfies ${\cal V}_{\mu}{\cal V}^{\mu}=1+{\cal O}(1/M^{2})$.
Let us remember that $H_{v}$ in Eq.~(\ref{eq:H_field}) was defined in the $v^{\mu}$-frame.
We consider the Lorentz boost from the $v^{\mu}$-frame to the ${\cal V}^{\mu}$-frame (${\cal V}^{\mu}$ is given in Eq.~(\ref{eq:original_four_velocity})) up to and including ${\cal O}(1/M)$.
Then, the Lorentz transformation of $H_{v}$ is given by
\begin{align}
{\cal H}_{v}(x) = H_{v}(x) + \frac{1}{2M} \left( i \vec{D} \hspace{-0.6em}/ \, H_{v}(x) - H_{v}(x) i\cev{D}\hspace{-0.6em}/ \, - 2v \!\cdot\! iD H_{v}(x) \right) + {\cal O}(1/M^2).
\end{align}
Note that ${\cal H}_{v}$ satisfies ${\cal V}\hspace{-0.6em}/\hspace{0.2em}{\cal H}_{v} = {\cal H}_{v}+{\cal O}(1/M^{2})$ and ${\cal H}_{v}{\cal V}\hspace{-0.6em}/ = - {\cal H}_{v}+{\cal O}(1/M^{2})$, hence $(1+{\cal V}\hspace{-0.6em}/ )/2$ operates like the projection operator $(1+v\hspace{-0.5em}/)/2$ for $H_{v}$ in Eq.~(\ref{eq:H_field}).
By the velocity rearrangement between $v^{\mu}$ and $w^{\mu}=v^{\mu}+q^{\mu}/M$ with $q^{\mu}$ much smaller than $m_{Q}$, ${\cal H}_{v}(x)$ is transformed as
\begin{align}
 {\cal H}_{w}(x) = e^{iq \cdot x} {\cal H}_{v}(x) + {\cal O}(1/M^{2}),
\end{align}
in a covariant form.

As for the  breaking of the heavy quark symmetry, we consider the form ${\mathrm Tr} \left( \bar{H}_{v} \Gamma H_{v} \cdots \right)$ with Dirac gamma matrices $\Gamma=i\gamma_{5}, \gamma^{\mu}\gamma_{5}, \sigma^{\mu\nu}$.
In fact, those $\Gamma$'s are not commutable with the spin operator (\ref{eq:spin_operator}).
We may choose $\Gamma=\gamma^{\mu}\gamma_{5}$ for the axial-vector coupling to the pion field. 

Following both the velocity rearrangement and the heavy quark spin breaking,
we can construct the effective Lagrangian up to and including ${\cal O}(1/M)$.
The interaction Lagrangian in Eq.~(\ref{eq:L_heavy_hadron_pion}) is extended as~\cite{Kitazawa:1993bk,Yasui:2013xr}
\begin{align}
 {\cal L}^{\mathrm{LO+NLO}}_{\pi H_{v} H_{v}}
&=
 \left( g+\frac{g_{1}}{M} \right) {\mathrm Tr} \bar{H}_{v} H_{v} \gamma_{\mu} \gamma_{5} A^{\mu} \nonumber \\
&
 + \frac{g}{2M}
 \left( {\mathrm{Tr}} v \!\cdot\! iD \bar{H}_{v} H_{v} \gamma_{\mu} \gamma_{5} A^{\mu} - {\mathrm{Tr}} \bar{H}_{v} v \!\cdot\! iD H_{v} \gamma_{\mu} \gamma_{5} A^{\mu} \right) \nonumber \\
&
 + \frac{g}{4M} \varepsilon_{\mu\nu\rho\sigma} \left( {\mathrm{Tr}} iD^{\nu} \bar{H}_{v} H_{v} \sigma^{\rho\sigma}A^{\mu} - {\mathrm{Tr}} \bar{H}_{v} iD^{\nu} H_{v} \sigma^{\rho\sigma} A^{\mu} \right) \nonumber \\
&
 + \frac{g_{2}}{M} {\mathrm{Tr}} \bar{H}_{v} \gamma_{\mu} \gamma_{5} H_{v} A^{\mu} + {\cal O}(1/M^{2}),
 \label{eq:L_heavy_hadron_pion_2}
\end{align}
where $g_{1}$ and $g_{2}$ are new coupling constants different from $g$.
Note that the first, second and third terms in the right-hand-side are related with each other with common coupling $g$ due to the invariance under the velocity-rearrangement.
Those three terms satisfy the heavy quark symmetry.
Only the last term breaks the heavy quark symmetry.

We have demonstrated the construction of the effective Lagrangian up to and including ${\cal O}(1/M)$ for the axial-vector coupling.
Furthermore, we may consider the vector current term of the pion field, as shown in Ref.~\cite{Kitazawa:1993bk}.
This scheme can be applied also for other cases including higher order terms of the pion momentum.

\paragraph{Heavy baryon effective theory}

So far, we have considered the heavy-light ($q\bar{Q}$ and $Q\bar{q}$) mesons.
Because the construction scheme of the effective field is quite general,
we may consider the other heavy hadrons with other brown muck spin~\cite{Falk:1991nq}.
Let us investigate the effective field of $qqQ$ ($\Lambda_{c}$, $\Sigma_{c}$, $\Sigma_{c}^{\ast}$, $\dots$, $\Lambda_{b}$, $\Sigma_{b}$, $\Sigma_{b}^{\ast}$, $\dots$) baryons.

For the brown muck spin $j=0$ ($\Lambda_{c}$ baryon and $\Lambda_{b}$ baryon), the effective field is trivially given by the spinor
\begin{align}
\psi_{v}=u_{h},
\end{align}
with $u_{h}$ being the heavy quark spinor satisfying $v\hspace{-0.5em}/\hspace{0.1em}u_{h}=u_{h}$ (cf.~Eq.~(\ref{eq:positive_projection})).
Here $\psi_{v}$ satisfies the condition $v\hspace{-0.4em}/\psi_{v} = \psi_{v}$.
In literature, $\psi_{v}$ is sometimes represented by $B_{\bar{\bf 3}}$ for the SU(3) ($u$, $d$, $s$) flavor anti-triplet representation for $qq$ in $qqQ$.

For the brown muck spin $j=1$ ($\Sigma_{c}$, $\Sigma_{c}^{\ast}$ baryons and $\Sigma_{b}$, $\Sigma_{b}^{\ast}$ baryons),
the effective field is given by the vector-spinor
\begin{align}
 \psi_{v}^{\mu} = A^{\mu} u_{h},
\end{align}
with $A^{\mu}$ being the brown muck axial-vector with $j^{\,{\cal P}}=1^{+}$ and $u_{h}$ being the heavy quark spinor with the condition $v\hspace{-0.5em}/\hspace{0.1em}u_{h}=u_{h}$.
Notice that the $v\hspace{-0.4em}/\psi_{v}^{\mu} = \psi_{v}^{\mu}$ as well as $v_{\mu}\psi_{v}^{\mu}=0$ are satisfied, in which the latter is induced from the condition $v_{\mu}A^{\mu}=0$ for the axial-vector field $A^{\mu}$.
The $\psi^{\mu}_{v}$ is a superposed field of the spin $3/2$ baryon and the spin $1/2$ baryon.
Let us decompose the $\psi^{\mu}_{v}$ into physical components $\psi_{v\,3/2}^{\mu}$ and $\psi_{v\,1/2}$ with spin $3/2$ and $1/2$, respectively.
For this purpose, we recall the condition for the spin $3/2$ field $\Psi^{\mu}$ in the Rarita-Schwinger formalism~\cite{Rarita:1941mf} (see also Ref.~\cite{Scherer:2012xha}),
\begin{align}
 \gamma_{\mu} \Psi^{\mu}=0
 \hspace{0.5em}
 \mathrm{and}
 \hspace{0.5em}
 \partial_{\mu} \Psi^{\mu}=0.
\end{align}
In the case of $\psi_{v\,3/2}^{\mu}$ in the $v$-frame, those two conditions are rewritten as
\begin{align}
 \gamma_{\mu} \psi_{v\,3/2}^{\mu}=0
 \hspace{0.5em}
 \mathrm{and}
 \hspace{0.5em}
 v_{\mu}  \psi_{v\,3/2}^{\mu}=0.
 \label{eq:RS_condition_2}
\end{align}
Then, we find the following form
\begin{align}
 \psi_{v\,3/2}^{\mu} = \left( g^{\mu}_{\nu} -\frac{1}{3} \left( \gamma^{\mu}+v^{\mu} \right) \gamma_{\nu}\right) \psi_{v}^{\nu},
\end{align}
satisfies Eq.~(\ref{eq:RS_condition_2}) as well as $v\hspace{-0.5em}/\hspace{0.1em} \psi_{v\,3/2}^{\mu}=\psi_{v\,3/2}^{\mu}$.
The spin 1/2 field $\psi_{v\,1/2}$ is given as the orthogonal component to $\psi_{v\,3/2}^{\mu}$;
\begin{align}
 \psi_{v\,1/2} = \frac{1}{\sqrt{3}} \gamma_{5} \gamma_{\nu} \psi_{v}^{\nu},
\end{align}
which satisfies $v\hspace{-0.5em}/\hspace{0.1em} \psi_{v\,1/2}=\psi_{v\,1/2}$.
As a result, we can express $\psi_{v}^{\mu}$ by $\psi_{v\,3/2}^{\mu}$ and $\psi_{v\,1/2}$ as
\begin{align}
 \psi_{v}^{\mu} = \psi_{v\,3/2}^{\mu} + \frac{1}{\sqrt{3}} \left( \gamma^{\mu} + v^{\mu} \right) \gamma_{5} \psi_{v\,1/2}.
 \label{eq:superfield_baryon}
\end{align}
In literature, $\psi_{v}^{\mu}$ is sometimes represented by $S_{{\bf 6}}^{\mu}$ in the SU(3) ($u$, $d$, $s$) flavor sextet representation.
With those effective fields, we can construct the interaction Lagrangian with light mesons.
The examples can be seen in Ref.~\cite{Liu:2011xc} and references therein.
The potential between a heavy baryon and a nucleon will be given in Sect.~\ref{sec:charm_baryons}.

\subsection{Theoretical approaches to finite density}
\label{sec:finite_density}

There are several theoretical approaches to investigate the properties of hadrons in nuclear medium.
First, we introduce the few-body calculation based on the Gaussian expansion method (Sect.~\ref{sec:fewbody_systems}).
In principle, the few-body calculation can be applied to large baryon numbers,
but it is not practical in many cases.
Instead, we introduce the treatment of the in-medium nucleon propagator used in the many-body calculation in infinitely extended nuclear matter (Sect.~\ref{sec:nuclear_matter}).
It is an important aspect that several quark and gluon condensates in nuclear medium become different from those in vacuum.
In the leading order of the low baryon number density, we can obtain the model-independent result for those changes (Sect.~\ref{sec:chiral_symmetry_at_finite_density}).
We introduce the QCD sum rule as a theoretical technique to connect the quark and gluon condensates and the hadron properties (Sect.~\ref{sec:QCDSR}).

\subsubsection{Few-body systems}
\label{sec:fewbody_systems}

In this section, we give a brief review of the
Gaussian expansion method which is one of the numerical techniques to solve accurately quantum few-body systems.
The few-body calculations often appear in many important problems of
the quantum theory.
In order to solve the few-body systems with high precision,
various methods have been developed so far~\cite{Kamada:2001tv}.
As one of the standard methods,
the variational method with
appropriate basis is widely used
to analyze bound and scattering states of the few-body systems.
In this method, the eigenvalues are obtained by diagonalizing the
Hamiltonian with the wave functions expanded by the basis
functions.
The calculation is iterated while changing the variational parameters,
and continued until the obtained values are converged.
The few-body system with a finite number of
$L^2$-class basis functions has a large number of degrees of freedom.
Therefore, it is important to choose the appropriate basis functions.

The Gaussian expansion method (GEM) 
proposed in Refs.~\cite{Kamimura:1988zz,Hiyama:2003cu,Hiyama:2012sma}
is the well-known method,
where
the wave functions
are expanded in terms of Gaussian basis functions.  
GEM
provides an efficient approach for few-body calculations,
because
(i) it can describe the physical behavior of wave functions
such as short-range correlations and long-range
asymptotic behavior, (ii) it is easy to perform the coordinate
transformation and the calculation of the matrix elements for
various interactions including non-central
force such as the spin-orbit force and tensor force.
The results by the GEM are equivalent to those by the 
Faddeev method giving accurate solutions for the three-body systems,
when the good convergence of the eigenenergy is obtained~\cite{Hiyama:2003cu}.
The GEM has been applied to various few-body problems such as 
molecular
physics~\cite{Kamimura:1988zz,Hiyama:2003cu,Kino:1993hi,Kino:1993hi195,PhysRevA.52.870,Hamahata:2001hi},
atomic physics~\cite{Kino:1999hi,Kino:2001hi138,PhysRevA.85.022502,Hiyama:2012cj}, 
nuclear
physics~\cite{Matsumoto:2003qw,Matsumoto:2004ck,Matsumoto:2005pd,Prog.Theor.Phys11620061Aoyama,Myo:2007vm,Funaki:2008gb,Yamanaka:2015qfa,Hiyama:2016nwn,Yamanaka:2016fjj},
astrophysics~\cite{Hamaguchi:2007mp,Kusakabe:2007fv,Kamimura:2008fx},
hypernuclear
physics~\cite{Hiyama:2003cu,KanadaEn'yo:2008wm,Jido:2008kp,Hiyama:2010zzb,Hiyama:2010zz,Yang:2011rp,Dote:2014ema},
and
other hadron few-body systems~\cite{Hiyama2006237,Segovia:2008zz,Ortega:2010qq,Yokota:2013sfa,Yamaguchi:2013hsa,Yoshida:2015tia,Maeda:2015hxa}.

\begin{figure}[t]
 \begin{center}
    \includegraphics[width=8cm,bb=0 0 783 232]{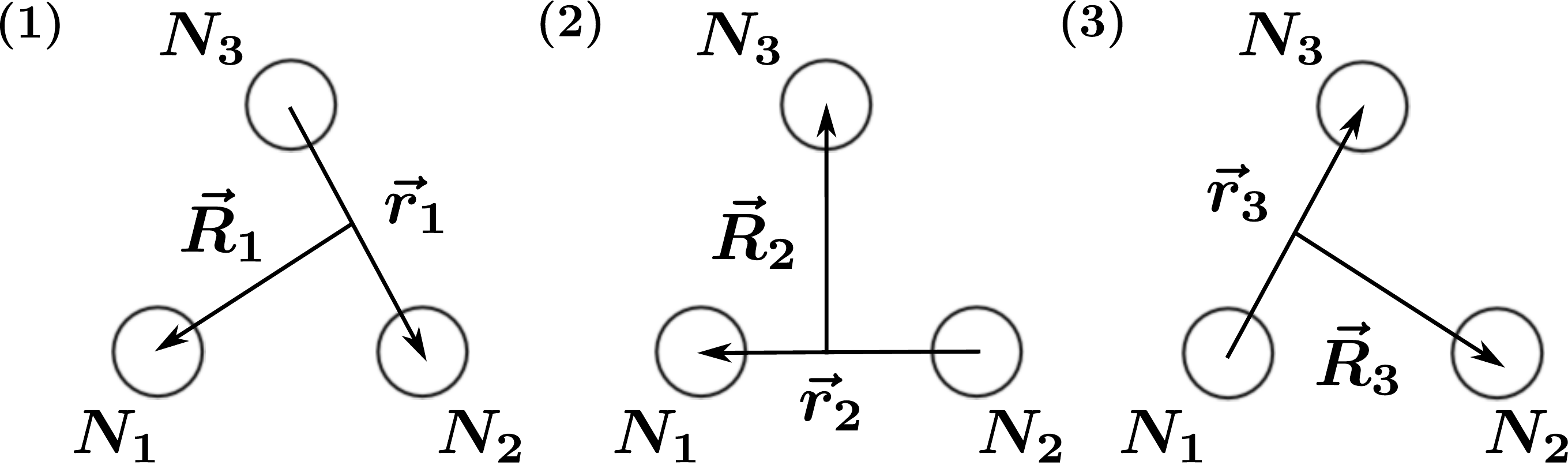}
  \caption{Jacobi coordinates of the three-body systems.}
  \label{fig:jacobi}
 \end{center}
\end{figure}

In order to describe the 
basis functions,
the 
Jacobi coordinates 
are employed.
In Fig.~\ref{fig:jacobi}, the Jacobi coordinates of the three-body
system are shown.
By introducing the three-channels of the Jacobi coordinates,
the function space becomes larger than the case with a single channel.
The basis function with three channels takes account of all two-particle
correlations.
If you employ only one Jacobi channel of them,
it is difficult 
to describe the two-particle correlations corresponding to the other Jacobi
channels.
The set of the Jacobi coordinates makes a good convergence even if the
orbital angular momentum considered in the calculation is restricted.

In Fig.~\ref{fig:jacobi}, the coordinates $\vec{r}_c,\vec{R}_c$ ($c=1,2,3$) and the center of mass
$\vec{G}$
are given by the
single-particle coordinates $x_i$  ($i=1,2,3$) and the mass $m_i$.
The coordinates for $c=1$ are written by
\begin{align}
 &\left(
 \begin{array}{c}
  \vec{r}_1 \\
  \vec{R}_1 \\
  \vec{G} \\
 \end{array}
 \right) =
 \left(
 \begin{array}{c}
  \vec{x}_2-\vec{x}_1 \\
  \displaystyle \vec{x}_3-\frac{m_1\vec{x}_1+m_2\vec{x}_2}{m_1+m_2} \\[2.5mm]
  \displaystyle \frac{m_1\vec{x}_1+m_2\vec{x}_2+m_3\vec{x}_3}{m_1+m_2+m_3} \\
 \end{array}
 \right) =
  \left(
 \begin{array}{ccc}
  -1&1&0 \\
  \displaystyle -\frac{m_1}{m_{12}}&\displaystyle -\frac{m_2}{m_{12}}&1 \\[2.5mm]
  \displaystyle \frac{m_1}{m_{123}}&\displaystyle
   \frac{m_2}{m_{123}}&\displaystyle \frac{m_3}{m_{123}} \\
 \end{array}
 \right)
 \left(
 \begin{array}{c}
  \vec{x}_1 \\
  \vec{x}_2 \\
  \vec{x}_3 \\
 \end{array}
 \right) \, ,
\end{align}
with $m_{12\cdots n}=m_1+m_2+\cdots+m_n$.
The coordinates for $c=2,3$ can be obtained in the similar way.

The wave function of the three-body system is given by a sum of the
rearrange channel amplitudes ($c=1,2,3$) as
\begin{align}
 \Psi_{JM}=&\sum_{c=1}^{3}\Phi^{c}_{JM}(\vec{r}_c,\vec{R}_c) \, .
\end{align}
The wave function of the channel $c$ is 
\begin{align}
 \Phi^{c}_{JM}(\vec{r}_c,\vec{R}_c)&=
 \sum_{nl_1,Nl_2,L}\sum_{s_{12}S,I_{12}}
 C^{(c)}_{nl_1,Nl_2,L,s_{12}S,I_{12}I}
 \left\{
 \left[\left[\phi^{(c)}_{nl_1m_1}(\vec{r}_c)\psi^{(c)}_{Nl_2m_2}(\vec{R}_c)\right]_{L}
 \right.\right. \notag\\
 &\times
 \left.\left.
 \left[\left[\chi_{s_1}\chi_{s_2}\right]_{s_{12}}\chi_{s_3}\right]_{S}
 \right]_{JM}
 \left[\left[\eta_{I_1}\eta_{I_2}\right]_{I_{12}}\eta_{I_3}\right]_{I}
 \right\}
 \, . \label{Eq:gaussian_expansion}
\end{align}
$l_1$ and $l_2$ stand for the relative orbital angular momenta
associated with the coordinates $\vec{r}_c$ and $\vec{R}_c$,
respectively.
$L$ is the total orbital angular momentum of the
three-body system.
$\chi_{s_i}$ ($\eta_{I_i}$) with $i=1,2,3$ is the spin (isospin)
function of
the particle with the spin $s_i$ (isospin $I_i$).
$s_{12}$ ($I_{12}$) is the spin (isospin) of two particles combined by
the relative coordinate $\vec{r}_c$, and $S$ ($I$) is the total spin
(isospin) of the
three-body system.
The (anti-)symmetrization is needed~\cite{Hiyama:2003cu}
when the system includes identical particles.
For the sum in Eq.~\eqref{Eq:gaussian_expansion},
all possible coupled channels are included to obtain
solutions with sufficiently good accuracy.  

The functions $\phi_{nl_1m_1}(\vec{r}\,)$ and
$\psi_{Nl_2m_2}(\vec{R}\,)$ are
expressed in terms of the Gaussian functions~\cite{Hiyama:2003cu} as
\begin{align}
 \phi_{nl_1m_1}(\vec{r}\,)&=\sqrt{\frac{2}{\Gamma(l_1+3/2)b^{2l_1+3}_n}}
 \exp{\left(-\frac{r^2}{2b^2_n}\right)}{\cal Y}_{l_1m_1}(r)
 \, , \\
 \psi_{N{l_2}m_2}(\vec{R}\,)&=\sqrt{\frac{2}{\Gamma({l_2}+3/2)B^{2l_2+3}_N}}
 \exp{\left(-\frac{R^2}{2B^2_N}\right)}{\cal Y}_{l_2m_2}(R)
 \, ,
\end{align}
where ${\cal Y}_{lm}(r)$ is a solid spherical harmonic ${\cal Y}_{lm}(r)=r^lY_{lm}(\hat{r})$.
The Gaussian ranges $b_n$ and $B_N$ are given by the form of geometric
series as
\begin{align}
 b_n&=b_1 a^{n-1}\, (n=1,\cdots,n_{max}), \\
 B_N&=B_1 A^{N-1} \, (N=1,\cdots,N_{max}),
\end{align}
where $(b_1,b_{n_{max}})$ and $(B_1,B_{N_{max}})$ 
(or $(b_1,a)$ and $(B_1,A)$) are the variational parameter.

The eigenvalues and coefficients $C^{(c)}_{nl_1,Nl_2,L,s_{12}S,I_tI}$ in
Eq.~\eqref{Eq:gaussian_expansion} are
obtained by solving the eigenvalue problem
\begin{align}
{\boldmath HC}=E{\boldmath NC} \, ,
 \label{eq:eigenvalueProb}
\end{align}
where the matrix elements of the Hamiltonian and norm are given by
\begin{align}
 H_{nn^\prime}=\langle \Phi_{JM,n}|H|\Phi_{JM,n^\prime}\rangle \, , \\
 N_{nn^\prime}=\langle \Phi_{JM,n}|1|\Phi_{JM,n^\prime}\rangle \, .
\end{align}

\subsubsection{Treatment of nuclear matter in hadron effective theories}
\label{sec:nuclear_matter}

In the framework of the hadron effective theory,
the fundamental degrees of freedom are hadrons.
The change of hadrons in nuclear matter is induced through the interaction with nucleons at finite baryon density.
Let us consider zero temperature.
For the Fermi momentum $k_{\mathrm{F}}$, the non-interacting nucleon propagator is given by
\begin{align}
 iS(p_{0},\vec{p}\,;k_{\mathrm{F}}) = (p\hspace{-0.4em}/+m)
 \left( \frac{i}{p^{2}-m^{2}+i\varepsilon} - 2\pi \theta(p_{0}) \delta(p^{2}-m^{2}) \theta(k_{\mathrm{F}}-|\vec{p}\,|) \right),
\label{eq:propagator_nnucleon_medium}
\end{align}
with $\varepsilon>0$.
The second term in the large parentheses is for the subtraction of the nucleon propagation with on-mass-shell and positive energy inside Fermi sphere due to the Pauli exclusion principle.
This is easily confirmed by performing the $p_{0}$-integration (on the complex $p_{0}$-plane) after multiplying a regular function of $p_{0}$.
Note that the covariance is lost due to the finite density.
An alternative way for introducing the nucleon propagator at finite density is to introduce the propagator
\begin{align}
 iS'(p_{0},\vec{p}\,;\mu) = \frac{i}{p\hspace{-0.4em}/-m + \mu \gamma^{0} + i\varepsilon'},
 \label{eq:propagator_nnucleon_medium_2}
\end{align}
with chemical potential $\mu$,
where $\varepsilon'$ is defined as $\varepsilon' > 0$ for $p_{0}>0$ and $|\vec{p}\,|\ge k_{\mathrm{F}}$ and for $p_{0}<0$, while $\varepsilon' < 0$ for $p_{0}>0$ and $|\vec{p}\,| < k_{\mathrm{F}}$.
It means that the poles for $p_{0}>0$ and $|\vec{p}\,| < k_{\mathrm{F}}$, namely holes inside the Fermi sphere, are regarded as the antiparticles in vacuum.
We notice 
 $\mu=\sqrt{k_{\mathrm{F}}^{2}+m^{2}}$ for free nucleons.
Eqs.~(\ref{eq:propagator_nnucleon_medium}) and (\ref{eq:propagator_nnucleon_medium_2}) give the same result.
In the latter formalism, we can easily consider the propagator at finite temperature by introducing the Matsubara sum for the $p_{0}$-integration~\cite{Bellac:2011kqa,Kapusta:2006pm}.

The hadrons in nuclear matter exhibit a complicated dynamics through the interaction with nucleons.
In principle, the dynamics should be solved as $N$-body problem with $N$ being the number of particles.
Of course, this is not possible as practical calculations.
In many cases, we can get important information by focusing on the two-body scattering of the hadron and the nucleon in nuclear matter.
In general, the two-body scattering is described by the coupled channel equation.
$T$-matrix of the two-body scattering is given by the Lippmann-Schwinger equation (or the Bethe-Salpeter equation),
which is schematically written as
\begin{align}
 T=V+VGT,
\end{align}
where $V$ is the interaction kernel and $G$ is the two-body propagator including the in-medium nucleon propagator (\ref{eq:propagator_nnucleon_medium}) or (\ref{eq:propagator_nnucleon_medium_2}).
Through the channel coupling, the hadrons acquire non-trivial behaviors in nuclear medium, such as change of spectral function, mass modification and change of decay modes.
Examples for heavy-light hadrons will be discussed in Sect.~\ref{sec:D_mesons}.
As a principle, we have to be careful that the many-body problem is not reduced to the two-body scattering in general cases, and there is a possibility that the $N$-body scattering could play some role.
For the three-body scattering, we need to consider the Faddeev equation.

\subsubsection{Chiral symmetry at finite density}
\label{sec:chiral_symmetry_at_finite_density}

\paragraph{Quark condensates in nuclear matter}

It is well known that chiral symmetry is dynamically broken in the QCD vacuum as discussed in Sect.~\ref{sec:chiral_symmetry}.
It is considered that this symmetry breaking is partially restored at finite baryon number density, such as in nuclear medium \cite{Cohen:1991nk,Drukarev:1991fs}, and the quark condensate deviates from the value in the QCD vacuum.
Let us investigate how the quark condensate changes at finite baryon number density.
This problem should be, of course, solved as many-body problem for general baryon number density.
However, as far as we focus on the lowest order of the finite baryon number density,
we can give a rigorous result which is model-dependence~\cite{Cohen:1991nk,Drukarev:1991fs}.

To start the discussion, we note that the light quark current mass $m_{q}$ enters into the QCD Hamiltonian density as a parameter,
\begin{align}
 {\cal H}_{\mathrm{mass}} = m_{q} \bar{\psi}\psi,
\end{align} 
for the two flavor ($\psi=(u,d)^{\mathrm{t}}$) and $m_{q}=m_{u}=m_{d}$ in the isospin limit.
Let us denote the state of the nuclear matter $| {\Omega}(m_{q}) \rangle$ for a notation representing the dependence on $m_{q}$.
The difference between the quark condensate in the nuclear matter and that in vacuum is expressed
\begin{align}
 \frac{\mathrm{d}}{\mathrm{d}m_{q}} \epsilon
&=
  \langle {\Omega}(m_{q}) | \bar{q}q | {\Omega}(m_{q}) \rangle
 -\langle {0}(m_{q}) | \bar{q}q | {0}(m_{q}) \rangle,
 \label{eq:HF_equation}
\end{align}
with
\begin{align}
 \epsilon \equiv \langle {\Omega}(m_{q}) | {\cal H}_{\mathrm{QCD}} | {\Omega}(m_{q}) \rangle
 -\langle {0}(m_{q}) | {\cal H}_{\mathrm{QCD}} | {0}(m_{q}) \rangle,
\end{align}
where ${\cal H}_{\mathrm{QCD}}$ is the full Hamiltonian of QCD.
Note that the vacuum state $| {0}(m_{q}) \rangle$ is dependent on the quark mass $m_{q}$, because $m_{q}$ is contained in the QCD Lagrangian as the parameter.
Here we used the Hellmann-Feynman theorem.\footnote{We consider the eigenstate $|\phi(\lambda)\rangle$ as an eigenstate of the Hamiltonian $H(\lambda)$ which is dependent on the parameter $\lambda$: $H(\lambda)|\phi(\lambda)\rangle=E(\lambda)|\phi(\lambda)\rangle$ ($E(\lambda)$ the eigenenergy) with the normalization condition $\langle\phi(\lambda)|\phi(\lambda)\rangle=1$. Then, 
\begin{align}
 \langle\phi(\lambda)|\frac{\mathrm{d}}{\mathrm{d}\lambda} H(\lambda)|\phi(\lambda)\rangle 
 = \frac{\mathrm{d}}{\mathrm{d}\lambda} \langle\phi(\lambda)|H(\lambda)|\phi(\lambda)\rangle. 
\end{align}}
In Eq.~(\ref{eq:HF_equation}),
$\epsilon$
  is the energy density of the nuclear matter, which is measured from the vacuum energy.
Let us represent it by
\begin{align}
\epsilon=m_{N}\rho+\delta \epsilon, 
\end{align}
with $\rho$ being the baryon number density and $\delta \epsilon$ being the binding energy of the nuclear matter.
At the low density, we may approximate $\epsilon$ as $\epsilon \simeq m_{N}\rho$ by neglecting $\delta \epsilon$.
Furthermore we introduce the quantity defined by
\begin{align}
 \sigma_{N}
&=
 \frac{1}{N_{f}} 
  \sum_{a} 
   \left(
    \langle N(m_{q}) | \left[Q_{A}^{a} , \left[ Q_{A}^{a}, H_{\mathrm{QCD}} \right] \right] | N(m_{q}) \rangle
  - \langle {0}(m_{q}) | \left[Q_{A}^{a} , \left[ Q_{A}^{a}, H_{\mathrm{QCD}} \right] \right] | {0}(m_{q}) \rangle
  \right),
\end{align}
with $N_{f}=2$, where $Q_{A}^{a}$ is the axial charge, $H_{\mathrm{QCD}}=\int \mathrm{d}^{3}\vec{x} \,\, {\cal H}_{\mathrm{QCD}}$ is the QCD Hamiltonian, and $|N(m_{q}) \rangle$ is the one nucleon state.
This can be rewritten as
\begin{align}
 \sigma_{N} 
&= m_{q} \int \mathrm{d}^{3}\vec{x} \left( \langle N(m_{q}) |\bar{q}q| N(m_{q}) \rangle - \langle {0}(m_{q}) |\bar{q}q| {0}(m_{q}) \rangle \right) \nonumber \\
&= m_{q} \frac{\mathrm{d}m_{N}}{\mathrm{d}m_{q}},
\end{align}
with $m_{N} = \langle N(m_{q}) | H_{\mathrm{QCD}} | N(m_{q}) \rangle$, where the Hellmann-Feynman theorem is used again.
Then, we obtain
\begin{align}
 \langle {\Omega}(m_{q}) | \bar{q}q | {\Omega}(m_{q}) \rangle
 -\langle {0}(m_{q}) | \bar{q}q | {0}(m_{q}) \rangle
\simeq
 \frac{\sigma_{N} \rho}{m_{q}}.
\end{align}
By using the GMOR relation~(\ref{eq:GMOR})\footnote{We note the difference of the notations: $\hat{m}=m_{q}$ and $|0\rangle=|{0}(m_{q})\rangle$.}, we obtain
\begin{align}
 \frac{\langle {\Omega}(m_{q}) | \bar{q}q | {\Omega}(m_{q}) \rangle}
 {\langle {0}(m_{q}) | \bar{q}q | {0}(m_{q}) \rangle}
\simeq
 1- \frac{\sigma_{N}}{f_{\pi}^{2}m_{\pi}^{2}}\rho.
\end{align}
This relation gives the change of the quark condensate in nuclear matter at baryon number density $\rho$.
The value of $\sigma_{N} = 40$-$45$ MeV is obtained from the analysis in the chiral perturbation theory~\cite{Buettiker:1999ap,Gasser:1990ce} (see also Ref.~\cite{Scherer:2012xha}).
When we use this value, we obtain
\begin{align}
 \frac{\langle {\Omega}(m_{q}) | \bar{q}q | {\Omega}(m_{q}) \rangle}
 {\langle {0}(m_{q}) | \bar{q}q | {0}(m_{q}) \rangle}
\simeq
 1- 0.3 \frac{\rho}{\rho_{0}},
 \label{eq:qqbarrho}
\end{align}
with the normal nuclear matter density $\rho_{0}=0.17$ fm$^{-3}$.
Hence, the quark condensate decreases by about 30 \% in normal nuclear matter.
Here we note that the above derivation is model-independent, and is valid in the limit of the low baryon number density, where only the leading order of $\rho$ should be dominant.
We have to be careful that no nucleon interaction is taken into account because its effect is completely neglected ($\delta \epsilon=0$).
Such effect can be included by higher order effect of $\rho$, which can be studied in nuclear many-body problems (see Ref.~\cite{Birse:1994cz,Cassing:1999es,Meissner:2001gz}).

\paragraph{Gluon condensates in nuclear matter}

Similarly, we can estimate the change of the gluon condensate in nuclear matter~\cite{Cohen:1991nk,Drukarev:1991fs}.
Here, without derivations, we quote on the result~\cite{Cohen:1991nk}
\begin{align}
\langle {\Omega}(m_{q}) | \frac{\alpha_{s}}{\pi} G^{a}_{\mu\nu} G^{a\mu\nu} | {\Omega}(m_{q}) \rangle
 - \langle {0}(m_{q}) | \frac{\alpha_{s}}{\pi} G^{a}_{\mu\nu} G^{a\mu\nu} | {0}(m_{q}) \rangle
\simeq
 -\frac{8}{9} \left( m_{N} - \sigma_{N} - y \right) \rho,
 \label{eq:gluonrho}
\end{align}
in the linear density approximation with $y$ the strangeness content in a nucleon
\begin{align}
 y
= m_{s} \int \mathrm{d}^{3}\vec{x} \left( \langle N(m_{q}) |\bar{s}s | N(m_{q}) \rangle - \langle {0}(m_{q}) |\bar{s}s | {0}(m_{q}) \rangle \right),
\end{align}
with the strange quark current mass $m_{s}$.

\subsubsection{QCD sum rules}
\label{sec:QCDSR}

In the QCD sum rule hadron properties such as masses and coupling constants are  
studied by two point correlation 
functions~\cite{Shifman:1978bx,Shifman:1978by} (see Refs.~\cite{Reinders:1984sr,Colangelo:2000dp,Ioffe:2005ym} for a review).  
In momentum space at $p = (\omega, \bm p)$ it is defined as  
\begin{align}
\Pi (p^2) = \int \mathrm{d}^4x\ e^{i \omega t - i \bm p \cdot \bm x} \langle \Omega| T(A(x) \bar B(0)) |\Omega\rangle,
\label{eq_def_correlationAB}
\end{align}
where $A$ and $B$ are appropriate hadronic current operators which create 
physical states  that we are interested  in.  
Here we suppress $m_q$ in the nuclear matter state $|\Omega\rangle$.  
The operators can be taken, for instance, 
$A, B = \bar q \gamma_\mu \vec \tau q$ for the isovector $\rho$ meson, where $\vec \tau$ is an isospin matrix, and 
$\bar s \gamma_\mu s$ for the $\phi$ meson.  

The correlation function is computed in two different ways.  
One is a phenomenological manner
by using a spectral function $\sigma$, the imaginary part of $\Pi$,   
which contains hadronic parameters such as masses and coupling constants,  
\begin{align}
\Pi (p^2) 
=
\int^\infty_{0} \mathrm{d}s\ \frac{\sigma(s)}{s-p^2 - i \epsilon}\, .
\label{Pi_phenomenological}
\end{align}
For example, for a simple case of one bound or resonant state at the mass $m_{res}$ with the residue $\lambda$ and a continuum 
starting from a threshold $s_{th}$, the spectral function can be parametrized as 
\begin{align}
\sigma(s) = \lambda \delta(s - m_{ res}) + f(s) \theta (s - s_{th}) .
\end{align}
The simplest choice of $f(s)$ is the phase space of physical multi-hadron states.

Another method, which is essential in the QCD sum rule, 
is the operator product expansion (OPE) of QCD.
The correlation function (\ref{eq_def_correlationAB}) 
is computed in the asymptotic (deep Euclidean) region by QCD, 
where the perturbation method is available. 
The Wilson's operator product expansion is performed 
at short distances~\cite{Wilson:1969zs}, 
\begin{align}
\lim_{x \to y} A(x) \bar B(y)  \to \sum_i C_i (x-y) {\cal O}_i ((x+y)/2)  , 
\label{eq_WilsonsOPE}
\end{align}
where ${\cal O}_i$ are local operators and $C_i (x-y) $
the Wilson's coefficients  which can be calculated perturbatively.  
The indices summed over $i$ are  ordered by the dimensions of the  operators ${\cal O}_i$.  
Examples of such operators for hadrons in the vacuum written in terms of the quark and gluon fields 
are given in Ref.~\cite{Reinders:1984sr}.
In this decomposition, the singularities at short distances, $x \to y$ are isolated by the 
Wilson coefficients $C_i (x-y)$.  
After taking the matrix elements by $|\Omega\rangle$, 
the operators are replaced by numerical values which characterize 
the non-perturbative dynamics of hadrons.
If $|\Omega\rangle$ is the vacuum, only Lorentz invariant matrix elements survive, 
while for a finite density matter,  non-Lorentz invariant ones can also survive.   

It is known that the asymptotic behaviors of 
$C_i$'s are determined up to logarithmic factor by the canonical 
dimensions of the operators 
$d_A, d_B$ and $d_i$, 
\begin{align}
C_i(x-y) 
\xrightarrow[|x-y| \ll 1/\Lambda_{\mathrm{QCD}}]{}
|x-y|^{d_i - d_A - d_B} 
(1 + {\cal O}(|x-y|\Lambda_{\mathrm{QCD}})) .
\end{align}
For instance, for $A = \bar \psi \gamma_\mu \psi$, $d_A=3$.  
Therefore, the terms of higher dimensional operators (with larger $d_i$) 
are expected to be suppressed in the deep Euclidean region, $x \to y$ or $|p^2| \to \infty$,  
by powers of $\Lambda_{\rm QCD}^2/|p^2|$.  

Let us consider the application of the QCD sum rules in the nuclear matter.
General features of the nuclear matter was studied in Refs.~\cite{Drukarev:1991fs,Drukarev:1988kd}.
In what follows, we overview the original work of Hatsuda and Lee~\cite{Hatsuda:1991ez} for the study of 
vector mesons in a nuclear matter at finite density.  
For definiteness and simplicity the longitudinal component of the 
correlation function is considered for the charged $\rho$ meson, $\rho^+$, 
\begin{align}
\lim_{\bm q \to 0} \Pi_L (\omega,  \bm p) 
=
\int \mathrm{d}^4x\ e^{i \omega t - i \bm p \cdot \bm x} 
\langle \Omega| T(\bar d \gamma_\mu u (x) \bar  u \gamma^\mu d(0)) |\Omega\rangle . 
\end{align}
In this manner we do not have to consider possible Lorentz index dependence of the correlation function.  
Because we are interested in the ground state of the nuclear matter with equal numbers of protons and neutrons,  
the baryon number density takes a finite value
\begin{align}
\langle \Omega| \frac{1}{N_c} (\bar u \gamma_\mu u + \bar d \gamma_\mu d)
|\Omega\rangle
=
g_{\mu 0} \rho ,
\label{eq_VVatdensity}
\end{align}
where $\rho$ is the  baryon number density whose value for the normal nuclear matter 
is  $\sim 0.17\; {\rm fm}^{-3}$.   
The expression of (\ref{eq_VVatdensity})  violates the Lorentz invariance.  
Having the nuclear matter as a kind of background, 
it is convenient to study the spectral function as a function of $\omega$ with zero momentum
$\bm p = 0$.  

Now the remaining and the most relevant task is to find the operators 
${\calO_i}$.  
For the case of vector mesons, the terms of non-vanishing Wilson coefficients include up to dimension six, 
\begin{align}
{\rm Four \; quark \; operators}&:
\bar q \Gamma_\mu \lambda^a q
\bar q \Gamma_\nu \lambda^a q, 
\nonumber \\
{\rm  Quark \; bilinears} &:
\bar q \gamma_\mu D_\nu q, 
\; \; \; 
\bar q \gamma_\mu D_\nu D_\alpha D_\beta q .
\end{align}
The link to the physics of finite density is to take the matrix elements of these operators in the nuclear matter, and relate them to the quark condensate 
$\langle \Omega | \bar q q | \Omega\rangle$ 
and the gluon condensate $\langle \Omega | G_{\mu \nu} G^{\mu \nu} | \Omega \rangle$
because they carry the most important density dependence. 
Manipulations can be done by using equations of motion or Fierz rearrangements.  
Moreover, for higher dimensional operators, vacuum saturation, or equivalently
mean field approximation is used.
For instance, the four quark condensate can be written as 
\begin{align}
\langle \Omega | (\bar q q)^2 | \Omega\rangle
\sim 
\langle \Omega | \bar q q | \Omega\rangle ^2 .
\end{align}
Now by using the $\rho$ dependence of various condensates as shown in 
Eqs.~(\ref{eq:qqbarrho}) and (\ref{eq:gluonrho}), we can extract 
the $\rho$ dependence of physical quantities.  
In addition, 
in the nuclear matter the spectral density (\ref{Pi_phenomenological}) 
receives contributions from the Landau damping due to nucleon collisions.
This effect is relevant only in low energy region ($\sim \delta(s)$) which is expected 
to play only a minor role for the hadron properties.  

Now the QCD sum rule is obtained by matching 
the phenomenological spectral density and the one of the OPE with 
appropriate density dependence in various condensates.
The phenomenological parameters are determined such that the two values 
are optimally equated.  
In principle, these quantities are asymptotically expanded in the deep Euclidean region 
in powers of $1/(-q^2)$.  
Practically, this can be done by the finite energy sum rule, or Borel improved sum rule.   

\begin{figure}[t]
\begin{center}
\includegraphics[width=0.3 \linewidth,bb=0 0 589 584]{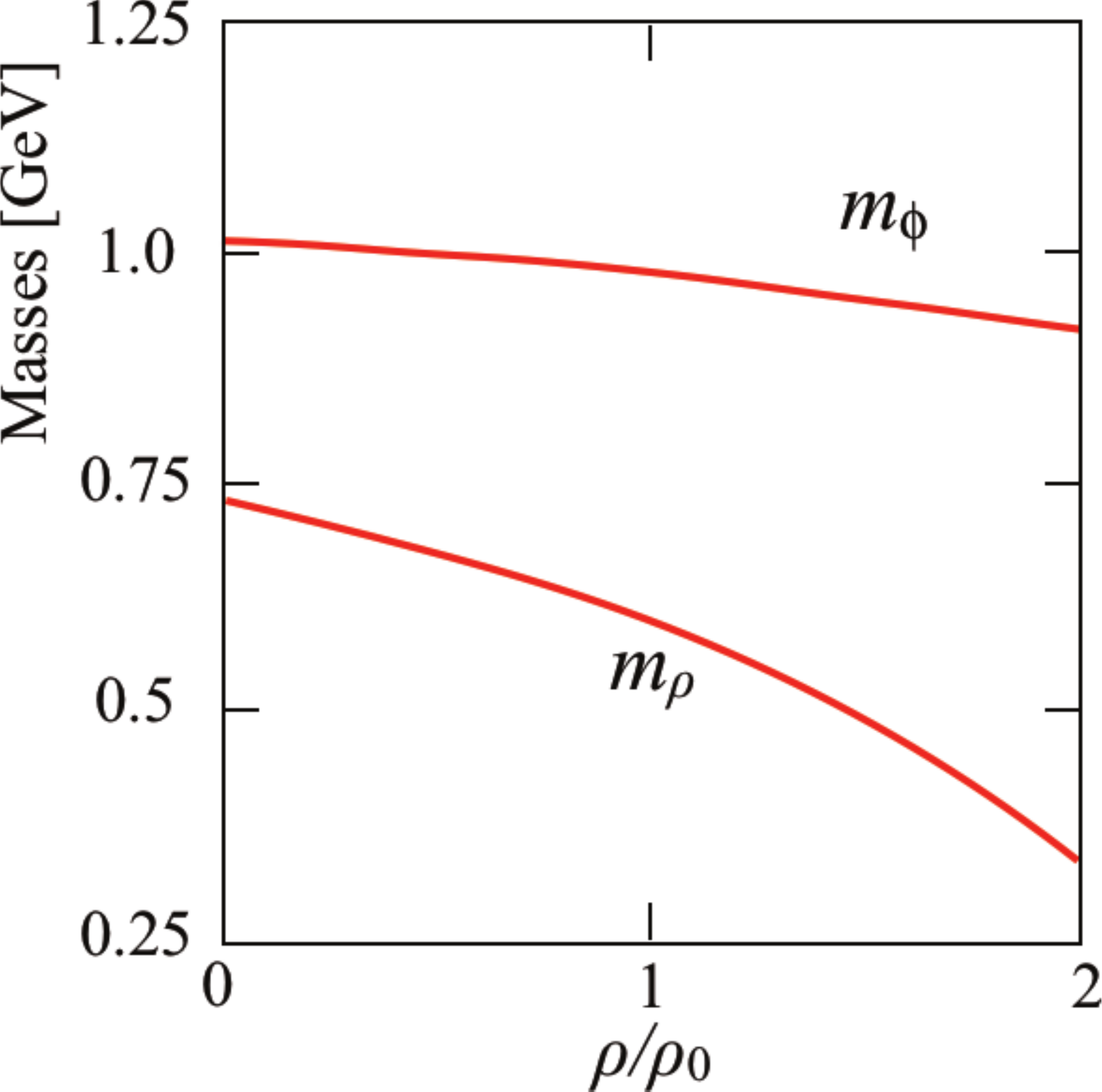}
\end{center}
\vspace{-5mm}
\caption{Masses of the $\rho$ and $\phi$ mesons at finite baryon densities $\rho$. The mass of $\phi$ is computed by using the strangeness content $y = 0.22$. This figure is made by using the results of Ref.~\cite{Hatsuda:1991ez}.   
}
\label{fig_sumrule}
\end{figure}

In Ref.~\cite{Hatsuda:1991ez}, the finite energy sum rule is used and the following results are obtained
\begin{align}
\lambda^2 
- s_{th}
\left( 1 - \frac{\alpha_S}{\pi} \right)
&=- \frac{2 \pi}{M_N} \rho ,
 \\
\lambda^2 m^2
- \frac{s_{th}^2}{2}
\left( 1 - \frac{\alpha_S}{\pi} \right)
&=
- Q_4 - 2 \pi^2A_2^{u+d}M_N \rho ,
\\
\lambda^2 m^4 
- \frac{s_{th}^3}{3}
\left( 1 - \frac{\alpha_S}{\pi} \right)
&=
- Q_6 - \frac{10}{3} \pi^2A_4^{u+d}M_N \rho ,
\end{align}
where 
\begin{align}
Q_4 &= \frac{\pi^2}{3} 
\langle \Omega | \frac{\alpha_S}{\pi} G^2 |\Omega \rangle ,
 \\
Q_6 &= \frac{896}{81} \pi^3 
\langle \Omega | \sqrt{\alpha_S} \bar qq |\Omega \rangle ^2 ,
\end{align}
and $A_n^{q}$ are given in terms of the quark distribution function
\begin{align}
A_n^{q} = 2 \int^1_0 \mathrm{d}x\ x^{n-1} 
(q(x, \mu^2) + (-1)^n\bar q(x, \mu^2) ) ,
\end{align}
at a scale $\mu^2$.  
The terms proportion to $\rho$ as well as the condensates at finite density 
determines the density dependence of the physical quantities.  
These equations imply that the mass of the vector meson $m$ decreases
at finite density.  
Detailed contents of quantitative discussions are found in Ref.~\cite{Hatsuda:1991ez}.  
Here we show their results in Fig.~\ref{fig_sumrule}, 
where we observe that the mass of $\rho$ decreases by about 120 MeV, 
and that of $\phi$ by 30 MeV at the normal nuclear matter density.  
Although the latter seems rather small, its experimental impact is important 
because the amount of the mass shift is well larger than the decay width  
as we discuss shortly
($\Gamma_\rho \sim 150$ MeV, and $\Gamma_\phi \sim 4$ MeV~\cite{Agashe:2014kda}).  

The change in the mass of hadrons under the change in the environment, 
such as density and temperature, is natural if they are composite systems.
In the present case, it is also related to the properties of the QCD vacuum.  
Theoretically, the predication was made in QCD by using the QCD sum rule.  
The results given in this section is exact up to the first order of $\rho$
which is the case of low density regions.  
In this regard, the QCD sum rule is useful and important to investigate the 
primary features of hadron properties.  
The present method is applicable also to the finite temperature, once we know 
the temperature dependence of various condensates which has been calculated in lattice simulations.  

Having this general argument given above, experimental effort has been made to observe 
the mass shift of hadrons in a nucleus of finite baryon density~\cite{Hayano:2008vn,Leupold:2009kz}.  
Experimentally, because of the rather wide decay width of the $\rho$ meson as we 
pointed out above,  it was attempted to observe a $\phi$ meson in proton-nucleus 
collisions.  
The result showed the mass spectrum with the strength spreading (broadening)
in the lower mass region (cf.~Refs.~\cite{Hayano:2008vn,Leupold:2009kz}.
However, the precise mechanism of this change has not been clarified yet.  
The difficulty lies in realizing a situation as close as 
to stationary condition, where the hadron stays inside a nucleus for sufficiently long time.  
This is a general problem that we have to solve whenever 
we would like to study hadron properties in a varied environment.  
Once this will be done, however, we expect to see the dynamical change in the vacuum 
which is the indication of the non-trivial structure of the QCD vacuum.

\section{Quarkonia}
\label{sec:charmonia}

\subsection{Quarkonium-nucleon interaction: gluon-exchange dominance}

Since 1980s, the interaction between a $J/\psi$ meson and a nucleon has been discussed as an unique inter-hadron interaction which is dominated by the gluon exchange~\cite{Brodsky:1989jd}.
Because $J/\psi$ is composed of $c\bar{c}$, the light quark-antiquark $q\bar{q}$ exchange between $J/\psi$ and nucleon is suppressed by the Okubo-Zweig-Iizuka (OZI) rule.
This property is quite different from other hadron-hadron interactions.

A charmonium can be considered as a small size system.
The interaction between $c$ and $\bar{c}$ is governed by the color Coulomb potential at short distance. 
While the color Coulomb potential generates a pointlike bound state for infinitely heavy quarks, in the actual charm quarks with a finite mass, the balance between the Coulomb potential and the kinetic energy determines the small but finite size of the $c\bar{c}$ system. This provides the ``color dipole" structure of the charmonia.
This dipole (multipole) structure makes the QCD van der Waals potential between $J/\psi$ and a nucleon as the perturbative interaction at short distance. 
This is analogous to the van der Waals potential in atomic physics where the interaction between electrically neutral atoms are induced by the instantaneous polarization.
At long distance, on the other hand, the $J/\psi N$ potential is dominated by the nonperturbative dynamics with multi-gluon exchanges, such as the pomeron exchange interaction.

In the early work by Brodsky, Schmidt and de T\'eramond~\cite{Brodsky:1989jd}, the $J/\psi N$ interaction has been investigated from the experimental data of the $\gamma N$ reaction.
By supposing the pomeron exchange, it is concluded that the $J/\psi N$ interaction is attractive.
Based on this interaction and a simple variational calculation, $J/\psi$ is shown to be bound in the nucleus with the atomic number $A\geq 3$.
However, the obtained binding energy is quite large (more than 100 MeV for $A=9$), because the saturation property of nuclear matter is not considered, though the overlapping of nucleons should provide a strong repulsion at short distances.
In a more realistic discussion with the saturation effect~\cite{Wasson:1991fb}, the binding energy is shown to be about 30 MeV for $A=200$.

To study the $J/\psi N$ interaction at high energies, the perturbative treatment of the QCD coupling constant $\alpha_s$ is performed by considering the color multipole expansion~\cite{Peskin:1979va,Bhanot:1979vb}.
Regarding a pair of $Q$ and $\bar{Q}$ as a bound state ($\phi$) by the color Coulomb potential, the QCD dipole and quadrupole moments are considered.
The color octet contributions are also included.
In the forward scattering of $J/\psi p$, the color dipole in $J/\psi$ interacts with gluons contained as partons in the proton.
The scattering between $J/\psi$ and gluon is given by the diagrams of absorbing and emitting the gluons. The scattering matrix is given by
\begin{align}
{\cal M}_{J/\psi g} = -\frac{g^{2}}{2N_{c}} \langle \phi | \, \vec{r} \!\cdot\! \vec{E} \frac{1}{H_{a}+\epsilon+iD_{0}} \vec{r} \!\cdot\! \vec{E} \, | \phi \rangle,
 \label{eq:multi_pole_Peskin}
\end{align}
where $g$ is the $QQg$ coupling constant, $N_{c}$ is the number of colors, $\phi$ represents the $\bar{Q}Q$ state, $\vec{r}$ is the distance between $\bar{Q}$ and $Q$, $\vec{E}$ is the electric gluon field, $H_{a}$ the color octet Hamiltonian, $\epsilon$ is the Coulomb binding energy of the $\bar{Q}Q$ state, and $D_{0}$ is the zeroth component of the covariant derivative.
Based on the dipole picture, the cross section of the elastic scattering is estimated as $\sigma^{J/\psi p}=3.6$-$5.0$ mb for $J/\psi p$ and $\sigma^{\Upsilon p}=0.9$-$1.2$ mb for $\Upsilon p$ as asymptotic values at the high scattering energy.
More information about the QCD van der Waals potential can be found in a review article~\cite{Kharzeev:1995ij}.
The above scheme is applicable to a variety of reaction processes of quarkonia with hadrons.

We have to notice that the $J/\psi N$ interaction relevant for the binding in nuclear matter occurs at the low energy, where the pomeron exchange may not be an appropriate picture.
In the same manner with Refs.~\cite{Peskin:1979va,Bhanot:1979vb}, the scattering length of the $\eta_{c} N$ interaction has been discussed in Ref.~\cite{Kaidalov:1992hd}.
From the multi-pole expansion~\cite{Peskin:1979va,Bhanot:1979vb} 
[cf.~Eq.~(\ref{eq:multi_pole_Peskin})], the scattering amplitude for a quarkonium $\phi$ and a nucleon $N$ is given by
\begin{align}
 T = \frac{4\pi}{3} M_{\phi} \langle \phi | r^{i} \frac{1}{H_{a}+\epsilon} r^{j} | \phi \rangle \langle N | \alpha_{s} E^{i} E^{j} | N \rangle.
 \label{eq:T_matrix_Peskin}
\end{align}
To estimate the first matrix element, we use the result of Ref.~\cite{Bhanot:1979vb} for the $1S$ wave function of the $c\bar{c}$ state
\begin{align}
 \langle \phi | r^{i} \frac{1}{H_{a}+\epsilon} r^{j} | \phi \rangle
=
\delta^{ij}
\frac{28}{27} \pi a^{3},
\end{align}
with $a$ being the size parameter of the $1S$ $c\bar{c}$ state.
The second matrix element can be evaluated by the energy-momentum tensor $\theta_{\mu\mu} = \frac{\beta(\alpha_{s})}{4\alpha_{s}} G^{a}_{\mu\nu} G^{a}_{\mu\nu}$ as
\begin{align}
 \alpha_{s} E^{i} E^{j} = \frac{2\pi}{b} \theta_{\mu\mu} + {\cal O}(\alpha_{s}),
\end{align}
with $b=9$ from the leading order term in the Gell-Mann--Low beta function $\beta({\alpha_{s}})$. The matrix element of the energy-momentum tensor is given by $\langle N | \theta_{\mu\nu} | N \rangle_{p_{1}=p_{2}=p}=2p_{\mu}p_{\nu}$ for the initial and final momenta $p_{1}$ and $p_{2}$.
Hence, the scattering amplitude for $\eta_{c}N$ at the threshold $p_{0}=m_{N}$ is given as
\begin{align}
 T = \frac{64\pi^{3}}{3^{6}} 7 M_{\eta_{c}} m_{N}^{2} a^{3}.
\end{align}
As a result, we obtain the scattering length\footnote{
Throughout this paper, we define the scattering length as $a_{s}=\lim_{k\to 0}f(k)$ where $f(k)$ is the elastic scattering amplitude of the two-body system with the momentum $k$. In the absence of a shallow bound state, the positive (negative) $a_{s}$ corresponds to the attractive (repulsive) interaction at threshold. Our convention is commonly used in hadron physics (meson-meson/meson-baryon scattering), while the convention with an opposite sign is often used in the nuclear and atomic physics. See also the discussion in section~\ref{sec:DNinteraction}.} 
\begin{align}
 a_{s} = \frac{T}{8\pi(M_{\eta_{c}}+m_{N})} = 0.05 \hspace{0.5em} \mathrm{fm}.
\end{align}
This attractive scattering length is used to estimate the binding energy of $\eta_{c}$ in nuclear matter~\cite{Kaidalov:1992hd}.
The result was about 3 MeV.

From the phenomenological viewpoint, the bound state approach of the Skyrmion model has been used in Refs.~\cite{Gobbi:1992cf,Gobbi:1993wu} where $\eta_{c}$ is bound in the Skyrmion as a nucleon.
This model is applicable also to the $\eta$ bound state in the Skyrmion.
The Skyrmion is a topologically stable object composed of pions~\cite{Adkins:1983ya}.
At first sight, it may seem that strange and charm quarks are irrelevant to the Skyrmion, because the Skyrmion obeys the SU(2) flavor symmetry for $u$ and $d$ quarks.
Recalling that pions are the adjoint representation of SU(2) flavor symmetry, however, we can extend the pion field to the adjoint representation of the SU(4) flavor symmetry for $u$, $d$, $s$ and $c$ quarks, and hence
we can treat $\eta$ as well as $\eta_{c}$ as the same multiplet as pions.
Because the mass of $\eta$ and $\eta_{c}$ is much heavier than that of pions, the ``bound-state" approach is adopted.
The obtained binding energy of the $\eta_{c}N$ state is about 1 GeV, and it is much larger than the values expected from other theoretical studies, such as the ``color dipole" picture~\cite{Kaidalov:1992hd}, the lattice QCD and the QCD sum rules discussed below.

First study of the charmonium-nucleon interaction on the lattice QCD simulation has been performed with the quenched approximation in Ref.~\cite{Yokokawa:2006td}.
Subsequently, several groups have investigated the $J/\psi N$ interaction~\cite{Liu:2008rza,Kawanai:2010ev,Kawanai:2010ru,Beane:2014sda}.
The results in Refs.~\cite{Liu:2008rza,Kawanai:2010ev,Kawanai:2010ru} indicate that the $J/\psi N$ interaction and the $\eta_{c}N$ interaction are attractive with a moderate strength.
In Ref.~\cite{Beane:2014sda}, the $J/\psi(\eta_{c}) N$ system is shown to be bound in the unphysical quark mass region. The results of the scattering lengths by the lattice QCD simulation are summarized in Table~\ref{table:scattering_length_charmonium_nucleon}.
It is interesting that those numbers are comparable with the result $a_{J/\psi N}=0.10 \pm 0.02$ fm obtained by the QCD sum rules for the forward scattering amplitude~\cite{Hayashigaki:1998ey}. In the low density approximation, this scattering length corresponds to the negative mass shift by 4-7 MeV in nuclear matter, as stated below.

\begin{table}[tbp]
\caption{Scattering length $a$ [fm] from the lattice QCD simulations. SAV indicates the spin averaged value, $(a_{1/2}+2a_{3/2})/3$. PSF and LLE indicate the different methods, phase shift formula and leading large-$L$ expansion, respectively, used in Ref.~\cite{Yokokawa:2006td}. The number with ``*" is calculated in this review. The scattering lengths of $J/\psi N$ with spin 1/2 and 3/2 are shown in the figure in Ref.~\cite{Kawanai:2010ru}, though the numbers were not given explicitly.}
\begin{center}
\begin{tabular}{|cc|cccc|}
\hline
 & spin & \cite{Yokokawa:2006td} PSF & \cite{Yokokawa:2006td} LLE  & \cite{Liu:2008rza} & \cite{Kawanai:2010ev,Kawanai:2010ru} \\
\hline 
$\eta_c N$    & 0  & $0.70 \pm 0.66$ & $0.39 \pm 0.14$ & 0.18 (9) & 0.25 \\
\hline
$J/\psi N$ &1/2 & $0.57 \pm 0.42$ & $0.35 \pm 0.15$ & $-0.05$(77) & --- \\
                    & 3/2 & $0.88 \pm 0.63$ & $0.43 \pm 0.16$ & 0.24(35) & --- \\
                    & SAV & $0.71 \pm 0.48$ & $0.39 \pm 0.14$  & 0.16 (70)* & 0.35 \\
\hline
\end{tabular}
\end{center}
\label{table:scattering_length_charmonium_nucleon}
\end{table}%

\subsection{Few-body systems}

Due to the $J/\psi N$ attraction, it is argued that the $J/\psi$-bound state can exist in the atomic nuclei~\cite{Brodsky:1989jd,Wasson:1991fb}.
In Ref.~\cite{Belyaev:2006vn}, the Faddeev calculations of $\eta_{c} d$ ($d$ is a deuteron) and $\eta_{c} ^{3}\mathrm{He}$ are performed for the Yukawa-type potential for $\eta_{c}N$ and the Paris potential for $NN$, where the $\eta_{c}N$ potential is parametrized according to Ref.~\cite{Brodsky:1989jd}.
Recently, a 
 few-body calculation has been performed for the $J/\psi(\eta_{c})$-nuclei~\cite{Yokota:2013sfa,Yokota:2014lma} by using the Gaussian expansion method explained in section~\ref{sec:fewbody_systems}.
Though the QCD van der Waals potential exists at short distance, the confinement force would provide a long range potential. In Refs.~\cite{Yokota:2013sfa,Yokota:2014lma} the authors use a Gaussian-type potential for the $J/\psi(\eta_{c})$-$N$ system, whose model parameters (potential depth and range) are related to the two-body scattering length.
By varying the value of the $J/\psi(\eta_{c})$-$N$ scattering length, various few-body systems such as $J/\psi(\eta_{c})$-$NN(d)$, $J/\psi(\eta_{c})$-$^{4}\mathrm{He}$ and $J/\psi$-$^{8}\mathrm{Be}$, where the $\alpha\alpha$ cluster structure for $^{8}\mathrm{Be}$ is considered, are examined.
The Minnesota potential is used for $NN$ as a phenomenological nuclear potential.
The $J/\psi(\eta_{c})$-$N$ scattering lengths necessary for the formation of the bound states are found to be $a_{J/\psi N}>0.95$ fm for $J/\psi(\eta_{c})$-$NN$ and $a_{J/\psi N}>0.24$ fm for $J/\psi$-$^{4}\mathrm{He}$~\cite{Yokota:2013sfa,Yokota:2014lma},\footnote{The present convention of the sign of the scattering length is different from the original ones in Refs.~\cite{Yokota:2013sfa,Yokota:2014lma}.} which can be compared with the results in the lattice QCD simulations in Table~\ref{table:scattering_length_charmonium_nucleon}.
To form a bound state of $J/\psi$-$^{8}\mathrm{Be}$, $a_{J/\psi N}>0.16$ fm is required.
The results in Refs.~\cite{Yokota:2013sfa,Yokota:2014lma} are consistent with that in Ref.~\cite{Belyaev:2006vn}.
By adopting the scattering lengths in Refs.~\cite{Kawanai:2010ev,Kawanai:2010ru}, bound states are found for $A\geq 4$ with the biding energies 0.5 MeV for $J/\psi(\eta_{c})$-$^{4}\mathrm{He}$ and 2 MeV for $J/\psi(\eta_{c})$-$^{8}\mathrm{Be}$.

\subsection{Nuclear matter}

The binding energy of the quarkonium in nuclear matter has been estimated by considering the scale anomaly (trace anomaly)~\cite{Luke:1992tm}.
For the Coulomb bound state of $\bar{Q}Q$, the interaction Lagrangian with the dimension seven operators is given by
\begin{align}
{\cal L}_{\mathrm{int}} = \sum_{v} \frac{1}{\Lambda_{Q}^{3}} (P_{v}^{\dag} P_{v} - V_{v\mu}^{\dag} V_{v}^{\mu}) (c_{E} {\cal O}_{E} + c_{B} {\cal O}_{B}),
\end{align}
where
${\cal O}_{E} \equiv - G^{\mu\alpha a} G_{\alpha}^{\nu a} v_{\mu} v_{\nu}$ and 
${\cal O}_{B} \equiv \frac{1}{2} G^{\alpha \beta a} G_{\alpha \beta}^{a} - G^{\mu\alpha a} G_{\alpha}^{\nu a} v_{\mu} v_{\nu}$
with the compositeness scale $\Lambda_{Q} \simeq \alpha_{s}(\Lambda_{Q})m_{Q}$ for a Coulomb bound state,
$P_{v}$ creates the pseudoscalar meson ($\eta_{c}$) and $V_{v\mu}$ creates the vector meson ($J/\psi$) with four velocity $v^{\mu}$.
This is the covariant form of the leading order of the multipole expansion.
In the rest frame $v^{\mu}=(1,\vec{0}\,)$, these operators reduce to ${\cal O}_{E}\to\vec{E}^{a} \!\cdot\! \vec{E}^{a}$ and ${\cal O}_{B}\to\vec{B}^{a} \!\cdot\! \vec{B}^{a}$, from which we see that $c_{E}$ and $c_{B}$ are also related to the Stark and Zeeman energies, respectively.
Because ${\cal O}_{E}$ and ${\cal O}_{B}$ are related to the twist-two gluon operator $G_{2\,g}^{\mu\nu}= \frac{1}{4} g^{\mu\nu} G^{\alpha \beta a}G_{\alpha \beta}^{a} - G^{\mu\alpha a}G_{\alpha}^{\nu a}$ and the energy momentum tensor $T_{\alpha}^{\alpha}=\frac{\beta(g)}{2g}G^{\alpha\beta a}G_{\alpha \beta}^{a}$, the forward scattering amplitude can be expressed as
\begin{align}
 {\cal M} = 2V_{2}(\Lambda_{Q}) \frac{c_{E}+c_{B}}{\Lambda_{Q}^{3}} M^{2} \left( \gamma^{2}-\frac{1}{4} \right)
 + 2M^{2} \frac{c_{E}-c_{B}}{\Lambda_{Q}^{3}} \frac{2\pi}{b_{Q} \alpha_{s}(\Lambda_{Q})},
\end{align}
with $\gamma=v^{0}$, where $\beta(g)=-b_{Q}\frac{g^{3}}{16\pi^{2}}$ is the Gell-Mann--Low beta function at the leading order.
The matrix elements of $G_{2\,g}^{\mu\nu}$ and $T_{\alpha}^{\alpha}$ are parametrized as
$\langle p | G_{2\,g}^{\mu\nu} | p \rangle = 2V_{2}(\mu) \left( p^{\mu}p^{\nu} - \frac{1}{4} g^{\mu\nu} p^{2} \right)$,
and $\langle p | T_{\alpha}^{\alpha} | p \rangle = 2M^{2}$ with the nucleon mass $M$.
The value of $V_{2}(\mu)$ indicates the gluon momentum fraction, which is estimated from the deep inelastic scattering at energy scale $\mu$.
Because the forward scattering amplitude is related to the energy shift of the $\bar{Q}Q$ due to the interaction with gluons in nuclear matter, we can obtain the binding energy of $\bar{Q}Q$ possessing Bohr radius $r_{B}$,
\begin{align}
 U_{\mathrm{binding}} = \frac{14}{27} \pi r_{B}^{3} \, \rho \left( V_{2}(\Lambda_{Q}) \left(\gamma^{2}-\frac{1}{4}\right) + \frac{2\pi}{b_{Q}\alpha_{s}(\Lambda_{Q})} \right),
\end{align}
at baryon number density $\rho$~\cite{Luke:1992tm}.
We quote the resulting binding energy in nuclear matter $U_{\mathrm{binding}}=8$-$11$ MeV for $J/\psi$ and $U_{\mathrm{binding}}=2$-$4$ MeV for $\Upsilon$.

In Ref.~\cite{Lee:2003jh}, a more explicit form of the mass shift of the quarkonium with the binding energy $\epsilon=2m_{c}-m_{J/\psi}$ is presented as
\begin{align}
 \Delta m_{J/\psi}
=
-\frac{1}{9}
\int \mathrm{d}k^{2} \left| \frac{\partial \psi(k)}{\partial \bf{k}} \right|^{2} \frac{k}{k^{2}/m_{c}+\epsilon}
\left\langle \frac{\alpha_{s}}{\pi} E^{2} \right\rangle_{\Omega}
\frac{\rho}{2m_{N}},
\label{eq:m_jpsi_Lee:2003jh}
\end{align}
with the quarkonium wave function $\psi(k)$ in momentum space.
This formula can be applied not only to the ground states but also to the excited charmonia.
The mass shift in Eq.~(\ref{eq:m_jpsi_Lee:2003jh}), reproducing the QCD second order Stark effect in Ref.~\cite{Luke:1992tm}, was given by the operator product expansion (OPE).
The formula (\ref{eq:m_jpsi_Lee:2003jh}) is analogous to the result by the multi-pole expansion in Eq.~(\ref{eq:T_matrix_Peskin}), although the derivations are given by different styles of calculation.
In the linear density approximation, the electric gluon condensate is given by
\begin{align}
\left\langle \frac{\alpha_{s}}{\pi} E^{2} \right\rangle_{\Omega}
=
\left(
 \frac{4}{9} m_{N} m_{N}^{0} + \frac{3}{2} m_{N}^{2} \frac{\alpha_{s}}{\pi} A_{2}^{g}
\right)
\frac{\rho}{2m_{N}},
\label{eq:electric_gluon}
\end{align}
with $m_{N}^{0}$ being the nucleon mass in the chiral limit and $A_{2}^{g}$ the second moment of the gluon distribution function in a nucleon (see also the discussion below).
Numerically, we have $\left\langle \frac{\alpha_{s}}{\pi} E^{2} \right\rangle_{\Omega} \simeq 0.5$ GeV$^{2}$,
and we find the mass shifts $\Delta m_{J/\psi}=-8$ MeV, $\Delta m_{\psi(3686)}=-100$ MeV and $\Delta_{\psi(3770)}=-140$ MeV for $J/\psi$, $\psi(3686)$ and $\psi(3770)$, respectively.
The correction by the $D$ meson loop effect is analyzed by a model Lagrangian with $D$ meson and nucleon interaction,
and it is found that the mass shifts are modified as $\Delta m_{J/\psi}=-5$ MeV, $\Delta m_{\psi(3686)}=-130$ MeV and $\Delta_{\psi(3770)}=-125$ MeV.
The binding energy of $J/\psi$ is consistent with the binding energy 3 MeV for $\eta_{c}$ estimated by the two-body scattering length from the multi-pole expansion in Ref.~\cite{Kaidalov:1992hd} as already discussed.

So far, the studies based on the QCD van der Waals potential and the scale anomaly provided the investigation about the binding energy of $J/\psi$ in nuclear matter. The relations between the in-medium property of hadrons and various condensates can be obtained in the QCD sum rules~\cite{Klingl:1998sr,Hayashigaki:1998ey}. As shown in Ref.~\cite{Klingl:1998sr}, the gluon condensate in nuclear matter ($\Omega$) at baryon number density $\rho$ is given in the linear density approximation as
\begin{align}
\left\langle \frac{\alpha_{s}}{\pi} G_{\mu\nu} G^{\mu\nu} \right\rangle_{\Omega}
=
\left\langle \frac{\alpha_{s}}{\pi} G_{\mu\nu} G^{\mu\nu} \right\rangle_{0}
-\frac{8}{9} m_{N}^{0} \rho,
\end{align}
in comparison with that in vacuum (0)
for the nucleon mass $m_{N}^{0} \simeq 750$ MeV in the chiral limit.
Notice that, in vacuum, only the scalar condensate $\left\langle \frac{\alpha_{s}}{\pi} G_{\mu\nu}G^{\mu\nu} \right\rangle$ gives the contribution up to dimension four.
In nuclear matter, additionally, the twist-two gluon operator $\left\langle \frac{\alpha_{s}}{\pi} G_{\mu\rho}G^{\rho}_{\nu} \right\rangle$ contributes.
The twist-two gluon operator is related to the gluon distribution function $G(x,\mu^{2})$ in a nucleon (cf.~Ref.~\cite{Luke:1992tm}):
\begin{align}
\langle N(p) | \frac{\alpha_{s}}{\pi} G^{\alpha \sigma} G_{\sigma}^{\beta} | N(p) \rangle
=
- \left( p^{\alpha} p^{\beta} - \frac{1}{4} g^{\alpha\beta} p^{2} \right) \frac{\alpha_{s}}{\pi} A_{2}^{g},
\end{align}
with the second momentum of the gluon distribution function
\begin{align}
A_{2}^{g}(\mu^{2}) = 2 \int_{0}^{1} \mathrm{d}x\, x\,G(x,\mu^{2}),
\end{align}
at energy scale $\mu$ corresponding to the charmonium mass.
Including the twist-two gluon operator contribution, the gluon condensate in  nuclear matter is given by
[cf.~Eq.~(\ref{eq:electric_gluon})]
\begin{align}
\left\langle \frac{\alpha_{s}}{\pi} G^{2} \right\rangle_{\Omega}
&=
\left\langle \frac{\alpha_{s}}{\pi} G^{2} \right\rangle_{0}
-
\left(
 \frac{8}{9} m_{N}^{0} + \frac{3}{2} m_{N} \frac{\alpha_{s}}{\pi} A_{2}^{g}
\right) \rho
\nonumber \\
&\simeq
\left\langle \frac{\alpha_{s}}{\pi} G^{2} \right\rangle_{0}
\left( 1-0.06\frac{\rho}{\rho_{0}} \right),
\end{align}
with the normal nuclear matter density $\rho_{0}$.
Quantitatively, the contribution from the twist-two gluon operator is found to be about 10 \%.
In the QCD sum rules, we can extract the information about hadron property (spectral function) from the OPE data (see Sect.~\ref{sec:QCDSR}).
As the results, the mass shifts $\Delta m_{J/\psi} \simeq -7$ MeV and $\Delta m_{\eta_{c}} \simeq -5$ MeV were obtained at normal nuclear matter density~\cite{Klingl:1998sr}.
Interestingly, those values are consistent with the previous studies based on the color dipole picture~\cite{Kaidalov:1992hd} and the scale anomaly~\cite{Luke:1992tm} in sign as well as in order of the magnitude.
The QCD sum rule analysis with dimension up to six is performed, and the resulting mass shift is $\Delta m_{J/\psi}=-4$ MeV~\cite{Kim:2000kj}.
The mass shifts of $J/\psi$ and $\eta_{c}$ in isospin asymmetric nuclear matter are investigated in the QCD sum rules by introducing the dilaton field for the scale anomaly~\cite{Kumar:2010hs}.

The mean-field approach has been discussed in Ref.~\cite{Tsushima:2011kh}.
The (virtual) intermediate states $D\bar{D}$, $D^{\ast}\bar{D}$, $D\bar{D}^{\ast}$, $D^{\ast}\bar{D}^{\ast}$ in the $J/\psi$ interact with nucleons in nuclear matter (cf.~Sect.~\ref{sec:D_mesons}).
By this virtual processes, $J/\psi$ can be bound state in nuclear matter.
Several bound states including excited states are found for $^{4}\mathrm{He}$, $^{12}\mathrm{C}$, $^{16}\mathrm{O}$, $^{40}\mathrm{Ca}$, $^{90}\mathrm{Zr}$ and $^{208}\mathrm{Pb}$ nuclei.

As discussed above, the mass spectroscopy of charmonium in nuclear matter provides us with new tools to investigate the properties of the QCD vacuum, such as the modification of the gluon condensate.
For this purpose, we have focused on the $J/\psi N$ interaction around the threshold.
If we increase the energy by about 100 MeV, the $J/\psi N$ state can be converted to various channels such as $\bar{D}^{(\ast)}\Lambda_{c}$ and $\bar{D}^{(\ast)}\Sigma_{c} ^{(\ast)}$.
Recently, it has been reported by LHCb that there are two resonances $P_{c}(4380)$ and $P_{c}(4450)$, the hidden charm pentaquarks, around the mass thresholds of the $\bar{D}\Sigma_{c}$ and $\bar{D}^{\ast}\Sigma_{c}$ channels, respectively~\cite{Aaij:2015tga}.
The resonance signals are observed in the invariant mass spectrum of the $J/\psi p$ system in the final states.
This observation indicates that the $J/\psi N$ scattering have non-trivial structures due to the existence of the resonances and the coupled-channel effects.
Hence it is an interesting subject to study the spectral functions of nuclear matter with $J/\psi$ in a wide range of the scattering energies.
For example, it has been discussed that the $J/\psi$ with a finite momentum in nuclear matter gives a large cross section of $J/\psi$-nucleus reaction due to the existence of $P_{c}$ resonances~\cite{Molina:2012mv}.

\section{Heavy-light mesons}
\label{sec:D_mesons}

To study the heavy-light mesons in nuclear medium, let us first recall its naming scheme which is somehow confusing. Heavy-light meson with charm $C=+1$ is called $D$ meson, and that with bottomness $B=+1$ is denoted by $B$. Because the charm quark $c$ (bottom quark $b$) has $C=+1$ ($B=-1$), the quark content of the $D$ meson ($B$ meson) is $c\bar{q}$ ($\bar{b}q$) with a light quark $q$. Thus, we obtain the quark contents of $D$, $B$ and their antiparticles as summarized in Table~\ref{tbl:DB}. In vacuum, the basic properties of $D$ and $\bar{D}$ are symmetric. On the other hand, their in-medium properties are quite different, because the nuclear medium is composed only of the light quarks. The light degrees of freedom make the heavy-light mesons sensitive to the partial restoration of chiral symmetry in nuclear medium. Moreover, the modification of the properties of $D$ and $\bar{D}$ also affects the properties of excited charmonia near the $\bar{D}D$ threshold, such as $\psi(3770)$~\cite{Hayashigaki:2000es,Friman:2002fs}. In the following, we summarize these rich phenomena of heavy-light mesons in nuclear medium.

\begin{table}[tbp]
\caption{Quark contents of heavy-light mesons.}
\begin{center}
\begin{tabular}{|cc|cc|}
\hline
\multicolumn{2}{|c|}{$D(c\bar{q})$} & \multicolumn{2}{c|}{$\bar{D}(\bar{c}q)$} \\
$D^{+}(c\bar{d})$ & $D^{0}(c\bar{u})$ 
& $\bar{D}^{0}(\bar{c}u)$ & $D^{-}(\bar{c}d)$ \\
\hline
\multicolumn{2}{|c|}{$\bar{B}(b\bar{q})$} & \multicolumn{2}{c|}{$B(\bar{b}q)$} \\
$\bar{B}^{0}(b\bar{d})$ & $B^{-}(b\bar{u})$ 
& $B^{+}(\bar{b}u)$ & $B^{0}(\bar{b}d)$ \\
\hline
\end{tabular}
\end{center}
\label{tbl:DB}
\end{table}%

\subsection{Two-body interaction with nucleon}
\label{sec:DNinteraction}

The most basic information to study the heavy-light meson in nuclei is the two-body interaction with a nucleon. Because the isospin of the $D/\bar{D}$ meson is $I=1/2$, there are two isospin components in the $DN/\bar{D}N$ system, $I=0$ and $I=1$. At present, there is no direct experimental data to constrain these interactions. The $DN/\bar{D}N$ interactions are thus constructed by generalizing some established framework of the hadron interactions. Here we introduce several approaches proposed for the $DN/\bar{D}N$ interaction.

There are two important remarks. First, the quark-antiquark annihilation is allowed in the $D+N\sim \bar{q}c+qqq$ system, while it does not exist in the $\bar{D}+N\sim \bar{c}q+qqq$ system. As a consequence, the $DN$ system couples with the lower energy $\pi\Sigma_{c}$ and $\pi\Lambda_{c}$ channels, as well as the charmed baryon resonances, $\Lambda_{c}^{*}$ and $\Sigma_{c}^{*}$. To study the $DN$ system, we need to solve a complicated coupled-channel problem. In contrast, the $\bar{D}N$ system has no coupled channels at lower energies so that the problem is much simpler than the $DN$ system. The $\bar{D}N$ system is in a exotic channel, whose quantum numbers ($C=-1$ and baryon number $B=1$) cannot be expressed by three valence quarks. Second, heavy quark symmetry relates the pseudoscalar $D$ meson with the vector $D^{*}$ meson. From the viewpoint of heavy quark symmetry, the $DN/\bar{D}N$ interaction should be considered together with the $D^{*}N/\bar{D}^{*}N$ channel. As we will see below, the inclusion of the vector meson channel plays an important role, especially in the $\bar{D}N$ system.

\paragraph{Contact interaction models with SU(4) symmetry}

As discussed in section~\ref{sec:chiral_effective_theory}, the scattering of the pseudoscalar mesons off the baryons in the flavor SU(3) sector can be well described by the resummation of the Weinberg-Tomozawa interaction~\cite{Hyodo:2011ur,Kaiser:1995eg,Oset:1997it,Oller:2000fj,Ikeda:2011pi,Ikeda:2012au}. The interaction kernel is given by
\begin{align}
   V_{ij}(\sqrt{s}) 
   =-\frac{ C_{ij}}{4f^{2}} (2\sqrt{s}-M_{i}-M_{j}),
   \label{eq:WTinteraction}
\end{align}
where $C_{ij}$ is the group theoretical coefficient. The scattering amplitude $T_{ij}(\sqrt{s})$ is given by the solution of the matrix equation
\begin{align}
   T(\sqrt{s})  = V(\sqrt{s})  + V(\sqrt{s}) G(\sqrt{s}) T(\sqrt{s}) .
   \label{eq:scatteringequation}
\end{align}
The loop function $G_{i}(\sqrt{s})$ contains a cutoff parameter (subtraction constant) which should be determined either by fitting to the experimental data, or by some theoretical arguments~\cite{Oller:2000fj,Lutz:2001yb,Hyodo:2008xr}. When there is a sufficient attraction in the interaction $V$, a bound state or a resonance can be dynamically generated, which is expressed by the pole of the scattering amplitude $T$. The success of chiral SU(3) dynamics with the Weinberg-Tomozawa interaction indicates that the SU(3) symmetric interaction is a good starting point for these channels, even though there is a sizable SU(3) breaking effect in the observed hadron masses. While the Weinberg-Tomozawa interaction~\eqref{eq:WTinteraction} is a contact interaction, it can be regarded as the zero-range limit of the t-channel vector meson exchange, with the help of the universal vector meson coupling and the Kawarabayashi--Suzuki--Riazuddin--Fayyazuddin (KSRF) relation~\cite{Kawarabayashi:1966kd,Riazuddin:1966sw}. With these observations, there are series of works which utilize the contact interaction model for the $DN/\bar{D}N$ systems, using the same form with Eq.~\eqref{eq:WTinteraction} under the larger flavor or spin-flavor symmetry.\footnote{This generalized framework is often referred to as the ``Weinberg-Tomozawa'' approach. However, the original Weinberg-Tomozawa theorem~\cite{Weinberg:1966kf,Tomozawa:1966jm} is derived from chiral symmetry. It is certainly not justified to start from the four-flavor chiral symmetry in QCD. When the spin symmetry is included, Eq.~\eqref{eq:WTinteraction} determines the interaction of the non-NG modes such as $\rho$ meson, whose dynamics cannot be constrained by chiral symmetry. It is therefore appropriate to regard this approach as a contact interaction model motivated by the vector meson exchange mechanism.}

The first study in this direction is performed in Ref.~\cite{Hofmann:2005sw}, by extending the flavor symmetry to SU(4) with the charm quark. The ground state $1/2^{+}$ baryons are in the $\bm{20}\in \bm{4}\otimes\bm{4}\otimes \bm{4}$ representation of SU(4), and the $0^{-}$ mesons are in the $\bm{15}\in \bm{4}\otimes\bm{\bar{4}}$ representation. The $\bm{20}$ representation contains four SU(3) multiplets: $\bm{8}=\{N,\Lambda,\Sigma,\Xi\}$ of the light baryons, $\bm{\bar{3}}=\{\Lambda_{c},\Xi_{c}\}$ and $\bm{6}=\{\Sigma_{c},\Xi_{c}^{\prime},\Omega_{c}\}$ of the heavy-light-light baryons, and $\bm{3}=\{\Xi_{cc},\Omega_{cc}\}$ of the heavy-heavy-light baryons. The $\bm{15}$ multiplet can be decomposed into  $\bm{8}=\{\pi,K,\eta_{8}\}$ of the light mesons, $\bm{\bar{3}}=\{D,D_{s}\}$ and $\bm{3}=\{\bar{D},\bar{D}_{s}\}$ of the heavy-light mesons, and $\bm{1}=\{\eta_{1}\}$. Physical $\eta$, $\eta^{\prime}$, and $\eta_{c}$ are expressed by the linear combinations of $\eta_{8}$ and $\eta_{1}$, together with the SU(4) singlet in $\bm{4}\otimes\bm{\bar{4}}$. We note that the vector $D^{*}$ mesons are not included in this model. From the SU(4) symmetry, the strengths of the diagonal $DN/\bar{D}N$ interactions [in the convention of Eq.~\eqref{eq:WTinteraction}] are found to be~\cite{Mizutani:2006vq}
\begin{align}
   C_{DN  DN}^{(I=0)}&=3 , 
   \quad C_{DN DN}^{(I=1)}=1 , 
   \label{eq:DNcontact} \\
   C_{\bar{D}N  \bar{D}N}^{(I=0)}&=0 , 
   \quad C_{\bar{D}N \bar{D}N}^{(I=1)}=-2, 
   \label{eq:DbarNcontact}
\end{align}
where the positive (negative) sign corresponds to the attractive (repulsive) interaction. Namely, the $DN(I=0)$ channel is strongly attractive, the $DN(I=1)$ channel is moderately attractive, the interaction vanishes in the $\bar{D}N(I=0)$ channel, and the $\bar{D}N(I=1)$ channel is repulsive. Interestingly, the strengths of the interactions in Eqs.~\eqref{eq:DNcontact} and \eqref{eq:DbarNcontact} are identical with those of the Weinberg-Tomozawa term in the $\bar{K}N/KN$ sector, because of the replacement of $c\leftrightarrow s$. The non-attractive nature of the $\bar{D}N$ interactions can be understood by the group theoretical property of $C_{ij}$, which is in most cases repulsive in the exotic channels~\cite{Hyodo:2006yk,Hyodo:2006kg}. By solving the scattering equation~\eqref{eq:scatteringequation}, quasi-bound states are dynamically generated in the $DN$ sector, while no state is generated in the $\bar{D}N$ sector, reflecting the interaction strengths in Eqs.~\eqref{eq:DNcontact} and \eqref{eq:DbarNcontact}. The quasi-bound states have the quantum number $J^{P}=1/2^{-}$, because the contact term gives the $s$-wave interaction. The analysis of the speed plot\footnote{The speed plot (absolute value of the energy derivative of the scattering amplitude) shows a peak at the energy of a narrow resonance and is related to the time delay (see, e.g., Ref.~\cite{Kelkar:2003pt}).} shows the existence of a narrow state around 2590 MeV (2620 MeV) in the $DN(I=0)$ channel [$DN(I=1)$ channel]~\cite{Hofmann:2005sw}. The $I=0$ state is identified by the negative parity $\Lambda_{c}(2595)$ state. The results are summarized in Tables~\ref{tbl:DNbound} and \ref{tbl:DbarNbound}. We also note that an exotic state is found around 2780 MeV in the $\bar{D}_{s}N$-$\bar{D}\Lambda$-$\bar{D}\Sigma$ coupled-channels. The quark content of this manifestly exotic state is $\bar{c}sqqq$, which corresponds to the pentaquark state discussed in the constituent quark model~\cite{Lipkin:1987sk}. In the $D_{s}N$ channel, a similar state with the $\bar{s}cqqq$ configuration is found at 2892 MeV. As we will see below, the existence of these states plays an important role for the study of $D_{s}/\bar{D}_{s}$ in nuclear matter.

\begin{table}[tbp]
\caption{Quasi-bound states with $J^{P}=1/2^{-}$ in the $DN$ channel in various models. All numbers are given in units of MeV, and rounded off at one MeV precision. Here we summarize the states found below the $DN$ threshold $\sim 2806$ MeV. Results of Ref.~\cite{Hofmann:2005sw} are those with the SU(4) breaking effect. Results of Ref.~\cite{Yamaguchi:2013ty} are those in the $\pi\rho\omega$ model, where the state below the $DN$ threshold is stable because the lower energy channels are not included. The pole position of the model in Ref.~\cite{Mizutani:2006vq} can be found in Ref.~\cite{GarciaRecio:2008dp}. }
\begin{center}
\begin{tabular}{|l|ll|}
\hline
Model & $DN(I=0)$ & $DN(I=1)$ \\
\hline
SU(4) contact~\cite{Hofmann:2005sw} 
  & $M_{R}=2593$, $\Gamma_{R}<1$  
  & $M_{R}=2620$, $\Gamma_{R}=1$  \\
  \hline
SU(4) contact~\cite{Mizutani:2006vq} 
  & $M_{R}=2595$, $\Gamma_{R}=2$   
  & $M_{R}=2661$, $\Gamma_{R}=37$   \\
  & $M_{R}=2625$, $\Gamma_{R}=103$  
  & $M_{R}=2695$, $\Gamma_{R}=153$  \\
  \hline
SU(8) contact~\cite{GarciaRecio:2008dp}
  & $M_{R}=2595$, $\Gamma_{R}=1$   
  & $M_{R}=2554$, $\Gamma_{R}=1$   \\
  & $M_{R}=2610$, $\Gamma_{R}=71$   
  & $M_{R}=2612$, $\Gamma_{R}=179$   \\
  &  
  & $M_{R}=2637$, $\Gamma_{R}=80$   \\
  \hline
Meson exchange~\cite{Haidenbauer:2010ch} 
  & $M_{R}=2594$, $\Gamma_{R}=6$   
  & $M_{R}=2793$, $\Gamma_{R}=12$   \\
  & $M_{R}=2603$, $\Gamma_{R}=126$   
  &   \\
  \hline
Pion exchange~\cite{Yamaguchi:2013ty}
  & $M_{R}=2724$ & None  \\
\hline
\end{tabular}
\end{center}
\label{tbl:DNbound}
\end{table}%

\begin{table}[tbp]
\caption{Bound states with $J^{P}=1/2^{-}$ in the $\bar{D}N$ channel in various models. All numbers are given in units of MeV, and rounded off at one MeV precision. Results of Ref.~\cite{Yamaguchi:2011xb} are those in the $\pi\rho\omega$ model. Here we summarize the states found below the $\bar{D}N$ threshold $\sim 2806$ MeV.}
\begin{center}
\begin{tabular}{|l|ll|}
\hline
Model & $\bar{D}N(I=0)$ & $\bar{D}N(I=1)$ \\
\hline
SU(4) contact~\cite{Hofmann:2005sw} 
  & None & None \\
SU(8) contact~\cite{Gamermann:2010zz} 
  & 2805  & None \\
Meson exchange~\cite{Haidenbauer:2007jq}
  & None & None \\
Pion exchange~\cite{Yamaguchi:2011xb}
  & 2804 & None \\
Chiral quark model~\cite{Carames:2012bd} & None & None \\
\hline
\end{tabular}
\end{center}
\label{tbl:DbarNbound}
\end{table}%

This framework for the $DN$ channel is further studied in Ref.~\cite{Mizutani:2006vq}. The interaction kernel is modified to be consistent with the zero-range limit, and a suppression factor $\kappa_{c}=(\bar{m}_{V}/\bar{m}^{c}_{V})^{2} \simeq 1/4$ is introduced in view of the underlying vector meson exchange mechanism, which partially breaks the SU(4) symmetry. From the analysis of the amplitude on the real axis, the quasi-bound states are found at qualitatively similar positions with Ref.~\cite{Hofmann:2005sw}, while the $I=1$ state becomes broader. The pole positions of this SU(4) model are calculated in Ref.~\cite{GarciaRecio:2008dp} as summarized in Table~\ref{tbl:DNbound}. Interestingly, there is an additional broad state in each sector. Note that such broad state may not be extracted in the analysis of the speed plot in Ref.~\cite{Hofmann:2005sw}. The appearance of two poles between the $\pi\Sigma_{c}$ and $DN$ thresholds in the $I=0$ sector is analogous to the double-pole structure of $\Lambda(1405)$ in the SU(3) sector~\cite{Jido:2003cb,Hyodo:2007jq}.
The effect of the finite range interaction is examined in Ref.~\cite{JimenezTejero:2009vq}. The use of the nonlocal interaction kernel gives qualitatively similar results with Ref.~\cite{Mizutani:2006vq}, but the generated resonances generally have a broader width. This is because the suppression of the charm exchange process is milder than the zero-range approximation.

\paragraph{Contact interaction models with SU(8) symmetry}

The SU(4) contact interaction approach does not include the $D^{*}N/\bar{D}^{*}N$ channel, in contrast to the requirement of the heavy quark symmetry. Moreover, because the SU(4) symmetry relates the charm quark with the light quarks, at first glance, it looks contradicting with the heavy quark symmetry which emerges in the heavy quark limit. However, we should recall that in the SU(4) approaches the flavor symmetry is mainly used to determine the unknown coupling constants of charm hadrons, and the symmetry is largely broken by the physical hadron masses in the loop functions. As long as the physical hadron masses are used, it is mandatory to couple the $D^{*}N/\bar{D}^{*}N$ channels with the $DN/\bar{D}N$ channels, from the viewpoint of heavy quark spin symmetry. 
In addition, the coupling constants within the heavy quark spin multiplets should follow the heavy quark spin symmetry.
The incorporation of the heavy quark spin symmetry is discussed in Ref.~\cite{GarciaRecio:2008dp} for the $DN$ sector and in Ref.~\cite{Gamermann:2010zz} for the $\bar{D}N$ sector, within the framework of spin-flavor SU(8) symmetry. As a generalization of the SU(6) approach~\cite{GarciaRecio:2005hy}, the SU(8) model combines spin SU(2) with flavor SU(4). 

The ground state baryon field $B$ is in the $\bm{120}\in \bm{8}\times\bm{8}\times \bm{8}$ representation of SU(8), and the mesons $M$ are in the $\bm{63}\in \bm{8}\times\bm{\bar{8}}$ representation. The $\bm{120}$ multiplet can be decomposed into the SU(4) representations as ${\bf 20_{2}} \oplus {\bf 20'_{4}}$ where the subscript represents the dimension of the SU(2) representation. The former corresponds to the $1/2^{+}$ baryons in the SU(4) approach (nucleon and its flavor partners), and the latter contains the $3/2^{+}$ baryons with SU(3) representations $\bm{10}\oplus\bm{6}\oplus\bm{3}\oplus\bm{1}$ ($\Delta$ and its flavor partners). The meson in the $\bm{63}$ representation consists of ${\bf 15_{1}} \oplus {\bf 15_{3}} \oplus {\bf 1_{3}}$, where ${\bf 15_{1}}$ is the pseudoscalar mesons ($\pi$ and its flavor partners) and ${\bf 15_{3}}$ and ${\bf 1_{3}}$ correspond to the vector mesons ($\rho$ and its flavor partners). Because the vector current is given by the adjoint representation ${\bf 63}$, the SU(8) symmetric contact interaction is constructed as 
\begin{align}
{\cal L}^{\mathrm{SU}(8)} \sim ((M^{\dag} \otimes M)_{{\bf 63}_{a}} \otimes (B^{\dag} \otimes B)_{\bf 63})_{\bf 1} .
 \label{eq:Lagrangian_SU8}
\end{align}
Since the ground state multiplet contains spin $3/2$ baryons and spin $1$ mesons, the two-body $s$-wave system can be not only in $1/2^{-}$ but also in $3/2^{-}$ and $5/2^{-}$. The coupling strengths in the normalization of Eq.~\eqref{eq:WTinteraction} in the spin $1/2$ channel are given by\footnote{As a convention, positive numbers in the diagonal components are attraction, while negative numbers are repulsion.}
\begin{align}
   C_{ij}^{DN(I=0)}
   &=
   \begin{pmatrix}
   3 & \sqrt{27} \\
   \sqrt{27} & 9
   \end{pmatrix} , 
   \quad C_{ij}^{DN(I=1)}
   = \begin{pmatrix}
   1 & \sqrt{\frac{1}{3}} & -\sqrt{\frac{32}{3}}\\
   \sqrt{\frac{1}{3}} & \frac{1}{3} & \sqrt{\frac{32}{9}} \\
   -\sqrt{\frac{32}{3}} & \sqrt{\frac{32}{9}} & \frac{32}{3}
   \end{pmatrix} , 
   \label{eq:DNcontactSU8} \\
   C_{ij}^{\bar{D}N(I=0)}
   &=
   \begin{pmatrix}
   0 & \sqrt{12} \\
   \sqrt{12} & -4
   \end{pmatrix} , 
   \quad 
   C_{ij}^{\bar{D}N(I=1)}
   =\begin{pmatrix}
   -2 & -\sqrt{\frac{16}{3}} & -\sqrt{\frac{32}{3}} \\
   -\sqrt{\frac{16}{3}} & \frac{2}{3} & -\sqrt{\frac{32}{9}} \\
   -\sqrt{\frac{32}{3}} & -\sqrt{\frac{32}{9}} & -\frac{2}{3}
   \end{pmatrix}
   , 
   \label{eq:DbarNcontactSU8}
\end{align}
where the basis states are given by
\begin{align}
\{ DN/\bar{D}N(^{2}\mathrm{S}_{1/2}), D^{*}N/\bar{D}^{\ast}N(^{2}\mathrm{S}_{1/2})\} ,
\end{align}
for $I=0$ and
\begin{align}
\{ DN/\bar{D}N(^{2}\mathrm{S}_{1/2}), D^{*}N/\bar{D}^{\ast}N(^{2}\mathrm{S}_{1/2}), D^{*}\Delta/\bar{D}^{*}\Delta(^{2}\mathrm{S}_{1/2})\} ,
\end{align}
for $I=1$ with the spectroscopic notation $^{2S+1}L_{J}$. The other coupled channels are not shown explicitly here. We note that the 11 components are identical with Eqs.~\eqref{eq:DNcontact} and \eqref{eq:DbarNcontact}, because the SU(4) model is in the subspace of the SU(8) model. We note that the contact interaction only gives the $s$-wave interaction. When the tensor force is included as in the pion-exchange model, the $d$-wave components also couples to the problem.

By solving the scattering equation, several states are dynamically generated in the $DN$ sector. The pole structure in the $DN(I=0)$ channel is similar to the SU(4) model; one narrow pole and one broad pole appear around 2600 MeV. However, by analyzing the coupling strengths, it is shown that the narrow pole strongly couples to the $D^{*}N$ channel~\cite{GarciaRecio:2008dp}. In the $DN(I=1)$ sector, three states are found below the $DN$ threshold, because of the larger model space of the SU(8) approach than the SU(4) one. Further extension of the contact interaction model to the bottom sector is discussed in Ref.~\cite{GarciaRecio:2012db} where the corresponding $1/2^{-}$ resonance of  $\Lambda_{b}(5912)$ is dynamically generated in the $\bar{B}N(I=0)$ sector. Note however that the binding energy of the $\Lambda_{c}(2595)$ [$\Lambda_{b}(5912)$] as the $DN$ ($\bar{B}N$) bound state is $\sim 200$ MeV ($\sim 300$ MeV), and the ``molecule'' picture may not be justified for such tightly bound system.

In the $\bar{D}N(I=0)$ sector, a shallow bound state is found at 2805 MeV (1 MeV below the $\bar{D}N$ threshold)~\cite{Gamermann:2010zz} in the coupled-channel $\bar{D}N$-$\bar{D}^{*}N$ system. At first glance, the origin of the attraction may not be clear, because the diagonal interactions are zero and repulsive in Eq.~\eqref{eq:DbarNcontactSU8}. As pointed out in Ref.~\cite{Gamermann:2010zz}, however, the attractive interaction arises by taking the linear combination of $(\sqrt{3}\ket{\bar{D}N}+\ket{\bar{D}^{*}N})/2$. In fact, this combination coincides with the spin-complex basis which is introduced to extract the heavy quark spin symmetry in multi-hadron systems~\cite{Yasui:2013vca,Yamaguchi:2014era}. The coupling strength in Eq.~\eqref{eq:DbarNcontactSU8} in the spin-complex basis is given as
\begin{align}
   C_{ij}^{\bar{D}N(I=0), {\rm SC}}
   &=
   \begin{pmatrix}
   -6 & 0 \\
   0 & 2
   \end{pmatrix}  ,
\end{align}
and the attraction $C_{22}^{\bar{D}N(I=0), {\rm SC}}=2$ is responsible for the bound state formation. The absence of the off-diagonal components in the spin-complex basis is a consequence of the heavy quark symmetry, because the contained spin-complexes are different. In this way, when the heavy quark symmetry is incorporated, a shallow bound state is supported in the $\bar{D}N(I=0)$ sector, which otherwise has no bound state. The SU(8) model also generates bound and resonance states in other spin-isospin sectors. For instance, a bound state of $N\bar{D}^{\ast}$ is found at 2922 MeV in the $I(J^{P})=0(3/2^-)$ channel. This state forms a HQS doublet with the $1/2^{-}$ state~\cite{Yamaguchi:2014era}. Although the $I(J^{P})=0(3/2^-)$ state is stable in this approach, it can decay into the $\bar{D}N$ channel in $d$ wave. Other states are found in the same way; a resonance with $M_{R}=2873$ MeV and $\Gamma_{R}=91$ MeV in the $I(J^{P})=1(1/2^-)$ channel, a resonance with $M_{R}=2979$ MeV and $\Gamma_{R}=32$ MeV in the $I(J^{P})=1(3/2^-)$ channel, a bound state at 3126 MeV in the $I(J^{P})=1(5/2^{-})$ channel, 
a bound state at 3126 MeV in the $I(J^{P})=2(1/2^-)$ channel, and a bound state at 3061 MeV in the $I(J^{P})=2(3/2^{-})$ channel. Note that some states have a large binding energy; the $I(J^{P})=1(5/2^{-})$ state is about 100 MeV bound from the $\Delta \bar{D}^{*}$ threshold. When the decay width of $\Delta$ is taken into account, the bound states acquire a width of several tens of MeV, while the binding energies are not very much affected.

\paragraph{Meson exchange models}

One of the traditional approaches to study the hadron-hadron interaction is the meson exchange potential~\cite{Machleidt:1987hj}. In the strangeness sector, the $\bar{K}N/KN$ interactions have been developed by the J\"ulich group~\cite{Buettgen:1990yw,MuellerGroeling:1990cw,Hoffmann:1995ie}. The J\"ulich meson exchange model is generalized to the charm sector in Refs.~\cite{Haidenbauer:2010ch,Haidenbauer:2007jq}.

The meson exchange model is first applied to the $\bar{D}N$ sector~\cite{Haidenbauer:2007jq}. As a generalization of the $KN$ potential~\cite{Buettgen:1990yw,Hoffmann:1995ie}, $\bar{D}$ and $N$ interact with each other through the exchange of the $\rho$, $\omega$, $\sigma$, and $a_{0}$ mesons. The box-diagrams with the intermediate $\bar{D}^{*}N$, $\bar{D}^{*}\Delta$, $\bar{D}\Delta$ states are included. The coupling constants are determined by the SU(4) symmetry. As a short range interaction, the quark interchange processes with one-gluon exchange are also considered. The scattering amplitude is obtained by solving the Lippmann-Schwinger equation with this potential. The existence of bound states is not reported. In the $\bar{D}N(I=0)$ sector, the long-range meson exchange is attractive, while the short-range quark-gluon exchange is repulsive. As a consequence, the phase shift changes from attractive around the threshold to repulsive at higher energies. The interaction in the $\bar{D}N(I=1)$ sector is found to be repulsive. 

The $DN$ sector is studied with the same mechanisms. The results are partly presented in Ref.~\cite{Haidenbauer:2008ff} and the detailed description is given in Ref.~\cite{Haidenbauer:2010ch}. In the $DN$ sector, however, the quark-gluon exchange is absent, because the antiquark in the $D$ meson cannot be exchanged with the quarks in the nucleon. Then the dominant vector meson exchange plays a similar role with the contact interaction in the SU(4) model. In fact, $DN(I=0)$ sector generates two poles near the $\pi\Sigma_{c}$ threshold, which can be identified with $\Lambda_{c}(2595)$. The quasi-bound state in the $DN(I=1)$ channel is obtained at higher energies than the contact interaction model, and identified as $\Sigma_{c}(2800)$ whose spin-parity is not yet determined experimentally. More detailed comparison with the SU(4)~\cite{Mizutani:2006vq} and SU(8)~\cite{GarciaRecio:2008dp} models can be found in Ref.~\cite{GarciaRecio:2008dp}.

In the meson exchange model, the $D^{*}N/\bar{D}^{*}N$ channel is included in the intermediate state of the box diagrams. However, the diagonal $D^{*}N\to D^{*}N/\bar{D}^{*}N\to \bar{D}^{*}N$ interaction is not included, and the role of the heavy quark symmetry is not very clear.

\paragraph{Pion exchange models with heavy quark symmetry}

The pion-exchange potential, accompanying the tensor force, is one of the most important ingredients in the $NN$ interaction.
As for the $D^{(\ast)}N$ and $ \bar{D}^{(\ast)}N$ interactions,
 the one-pion exchange becomes possible to exist when the heavy quark symmetry is adopted; 
the one-pion exchange is provided by the mixing of $DN/\bar{D}N$ and $D^{*}N/\bar{D}^{\ast}N$.
It is important that the proximity of $DN/\bar{D}N$ and $D^{\ast}N/\bar{D}^{\ast}N$ channels ($m_{D^{\ast}}-m_{D}\simeq 140$ MeV) enhances the strength of the pion exchange potential through the off-diagonal channel coupling, while there is no pion exchange in the diagonal $DN/\bar{D}N$ channel due to the parity conservation.\footnote{In contrast to the $\Lambda N$-$\Sigma N$ coupling in hypernuclei which is the mixing of different isospin states, the $\bar{D}N$-$\bar{D}^{\ast}N$ coupling is the mixing of different spin states.}
The mixing effect becomes more enhanced in the $\bar{B}^{(\ast)}N/B^{(\ast)}N$ case.
In the light quark sector, in contrast, the pion exchange does not play an important role, where the mass splitting of the spin zero and spin one states is large; $m_{K^{\ast}}-m_{K}\simeq 350$ MeV and $m_{\rho}-m_{\pi}\simeq 640$ MeV.

The pion-exchange model is intensively applied to the exotic $\bar{D}N$ sector where the coupling to lower energy channels is absent.
In Ref.~\cite{Cohen:2005bx}, the pion-exchange model is considered to examine the properties of the stable heavy pentaquark bound state, which is shown to exist in the heavy quark limit combined with the large $N_{c}$ limit\footnote{There are early studies of Skyrmions where a $D^{(\ast)}/\bar{D}^{(\ast)}$ meson is bound in the bound-state approach~\cite{Riska:1992qd,Oh:1994np,Oh:1994yv}.}.
In this study, however, the coupling of different angular momentum states via the tensor force is not completely considered. The full calculation including the tensor force is performed in Ref.~\cite{Yasui:2009bz}. By an analogy with the nucleon-nucleon interaction, the tensor force plays an important role in the pion exchange. In fact, the tensor force is known to significantly affect the nuclear structure including the neutron rich nuclei (see Refs.~\cite{Myo:2007vm,Otsuka:2005zz,Myo:2015rbv}). The tensor force strongly works through the mixing of the state with angular momentum $L$ and those with $L\pm2$. In the present case, it is important that there is the mixing of $\bar{D}N$ and $\bar{D}^{\ast}N$, in addition to the angular momentum mixing.
All the relevant channels for the $J\le 3/2$ (5/2) systems with the negative (positive) parity are listed in Table~\ref{table:DbarN_all}.

\begin{table}[tbp]
\centering
\caption{Various coupled channels for a 
given quantum number $J^P$ for negative/positive parity $P = \mp1$~\cite{Yasui:2009bz,Yamaguchi:2011xb,Yamaguchi:2011qw}. }
\begin{tabular}{ c  | c c c c}
\hline
$J^P$ &  \multicolumn{4}{c }{channels} \\
\hline
$1/2^-$ &  $\bar{D}N(^2\mathrm{S}_{1/2})$ & $\bar{D}^*N(^2\mathrm{S}_{1/2})$ & 
     $\bar{D}^*N(^4\mathrm{D}_{1/2})$ & \\
$3/2^-$ &  $\bar{D}N(^2\mathrm{D}_{3/2})$ &  $\bar{D}^*N(^4\mathrm{S}_{3/2})$ & 
     $\bar{D}^*N(^4\mathrm{D}_{3/2})$ & $\bar{D}^*N(^2\mathrm{D}_{3/2})$ \\
\hline
$1/2^+$ &$\bar{D}N(^2\mathrm{P}_{1/2})$&$\bar{D}^\ast N(^2\mathrm{P}_{1/2})$&$\bar{D}^\ast N(^4\mathrm{P}_{1/2})$& \\
$3/2^+$ & $\bar{D}N(^2\mathrm{P}_{3/2})$&$\bar{D}^\ast N(^2\mathrm{P}_{3/2})$&$\bar{D}^\ast
	      N(^4\mathrm{P}_{3/2})$&$\bar{D}^\ast N(^4\mathrm{F}_{3/2})$ \\
$5/2^+$&$\bar{D}N(^2\mathrm{F}_{5/2})$&$\bar{D}^\ast N(^4\mathrm{P}_{5/2})$&$\bar{D}^\ast
	      N(^2\mathrm{F}_{5/2})$&$\bar{D}^\ast N(^4\mathrm{F}_{5/2})$ \\
\hline
\end{tabular}
\label{table:DbarN_all}
\end{table}

For instance, in the $J^{P}=1/2^{-}$ and $3/2^{-}$ sectors relevant coupled-channels are given by
\begin{align}
\{ \bar{D}N(^{2}\mathrm{S}_{1/2}), \bar{D}^{\ast}N(^{2}\mathrm{S}_{1/2}), \bar{D}^{\ast}N(^{4}\mathrm{D}_{1/2}) \} ,
\end{align}
and
\begin{align}
\{ \bar{D}N(^{2}\mathrm{D}_{3/2}), \bar{D}^{\ast}N(^{4}\mathrm{S}_{3/2}), \bar{D}^{\ast}N(^{4}\mathrm{D}_{3/2}), \bar{D}^{\ast}N(^{2}\mathrm{D}_{3/2}) \} ,
\end{align}
respectively. The coupled-channel Hamiltonian of the pion-exchange model is given by
\begin{align}
H_{1/2^-} &=
\left(
\begin{array}{ccc}
 K_0 & \sqrt{3} \, {C} & -\sqrt{6} \, {T}  \\
\sqrt{3} \, {C} & K_0-2 \, {C} + \Delta M & -\sqrt{2} \, {T} \\
-\sqrt{6} \, {T} & -\sqrt{2} \, {T} & K_2 + ({C} - 2\, {T}) + \Delta M
\end{array}
\right), \label{eq:OPEP_DbarN_1/2} \\
H_{3/2^-} &=
\left(
\begin{array}{cccc}
 K_2 & \sqrt{3}\, {T} & -\sqrt{3} \, {T} & \sqrt{3}\,{C} \\
\sqrt{3}\,{T} &K_0 + {C} + \Delta M & 2\,{T} & {T} \\
-\sqrt{3}\,{T} & 2\,{T} & K_2 + {C} + \Delta M & -{T} \\
\sqrt{3}\,{C} & {T} & -{T} & K_2 -2\,{C} + \Delta M
\end{array}
\right), \label{eq:OPEP_DbarN_3/2} 
\end{align}
where the kinetic term is $K_{\ell}=-( \partial^2 / \partial r^2 + (2/r)\partial / \partial r - \ell(\ell+1)/r^2)/2\mu$ with the reduced mass $\mu$, the central and tensor potentials are ${C}=\kappa\, C(r;m_\pi)$ and ${T} = \kappa\, T(r;m_\pi)$, respectively, with $r$ the distance between $\bar{D}^{(\ast)}$ and N. The coupling constant is given by $\kappa = (g g_{\pi NN}/\sqrt{2} m_N f_{\pi})(\vec{\tau}_{P} \!\cdot\! \vec{\tau}_N/3)$ with $g$ the pion coupling of the $D$ meson in Eq.~\eqref{eq:L_heavy_hadron_pion} and $g_{\pi NN}$ the pion-nucleon coupling constant. The spatial dependences of the central and tensor potentials are given by
\begin{align}
C(r;m) &= \int \frac{\mathrm{d}^{3}\vec{q}}{(2\pi)^{3}} \frac{m^{2}}{|\vec{q}\,|^{2}+m^{2}} e^{i\vec{q}\cdot\vec{r}} F(\Lambda_{P},\vec{q}\,) F(\Lambda_{N},\vec{q}\,), \label{eq:central_C} \\
T(r;m) S_{12}(\hat{r}) &= \int \frac{\mathrm{d}^{3}\vec{q}}{(2\pi)^{3}} \frac{-|\vec{q}\,|^{2}}{|\vec{q}\,|^{2}+m^{2}} S_{12}(\hat{q}) e^{i\vec{q}\cdot\vec{r}} F(\Lambda_{P},\vec{q}\,) F(\Lambda_{N},\vec{q}\,), \label{eq:tensor_T}
\end{align}
where $S_{12}(\hat{x}) = 3 (\vec{\cal O}_{1} \cdot
\hat{x})(\vec{\sigma}_{2} \cdot \hat{x}) - \vec{\cal O}_{1} \cdot
\vec{\sigma}_{2}$
with $\hat{x}=\vec{x}/|\vec{x}\,|$, and ${\cal O}_{1}$ the polarization vector $\vec{\varepsilon}^{\,(\lambda)}$ or $\vec{\varepsilon}^{\,(\lambda)\dag}$ with helicity $\lambda$ (spin-one operator $\vec{T}$) 
from the $\pi \bar{D}\bar{D}^{\ast}$ (the $\pi \bar{D}^{\ast}\bar{D}^{\ast}$) vertex (the subscript 1 for $P^{(\ast)}$ and the subscript 2 for $N$) \cite{Yasui:2009bz}.
The hadronic form factor is $F(\Lambda,\vec{q}\,)=(\Lambda^{2}-m^{2})/(\Lambda^{2}+|\vec{q}\,|^2)$ with a cutoff $\Lambda$.
The energy is measured from the $\bar{D} N$ threshold and the $\Delta M=M_{\bar{D}^{\ast}}-M_{\bar{D}}$ is the mass difference of the $\bar{D}N$ and $\bar{D}^{\ast}N$ thresholds.
The cutoff $\Lambda_P$ 
is determined by the ratio of the matter radii of the heavy meson $P$ and
the nucleon $N$, namely $\Lambda_P/\Lambda_N=r_N/r_P$~\cite{Yasui:2009bz}.
We note that in the central potential in Eqs.~(\ref{eq:central_C}) and (\ref{eq:tensor_T})
 the contact term contribution 
has been neglected. This is because the cutoff $\Lambda_{N}$ of the nucleon is fixed 
to reproduce the deuteron binding energy, without including the contact term in the 
nucleon-nucleon potential. Similarly, the contact term of the heavy meson-nucleon 
potential is also neglected.

By calculating the eigenvalues of this Hamiltonian, a shallow bound state (a few MeV binding) is obtained below the $\bar{D}N$ threshold in the $I(J^{P})=0(1/2^{-})$ channel, while no bound state is found in other channels. Inclusion of the $\rho$ and $\omega$ exchanges does not change the qualitative results. Interestingly, the results of the bound state quantitatively agree with the SU(8) contact model (see Table~\ref{tbl:DbarNbound}), although the driving mechanism is different; the tensor force is crucial for the bound state in the pion-exchange model. On the other hand, from the comparison with the SU(4) contact model and the meson exchange model, the bound state below the $\bar{D}N$ threshold is obtained only when the heavy quark symmetry is incorporated in the framework, because the channel coupling of $\bar{D}N$ and $\bar{D}^{\ast}N$ gives an effective attraction.

\begin{figure}[tbp]
\begin{center}
\includegraphics[width=11cm,bb=0 0 548 289]{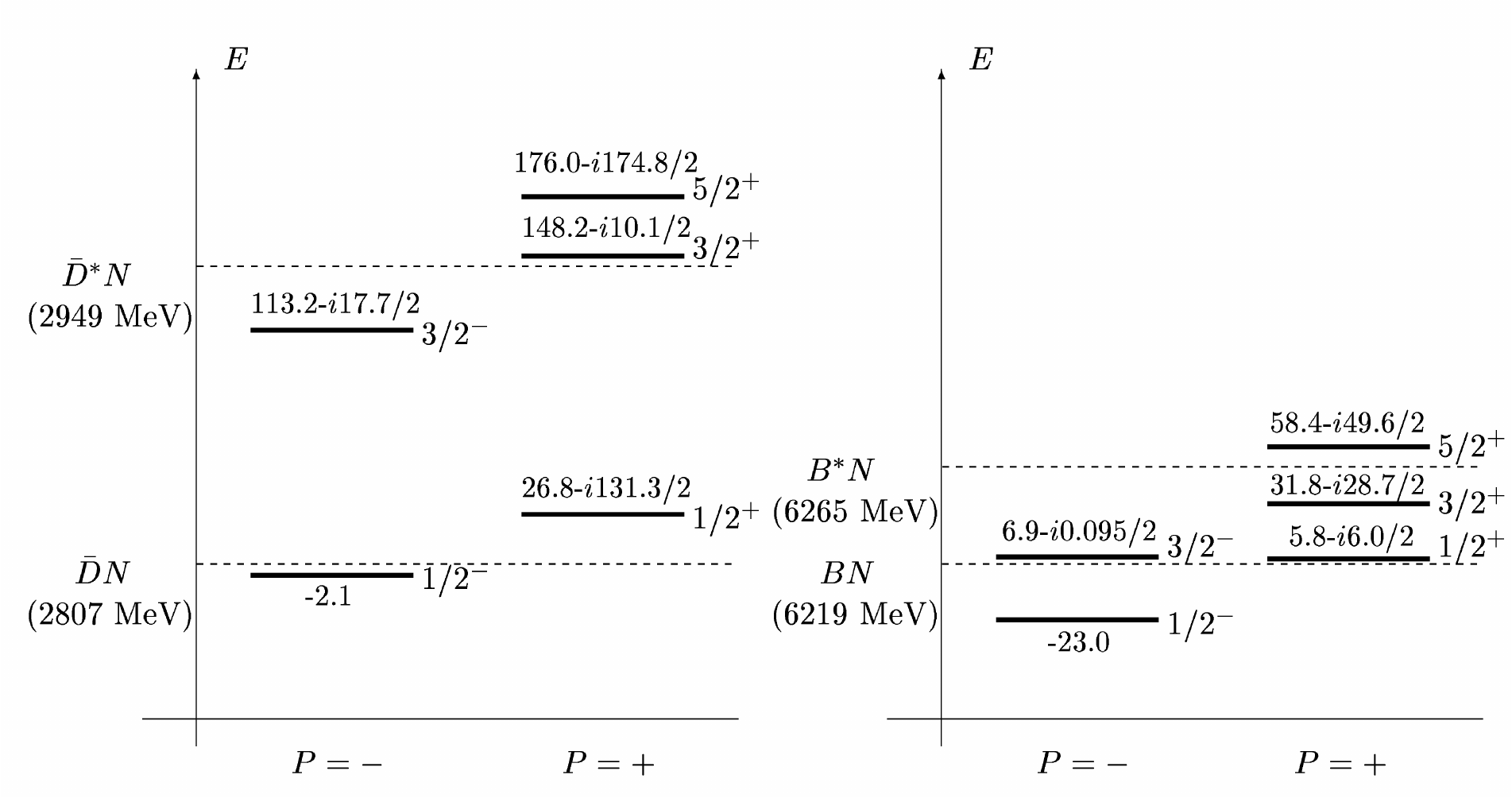}
\caption{Energy levels of $\bar{D}N$-$\bar{D}^{\ast}N$ bound/resonant states (left) and their counterpart in the bottom sector (right) in the pion-exchange model~\cite{Yamaguchi:2011qw}. }
\label{fig:Yamaguchi:2011qw}
\end{center}
\end{figure}

Other $I(J^{P})$ sectors are further studied in Refs.~\cite{Yamaguchi:2011xb,Yamaguchi:2011qw}, finding various resonances not only in the negative parity channels but also in the positive parity ones. Of particular importance is the resonance in the $I(J^{P})=0(3/2^{-})$ channel.
Interestingly, the main channel for attraction in this resonance is provided by the $\bar{D}^{\ast}N(^{4}\mathrm{S}_{3/2})$ channel.
The decay to the lowest threshold $\bar{D}N$ is much suppressed, because the decay channel $\bar{D}N(^{2}\mathrm{D}_{3/2})$ contains the $d$-wave.
Hence this is regarded as the Feshbach resonance.
 In contrast to the SU(8) contact interaction model, the decay of the $I(J^{P})=0(3/2^{-})$ state into the $\bar{D}N$ channel occurs in the pion-exchange model, thanks to the tensor force. The bottom sector can also be analyzed by replacing $\bar{D}$ by $B$. The results of the bound/resonant states are summarized in Fig.~\ref{fig:Yamaguchi:2011qw}. 

The $I(J^{P})=0(1/2^{-})$ and $I(J^{P})=0(3/2^{-})$ states become degenerate in mass in the heavy quark limit~\cite{Yasui:2013vca,Yamaguchi:2014era}.
To see this, let us rewrite the Hamiltonian by performing the transformation from the particle basis to the spin-complex basis.
From the spin rearrangement, we change $NP^{(\ast)}$ with $P=(q\bar{Q})_{\mathrm{spin}\,0}$ and $P^{\ast}=(q\bar{Q})_{\mathrm{spin}\,1}$ into the form $[Nq]\bar{Q}$ with the spin-complex $[Nq]$, and define the spin-complex wave function as
\begin{align}
\left\{ | [Nq]^{(0,\mathrm S)}_{0^+} \, \bar{Q} \rangle_{1/2^-}, | [Nq]^{(1, {\mathrm S})}_{1^+} \bar{Q} \rangle_{1/2^-}, | [Nq]^{(1,{\mathrm D})}_{1^+} \bar{Q} \rangle_{1/2^-} \right\},
\end{align}
and 
\begin{align}
\left\{ | [Nq]^{(1, {\mathrm S})}_{1^+} \bar{Q} \rangle_{3/2^-}, | [Nq]^{(1, {\mathrm D})}_{1^+} \bar{Q} \rangle_{3/2^-},  | [Nq]^{(0, {\mathrm D})}_{2^+}  \bar{Q} \rangle_{3/2^-}, | [Nq]^{(1,{\mathrm D})}_{2^+} \bar{Q} \rangle_{3/2^-} \right\},
\end{align}
  respectively, for $1/2^{-}$ and $3/2^{-}$. 
The notation $[Nq]^{(S,L)}_{j^{\cal P}}$ denotes the spin-complex composed of a nucleon and a quark (in $P^{(\ast)}$ meson) with total spin $S$, angular momentum $L$, and total angular momentum and parity $j^{\cal P}$.
The particle basis $\{ |P^{(\ast)}N(^{2S+1}L_{J})\rangle \}$ and the spin-complex basis $\{ | [Nq]^{(S,L)}_{j^{\cal P}} \bar{Q} \rangle_{J^{P}} \}$ are related by unitary matrix, $U$ for $1/2^-$ and $U'$ for $3/2^-$:
\begin{align}
\left(
\begin{array}{c}
 | PN(^{2}{\mathrm S}_{1/2}) \rangle \\
 | P^{\ast}N(^{2}{\mathrm S}_{1/2}) \rangle \\
 | P^{\ast}N(^{2}{\mathrm D}_{1/2}) \rangle
\end{array}
\right) = U
\left(
\begin{array}{c}
 | [Nq]^{(0,\mathrm S)}_{0} \, \bar{Q} \rangle_{1/2^-} \\
 | [Nq]^{(1, {\mathrm S})}_{1} \bar{Q} \rangle_{1/2^-} \\
 | [Nq]^{(1,{\mathrm D})}_{1} \bar{Q} \rangle_{1/2^-}
\end{array}
\right),
\hspace{0.5em}
\mathrm{with}
\hspace{0.5em}
U=
\left(
\begin{array}{ccc}
 -\frac{1}{2} & \frac{\sqrt{3}}{2} & 0  \\
 \frac{\sqrt{3}}{2} & \frac{1}{2} & 0  \\
 0 & 0 & -1   
\end{array}
\right),
\end{align}
for $1/2^{-}$, and
\begin{align}
\left(
\begin{array}{c}
 | PN(^{2}{\mathrm D}_{3/2}) \rangle \\
 | P^{\ast}N(^{4}{\mathrm S}_{3/2}) \rangle \\
 | P^{\ast}N(^{4}{\mathrm D}_{3/2}) \rangle \\
 | P^{\ast}N(^{2}{\mathrm D}_{3/2}) \rangle
\end{array}
\right) = U'
\left(
\begin{array}{c}
 | [Nq]^{(1, {\mathrm S})}_{1} \bar{Q} \rangle_{3/2^-} \\
 | [Nq]^{(1, {\mathrm D})}_{1} \bar{Q} \rangle_{3/2^-} \\
 | [Nq]^{(0, {\mathrm D})}_{2} \bar{Q} \rangle_{3/2^-} \\
 | [Nq]^{(1, {\mathrm D})}_{2} \bar{Q} \rangle_{3/2^-} 
\end{array}
\right),
\hspace{0.5em}
\mathrm{with}
\hspace{0.5em}
U'=
\left(
\begin{array}{cccc}
 0 & \frac{\sqrt{6}}{4} & \frac{1}{2} & \frac{\sqrt{6}}{4} \\
 1 & 0 & 0 & 0 \\
 0 & \frac{1}{\sqrt{2}} & 0 & -\frac{1}{\sqrt{2}} \\
 0 & \frac{1}{2\sqrt{2}} & - \frac{\sqrt{3}}{2} & \frac{1}{2\sqrt{2}}
\end{array}
\right),
\end{align}
for $3/2^{-}$.
We note that the $s$-wave channels, $PN(^{2}\mathrm{S}_{1/2})$, $P^{\ast}N(^{2}\mathrm{S}_{1/2})$ and $P^{\ast}N(^{4}\mathrm{S}_{3/2})$, have been discussed in the contact interaction with SU(8) symmetry. 
Let us rewrite the Hamiltonians $\tilde{H}_{1/2^-}$ and $\tilde{H}_{3/2^-}$, which are defined by Eqs.~(\ref{eq:OPEP_DbarN_1/2}) and (\ref{eq:OPEP_DbarN_3/2}) by setting $\Delta M=0$ in the heavy quark limit, in terms of the spin-complex basis.
They are transformed as
\begin{align}
 U^{-1} \tilde{H}_{1/2^-} U 
&=
\left(
\begin{array}{c|cc}
 K_{0} \!-\! 3\,{C} & 0 & 0 \\
 \hline
 0 & K_{0} \!+\! {C} & -2\sqrt{2} \,{T} \\
 0 & -2\sqrt{2} \,{T} & K_{2} \!+\!  ({C} \!-\! 2\,{T})
\end{array}
\right) \nonumber \\
&=
{\mathrm{diag}}( \tilde{H}_{1/2^-}^{(0^+)},\tilde{H}_{1/2^-}^{(1^+)} ), \\
  U'^{-1} \tilde{H}_{3/2^-} U' 
&=
\left(
\begin{array}{cc|cc}
 K_{0} \!+\! {C} & 2\sqrt{2}\,{T} & 0 & 0 \\
 2\sqrt{2}\,{T} & K_{2} \!+\! ({C} \!-\! 2\,{T}) & 0 & 0 \\
\hline
 0 & 0 & K_{2} \!-\! 3\,{C} & 0 \\
 0 & 0 & 0 & K_{2} \!+\! ({C} \!+\! 2\,{T})
\end{array}
\right) \nonumber \\
&=
{\mathrm{diag}}( \tilde{H}_{3/2^-}^{(1^+)},\tilde{H}_{3/2^-}^{(2^+)} ),
\end{align}
with the block-diagonal forms.\footnote{
We notice that the off-diagonal terms in $\tilde{H}^{(2^{+})}_{3/2^{-}}$ vanish because the one pion-exchange potential is used. 
When other interactions are employed,
 the off-diagonal terms in $H_{J^P}^{(j^{\cal P})}$ may exist in general.
In our model space, we consider only $Nq$ for the spin-complex.}
Even when the other components such as $\Delta q$, $N \pi q$ are considered, 
the Hamiltonians are block-diagonalized with the spin-complex basis,
 as far as the heavy quark symmetry is respected.
From the transformed Hamiltonian with the spin-complex basis,
we immediately find that $\tilde{H}_{1/2^-}^{(1^+)}$ and $\tilde{H}_{3/2^-}^{(1^+)}$ have exactly the same eigenvalues.
This is reasonable because the same spin-complex with $1^{+}$ is contained in both $1/2^-$ and $3/2^-$ states.
Hence we confirm that the $1/2^{-}$ and $3/2^{-}$ states should be degenerate in the heavy quark limit, as indicated by the small mass difference between $1/2^{-}$ and $3/2^{-}$ states, as shown in Fig.~\ref{fig:Yamaguchi:2011qw}.
The small mass differences in the other channels can be explained analogously.

The non-exotic $DN$ sector is studied in Ref.~\cite{Yamaguchi:2013ty}. To concentrate on the molecular states around the $DN$ threshold, the lower energy $\pi\Sigma_{c}$ channels are not included. In the $I(J^{P})=0(1/2^{-})$ sector, the pion exchange gives a bound state around 14 MeV below the $DN$ threshold. The binding energy further increases to $\sim 80$ MeV by the $\rho$ and $\omega$ exchanges. No bound state is found in the $I=1$ sector. In contrast to the $\bar{D}N$ sector, the correspondence with the results of other models is not clear (see Table~\ref{tbl:DNbound}), partly because of the difference of the model space.

\paragraph{Chiral quark models}

In the chiral constituent quark models, the short range part of the hadron interactions can be studied. This model is successfully applied to the nuclear forces~\cite{Valcarce:2005em}, and in particular, to the $\Delta \Delta$ channel~\cite{Garcilazo:1997mf,Mota:1999qp,Valcarce:2001in} which is related to the resonance structure in the $pn\rightarrow d\pi^{0}\pi^{0}$ reaction~\cite{Bashkanov:2008ih}, known as the Abashian-Booth-Crowe (ABC) effect~\cite{Abashian:1960zz,Booth:1961zz}.

In Refs.~\cite{Carames:2012bd,Fontoura:2012mz,Carames:2014cna}, the $\bar{D}N$ scattering is studied in the chiral quark models. The $\bar{D}N$ interaction is constructed by the short range quark exchange mechanism, together with the long range meson exchanges. In the $(I,J^{P})=(0,1/2^{-})$ sector, the quark exchange works repulsively. As a consequence, no bound state is found in the $(I,J^{P})=(0,1/2^{-})$ channel.
A bound state is found as in the pion-exchange model~\cite{Yasui:2009bz,Yamaguchi:2011qw} if the quark exchange is turned off.
However, the repulsion by the quark exchange cancels partly the attraction of the pion exchange, 
though the total interaction is still attractive in the full model. On the other hand, the quark exchange is attractive in the $(I,J^{P})=(1,5/2^{-})$ sector. This attraction supports a $\Delta \bar{D}^{\ast}$ bound state with a binding energy of 3.87 MeV. The state can be observed as a $d$-wave resonance in the $N\bar{D}$ system. The mechanism of the binding of the high-spin state is similar to that of the $\Delta\Delta$ resonance in spin 3~\cite{Carames:2014cna}. The appearance of the shallow bound state in the $(I,J^{P})=(1,5/2^{-})$ sector is in clear contrast with the SU(8) contact interaction model  where a bound state is found about 100 MeV below the threshold.

\paragraph{Comparison of different models around threshold}

We have introduced several different approaches to the $DN/\bar{D}N$ interaction. Before closing this section, we would like to compare the prediction of different models in the low-energy limit, using the scattering lengths of the $DN/\bar{D}N$ system. The scattering length is defined by the two-body amplitude at the threshold energy. This is an observable quantity, and is accessible in principle by the lattice QCD. 
Moreover, the leading contribution to the mass of the $D/\bar{D}$ meson in the nuclear medium can be estimated by the scattering length.

To avoid confusion, let us recall the properties of the scattering length. We first show the convention of the scattering length in this paper:
\begin{align}
   a_{DN/\bar{D}N}
   &=
   \lim_{k\to 0}f_{DN/\bar{D}N}(k) ,
   \label{eq:scatteringlength}
\end{align}
where $f_{DN/\bar{D}N}(k)$ is the elastic scattering amplitude of the $DN/\bar{D}N$ channel with the momentum $k$.\footnote{In the $k\to 0$ limit, all the partial waves other than $l=0$ vanish, and the scattering amplitude is independent of the scattering angle. Our convention is commonly used in hadron physics (meson-meson/meson-baryon scattering), while the convention with an opposite sign is often used in the nuclear and atomic physics.} Because the total cross section at the threshold is given by $4\pi a_{DN/\bar{D}N}^{2}$, the scattering length determines the strength of the two-body scattering process in the zero momentum limit. In the above convention, the positive scattering length represents the attractive scattering at threshold (increasing phase shift), and the negative scattering length corresponds to the repulsive scattering (decreasing phase shift). We note that the attractive/repulsive nature of the scattering length does not always correspond to the property of the potential. If a shallow bound state is generated by an attractive potential, the scattering length is large and negative. In other words, the attractive scattering length indicates the absence of a shallow bound state near the threshold. While the $\bar{D}N$ scattering length is real, the $DN$ scattering length has an imaginary part, which expresses the inelastic scattering effect to the open $\pi\Sigma_{c}$ and $\pi\Lambda_{c}$ channels. For reference, the recent determination of the $\pi N$ scattering lengths gives $a_{\pi N}^{I=1/2}=169.8\times 10^{-3}M_{\pi}^{-1}=0.243$ fm and $a_{\pi N}^{I=3/2}=-86.3\times 10^{-3}M_{\pi}^{-1}=-0.123$ fm~\cite{Hoferichter:2015hva}. As an example of the channel with near-threshold quasi-bound state, the $\bar{K}N$ scattering length is found to be $a_{\bar{K}N}^{I=0}=-1.39+i\ 0.85$ fm~\cite{Ikeda:2011pi,Ikeda:2012au,Kamiya:2015aea} where the $\Lambda(1405)$ resonance lies below the $\bar{K}N$ threshold.

We summarize the the $DN$ scattering lengths in the SU(4) contact interaction models~\cite{Hofmann:2005sw,Mizutani:2006vq}, the SU(8) contact interaction model~\cite{GarciaRecio:2008dp}, and the meson exchange model~\cite{Haidenbauer:2010ch,Haidenbauer:2008ff} in Table~\ref{tbl:DNslength}. In the $I=0$ sector, all models give a moderately repulsive scattering length, except for the SU(8) contact interaction model which predicts very weakly attractive value. It is remarkable that the imaginary part is very small in all cases, indicating the transition to the $\pi\Sigma_{c}$ channel is suppressed. In the $I=1$ sector, the results are more scattered. The imaginary part is very small in Refs.~\cite{Hofmann:2005sw,GarciaRecio:2008dp} while it is sizable in Refs.~\cite{Mizutani:2006vq,Haidenbauer:2010ch}. A large negative scattering length of $DN(I=1)$ of the meson exchange model~\cite{Haidenbauer:2010ch} is a consequence of the shallow quasi-bound state which corresponds to $\Sigma_{c}(2800)$.

\begin{table}[tbp]
\caption{Scattering lengths in the $DN$ channel in various models. The isospin averaged scattering length $a_{D}$ is defined in Eq.~\eqref{eq:averagedslength}. All numbers are given in units of fm. The negative (positive) scattering length corresponds to the repulsive (attractive) scattering at threshold. When a shallow bound state exists, the scattering length becomes negative with a large magnitude. The results of Ref.~\cite{Hofmann:2005sw} are given in Ref.~\cite{Lutz:2005vx} where the imaginary parts are found to be negligible. The results of Refs.~\cite{Mizutani:2006vq,GarciaRecio:2008dp} are shown in Ref.~\cite{Haidenbauer:2010ch}.  }
\begin{center}
\begin{tabular}{|l|lll|}
\hline
Model & $a_{DN}^{I=0}$ & $a_{DN}^{I=1}$ & $a_{D}$\\
\hline
SU(4) contact~\cite{Hofmann:2005sw} 
  & $-0.43$  
  & $-0.41$ 
  & $-0.42$ \\
SU(4) contact~\cite{Mizutani:2006vq} 
  & $-0.57+i\ 0.001$  
  & $-1.47+i\ 0.65$  
  & $-1.25+i\ 0.49$  \\
SU(8) contact~\cite{GarciaRecio:2008dp}
  & $0.004+i\ 0.002$  
  & $0.33+i\ 0.05$ 
  & $0.29+i\ 0.038$  \\
Meson exchange~\cite{Haidenbauer:2010ch} 
  & $-0.41+i\ 0.04$    
  & $-2.07+i\ 0.57$  
  & $-1.66+i\ 0.44$   \\
\hline
\end{tabular}
\end{center}
\label{tbl:DNslength}
\end{table}%

As shown in Table~\ref{tbl:DbarNslength}, the $\bar{D}N$ scattering lengths are calculated in the SU(4) contact interaction model~\cite{Hofmann:2005sw},  the meson exchange model~\cite{Haidenbauer:2007jq}, the pion-exchange model~\cite{Yamaguchi:2011xb}, and the chiral quark model~\cite{Fontoura:2012mz}. The scattering lengths are in general not very large and comparable with the $\pi N$ sector, except for the pion-exchange model where the $I=0$ scattering length is enhanced by the near-threshold bound state (see Table~\ref{tbl:DbarNbound}). We comment that the pion-exchange model without the $\rho$ and $\omega$ exchanges provides an attractive scattering length ($a^{I=1}_{\bar{D}N}=0.22$ fm). However, the $\rho$ and $\omega$ exchanges in the diagonal component lead to the repulsive scattering length as shown in Table~\ref{tbl:DbarNslength}.

\begin{table}[tbp]
\caption{Scattering lengths in the $\bar{D}N$ channel in various models. 
The isospin averaged scattering length $a_{\bar{D}}$ is defined in Eq.~\eqref{eq:averagedslength}. All numbers are given in units of fm. The negative (positive) scattering length corresponds to the repulsive (attractive) scattering at threshold. When a shallow bound state exists, the scattering length becomes negative with a large magnitude. Results of Ref.~\cite{Yamaguchi:2011xb} are those in the $\pi\rho\omega$ model. The results of Ref.~\cite{Hofmann:2005sw} are given in Ref.~\cite{Lutz:2005vx}.}
\begin{center}
\begin{tabular}{|l|lll|}
\hline
Model & $a_{\bar{D}N}^{I=0}$ & $a_{\bar{D}N}^{I=1}$ & $a_{\bar{D}}$ \\
\hline
SU(4) contact~\cite{Hofmann:2005sw} 
  & $-0.16$  
  & $-0.26$ 
  & $-0.24$ \\
Meson exchange~\cite{Haidenbauer:2007jq} 
  & $0.07$    
  & $-0.45$ 
  & $-0.32$   \\
Pion exchange~\cite{Yamaguchi:2011xb}
  & $-4.38$  & $-0.07$ & $-1.15$ \\
Chiral quark model~\cite{Fontoura:2012mz} & $0.03$-0.16 & $0.20$-0.25 
  & $0.16$-0.23 \\
\hline
\end{tabular}
\end{center}
\label{tbl:DbarNslength}
\end{table}%

The $DN/\bar{D}N$ scattering length can be used to estimate the mass shift of the $D/\bar{D}$ in the nuclear medium. Under the linear density approximation~\cite{Dover:1971hr}, the mass shift of the $D/\bar{D}$ meson in the symmetric nuclear matter is given by
\begin{align}
\Delta m_{D/\bar{D}} = -2\pi \frac{M_{N}+m_{D}}{M_{N}m_{D}} \rho_{N} a_{D/\bar{D}},
\label{eq:scattering_massshift}
\end{align}
with the nucleon mass $M_{N}$, the $D$ meson mass $m_{D}$, and the normal nuclear matter density $\rho_{N}$. The isospin averaged scattering length is defined as
\begin{align}
  a_{D/\bar{D}} =\frac{a^{I=0}_{DN/\bar{D}N}+3a^{I=1}_{DN/\bar{D}N}}{4} 
  \label{eq:averagedslength} .
\end{align}
We see that the attractive scattering length $a_{D/\bar{D}}>0$ (repulsive scattering length $a_{D/\bar{D}}<0$) induces the decrease (increase) of the $D/\bar{D}$ mass in nuclear matter. In Tables~\ref{tbl:DNslength} and \ref{tbl:DbarNslength}, we show the results of the averaged scattering lengths~\eqref{eq:averagedslength}. We note that the the scattering length in the $I=1$ channel is important for the in-medium property of the $D/\bar{D}$ meson, because of the larger weight in Eq.~\eqref{eq:averagedslength}.
We however remind that Eq.~\eqref{eq:scattering_massshift} is a simple estimation, and more detailed analysis of the mass shift will be discussed in section~\ref{sec:Dinmatter}. In Ref.~\cite{Hayashigaki:2000es}, the two-body scattering length is evaluated by the QCD sum rules, in order to study the in-medium modification of the $D$ meson mass. The averaged scattering length of $D$ and $\bar{D}$ is estimated as $(a_{D}+a_{\bar{D}})/2 = 0.72 \pm 0.12 \hspace{0.5em} \mathrm{fm}$. This suggests the decrease of the averaged mass of $D$ and $\bar{D}$ by about $\simeq -48 M \pm 8$ MeV, while the later studies indicate the increase of the $D$ meson masses. Again, thorough discussion on the mass shift will be given in section~\ref{sec:Dinmatter}.

\subsection{Few-body systems}

We have seen several studies with an attractive $DN/\bar{D}N$ interaction, some of which predict a (quasi-)bound state below the threshold. These observations suggest the possible formation of a bound state of $D/\bar{D}$ with a few nucleons. If it exists, detailed investigations of the few-body systems will provide another clue to understand the two-body interaction, as in the case of the hypernuclei and the $\Lambda N$ interaction.

In Ref.~\cite{Bayar:2012dd}, the $DNN$ three-body system is studied with the $DN$ interaction of Ref.~\cite{Mizutani:2006vq} where the $DN$ system has a quasi-bound state of $\Lambda_{c}(2595)$ in the $I=0$ channel. The three-body system is solved by two techniques: a variational calculation based on the Gaussian expansion method as explained in Sect.~\ref{sec:fewbody_systems} and the fixed center approximation (FCA) to the Faddeev equation as developed in Refs.~\cite{Bayar:2011qj,Bayar:2012rk}. In both methods, a narrow quasi-bound state of $DNN$ is found around 3500 MeV in the $I(J^{P})=1/2(0^{-})$ channel. The $I(J^{P})=1/2(1^{-})$ channel is unbound with respect to the $\Lambda_{c}(2595)N$ threshold. The width from the two-body absorption process $DNN\to \Lambda_{c}N$ is evaluated in the FCA calculation and found to be several tens of MeV. By analyzing the wave function obtained in the variational calculation, it is found that the $DN(I=0)$ pair in the quasi-bound $DNN$ system has a similar structure with the $\Lambda_{c}(2595)$ in free space. This is a characteristic feature observed in the $\bar{K}NN$ quasi-bound state~\cite{Dote:2008in,Dote:2008hw}. The FCA approach is also applied to other three-body systems with the $D$ meson, such as $NDK$, $\bar{K}DN$, $ND\bar{D}$ systems~\cite{Xiao:2011rc}, showing the existence of several quasi-bound states.

\begin{figure}[tbp]
\begin{center}
\includegraphics[width=9cm,bb=0 0 1459 1277]{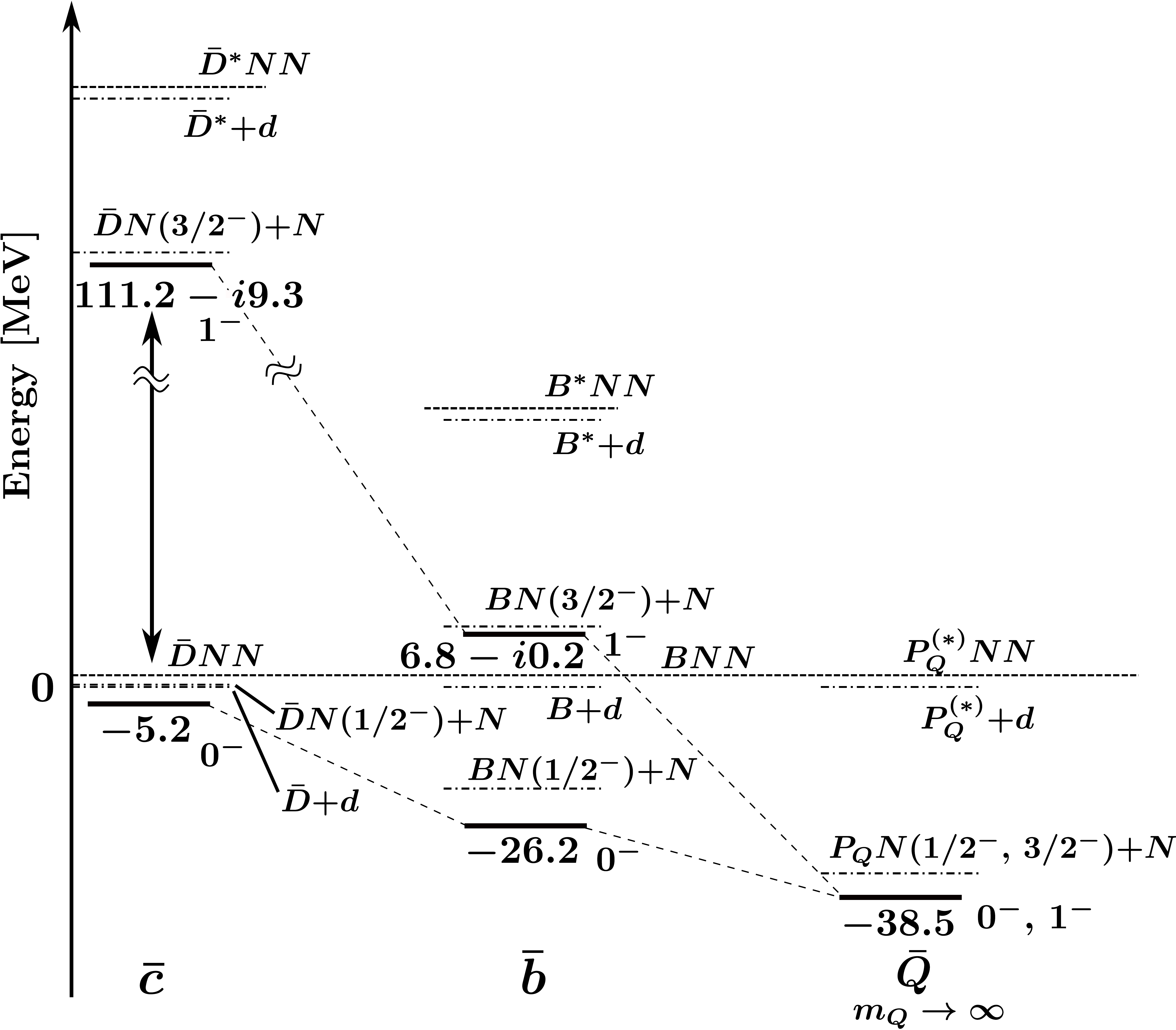}
\caption{Energy levels of $\bar{D}^{(\ast)}NN$, $B^{(\ast)}NN$ and  $P^{(\ast)}NN$ with $I=1/2$ and $J^{P}=0^{-}$ and $1^{-}$ (solid lines)~\cite{Yamaguchi:2013hsa}. The complex energies for resonances are given as $E_{re}-i\Gamma/2$, where $E_{re}$ is a resonance energy and $\Gamma/2$ is a half decay width. Thresholds (subthresholds) are denoted by dashed (dash-dotted) lines.}
\label{fig:threebody}
\end{center}
\end{figure}

The $\bar{D}NN$-$\bar{D}^{\ast}NN$ system is studied in Ref.~\cite{Yamaguchi:2013hsa} which indicates the existence of bound and resonance states in the three-body system. The $\bar{D}N$-$\bar{D}^{\ast}N$ interaction is given by the pion-exchange potential~\cite{Yasui:2009bz,Yamaguchi:2011xb,Yamaguchi:2011qw}, and the $NN$ interaction is adopted by the Argonne $v_{8}'$ (AV8') interaction~\cite{Pudliner:1997ck}. 
The Argonne $v_{8}'$ interaction includes the tensor force explicitly, as the pion-exchange potential is essential in nuclei. As a result, in the $I(J^{P})=1/2(0^{-})$ channel, a bound state is found at 5.2 MeV below the $\bar{D}NN$ threshold. In the $I(J^{P})=1/2(1^{-})$ channel, a resonance state is found at 111.2 MeV above the $\bar{D}NN$ threshold, with a 18.6 MeV decay width. As in the case of the $\bar{D}N$-$\bar{D}^{\ast}N$ system, it is found that the tensor force plays an important role in the $\bar{D}NN$-$\bar{D}^{\ast}NN$ system. The binding energy mostly comes from the tensor force in the $\bar{D}N$-$\bar{D}^{\ast}N$ system, while the central force is dominant in the $NN$ pair rather than the tensor force. The energy levels of the three-body $\bar{D}NN$-$\bar{D}^{\ast}NN$ system are summarized in Fig.~\ref{fig:threebody}, together with the $BNN$-$B^{*}NN$ system and $PNN$-$P^{*}NN$ which represents the $m_{Q}\to \infty$ limit. It is shown that the $I(J^{P})=1/2(1^{-})$ resonance in the charm sector degenerates with the bound $I(J^{P})=1/2(0^{-})$ state in the heavy quark limit. Thus, these states form a heavy quark spin doublet as a consequence of the heavy quark symmetry in the formulation~\cite{Yasui:2013vca,Yamaguchi:2014era}. 

It is interesting to note that the lowest energy state is found to be the state with total spin $J=0$ in both the $DNN$ and $\bar{D}NN$ systems. In the dominant $s$-wave $DNN/\bar{D}NN$ component in the $J=0$ state, the two nucleons are combined into the ${}^{1}S_{0}$ state. On the other hand, in the $NN$ system without $D/\bar{D}$, the lowest energy state is the bound deuteron in the ${}^{3}S_{1}$ channel, not the unbound ${}^{1}S_{0}$ channel. This means that, by adding $D/\bar{D}$, the the lowest energy configuration of the two-nucleon system changes from ${}^{3}S_{1}$ to ${}^{1}S_{0}$. The reason is attributed to the stronger $DN/\bar{D}N$ attraction in the $I=0$ channel than that in the $I=1$ channel. By analyzing the isospin decomposition, it is found that the $I(J^{P})=1/2(0^{-})$ channel has larger fraction of $I=0$ $DN/\bar{D}N$ pair than the $I(J^{P})=1/2(1^{-})$ channel~\cite{Bayar:2012dd}. This is analogous to the $\bar{K}NN$ system~\cite{Dote:2008in,Dote:2008hw} which also favors the $I(J^{P})=1/2(0^{-})$ state as the ground state. In this way, the injection of $D/\bar{D}$ causes the \textit{structure transition} of the two-body correlation of nucleons. A thorough investigation of the few-body systems will elucidate the property of the hadronic interactions inside the system.

\subsection{Nuclear matter}\label{sec:Dinmatter}

In general, the nuclear medium effect can shift the mass of the heavy-light meson, and can broaden the width of the spectrum. In some cases, an additional mode arises from the particle-hole type excitation in the same quantum numbers with the heavy-light meson. All these effects are included in the in-medium spectral function $S_{D/\bar{D}}(q_{0},\vec{q};\rho)$, which is related to the in-medium propagator $D_{D/\bar{D}}(q_{0},\vec{q};\rho)$ and the in-medium self-energy $\Pi_{D/\bar{D}}(q_{0},\vec{q};\rho)$ as
\begin{align}
  S_{D/\bar{D}}(q_{0},\vec{q};\rho) = -\frac{1}{\pi} \text{Im } D_{D/\bar{D}}(q_{0},\vec{q};\rho),
  \label{se:spectralfunction}
\end{align}
and
\begin{align}
  D_{D/\bar{D}}(q_{0},\vec{q};\rho)
  =
  \frac{1}{q_{0}^{2}-\vec{q}^{\,2}-m^{2}-\Pi_{D/\bar{D}}(q_{0},\vec{q};\rho)} ,
  \label{eq:propagator}
\end{align}
where $q$ is the four-momentum of the hadron, $m$ is the mass of the meson in vacuum, and $\rho$ is the nuclear matter density. A useful quantity for the application to the finite nuclei is the optical potential, which is also related to the self-energy as 
\begin{align}
   V_{\mathrm{opt}\ D/\bar{D}}(r,q^{0}) 
   = \frac{1}{2q^{0}} \Pi_{D/\bar{D}}(q^{0},\vec{q}=\vec{0},\rho(r)),
   \label{eq:opticalpotential}
\end{align}
where $\rho(r)$ is the density of the nucleus under the local density approximation. Thus, the in-medium self-energy $\Pi_{D/\bar{D}}(q_{0},\vec{q},\rho)$ is an essential quantity and is evaluated in various approaches. For instance, $\Pi_{D/\bar{D}}(q_{0},\vec{q},\rho)$ can be evaluated by the scalar and vector mean fields in nuclear matter. Hadronic interaction models and QCD sum rules are applied to evaluate $\Pi_{D/\bar{D}}(q_{0},\vec{q};\rho)$ through the forward scattering amplitude of the $DN/\bar{D}N$ system which is related to the self-energy at low densities [See Eq.~\eqref{eq:scattering_massshift}]. In the following, we overview the results of various approaches for the in-medium properties of the heavy-light mesons. Hereafter, the normal nuclear matter density is denoted as $\rho_{0}\sim 0.17 \text{ fm}^{-3}$.

\paragraph{Quark-meson coupling (QMC) model}

The first study of the $\bar{D}/D$ meson in nuclei is carried out in Ref.~\cite{Tsushima:1998ru} based on the quark-meson coupling (QMC) model~\cite{Guichon:1987jp}. The QMC model can describe the pion emission decays of the charmed baryons~\cite{Ivanov:1998qe}. This QMC model is used to study the modification of charm hadrons in nuclear matter. It is noted that the scalar meson ($\sigma$) exchange plays an important role for the relation between the modification of the hadron masses and the change of the chiral condensate in nuclear medium. The wave function of the light quarks ($u$, $d$, $\bar{u}$ and $\bar{d}$) and the charm quarks ($c$ and $\bar{c}$) inside the bag ($|\vec{x}\,|\le R$ with the bag radius $R$) follows the Dirac equations
\begin{align}
 \left( i\gamma  \!\cdot\! \partial  - \left( m_{q}-V_{\sigma}^{q}(\vec{x}\,) \right) - \gamma^{0} \left( V_{\omega}^{q}(\vec{x}\,) \pm    
\frac{1}{2} V_{\rho}^{q}(\vec{x}\,) \right) \right)
\psi_{u,d}(x)&=0, \\
 \left( i\gamma  \!\cdot\! \partial  - \left( m_{q}-V_{\sigma}^{q}(\vec{x}\,) \right) + \gamma^{0} \left( V_{\omega}^{q}(\vec{x}\,) \pm \frac{1}{2} V_{\rho}^{q}(\vec{x}\,) \right) \right)
\psi_{\bar{u},\bar{d}}(x)&=0, \\
 \left( i\gamma  \!\cdot\! \partial - m_{c} \right) \psi_{c,\bar{c}}(x)
 &=0,
\end{align}
where 
\begin{align}
 V_{\sigma}^{q}(\vec{x}\,) = g_{\sigma}^{q} \sigma(\vec{x}\,), \quad
 V_{\omega}^{q}(\vec{x}\,) = g_{\omega}^{q} \omega(\vec{x}\,), \quad
 V_{\rho}^{q}(\vec{x}\,) = g_{\rho}^{q} \rho(\vec{x}\,),
\end{align}
and $g_{M}^{q}$ is the coupling constant of the meson $M$ and the light quark $q$. The mean field of the meson $M(\vec{r}\,)$ is obtained by self-consistently solving the set of nonlinear equations of mesons and nucleon (the bag including quarks) as in the Serot-Walecka model~\cite{Serot:1984ey,Serot:1997xg} (See Eqs.~(23)-(30) in Ref.~\cite{Saito:1996sf}). The Coulomb interaction is included in this procedure. While the coupling constants are basically determined by the nuclear matter properties, it is found that the $\omega$ meson potential should be phenomenologically modified as $\tilde{V}_{\omega}^{q}=(1.4)^{2}V_{\omega}^{q}$ for the $K^{+}$ meson~\cite{Tsushima:1997df}. In Ref.~\cite{Tsushima:1998ru}, both $\tilde{V}_{\omega}^{q}$ and $V_{\omega}^{q}$ are used to examine the $D$ meson bound states in nuclei.

\begin{figure}[tbp]
 \begin{minipage}{0.5\hsize}
   \centering
   \includegraphics[scale=0.4,bb=0 0 488 374]{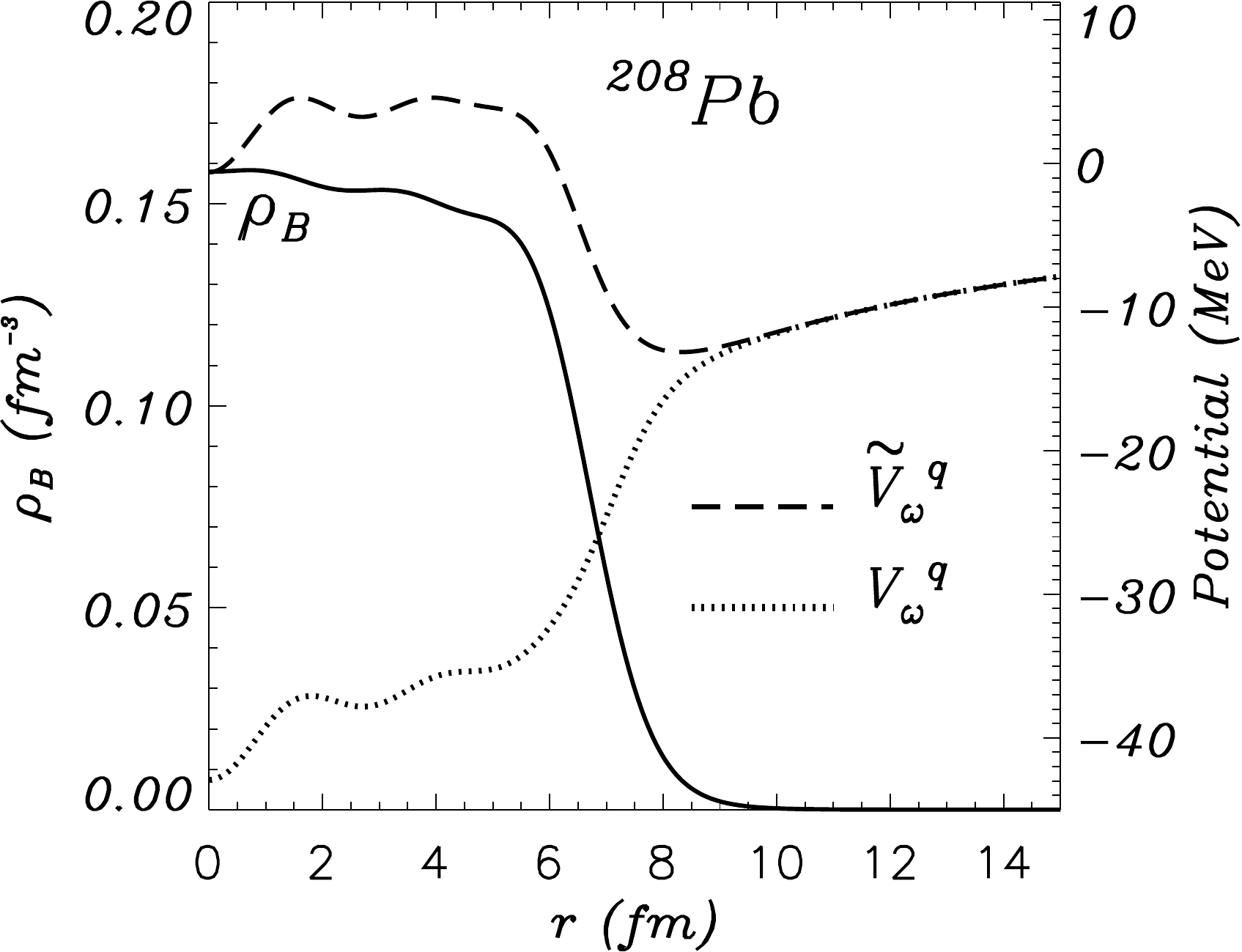}
 \end{minipage}
 \begin{minipage}{0.5\hsize}
    \centering
   \includegraphics[scale=0.37,bb=0 0 564 459]{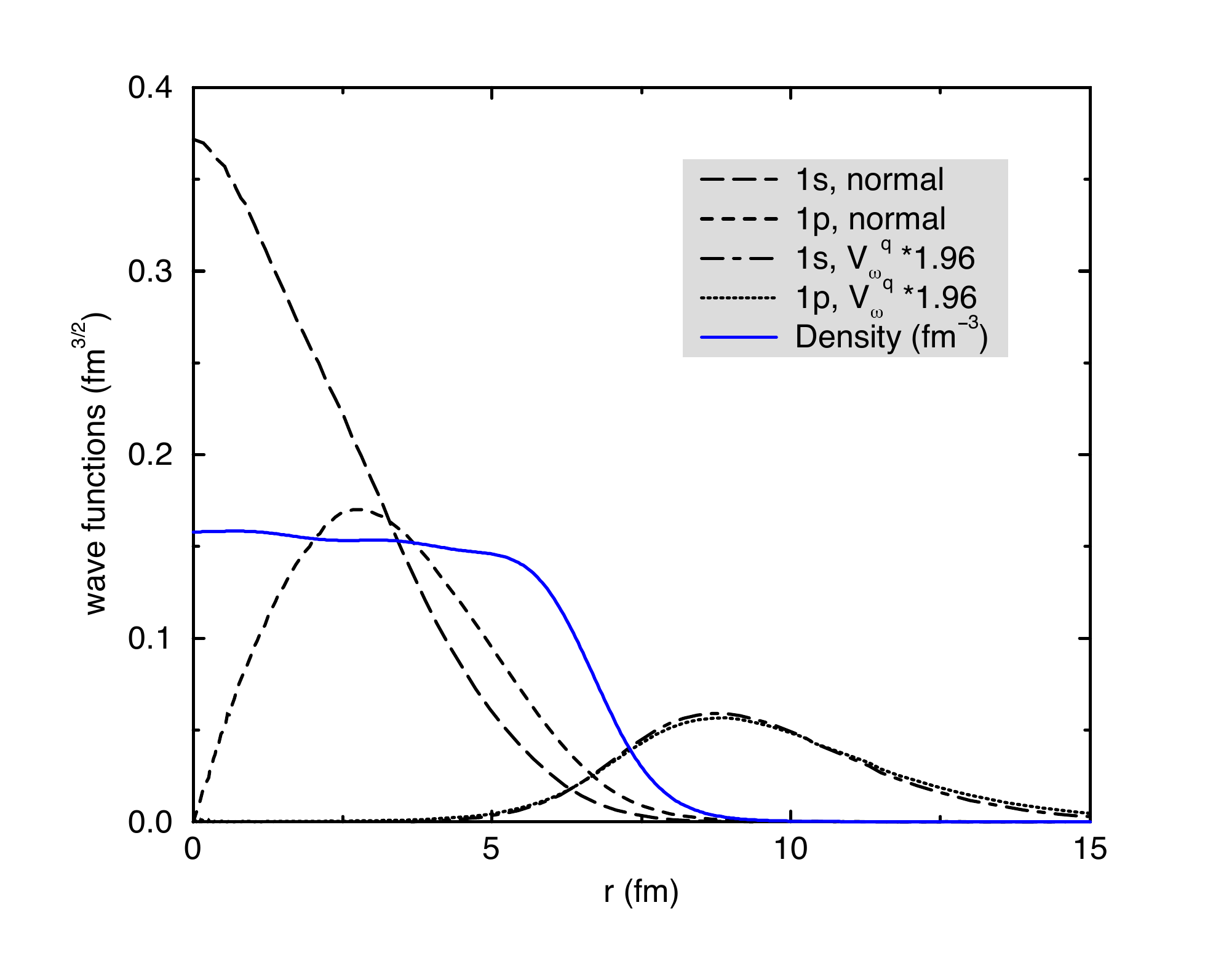}
 \end{minipage}
  \caption{The mean-field potentials (left) and the wave functions (right) of the bound $D^{-}$ ($\bar{D}$) meson in $^{208}\mathrm{Pb}$ in the QMC model \cite{Tsushima:1998ru}. In the left panel, the solid line represents the nuclear density distribution, the dotted line stands for the sum of the scalar, vector and Coulomb potentials for the $D^{-}$ meson, and the dashed line shows the total potential with the modified vector potential $\tilde{V}_{\omega}^{q}$. In the right panel, solid line represents the nuclear density distribution, the long dashed (short dashed) line represents the wave function of the $D^{-}$ meson in the 1s (1p) level, and the dashed-dotted (dotted) line shows the wave function of the $D^{-}$ meson in the 1s (1p) level with the modified vector potential $\tilde{V}_{\omega}^{q}$.}
  \label{fig:Tsushima:1998ru}
\end{figure}

In the left panel of Fig.~\ref{fig:Tsushima:1998ru}, the mean field potential with $V_{\omega}^{q}$ ($\tilde{V}_{\omega}^{q}$) for the $D^{-}$ meson in $^{208}\mathrm{Pb}$ is shown by the dotted (dashed) line. The attractive nature of the potential with $V_{\omega}^{q}$ is evident. The modification of the $\omega$ potential $V_{\omega}^{q}\to \tilde{V}_{\omega}^{q}$ largely cancels the attraction at small $r$, but still an attractive pocket exists at around $r\sim 8$ fm. By solving the Schr\"odinger-like equation for the $D^{-}$ meson, the bound state is found with 35 MeV binding (10 MeV binding) for the $V_{\omega}^{q}$ ($\tilde{V}_{\omega}^{q}$) case. Corresponding $D^{-}$ wave functions are shown in the right panel of Fig.~\ref{fig:Tsushima:1998ru}. It should be noted that the binding is caused mainly by the Coulomb interaction. In fact, the neutral $\bar{D}^{0}$ is unbound for the potential with $\tilde{V}_{\omega}^{q}$ which indicates that the $\bar{D}$ meson feels repulsion in the nuclear medium. In other words, the $D^{-}$ meson could be bound in the heavy nucleus due to the Coulomb force, even though the strong interaction is repulsive as a whole. In contrast to the $\bar{D}$ meson, the $\omega$ potential works with an opposite sign for the $D$ meson, leading to the binding energy of the $D^{0}$ meson of about 100 MeV. Note however that the channel coupling to the various decay modes, such as $DN\rightarrow Y_{c}\pi$, should be considered in the $D^{0}$ meson case. 

\begin{figure}[tbp]
\begin{center}
\includegraphics[scale=0.3,bb=0 0 529 621]{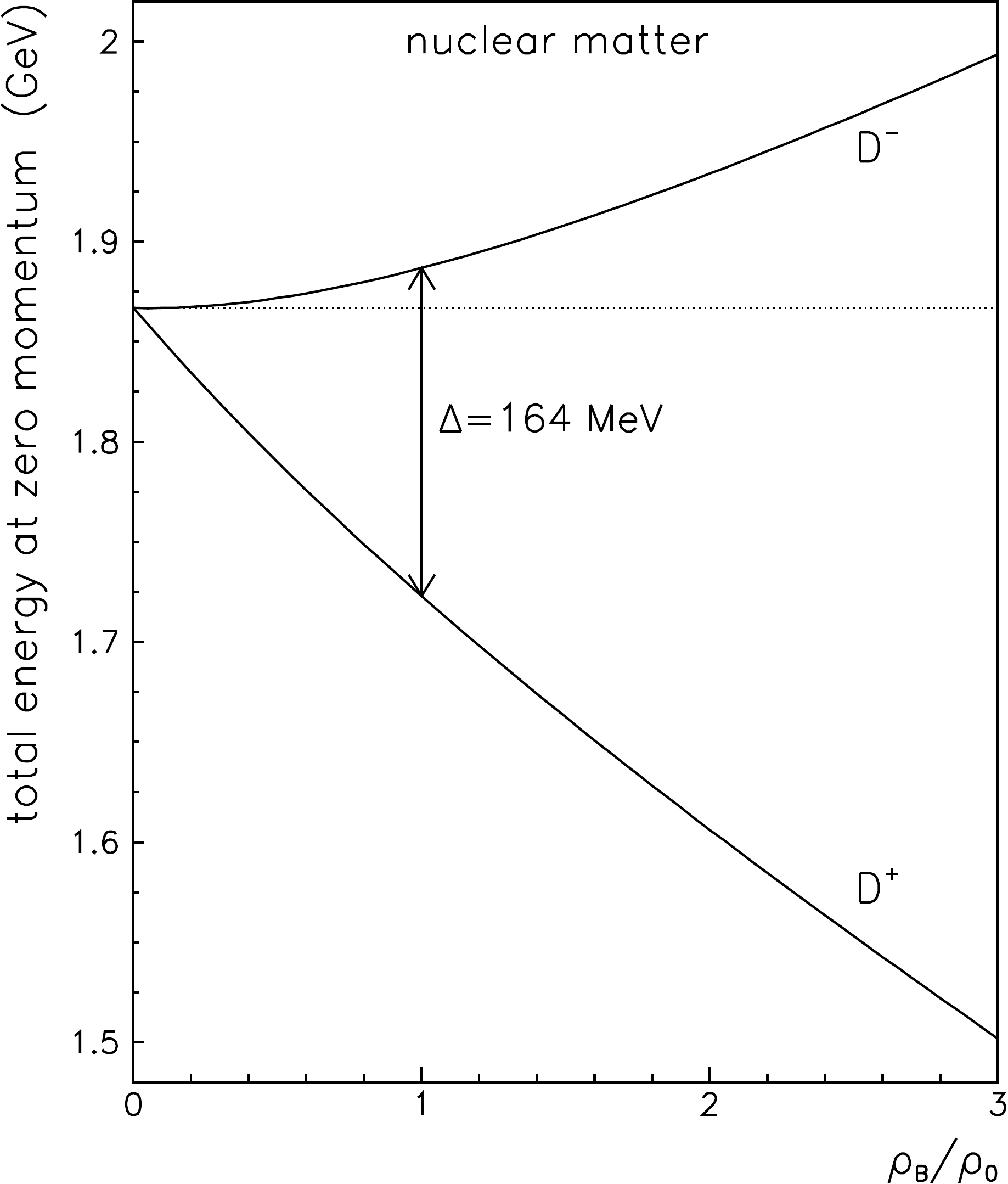}
\caption{Masses of $D^{-}$ ($\bar{D}$) meson and $D^{+}$ ($D$) meson in the QMC model~\cite{Sibirtsev:1999js}.}
\label{fig:Sibirtsev:1999js}
\end{center}
\end{figure}

The in-medium mass of the $D/\bar{D}$ meson is calculated in Ref.~\cite{Sibirtsev:1999js}. The modified $\tilde{V}_{\omega}^{q}$ is used for the $\omega$ potential. As shown in Fig.~\ref{fig:Sibirtsev:1999js}, the $D$ meson ($\bar{D}$ meson) feels attractive (repulsive) interaction in the nuclear medium. These behaviors are similar to those of the $\bar{K}/K$ meson studied in Ref.~\cite{Tsushima:1997df}. We note that the mixing of $D$ and $D^{*}$ is not included.

\paragraph{Nuclear mean-field approach}

In Refs.~\cite{Mishra:2003se,Mishra:2008cd,Kumar:2010gb,Kumar:2011ff}, the in-medium $D/\bar{D}$ meson is studied by the SU(4) generalization of the SU(3) mean-field approach~\cite{Mishra:2008kg,Mishra:2008dj}. The study is successively generalized to include the isospin asymmetry~\cite{Mishra:2008cd}, finite temperature~\cite{Kumar:2010gb}, and the strange hadronic matter environment~\cite{Kumar:2011ff}. The interaction Lagrangian of the $D/\bar{D}$ meson consists of the direct coupling to the nucleon (the Weinberg-Tomozawa vector-type term and the scalar-type term) and the coupling to the scalar-isoscalar mesons [$\sigma=f_{0}(600)$, $\zeta=f_{0}(980)$] and the scalar-isovector meson [$\delta=a_{0}(980)$].\footnote{While the coupling to the vector $\omega$ meson is considered in Ref.~\cite{Mishra:2003se}, it is omitted in the later studies~\cite{Mishra:2008cd,Kumar:2010gb,Kumar:2011ff} to avoid the double counting with the Weinberg-Tomozawa term.} The $D/\bar{D}$ meson self-energy $\Pi(\omega,|\vec{k}\,|)$ is evaluated by the mean-field approximation,\footnote{This self-energy $\Pi(\omega,|\vec{k}\,|)$ has an opposite sign from $\Pi_{D/\bar{D}}(q_{0},\vec{q};\rho)$ which is used in the other part of this subsection.} i.e., the nucleon and scalar meson fields are replaced by their expectation values in nuclear matter. The effect of the gluon condensate is considered through the trace anomaly with the ``dilaton'' field $\chi$~\cite{Cohen:1991nk,Kumar:2010hs}:
\begin{align}
 \left\langle \frac{\beta_{\mathrm{QCD}}}{2g} G^{a}_{\mu} G^{\mu\nu a} \right\rangle
 + \sum_{i} m_{i} \bar{q}_{i}q_{i}
=-(1-d) \chi^{4},
\end{align}
where the ``dilaton'' field couples to the scalar mesons. In Refs.~\cite{Mishra:2008cd,Kumar:2010gb,Kumar:2011ff}, the in-medium dispersion relation is obtained by solving the pole condition
\begin{align}
-\omega^{2} + \vec{k}^{\,2} + m_{D}^{2} - \Pi(\omega,|\vec{k}\,|) = 0 .
\end{align}
The in-medium mass is defined as $\omega(\vec{k}=\vec{0})$. Note that the mass shift is real for both $D$ and $\bar{D}$ in the mean-field approximation, because the self-energy does not have an imaginary part. In the symmetric nuclear matter, a negative mass shift is observed for both  $D$ and $\bar{D}$. The $D$ meson mass drops faster than the $\bar{D}$ meson. This is because the vector-type Weinberg-Tomozawa interaction gives attraction for the $D$ meson, while it is repulsive for the $\bar{D}$ meson. In the isospin asymmetric nuclear matter with $\eta=(\rho_{n}-\rho_{p})/2\rho_{B}\neq0$, the in-medium properties of the $D^{-}(\bar{c}d)$ and $\bar{D}^{0}(\bar{c}u)$ [$D^{0}(c\bar{u})$ and $D^{+}(c\bar{d})$] are different, reflecting the third component of the isospin. The difference is mainly caused by the Weinberg-Tomozawa contact interaction, and the contribution from the scalar-isovector exchange is small. In the neutron-rich nuclear matter ($\eta  > 0$; $n_{d}>n_{u}$), the masses of both $\bar{D}$ mesons ($D^{-}$ and $\bar{D}^{0}$) increases. On the other hand, for the $D$ mesons, the mass of $D^{0}$ increases but the mass of $D^{+}$ decreases (see also Ref.~\cite{Yasui:2012rw} for the masses of the $\bar{D}$ and $B$ mesons in the asymmetric nuclear matter in the pion-exchange model). It is shown that the $D$ meson is more sensitive to the isospin asymmetry than the $\bar{D}$ meson.

\paragraph{Contact interaction models with SU(4) symmetry}

There are several works studying the in-medium properties of the $D/\bar{D}$ meson using the contact interaction models introduced in section~\ref{sec:DNinteraction}. The common strategy is to include the Pauli blocking effect for the nucleon propagator and to treat the $D/\bar{D}$ meson self-energy in the ``self-consistent'' framework~\cite{Lutz:1997wt,Ramos:1999ku}. Let us briefly demonstrate this procedure for the $\bar{D}$ meson as an example. Ignoring the nucleonic correlation effects in nuclear matter, the in-medium self-energy of the $\bar{D}$ meson is given by the in-medium T-matrix as
\begin{align}
   \Pi_{\bar{D}}(q_{0},\vec{q};\rho)
   = \int \frac{{\mathrm d}^{3}p}{(2\pi)^{3}}
   n(\vec{p},\rho) \left( T_{\bar{D}N}^{(I=0)}(P_{0},\vec{P};\rho) 
   + 3T_{\bar{D}N}^{(I=1)}(P_{0},\vec{P};\rho) \right),
   \label{eq:Dbarselfenergy}
\end{align}
where $n(\vec{p},\rho)$ is the nucleon occupation (1 for the nucleons below the Fermi momentum and 0 for those above the Fermi momentum), $P_{0}=q_{0}+E_{N}(\vec{p})$ and $\vec{P}=\vec{q}+\vec{p}$. The in-medium T-matrix $T_{\bar{D}N}(P_{0},\vec{P};\rho) $ is obtained by solving the scattering equation in medium
\begin{align}
   T_{\bar{D}N}(P_{0},\vec{P};\rho) 
   =V+V\tilde{G}_{\bar{D}N}(P_{0},\vec{P};\rho) T_{\bar{D}N}(P_{0},\vec{P};\rho),
   \label{eq:inmediumamplitude}
\end{align}
where $V$ is the interaction kernel used to calculate the vacuum T-matrix in Eq.~\eqref{eq:scatteringequation}. The medium effect is included in the loop function
\begin{align}
   \tilde{G}_{\bar{D}N}(P_{0},\vec{P};\rho)
   &= i\int \frac{{\mathrm d}q_{0}{\mathrm d}^{3}q}{(2\pi)^{3}}
   D_{\bar{D}}(q_{0},\vec{q};\rho)
   S_{N}(P_{0}-q_{0},\vec{P}-\vec{q};\rho),
\end{align}
where $S_{N}(p_{0},\vec{p};\rho)$ is the in-medium nucleon propagator 
with Pauli blocking effect [cf. Eq.~\eqref{eq:propagator_nnucleon_medium}], and $D_{\bar{D}}(q_{0},\vec{q};\rho)$ is the $\bar{D}$ propagator which is related to the self-energy through Eq.~\eqref{eq:propagator}. In this way, The right-hand-side of Eq.~\eqref{eq:Dbarselfenergy} also depends on the $\bar{D}$ meson self-energy $\Pi_{\bar{D}}(q_{0},\vec{q},\rho)$. Thus, by solving the set of equations self-consistently, $\Pi_{\bar{D}}(q_{0},\vec{q},\rho)$ is determined. In the case of the $D$ meson, we should note that the $DN$ channel couples with $\pi Y_{c}$ channels. Medium modifications can also be applied to the other hadrons such as pions in the coupled channels.

\begin{figure}[tbp]
\begin{center}
\includegraphics[scale=0.9,bb=0 0 335 247]{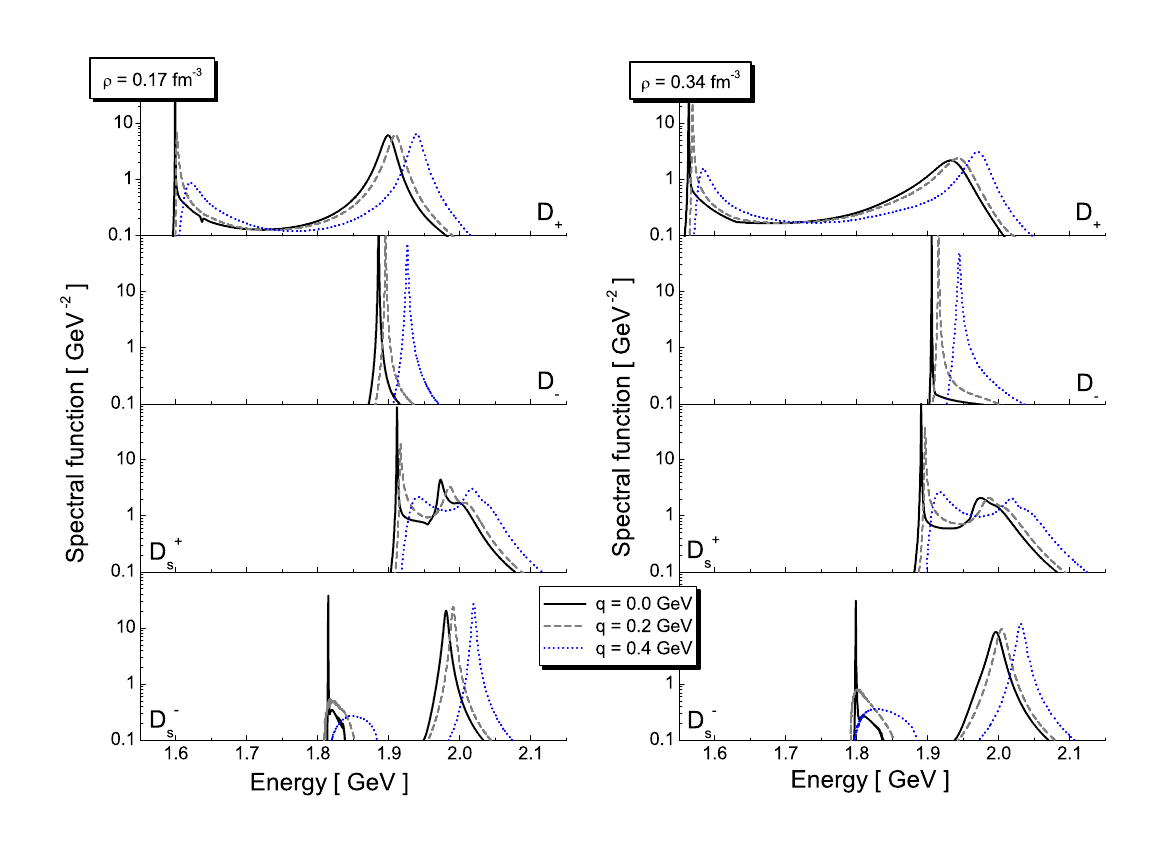}
\caption{Spectral functions of the $D$, $\bar{D}$, $D_{s}$ and $\bar{D}_{s}$ mesons at $\rho=0.17$ fm$^{-3}$ (left) and $0.34$ fm$^{-3}$ (right) in the SU(4) contact interaction model~\cite{Lutz:2005vx}.}
\label{fig:Lutz:2005vx}
\end{center}
\end{figure}

In Ref.~\cite{Lutz:2005vx}, the medium effect for the $D/\bar{D}$ meson is studied within the self-consistent framework developed in Ref.~\cite{Lutz:2001dq}, together with  the SU(4) contact interaction model of Ref.~\cite{Hofmann:2005sw}. As for the $\bar{D}$ meson, a simple estimation using the low-density theorem~\eqref{eq:scattering_massshift}, together with the scattering lengths shown in Table~\ref{tbl:DbarNslength}, leads to the mass shift of the $\bar{D}$ meson at nuclear density $\rho$ as $\Delta m_{\bar{D}} \simeq 17 \rho/\rho_{0}$ MeV, where $\rho_{0}$ is the normal nuclear matter density. The spectral function of the full self-consistent treatment is shown in Fig.~\ref{fig:Lutz:2005vx}. The corresponding value of the mass shift at $\rho=\rho_{0}$ is found to be 18 MeV, in good agreement with the above estimation. The spectral function of the $D$ meson is also shown in Fig.~\ref{fig:Lutz:2005vx}. The $D$ meson mode is pushed up by about 32 MeV from the free-space mass ($\sim 1867$ MeV), in accordance with the repulsive scattering length $a_{D}<0$ in Table~\ref{tbl:DNslength}. Another branch appears at lower energies, corresponding to the resonance-hole modes associated with the $I=0$ and $I=1$ resonances in the model of Ref.~\cite{Hofmann:2005sw} (see Table~\ref{tbl:DNbound}). The properties of the charm-strange mesons are also discussed. As mentioned in the previous subsection,
 there exists a $D_{s}N$ ($\bar{D}_{s}N$) quasi-bound state at 2892 (2780) MeV. As a consequence, the spectral functions of both the $D_{s}$ and $\bar{D}_{s}$ mesons (see Fig.~\ref{fig:Lutz:2005vx}) show two-peak structure which stems from the combination of the original $D_{s}/\bar{D}_{s}$ mode and the resonance-hole mode.

\begin{figure}[tb]
\begin{center}
\includegraphics[scale=1.1]{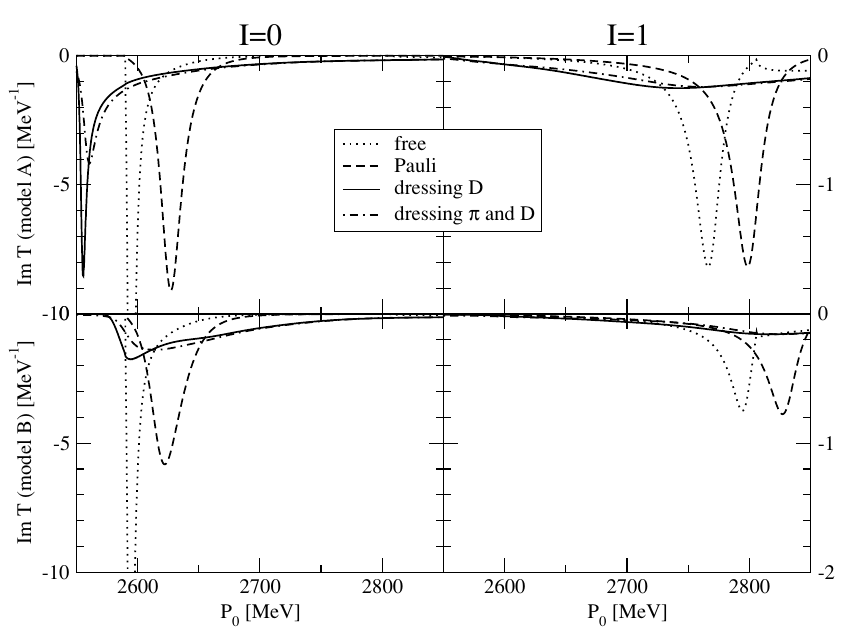}
\caption{Imaginary part of the in-medium $DN$ scattering amplitude $T_{\bar{D}N}(P_{0},\vec{P}=\vec{0};\rho)$ for I=0 (left) and I=1 (right) in model A (top) and in model B (down) at $\rho=\rho_{0}$, based on the SU(4) contact interaction model~\cite{Mizutani:2006vq}.}
\label{fig:Mizutani:2006vq}
\end{center}
\end{figure}

In Ref.~\cite{Mizutani:2006vq}, the constructed SU(4) contact interaction model is also applied to study the medium effect of the $D$ meson. In this study, medium modification of the pion propagator in the $\pi Y_{c}$ loops is also considered. In addition to the vector-type contact interaction, the scalar-isoscalar interaction is also considered to examine the model dependence. By fitting the $\Lambda_{c}(2595)$ resonance in the $I=0$ sector, they construct two models in which the scalar interaction is switched on (model A) and switched off (model B). The imaginary part of the in-medium $DN$ scattering amplitude $T_{DN}(P_{0},\vec{P}=\vec{0};\rho)$ at $\rho=\rho_{0}$ is shown in Fig.~\ref{fig:Mizutani:2006vq} with various medium effects applied. In both models, it is found that the Pauli blocking shifts the $\Lambda_{c}(2595)$ peak to the higher energy region, and the $D$ meson dressing pushes it down, as in the case of the $\Lambda(1405)$ in the $\bar{K}N$ sector~\cite{Lutz:1997wt,Ramos:1999ku}. The $D$ meson spectral function shows the two-peak structure, as observed in Ref.~\cite{Lutz:2005vx}.


\begin{figure}[tbp]
\begin{center}
\includegraphics[scale=0.3,bb=0 0 679 492]{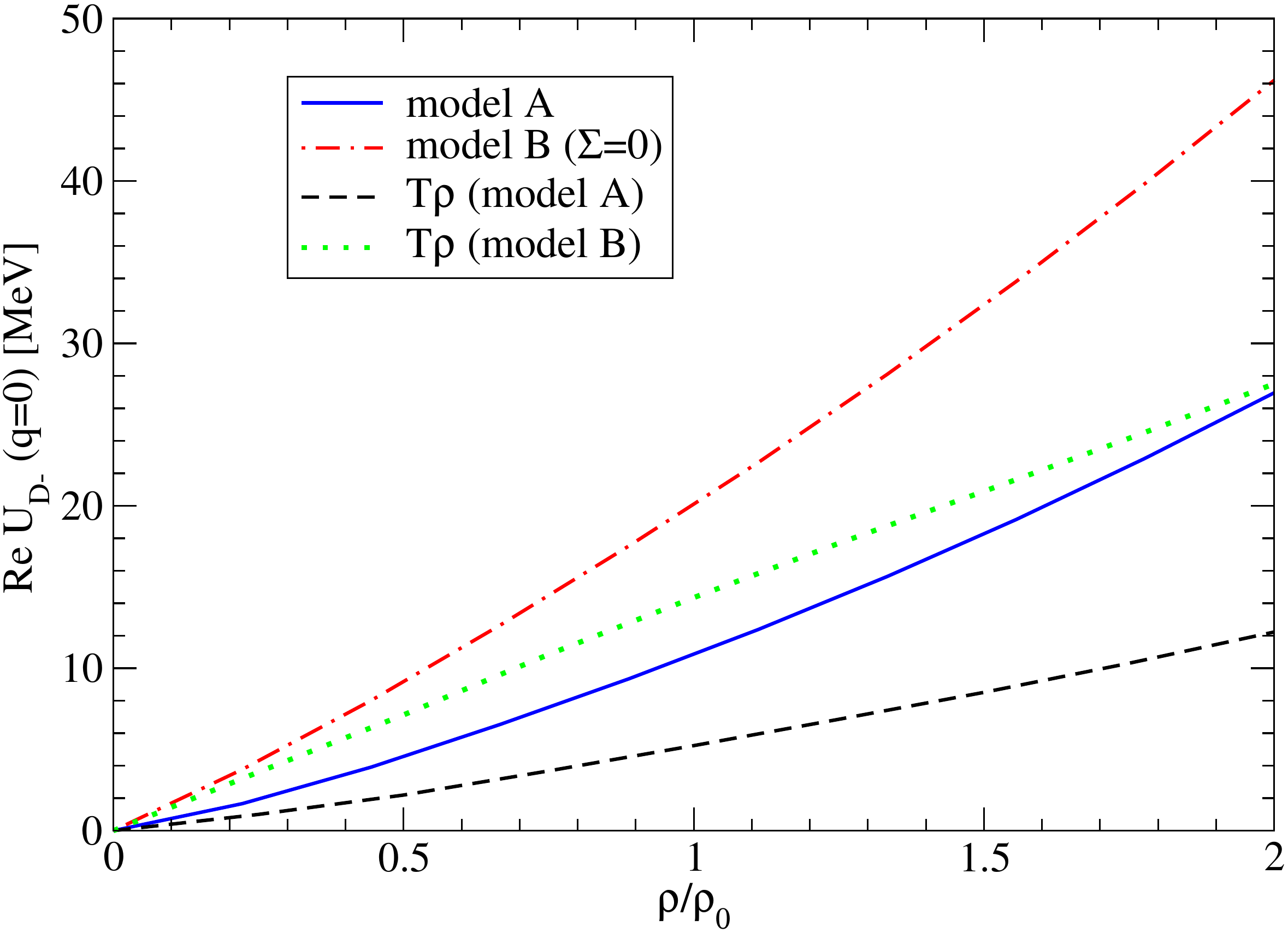}
\caption{The mass modification of the $\bar{D}$ meson in nuclear matter in the SU(4) contact interaction model~\cite{Tolos:2007vh}.}
\label{fig:Tolos:2007vh}
\end{center}
\end{figure}

In Ref.~\cite{Tolos:2007vh}, the same SU(4) contact interaction model is used to study the medium modification of both $D$ and $\bar{D}$. The mean-field binding for the baryons is considered by the Walecka-type $\sigma$-$\omega$ model~\cite{Kapusta:2006pm} and the finite temperature effect is also included. The results of the $D$ meson spectral function are similar to those of Ref.~\cite{Mizutani:2006vq} at zero temperature. On the other hand, at $T=100$ MeV, the $\Lambda_{c}(2595)$ peak in the in-medium T-matrix is smeared out and the resonance-hole mode in the $D$ meson spectral function disappears. In Fig.~\ref{fig:Tolos:2007vh}, the mass shift of the $\bar{D}$ meson is shown with (model A) and without (model B) the scalar-isoscalar interaction, estimated by the optical potential~\eqref{eq:opticalpotential}. The mass shift of $\bar{D}$ is repulsive, irrespective to the choice of the parameter set. It is shown that the low-density approximation (denoted as $T\rho$) quantitatively deviates from the self-consistent calculation.

\begin{figure}[tb]
\begin{center}
\includegraphics[height=5cm, scale=0.7,bb=3 49 391 229]{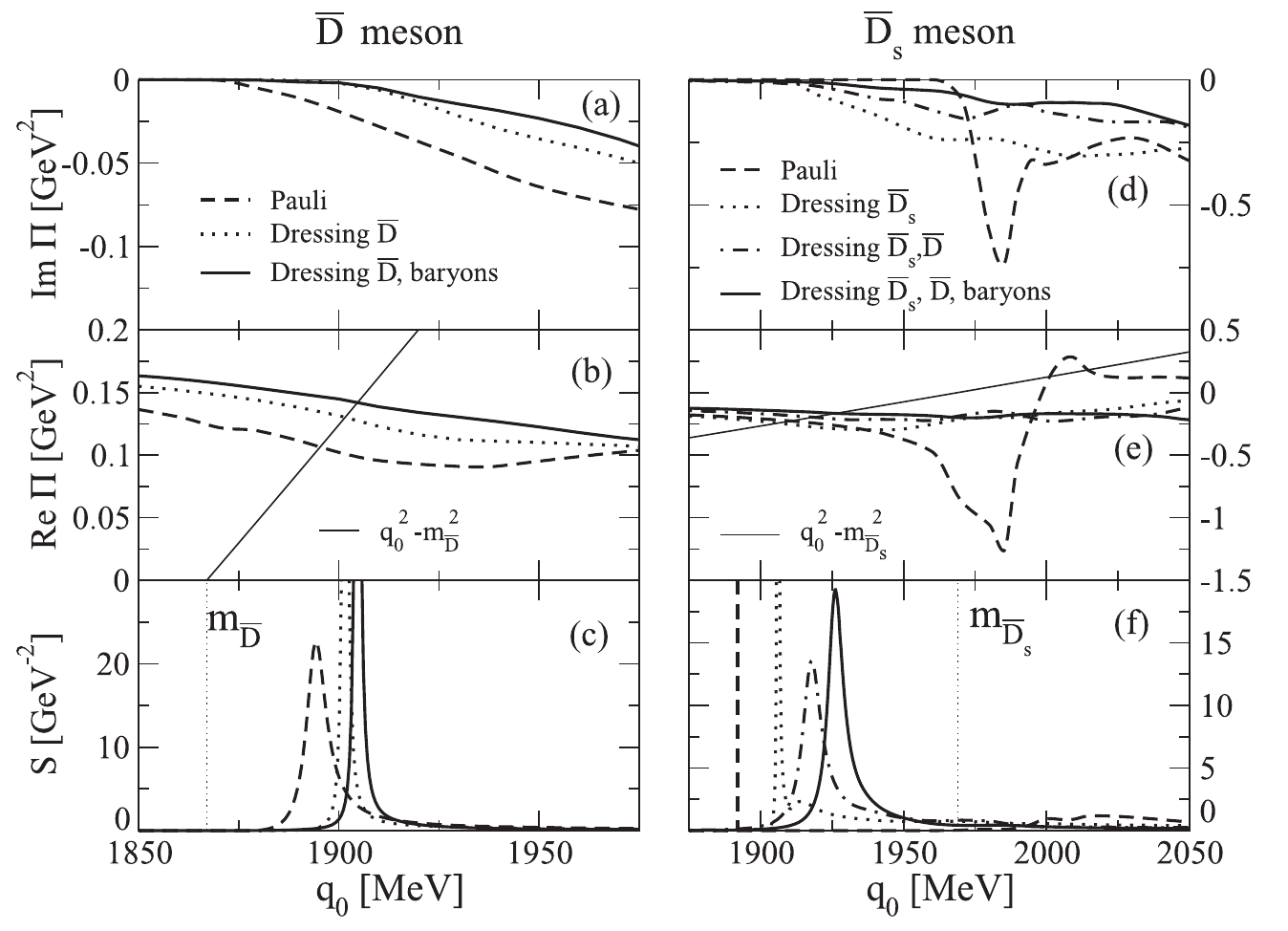}
\vspace*{1.5cm}
\caption{The self-energy and the spectral function of $\bar{D}$ (left) and $\bar{D}_{s}$ (right) mesons  at $\rho=\rho_{0}$ in the SU(4) contact interaction model \cite{JimenezTejero:2011fc}.}
\label{fig:JimenezTejero:2011fc_Fig4}
\end{center}
\end{figure}

The medium effect for $D/\bar{D}$ and $D_{s}/\bar{D}_{s}$ is studied in Ref.~\cite{JimenezTejero:2011fc} with the nonlocal interaction model~\cite{JimenezTejero:2009vq}. In this paper, the interplay between  $D/\bar{D}$ and $D_{s}/\bar{D}_{s}$ is investigated. Because the $DN$ channel couples with the $D_{s}Y$ channel, the medium modification of $D_{s}$ has an influence on the in-medium spectrum of $D$ through the coupled-channel effect. In fact, it is explicitly demonstrated in the self-consistent approach that the $D_{s}$ meson dressing has a nonnegligible effect on the $D$ meson self-energy in the nuclear medium. The same thing happens for $\bar{D}$ and $\bar{D}_{s}$; the $\bar{D}$ dressing affect the $\bar{D}_{s}$ spectrum because the $\bar{D}_{s}N$ channel couples with the $\bar{D}\Lambda$ channel. As shown in the right panel of Fig.~\ref{fig:JimenezTejero:2011fc_Fig4}, the dressing $\bar{D}$ modifies the spectral function of $\bar{D}_{s}$. The mass of $\bar{D}$ increase by about 35 MeV. The spectrum of $\bar{D}_{s}$ shows only one distinct peak, which is different from that in Ref.~\cite{Lutz:2005vx} where two peaks appear around the vacuum $\bar{D}_{s}$ mass.

\paragraph{Contact interaction models with SU(8) symmetry}

The contact interaction model with SU(8) symmetry~\cite{GarciaRecio:2008dp,Gamermann:2010zz} is utilized in Refs.~\cite{Tolos:2009nn,GarciaRecio:2010vt} for the $D$ meson and in Ref.~\cite{GarciaRecio:2011xt} for the $\bar{D}$ meson. Because of the heavy quark symmetry, the medium modification of the $D^{*}/\bar{D}^{*}$ meson is considered simultaneously. The in-medium self-energy of the $\bar{D}^{*}$ meson is given by the $\bar{D}N$ and $\bar{D}^{*}N$ T-matrices as
\begin{align}
 \Pi_{\bar{D}^{\ast}}(q^{0},\vec{q}; \rho)
  &=
  \int \frac{\mathrm{d}^{3}p}{(2\pi)^{3}}n(\vec{p},\rho)
  \left( 
  \frac{1}{3}T_{\bar{D}^{\ast}N}^{(I=0,J=1/2)}(P^{0},\vec{P};\rho) 
  +   T_{\bar{D}^{\ast}N}^{(I=1,J=1/2)}(P^{0},\vec{P};\rho) \right. \nonumber\\*
  &\quad \left. +
  \frac{2}{3}T_{\bar{D}^{\ast}N}^{(I=0,J=3/2)}(P^{0},\vec{P};\rho) 
  + 2T_{\bar{D}^{\ast}N}^{(I=1,J=3/2)}(P^{0},\vec{P};\rho)
  \right).
\end{align}
We note that the decay $\bar{D}^{\ast} \rightarrow \bar{D}\pi$ and the in-medium decay $\bar{D}^{\ast} \rightarrow \bar{D}NN^{-1}$ with nucleon $N$ and hole $N^{-1}$ are not considered.
The latter is investigated in pion-exchange interaction~\cite{Yasui:2012rw}.

As shown in Ref.~\cite{Tolos:2009nn}, the in-medium spectral function of the $D$ meson is qualitatively similar with that in the SU(4) model, but 
richer spectrum of resonances in the SU(8) model (see Table~\ref{tbl:DNbound}) induces various resonance-hole excitations. On the other hand, there is a qualitative difference from the SU(4) models in the mass shift. The mass shift of the $D$ meson is studied by the optical potential. In contrast to the SU(4) models, the mass of the $D$ meson decreases at finite density, reflecting the attractive scattering length $a_{DN}$ as shown in Table~\ref{tbl:DNslength}. Moreover, the imaginary part is shown to be small, which is also indicated by the small imaginary part of $a_{DN}$. On the other hand, the mass of the $D^{*}$ meson is shown to increase in the nuclear medium.

\begin{figure}[tbp]
\begin{center}
\includegraphics[scale=0.3,bb=0 0 792 612]{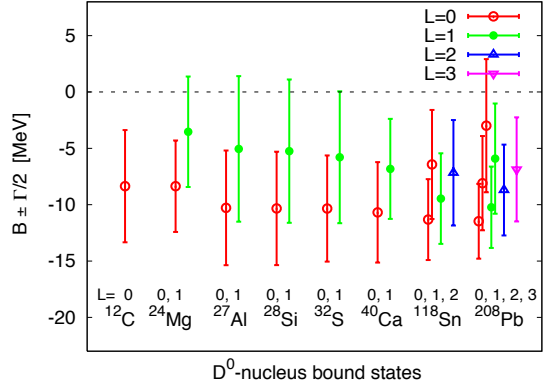}
\caption{Binding energies and widths for different $D^{0}$-nucleus states obtained using the strong energy-dependent $D$-potential with the SU(8) contact interaction model~\cite{GarciaRecio:2010vt}. The binding energy is defined as $E=B<0$ for the bound state.}
\label{fig:GarciaRecio:2010vt}
\end{center}
\end{figure}

This behavior of the $D$ meson in nuclear matter in the SU(8) model (attractive with small imaginary part) suggests the bound state formation in a finite nucleus. This possibility is pursued in Ref.~\cite{GarciaRecio:2010vt}. To calculate the $D$ bound states in finite nuclei, the energy-dependent optical potential for the $D$ meson is constructed with the local density approximation as in Eq.~\eqref{eq:opticalpotential} with the in-medium self-energy obtained in Ref.~\cite{Tolos:2009nn}. Solving the Schr\"odinger equation, the neutral $D^{0}$ bound nuclei are found. The states with widths smaller than the binding energies ($\Gamma/2<|B|$) are shown in Fig.~\ref{fig:GarciaRecio:2010vt}, which are advantageous in experimental identification. We note that the $DNN$ quasi-bound state found in Ref.~\cite{Bayar:2012dd} is qualitatively different from the $D$ nucleus in Ref.~\cite{GarciaRecio:2010vt}. The former is driven by the $DN$ attraction far from the threshold which forms $\Lambda_{c}(2595)$, while the latter is induced by the attraction of the $DN$ system near the threshold in the SU(8) model. For the positively charged $D^{+}$ meson, the Coulomb repulsion reduces the binding energies, and no state with $\Gamma/2<|B|$ is found.

\begin{figure}[tbp]
\begin{center}
\includegraphics[scale=0.8,bb=0 0 360 252]{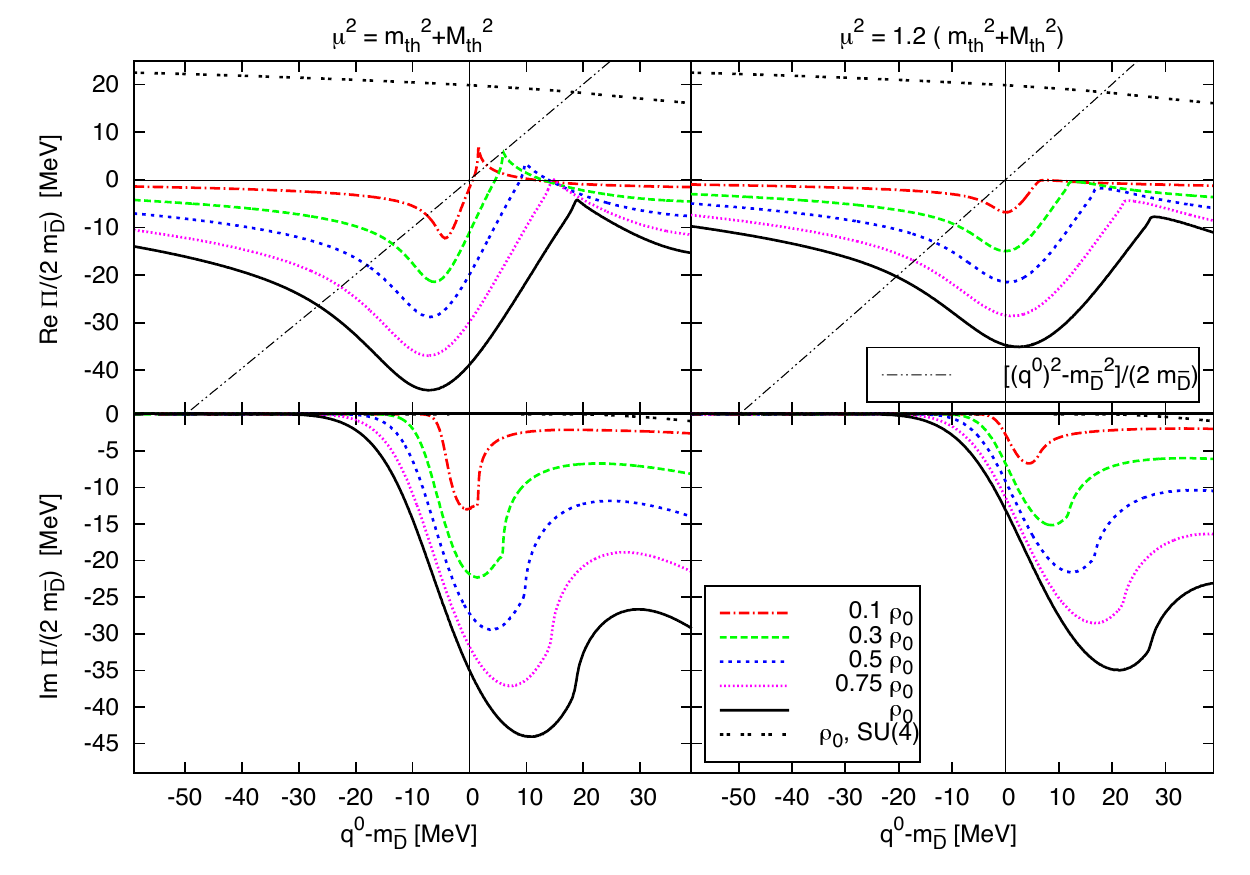}
\caption{Left panels show the real and imaginary parts of the $\bar{D}$ self-energy over $2m_{\bar{D}}$, at $\bm{q}=0$, as functions of the meson energy $q^{0}-m_{\bar{D}}$ for different densities and $\alpha=1$. Right panels show the same for $\alpha=1.2$. The oblique line is the function $[(q^{0})^{2}-m_{\bar{D}}^{2}]/(2m_{\bar{D}})$. The figure is taken from SU(8) contact interaction model of Ref.~\cite{GarciaRecio:2011xt}.}
\label{fig:GarciaRecio:2011xt_Pi}
\end{center}
\end{figure}

The result of the SU(8) model for the $\bar{D}$ meson is essentially different from the SU(4) models. As explained in section~\ref{sec:DNinteraction}, the $\bar{D}N$-$\bar{D}^{\ast}N$ system supports a bound state below the threshold, as a consequence of the coupled-channel effect~\cite{Gamermann:2010zz}. In Ref.~\cite{GarciaRecio:2011xt}, the bound state with $I(J^{P})=0(1/2^{-})$ is called $X(2805)$. The in-medium self-energy of the $\bar{D}$ meson is shown in Fig.~\ref{fig:GarciaRecio:2011xt_Pi} ($\alpha$ specifies the subtraction point). The energy of the in-medium mode is read off from the cross point of the real part of the self-energy Re $\Pi/(2m_{\bar{D}})$ with the oblique line representing $[(q^{0})^{2}-m_{\bar{D}}^{2}]/(2m_{\bar{D}})$. For instance, at $\rho=\rho_{0}$, the mode is found about $20$-$27$ MeV below the free $\bar{D}$ mass, depending on the choice of the subtraction constant. This mode can be regarded as the mixture of the in-medium $\bar{D}$ meson and the $X(2805)$-hole mode.
The imaginary part of the self-energy comes from the nucleon-hole pair creation. We note that the real part of the self-energy of the SU(4) model is always positive in the figure, so that the $\bar{D}$ meson is not bound, in accordance with Ref.~\cite{JimenezTejero:2011fc}. In addition, the strong energy dependence of the near-threshold amplitude indicates that the low-density ($T\rho$) approximation breaks down at very small density.

\begin{figure}[tbp]
 \begin{minipage}{0.5\hsize}
   \centering
   \includegraphics[scale=0.55,bb=0 0 360 252]{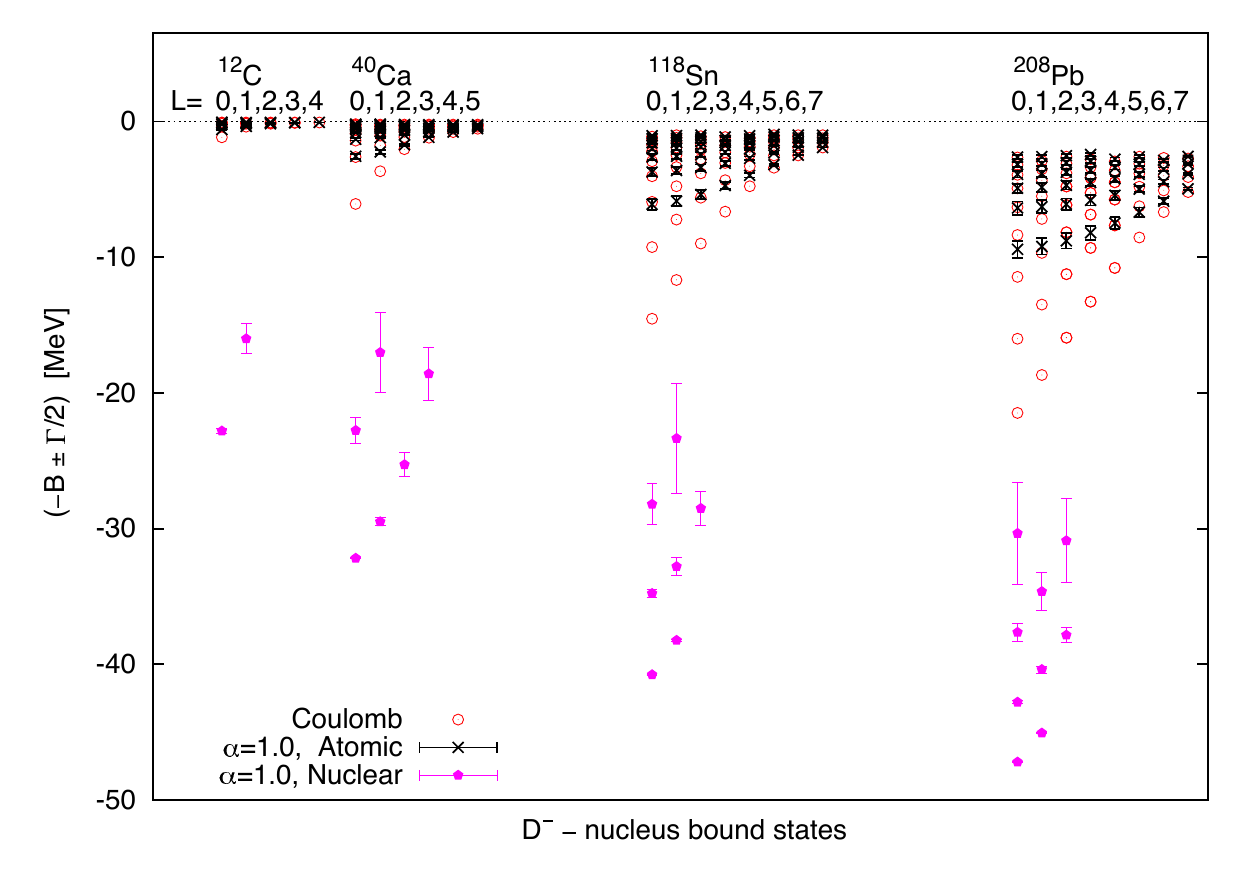}
 \end{minipage}
 \begin{minipage}{0.5\hsize}
    \centering
   \includegraphics[scale=0.55,bb=0 0 360 252]{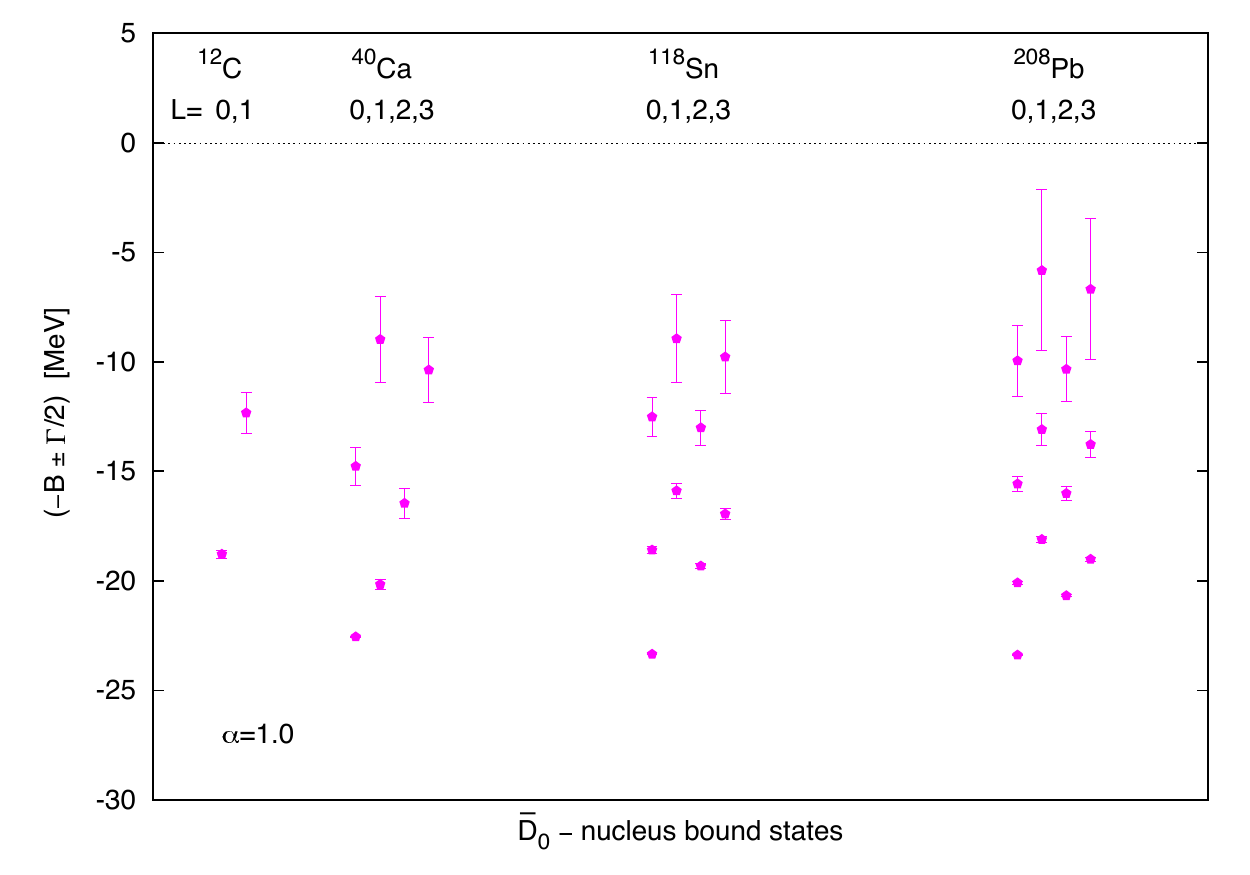}
 \end{minipage}
  \caption{The mass spectrum of $D^{-}$ (left) and $\bar{D}^{0}$ (right) mesons in atomic nuclei in the SU(8) contact interaction model~\cite{GarciaRecio:2011xt}. The binding energy is defined as $E=-B<0$ for the bound state.}
  \label{fig:GarciaRecio:2011xt_tolos}
\end{figure}

The $\bar{D}$ bound states in finite nuclei is studied by the optical potential from the self-energy $\Pi_{\bar{D}}$ shown in Fig.~\ref{fig:GarciaRecio:2011xt_Pi}. The Schr\"odinger equation for the $\bar{D}$ meson wave function $\Psi$ is written as
\begin{align}
 \left( -\frac{\vec{\nabla}^{2}}{2m_{\mathrm{red}}} +V_{\mathrm{Coul}}(r) + V_{\mathrm{opt}}(r,q^{0}) \right) \Psi
 =(-B-i\Gamma/2) \Psi,
\end{align}
where $m_{\mathrm{red}}$ is the reduced mass of the $\bar{D}$-nucleus system and $V_{\mathrm{Coul}}(r)$ is the Coulomb potential which acts only for $D^{-}$. Several bound states are obtained as shown by the filled symbols in Fig.~\ref{fig:GarciaRecio:2011xt_tolos}. In addition, for the $D^{-}$ case, there are atomic bound states indicated by crosses. Compared with the pure Coulombic ones (open circles), the energy levels of the $D^{-}$ atoms are shifted by the repulsive strong interaction effect due to the existence of the nuclear bound states. This is in accordance with the negative scattering length in Table~\ref{tbl:DbarNslength}.

\paragraph{Pion exchange models with heavy quark symmetry}

\begin{figure}[tbp]
\begin{center}
\includegraphics[scale=0.25,bb=0 0 842 595]{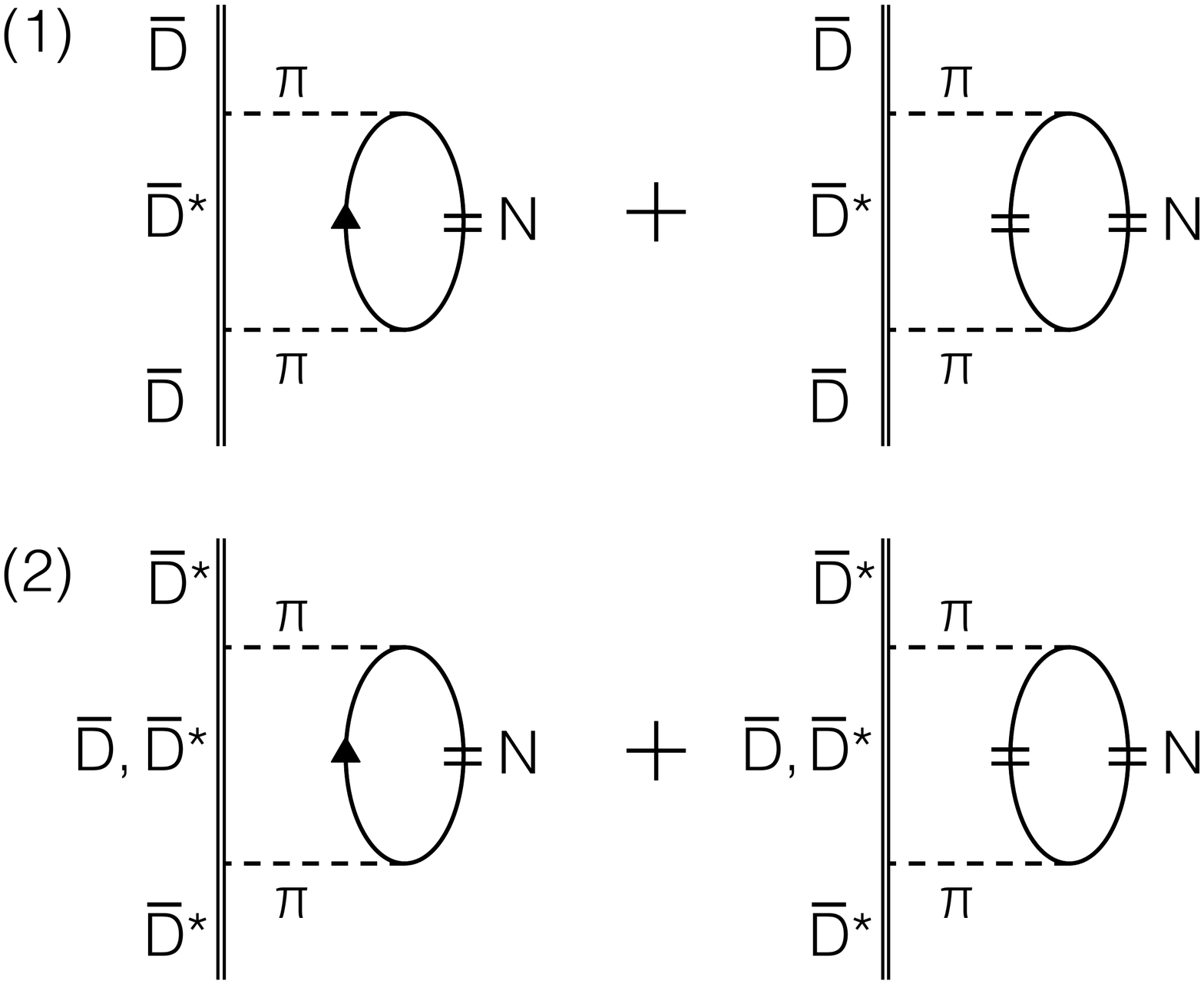}
\caption{Diagrams for the self-energies for $\bar{D}$ (top) and $\bar{D}^{*}$ (bottom) used in Ref.~\cite{Yasui:2012rw}. The line with $\parallel$ indicates the medium part, i.e. the second term in Eq.~(\ref{eq:propagator_nnucleon_medium}).}
\label{fig:Yasui:2012rw_diagram}
\end{center}
\end{figure}

The in-medium self-energies of $\bar{D}^{(\ast)}$ ($B^{(\ast)}$) is considered with the pion-exchange interaction as the longest force~\cite{Yasui:2012rw}. In the lowest order of the coupling constant of the $\bar{D}^{(\ast)}\bar{D}^{(\ast)}\pi$ vertex, the relevant diagrams are those shown in Fig.~\ref{fig:Yasui:2012rw_diagram}. The nucleon-hole excitation was included in the pion loop by the $NN\pi$ pseudovector coupling of Eq.~\eqref{eq:Yukawa}. The expression of the self-energy for the $\bar{D}$ meson is expressed by
\begin{align}
\Sigma_{\bar{\mathrm{D}}}(k_{\mathrm{F}}) =
 \Sigma_{\bar{\mathrm{D}}}^{(\bar{\mathrm{D}}^{\ast})}(k_{\mathrm{F}})
 + \Sigma_{\bar{\mathrm{D}}\,\mathrm{Pauli}}^{(\bar{\mathrm{D}}^{\ast})}(k_{\mathrm{F}}),
\end{align}
and that for the $\bar{D}^{\ast}$ meson is by
\begin{align}
\Sigma_{\bar{\mathrm{D}}^{\ast}}(k_{\mathrm{F}}) =
 \Sigma_{\bar{\mathrm{D}}^{\ast}}^{(\bar{\mathrm{D}}^{\ast})}(k_{\mathrm{F}})
 + \Sigma_{\bar{\mathrm{D}}^{\ast}}^{(\bar{\mathrm{D}})}(k_{\mathrm{F}})
 + \Sigma_{\bar{\mathrm{D}}^{\ast}\,\mathrm{Pauli}}^{(\bar{\mathrm{D}}^{\ast})}(k_{\mathrm{F}})
 + \Sigma_{\bar{\mathrm{D}}^{\ast}\,\mathrm{Pauli}}^{(\bar{\mathrm{D}})}(k_{\mathrm{F}}),
\end{align}
which correspond to (1) and (2), respectively, in Fig.~\ref{fig:Yasui:2012rw_diagram}.
The terms with superscript with the parentheses indicate the self-energy with the corresponding intermediate states, and the terms with ``Pauli" indicate the second diagram in (1) and (2).
The explicit equation forms are found in Ref.~\cite{Yasui:2012rw}.
It is important to note that there exist intermediate $\bar{D}^{\ast}$ states in the self-energy of $\bar{D}$ meson, because this intermediate state allows the $\pi$ exchange contribution for the $\bar{D}$ meson self-energy (Fig.~\ref{fig:Yasui:2012rw_diagram}~(1)).
The existence of the intermediate $\bar{D}^{\ast}$ states is important thanks to the small mass splitting between $\bar{D}$ and $\bar{D}^{\ast}$ proportional to the inverse of the heavy meson mass.
The same forms are applied to the $B$ and $B^{\ast}$ mesons by replacing their masses.
We note that the nonperturbative effect, such as the formation of the $\bar{D}N$ bound state in Ref.~\cite{Yasui:2009bz}, is not included in this scheme.

\begin{figure}[tbp]
 \begin{minipage}{0.5\hsize}
   \centering
   \includegraphics[scale=0.8,bb=-50 0 310 252]{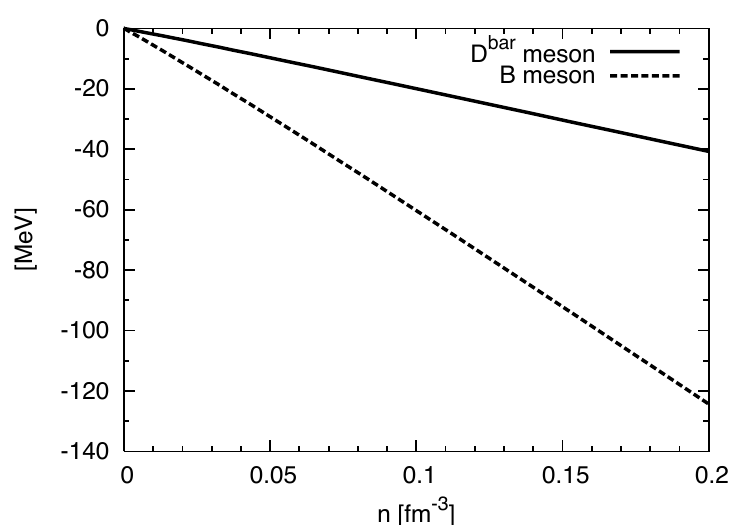}
 \end{minipage}
 \begin{minipage}{0.5\hsize}
    \centering
   \includegraphics[scale=0.8,bb=-50 0 310 252]{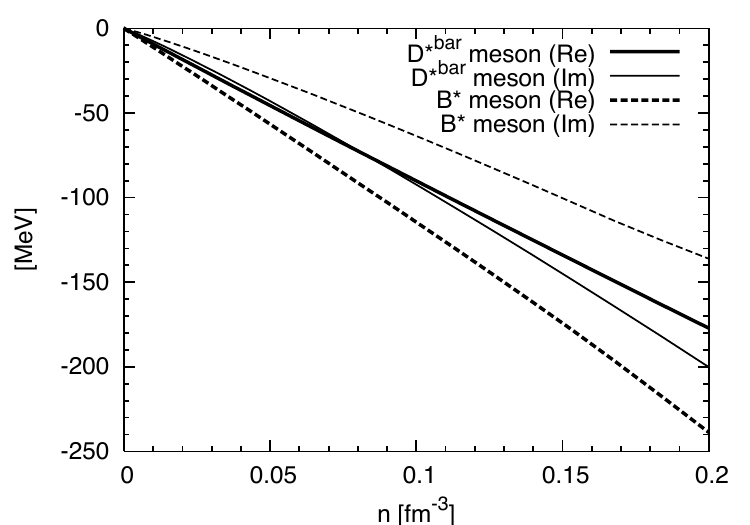}
 \end{minipage}
  \caption{The in-medium self-energies of $\bar{D}$ (left) and $\bar{D}^{\ast}$ (right) mesons in nuclear matter with baryon number density $n$, calculated by pion-exchange interaction~\cite{Yasui:2012rw}.}
  \label{fig:Yasui:2012rw}
\end{figure}

In this calculation, the first term in the self-energy $\Sigma_{\bar{\mathrm{D}}}^{(\bar{\mathrm{D}}^{\ast})}(k_{\mathrm{F}})$ of the $\bar{D}$ ($B$) meson has no imaginary part, unless the Fermi momentum exceeds the critical value $k_{F}^{\mathrm{cr}} = \sqrt{2\left(1+m_{N}/M\right)} \Delta$ with $M=(M_{P}+3M_{P^{\ast}})/4$ and  $\Delta = M_{P^{\ast}}-M_{P}$ ($P^{(\ast)}=\bar{D}^{(\ast)}$, $B^{(\ast)}$).
In the case of the $\bar{D}$ and $B$ mesons, the critical density $n_{\mathrm{cr}}=2k_{F}^{\mathrm{cr}\,2}/3\pi^{2}$ is larger than the normal nuclear density.
In any case, this imaginary part can be always canceled by the second term $\Sigma_{\bar{\mathrm{D}}\,\mathrm{Pauli}}^{(\bar{\mathrm{D}}^{\ast})}(k_{\mathrm{F}})$, as it should be.
In contrast to the $\bar{D}$ ($B$) meson, the self-energy of the $\bar{D}^{\ast}$ ($B^{\ast}$) meson always has an imaginary part, as it decays to the $\bar{D}$ ($B$) meson and the nucleon-hole pairs, $NN^{-1}$, even at the small baryon number density. It is interesting to compare the decay property of the $B^{\ast}$ meson with that of $B^{\ast}$ in vacuum. In the latter, $B^{\ast}$ cannot decay to $B\pi$ due to the closed mass threshold. In the former, however, $B^{\ast}$ decays to $B$ meson and the nucleon-hole pairs, because the minimum energy cost of nucleon-hole pairs is infinitely small due to the instability of the Fermi surface.
By calculating the self-energies, it is found that the mass of the $\bar{D}$ meson decreases at finite baryon number density and  the $\bar{D}^{\ast}$ meson mass also decreases, as shown in Fig~\ref{fig:Yasui:2012rw}. At the same time, the imaginary part of the self-energy causes the width broadening of the $\bar{D}^{\ast}$ meson.

Let us consider the above result in the heavy quark limit~\cite{Yasui:2013vca}.
In this limit, the self-energies of $\bar{D}$ and $\bar{D}^{\ast}$ mesons in Fig.~\ref{fig:Yasui:2012rw} can be compactly expressed as
\begin{align}
 -i \Sigma_{P} = 
 -\left( \frac{2g}{\sqrt{2}f_{\pi}} \right)^{2}
 \int \frac{\mathrm{d}^{4}k}{(2\pi)^{4}} \frac{k^{2}-(v\!\cdot\!k)^{2}}{2v\!\cdot\!(-k)+i\varepsilon}
 \left( \frac{1}{k^{2}-m_{\pi}^{2}+i\varepsilon} \right)^{2}
 \sum_{a,b} \tau^{a} \Pi_{\pi}^{ab}(k)\tau^{b},
\end{align}
for $P$ ($=\bar{D}$) meson and
\begin{align}
 -i \Sigma_{P^{\ast}} = -i\Sigma_{P^{\ast}}^{(P)}-i\Sigma_{P^{\ast}}^{(P^{\ast})},
\end{align}
with $\Sigma_{P^{\ast}}^{(P)}=(1/3)\Sigma_{P}$ and $\Sigma_{P^{\ast}}^{(P^{\ast})}=(2/3)\Sigma_{P}$ for $P^{\ast}$ ($=\bar{D}^{\ast}$) meson,
where the superscripts indicate the contained intermediate states.
Both $P$ and $P^{\ast}$ are defined in the heavy quark limit.
In this limit, $P^{\ast}$ meson has no imaginary part because of the mass degeneracy between $P$ and $P^{\ast}$ mesons: $M_{P}=M_{P}^{\ast}$.
The coupling constant $g$ was defined in Eq.~(\ref{eq:L_heavy_hadron_pion}), and $\Pi_{\pi}^{ab}(k)$ with isospin indices $a,b=1,2,3$ is the pion self-energy including nucleon-hole pairs in nuclear matter.
Here, the factors $1/3$ and $2/3$ included in $\Sigma_{P^{\ast}}^{(P)}$ and $\Sigma_{P^{\ast}}^{(P^{\ast})}$ for $\Sigma_{P^{\ast}}$ are important.
In fact, those factors can be understood in an intuitive way.
In the heavy quark limit, all the physical properties of $P$ and $P^{\ast}$ mesons are the same except for the spin degrees of freedom.
As for $\Sigma_{P}$, the intermediate $P^{\ast}$ states have three degrees of freedom due to spin one (cf.~ Fig.~\ref{fig:Yasui:2012rw}).
As for $\Sigma_{P^{\ast}}$, on the other hand, the intermediate $P$ states have one degree of freedom due to spin zero, and the intermediate $P^{\ast}$ states have two degree of freedom due to spin one.
Notice that the latter number
 is reduced from three to two, because there are antisymmetric tensor (helicity-flipping term) for the spin in the $\pi P^{\ast}P^{\ast}$ vertex in Eq.~(\ref{eq:L_heavy_hadron_pion}).
As a result, we obtain the same self-energies for $P$ and $P^{\ast}$ in nuclear matter in the heavy quark limit.
This result is consistent with the heavy quark symmetry.
The light spin-complex, which is defined in Sect.~\ref{sec:heavy_quark_symmetry}, is composed of $qNN^{-1}$ with $q$ the light quark in $P^{(\ast)}$$(=q\bar{Q})$ and $NN^{-1}$ nucleon-hole pairs.
Though the present result is given at the lowest order for the interaction, it will be straightforwardly extended to higher orders.

The isospin asymmetric nuclear matter is also investigated with the asymmetry parameter $\delta=(n_{n}-n_{p})/(n_{n}+n_{p})$~\cite{Yasui:2012rw}. For $\delta > 0$, namely, $n_{d}>n_{u}$ for $u$, $d$ quark number density, the $\bar{D}^{0}(\bar{c}u)$ mass becomes smaller, while the $D^{-}(\bar{c}d)$ mass become larger.
This is in contrast to the result in the mean-field approach~\cite{Mishra:2008cd,Kumar:2010gb}. To have more precise discussion, however, it will be necessary to include also the Weinberg-Tomozawa interaction for $NN \pi \pi$.

As discussed in Sect.~\ref{sec:heavy_quark_symmetry}, the $\lambda_{1}$ and $\lambda_{2}$ coefficients in the $1/m_{Q}$ expansion of the masses of the charm and bottom hadrons reflect the effects of the electric and magnetic gluons inside the charmed hadrons, respectively. By calculating the in-medium hadron masses, it is possible to study the change of the gluonic components through the modifications of the $\lambda_{1}$ and $\lambda_{2}$ in the nuclear matter. For this purpose, the $1/m_{Q}$ corrections should be introduced in the effective theory, as described in Sect.~\ref{sec:heavy_hadron_effective_theory}.
In Ref.~\cite{Yasui:2013iga}, the values of $\lambda_{1}(\rho)$ and $\lambda_{2}(\rho;m_{Q})$ at the baryon number density $\rho$ are calculated at the pion one-loop level.
It is important to include systematically the $1/M$ corrections with $M$ being the averaged mass of $P$ and $P^{\ast}$.
This is realized by using the $1/M$ corrections to the $P^{(\ast)}P^{(\ast)}\pi$ vertices as presented in Eq.~(\ref{eq:L_heavy_hadron_pion_2}).
When compared with the vacuum values, the coefficients at $\rho=\rho_{0}$ are obtained as
\begin{align}
  \frac{\lambda_{1}(\rho_{0})}{\lambda_{1}(0)}
  &=1.28\textrm{-}1.20 ,\quad
  \frac{\lambda_{2}(\rho_{0};m_{c})}{\lambda_{2}(0;m_{c})}
  =0.74\textrm{-}0.89,\quad
  \frac{\lambda_{2}(\rho_{0};m_{b})}{\lambda_{2}(0;m_{b})}
  =0.35\textrm{-}0.73 .
\end{align}
The increase of $\lambda_{1}$ indicates that the coupling of the electric gluon to the heavy quark in the heavy-light meson are enhanced in the nuclear medium, while the decrease of $\lambda_{2}$ shows the suppression of that of the magnetic gluon. For reference, the strength of the electric/magnetic gluon condensate is studied at finite temperature, using the QCD sum rule analysis of the lattice QCD data~\cite{Morita:2007hv,Lee:2008xp}. It is discussed that the increase of the electric condensate near the critical temperature leads to the decrease of the mass of $J/\psi$ through the Stark effect, while the magnetic condensate hardly changes around the critical temperature $T_{\mathrm c}$.

\paragraph{QCD sum rules}

The QCD sum rules have been used to study the in-medium properties for the light mesons such as $\rho$, $\omega$ and $\phi$~\cite{Hatsuda:1991ez}. The application to the heavy-light meson is firstly done in Refs.~\cite{Hayashigaki:2000es,Morath:1999cv}. The in-medium two-point correlation function of the $D/\bar{D}$ meson is given by 
\begin{align}
 \Pi_{\mathrm{PS}}^{\mathrm{NM}}(q)
 =
 i \int \mathrm{d}^{4}x e^{iq\cdot x}
 \langle T J_{5}(x) J_{5}^{\dag}(0) \rangle_{\mathrm{NM}(\rho_{N})},
 \label{eq:Trhosumrule}
\end{align}
where $\rho_{N}$ is the nuclear matter density and $q^{\mu}=(q^{0},\vec{q})$ is the momentum carried by the interpolating current $J_{5}(x)=J_{5}^{\dag}(x)=\left( \bar{c} i\gamma_{5} q(x) + \bar{q} i\gamma_{5} c(x) \right)/2$.
Here,
the mass of the in-medium $D/\bar{D}$ meson is evaluated by introducing the density dependence in the condensates.
In addition to the linear density approximation, in Ref.~\cite{Hayashigaki:2000es}, Eq.~(\ref{eq:Trhosumrule}) is approximated by
\begin{align}
 \Pi_{\mathrm{PS}}^{\mathrm{NM}}(q)
  \simeq
 \Pi_{\mathrm{PS}}^{0}(q) + \frac{\rho_{N}}{2M_{N}} T_{\mathrm{PS}}(q),
  \label{eq:Hayashigaki}
\end{align}
where $\Pi_{\mathrm{PS}}^{0}(q)$ is the vacuum correlation function and $T_{\mathrm{PS}}(q)$ is the forward scattering amplitude of a $D/\bar{D}$ meson and a nucleon in vacuum, defined as~\cite{Koike:1996ga}
\begin{align}
 T_{\mathrm{PS}}(q) = i \int \mathrm{d}^{4} x e^{iq\cdot x}
 \langle N(p) | \mathrm{T} J_{5}(x) J^{\dag}_{5}(0) | N(p) \rangle.
\end{align}
By applying the Borel transformation to $T_{\mathrm{PS}}(q)$ and comparing the OPE side with the phenomenological side, the modification of the $D/\bar{D}$ meson in medium is calculated. 
Up to the dimension four OPE, the relevant condensates are $\langle \bar{q}q \rangle$,
$\langle \frac{\alpha_{s}}{\pi}G^{2} \rangle$,
$\langle q^{\dag} i \vec{D}_{0} q \rangle$ and
$\langle \frac{\alpha_{s}}{\pi} \left( \frac{(vG)^{2}}{v^{2}}-\frac{G^{2}}{4} \right) \rangle$. In this case, the difference of the $D$ and $\bar{D}$ does not appear in the OPE side. As a result, the negative mass shift $\Delta m = -48 \pm 8$ MeV is obtained for both the $D$ and $\bar{D}$ mesons at normal nuclear matter density. Namely, the mass of the $D/\bar{D}$ meson decreases about 50 MeV in the nuclear medium. This attraction is comparable with that from the QMC model in Refs.~\cite{Sibirtsev:1999js,Tsushima:1998ru}. The negative mass shifts are also obtained by the improved analysis. In Ref.~\cite{Azizi:2014bba}, the mass shift is obtained as $\Delta m_{D}=-46 \pm 7$ MeV for the $D$ meson, and $\Delta m_{B}=-242 \pm 62$ MeV for the $B$ meson. Larger mass shift is suggested in Ref.~\cite{Wang:2015uya}; $\Delta m_{D}=-72$ MeV, $\Delta m_{B}=-478$ MeV, $\Delta m_{D^{\ast}}=-102$ MeV, and $\Delta m_{B^{\ast}}=-687$ MeV. In all cases, the medium modification acts as an attractive effect. We note that Refs.~\cite{Azizi:2014bba,Wang:2015uya} use the low density approximation of Eq.~\eqref{eq:Trhosumrule} to study the medium effect.

\begin{figure}[tbp]
 \begin{minipage}{0.5\hsize}
   \centering
   \includegraphics[scale=0.23,bb=0 0 792 612]{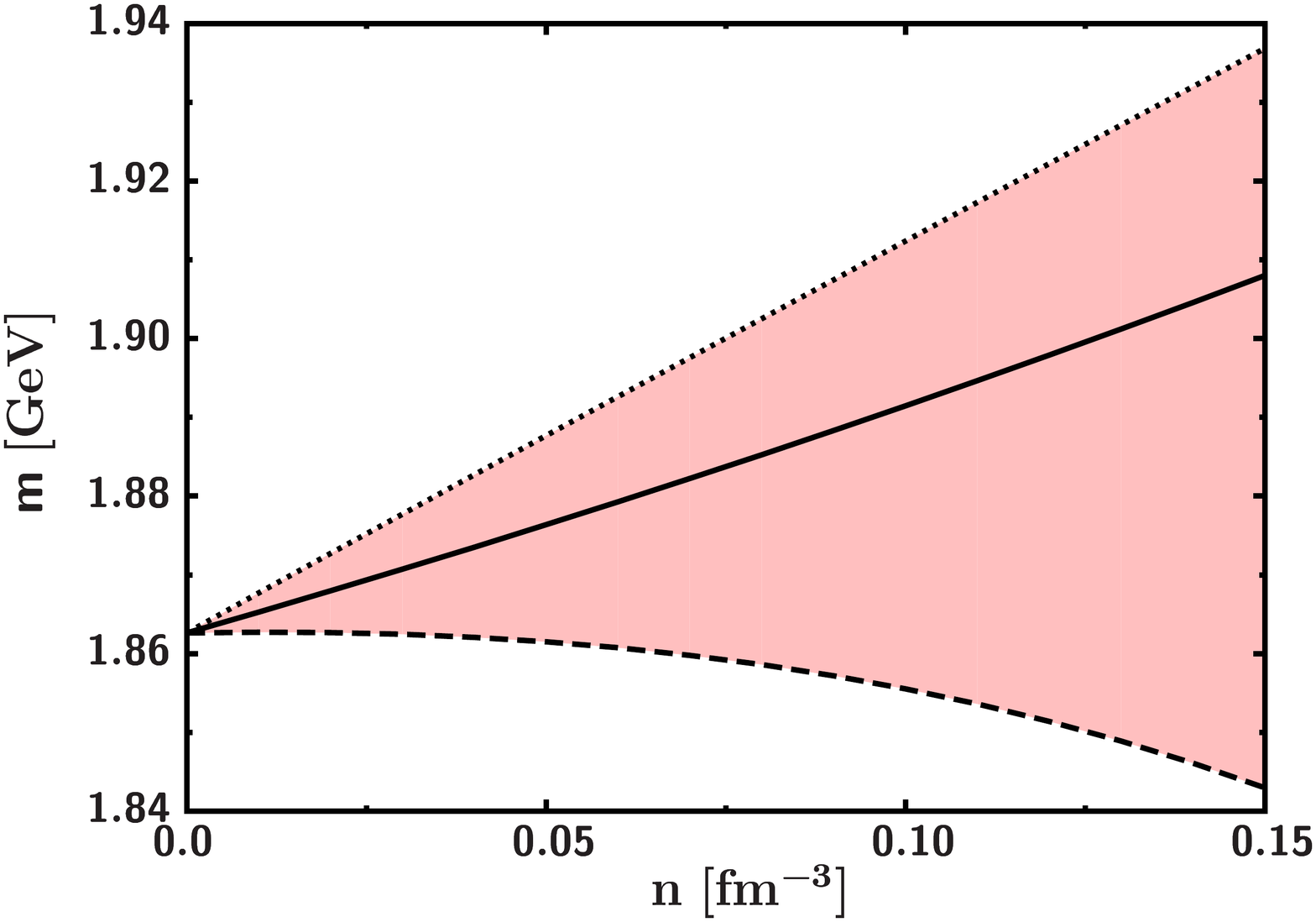}
 \end{minipage}
 \begin{minipage}{0.5\hsize}
    \centering
   \includegraphics[scale=0.23,bb=0 0 792 612]{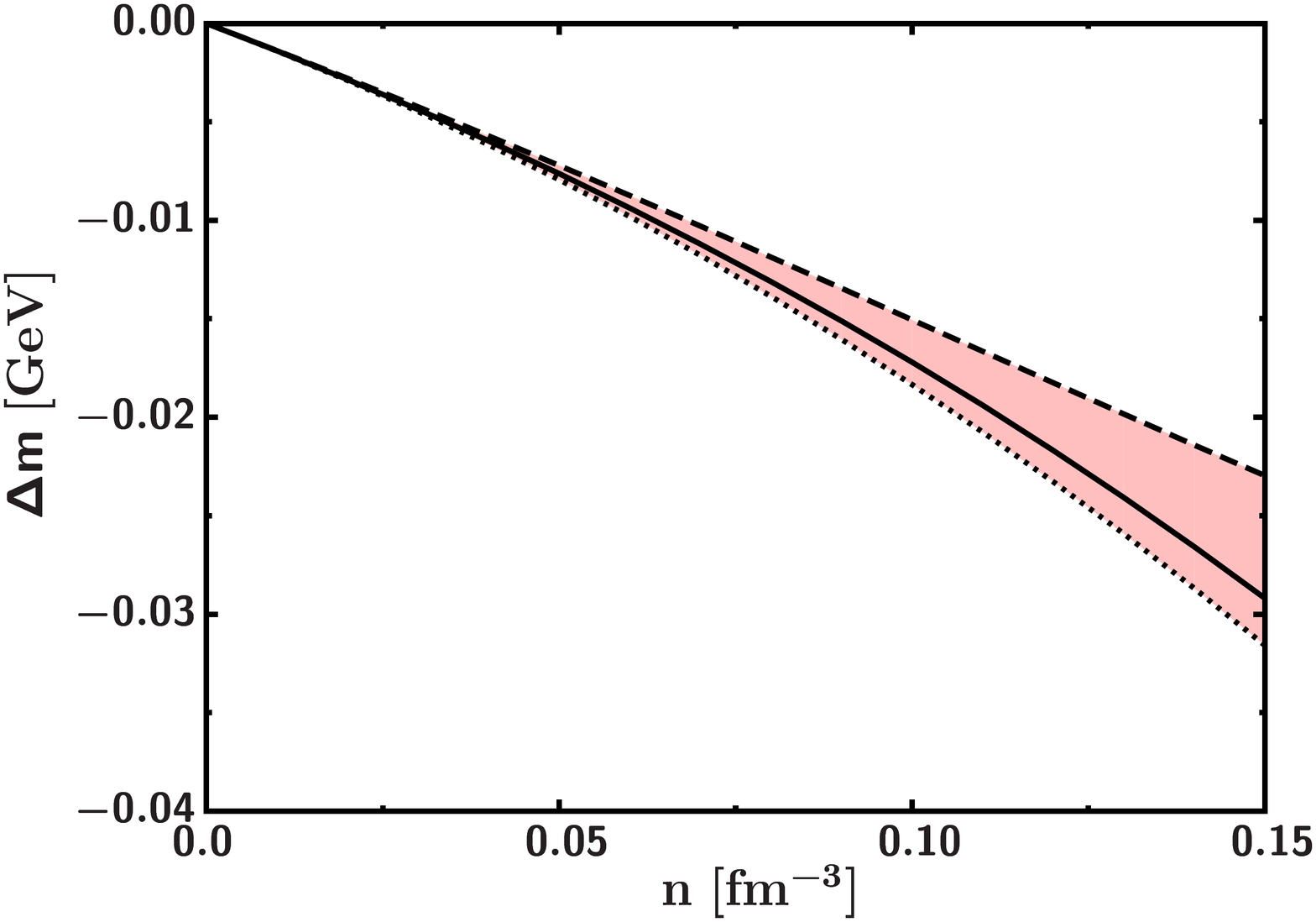}
 \end{minipage}
  \caption{The averaged mass $m=(m_{D}+m_{\bar{D}})/2$ and the mass splitting $\Delta m = (m_{D}-m_{\bar{D}})/2$ in nuclear matter with density $n$ in the QCD sum rule approach~\cite{Hilger:2008jg}.}
  \label{fig:Hilger:2010zb_D}
\end{figure}

On the other hand, the repulsive mass shift is predicted in Refs.~\cite{Hilger:2008jg,Hilger:2010zb} where the dimension five operator $\langle \bar{q} g \sigma G q \rangle$ and the $q_{0}$-odd operators 
$\langle q^{\dag}q \rangle$, $\langle q^{\dag} \vec{D}^{2}_{0} q \rangle$, $\langle q^{\dag} g \sigma G q\rangle$ are included in addition to the operators used in Ref.~\cite{Hayashigaki:2000es}.\footnote{Notice that $q_{0}$-odd terms do not appear in $\bar{q}q$ meson with $q=u, d$, such as $\rho$, $\omega$ mesons, as well as in quarkonia, such as $\phi$ and $J/\psi$, in isospin-symmetric nuclear matter.} 
In contrast to the assumption in Eq.~(\ref{eq:Hayashigaki}), the original form of the correlation function
\begin{align}
  \Pi_{\mathrm{PS}}^{\mathrm{NM}}(q) = \Pi_{\mathrm{PS}}^{\mathrm{NM},\mathrm{even}}(q_{0}^{2}) + q_{0} \Pi_{\mathrm{PS}}^{\mathrm{NM},\mathrm{odd}}(q_{0}^{2}),
\end{align}
with the superscripts indicate the $q_{0}$-even/odd operators is used.
In this case, the $q_{0}$-odd terms cause the splitting of $D$ and $\bar{D}$.
As a result, the averaged mass of $D$ and $\bar{D}$ mesons $m=(m_{D}+m_{\bar{D}})/2$ increases by $+45$ MeV at the normal nuclear matter density (see Fig.~\ref{fig:Hilger:2010zb_D}). 
By introducing the contribution of both $D^{+}$ and $D^{-}$ in the phenomenological side, the mass difference is extracted as $m_{D}-m_{\bar{D}}=-60$ MeV, indicating $m_{D}<m_{\bar{D}}$. The same tendency is observed in the bottom sector where 
$m=(m_{B}+m_{\bar{B}})/2$ increases about $+60$ MeV and $m_{\bar{B}}-m_{B}=-130$ MeV. In the case of the charm-strange mesons, the averaged mass $m=(m_{D_{s}}+m_{\bar{D}_{s}})/2$ increases by about $+30$ MeV, while $m_{D_{s}}-m_{\bar{D}_{s}}=+25$ MeV, showing $m_{D_{s}}>m_{\bar{D}_{s}}$ as opposed to the $D/\bar{D}$ case. For the detailed discussion on the sum rules with the mixed condensates and the four-quark condensates, see Refs.~\cite{Zschocke:2011aa,Buchheim:2014rpa}.

\begin{figure}[tbp]
\begin{center}
\includegraphics[scale=0.6,bb=0 0 360 252]{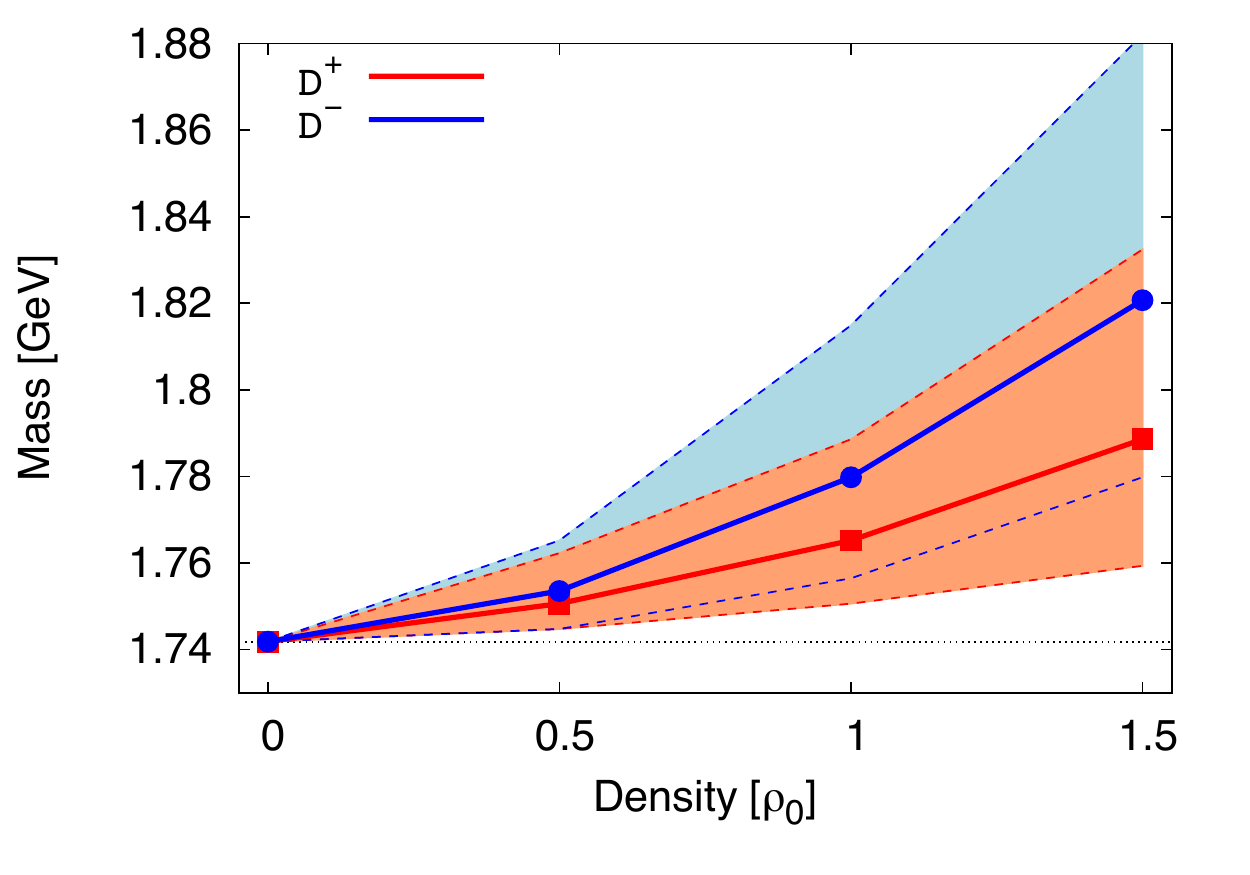}
\caption{Masses of $D^{-}$ ($\bar{D}$) and $D^{+}$ ($D$) meson at finite baryon density~\cite{Suzuki:2015est}. The unit of the horizontal axis is given by the saturation nuclear matter density $\rho_{0}$.}
\label{fig:Suzuki:2015est}
\end{center}
\end{figure}

In Ref.~\cite{Suzuki:2015est}, the in-medium spectral function of the $D/\bar{D}$ meson is extracted with the maximum entropy method~\cite{Gubler:2010cf}. In this method, the spectral function can be obtained without any assumption of its explicit form. The charge-conjugation projection is performed at the OPE level. The results are qualitatively similar to those in Ref.~\cite{Hilger:2008jg}, i.e., the $D/\bar{D}$ mass increases in the nuclear medium, as shown in Fig.~\ref{fig:Suzuki:2015est}. Quantitatively, the amount of the mass modification and the magnitude of the mass splitting are milder than those in Ref.~\cite{Hilger:2008jg}. Possible origin of the difference of the qualitative conclusions (attraction or repulsion) in the QCD sum rule approaches may be related with the choice of the Borel window which is used in the analysis (see e.g. Sect.~IV in Ref.~\cite{Suzuki:2015est} for more details). 

In Refs.~\cite{Hilger:2008jg,Hilger:2010zb,Suzuki:2015est} the mass of the $D/\bar{D}$ meson is predicted to increase. This is in contrast to the light mesons, whose masses in general decrease. An explanation of these behavior is given in Ref.~\cite{Park:2016xrw}, based on the constituent quark model with the linear confining potential. When chiral symmetry is partially restored in the nuclear medium, the constituent mass of the light quarks should decrease. It is shown that the light-light meson mass decreases along with the reduction of the constituent quark mass, while the heavy-light meson mass increases.

\paragraph{NJL model}

Nambu--Jona-Lasinio (NJL) model~\cite{Nambu:1961tp,Nambu:1961fr} describes the dynamical chiral symmetry breaking, and has been applied to various phenomena in the light quark sectors in finite temperature and density (see review articles~\cite{Klimt:1989pm,Vogl:1989ea,Klevansky:1992qe,Hatsuda:1994pi}). The generalization to the charm quark sector is done in Refs.~\cite{Ebert:1994tv,Ebert:1996vx} by utilizing the SU(4) symmetry.

\begin{figure}[tbp]
\begin{center}
\includegraphics[scale=0.3,bb=0 0 613 552]{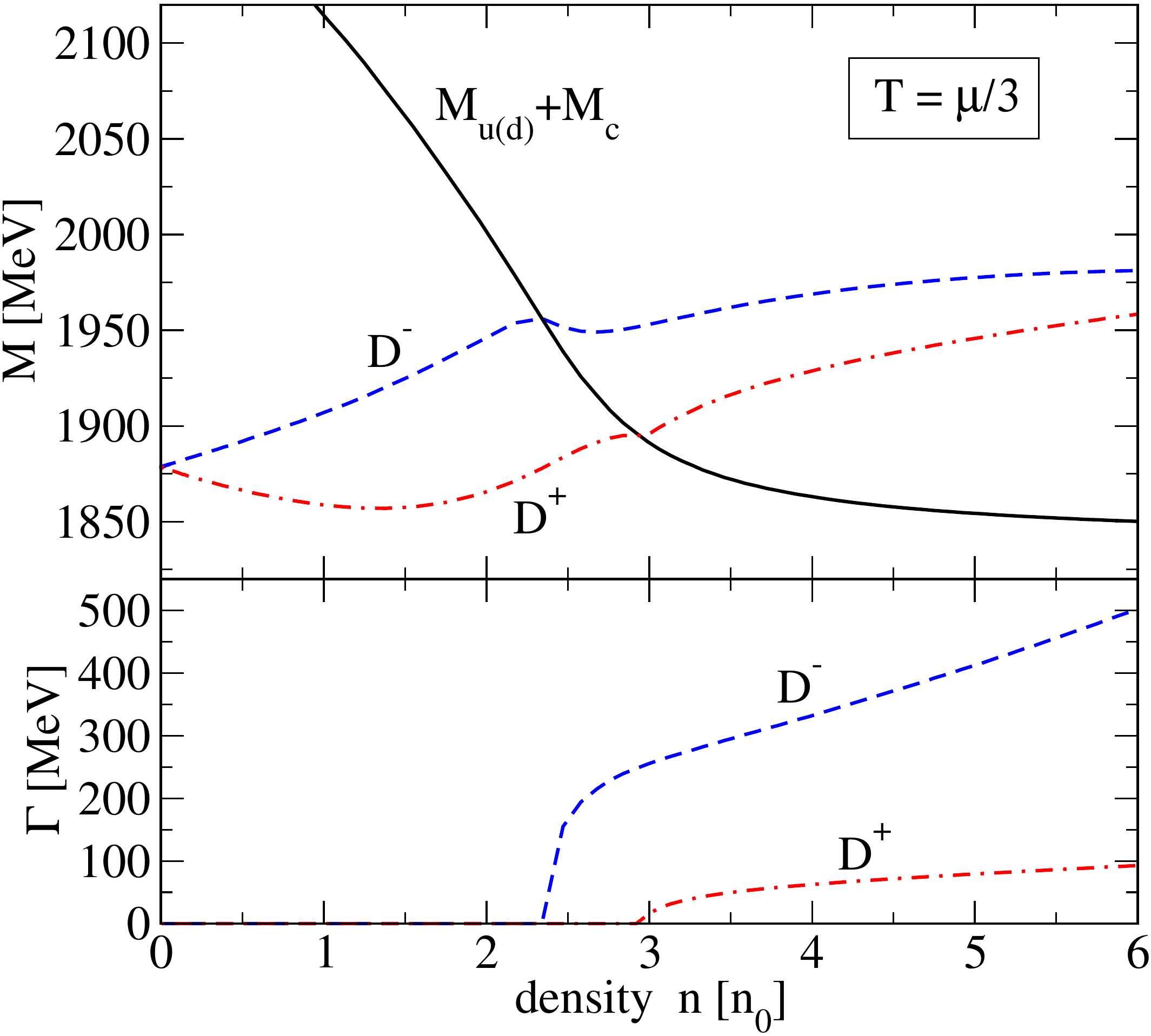}
\caption{Masses of $D^{-}$ ($\bar{D}$) and $D^{+}$ ($D$) meson at finite baryon density and temperature in the PNJL model \cite{Blaschke:2011yv}. The unit of the horizontal axis is given by the normal nuclear matter density $n_{0}=0.16$ fm$^{-3}$.}
\label{fig:Blaschke:2011yv}
\end{center}
\end{figure}

The in-medium properties of the heavy-light mesons are studied in Ref.~\cite{Blaschke:2011yv} where the Polyakov loop potential is also incorporated. The Lagrangian is given by
\begin{align}
{\cal L}_{\mathrm{PNJL}}
=
\bar{q} (i \gamma^{\mu} D_{\mu} + \hat{m}) q
+ G_{S} \sum_{a=0}^{15} \left( (\bar{q}\lambda^{a}q)^{2} + (\bar{q}i\gamma_{5}\lambda^{a}q)^{2} \right)
- {\cal U}(\Phi[A],\bar{\Phi}[A];T),
\end{align}
where $\Phi$ is the Polyakov loop and $q=(u,d,s,c)^{t}$ is the four-component quark field.
The light quark masses are dynamically generated by the gap equation.
The mass of the pseudoscalar meson $M_{P}$ is obtained by solving the gap equation 
\begin{align}
 1-2G_{S} \Pi^{ij}(P_{0}=M_{P},\vec{P}=0)=0,
\end{align}
with the polarization operators
\begin{align}
 \Pi_{ij}(P) = 
 iN_{c}
 \int \frac{\mathrm{d}^{4}p}{(2\pi)^{4}}
 \mathrm{tr}_{\mathrm{D}}
 \left[
  S_{i}(p) i\gamma_{5} S_{j}(p+P) i\gamma_{5}
 \right].
\end{align}
The masses of the $D^{-}$ ($\bar{D}$) and $D^{+}$ ($D$) mesons at finite density are shown in Fig.~\ref{fig:Blaschke:2011yv} with the finite temperature $T=\mu/3$. At low density, the mass of the $D^{-}$ ($\bar{D}$) meson increases, while that of the $D^{+}$ ($D$) meson decreases. This can be understood by the effective repulsion by the Pauli principle; the light quark inside $\bar{D}$ feels the Pauli blocking effect at finite density, whereas neither the charm quark nor the light antiquark are affected. At sufficiently large baryon density, the decay processes such as $\bar{D} \rightarrow q + \bar{c}$ or $D \rightarrow \bar{q}+c$ can occur and the decay width $\Gamma$ appears as shown in Fig.~\ref{fig:Blaschke:2011yv}. We should note the possible cutoff dependence of the threshold effect, as in the study of the vector mesons in vacuum~\cite{Takizawa:1991mx}.

\subsection{New aspects of heavy-light mesons in nuclear matter}

\paragraph{Partial restoration of broken chiral symmetry}

As discussed in Sect.~\ref{sec:chiral_symmetry}, the chiral condensate is one of the most important quantities to determine the properties of hadrons.
In the linear representation of chiral symmetry, the mass difference between chiral partners of heavy-light meson, $\Delta M$, is proportional to the pion decay constant, and hence it is proportional to the chiral condensate $\langle \bar{q}q \rangle$~\cite{Bardeen:2003kt}:
\begin{align}
 \Delta M  \propto f_{\pi}  \propto \langle \bar{q}q \rangle.
\end{align}
Thus, we can probe the chiral condensate in nuclear medium by measuring the $\Delta M$ for the heavy hadron.
In the following, we summarize the studies of the mass modifications of the chiral partners of the heavy-light pseudoscalar and scalar mesons, such as $D(0^{-})$ and $D_{0}^{\ast}(0^{+})$ as well as $D_{s}(0^{-})$ and $D_{s0}^{\ast}(0^{+})$.

There are several works based on the hadron dynamics.
In Ref.~\cite{Tolos:2009ck}, the in-medium properties (mass and decay width) of scalar mesons, $D_{0}^{\ast}(2400)$ and $D_{s0}^{\ast}(2317)$, was studied by investigating the T-matrix from the Lippmann-Schwinger equation based on the SU(8) flavor-spin symmetry.
It turned out that their in-medium decay width of $D_{s0}^{\ast}(2317)$ changes to be 100 MeV, which is much larger than the value ($<3.8$ MeV) in vacuum~\cite{Agashe:2014kda}.
The in-medium decay width of $D_{0}^{\ast}(2400)$ meson is also enhanced.
The reason of the large in-medium decay widths would be due to the strong absorption process via $DN$ and $DNN$ loops.
On the other hand, the mass modifications were small.
However, we have to be careful if the present result can be used for the partial restoration of the broken chiral symmetry in nuclear medium, because the non-linear representation is used (cf.~Sect.~\ref{sec:chiral_effective_theory}).

In contrast, the mass modifications of $D$, $D^{\ast}$, $D_{s}$, $D_{s}^{\ast}$ mesons were investigated at finite temperature in the linear-sigma model~\cite{Sasaki:2014asa,Sasaki:2014wma}.
We would like to introduce this study, because the discussion is based essentially on the change of chiral condensate in medium, and can be applied straightforwardly to nuclear medium.
The author focused especially on the parity doublets, $(0^{-},1^{-})$ and $(0^{+},1^{+})$.
Masses of $(0^{-},1^{-})$ states and $(0^{+},1^{+})$ states should coincide with each other when chiral symmetry recovers completely.\footnote{The $0^{-}$ and $1^{-}$ ($0^{+}$ and $1^{+}$) states should be closer to each other in mass, because of the approximate heavy quark symmetry (cf.~Sect.~\ref{sec:heavy_quark_symmetry}).}
In general, however, it is difficult to conclude if the masses of both parity + and - states become smaller, or if both of them become larger, or if the mass of parity + becomes larger and the mass of parity - becomes smaller.
In Ref.~\cite{Sasaki:2014asa}, the effective Lagrangian with SU(3) ($u$, $d$, $s$) flavor symmetry, the U(1)$_{\mathrm{A}}$ breaking and the heavy quark symmetry were considered~\cite{Nowak:1992um,Bardeen:1993ae} (see also Ref.~\cite{Bardeen:2003kt}).
We consider the $D=(D_{q},D_{q},D_{s})$ meson in the linear representation
\begin{align}
 {\cal H}_{L,R}(x) = \frac{1}{\sqrt{2}} \Bigl( G(x) \pm i H(x) \gamma_{5} \Bigr),
 \label{eq:H_LR}
\end{align}
with $H_{v}$ defined in Eq.~(\ref{eq:H_field}) for $(0^{-},1^{-})$ mesons and $G_{v}(x)$ defined by
\begin{align}
 G_{v}(x) = \frac{1+v\hspace{-0.5em}/}{2} \left( -iD^{\ast}_{v\,\mu}(x) \gamma^{\mu} \gamma_{5} + D_{v}(x) \right),
 \label{eq:G_field}
\end{align}
for $(0^{+},1^{+})$ mesons.
We consider also the light meson field $\Sigma = \left( \sigma^{a}+i\pi^{a} \right)T^{a}$ ($a=0,\dots,8$).\footnote{U(1)$_{\mathrm{V}}$ symmetry is included.}
Then, the heavy-light meson field ${\cal H}_{L,R}(x)$ and the light meson field $\Sigma$ are transformed as
\begin{align}
 {\cal H}_{L,R}(x) \rightarrow S {\cal H}_{L,R}(x) g_{L,R}^{\dag},
\end{align}
and
\begin{align}
 \Sigma \rightarrow g_{L}^{\dag} \Sigma g_{R},
\end{align}
according to the heavy quark spin transformation $S$ and chiral transformation $g_{L,R}$.
Then, we can construct the effective Lagrangian for the heavy-light meson ${\cal H}_{L,R}$ and the light meson $\Sigma$.
The effective Lagrangian of the light meson $\Sigma$ is given by the linear sigma model.
The effective Lagrangian concerning the heavy-light meson is given by the coupling to the light meson
\begin{align}
 {\cal L}_{\mathrm{HL}}
&=
\frac{1}{2} \mathrm{Tr}
\left( \bar{\cal H}_{L} v\!\cdot\!i\partial {\cal H}_{L} + \bar{\cal H}_{R} v\!\cdot\!i\partial {\cal H}_{R} \right)
+ \frac{m_{0}}{2} \mathrm{Tr} \left( \bar{\cal H}_{L} {\cal H}_{L} + \bar{\cal H}_{R} {\cal H}_{R} \right)
\nonumber \\
&
+
\frac{g_{\pi}}{4} \mathrm{Tr} \left( \Sigma^{\dag} \bar{\cal H}_{L} {\cal H}_{R}  + \Sigma \bar{\cal H}_{R} {\cal H}_{L} \right)
-i\frac{g_{A}}{2f_{\pi}}
\mathrm{Tr} \left( \gamma_{5}\left( \partial\hspace{-0.5em}/\hspace{0.1em}\Sigma^{\dag} \right) \bar{\cal H}_{L} {\cal H}_{R} - \gamma_{5} \left( \partial\hspace{-0.5em}/\hspace{0.1em}\Sigma \right) \bar{\cal H}_{R} {\cal H}_{L} \right),
\label{eq:linear_HL}
\end{align}
and by four-point interaction of the heavy-light meson (see Ref.~\cite{Sasaki:2014asa} for more details).
The four-point interaction is adopted for realizing the condensate of $D_{s}$ mesons at finite chemical potentials of strangeness and charm.

\begin{figure}[tbp]
\begin{center}
\includegraphics[scale=1.0,bb=0 0 174 121]{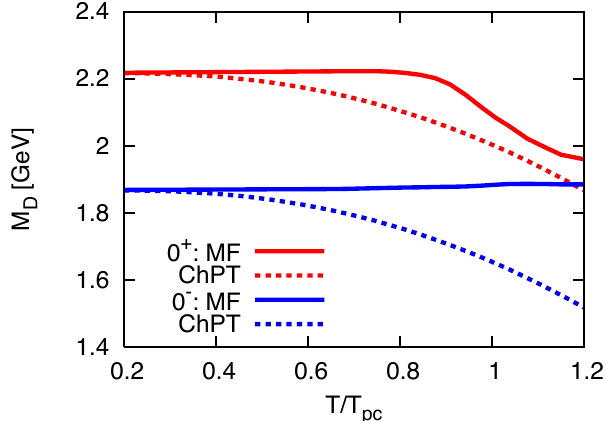}
\caption{Masses of non-strange charm meson in the mean-field model (solid lines) and the corresponding result with one-loop in the chiral-perturbation theory \cite{Sasaki:2014asa}.}
\label{fig:Sasaki:2014asa}
\end{center}
\end{figure}

We consider the SU(3) flavor ``$\sigma$" meson ($J^{P}=0^{+}$), the singlet $\sigma_{0}$ and the octet $\sigma_{8}$, in the linear representation.
They are transformed to a non-strange scalar $\sigma_{q}$  meson, a strange scalar $\sigma_{s}$ meson
\begin{align}
\left(
\begin{array}{c}
 \sigma_{q} \\
 \sigma_{s}
\end{array}
\right)
=
\frac{1}{\sqrt{3}}
\left(
\begin{array}{cc}
 \sqrt{2} & 1 \\
 1 & -\sqrt{2}
\end{array}
\right)
\left(
\begin{array}{c}
 \sigma_{0} \\
 \sigma_{8}
\end{array}
\right),
\end{align}
as a mixing of $\sigma_{0}$ and $\sigma_{8}$.
Their condensates in vacuum are given by
\begin{align}
 \langle \sigma_{q} \rangle &= f_{\pi},  \\
 \langle \sigma_{s} \rangle &= \frac{1}{\sqrt{2}} \left( 2f_{K} -  f_{\pi} \right),  
\end{align}
with the pion and kaon decay constants, $f_{\pi}$ and $f_{K}$.
The ground state of the heavy-light meson systems is given by the stationary condition for the thermodynamics potential $\Omega$,
\begin{align}
 \frac{\partial \Omega}{\partial \sigma_{q}} =  \frac{\partial \Omega}{\partial \sigma_{s}}
= \frac{\partial \Omega}{\partial D_{q}} =  \frac{\partial \Omega}{\partial D_{s}} = 0.
\end{align}
As a result, it turned out that the masses of $J^{P}=0^{-}$ and $0^{+}$ states for $D$ mesons are almost constant at low temperature, and only $0^{+}$ states becomes lighter around the critical temperature (Fig.~\ref{fig:Sasaki:2014asa}).
In addition, the mass difference between $J^{P}=0^{-}$ and $0^{+}$ states for $D$ mesons, $\delta M_{D}$, is almost the same as that between $J^{P}=0^{-}$ and $0^{+}$ states for $D_{s}$ mesons, $\delta M_{D_{s}}$.
This result suggests that the restoration of the broken chiral symmetry is almost the same both for $u$, $d$ quarks and $s$ quark.
This is different from the result in SU(4) flavor symmetry, namely that $\delta M_{D}$, is much smaller than $\delta M_{D_{s}}$~\cite{Roder:2003uz}.

In contrast to the linear representation, we may consider the non-linear representation at low temperature, because the heavy-light meson with positive parity can be eliminated due to the large mass~\cite{Sasaki:2014asa}.
In this setting, the mass difference between $J^{P}=0^{-}$ and $0^{+}$ states for $D$ mesons is investigated, where the pion one-loop effect is considered based on the chiral perturbation theory (e.g. Ref.~\cite{Harada:2003kt}).
Then, it is shown that the masses of both of $J^{P}=0^{-}$ and $0^{+}$ states become small (Fig.~\ref{fig:Sasaki:2014asa}).
The restoration of the broken chiral symmetry for the heavy-light mesons is also studied by the chiral susceptibility~\cite{Sasaki:2014wma}.

The mass degeneracy of $0^{-}$ and $0^{+}$ states is also discussed by Skyrmion crystal~\cite{Suenaga:2014sga}.
Because the $H_{v}(x)$ field as well as the $G_{v}(x)$ field couple to the pion field composing the Skyrmion,
their mass shifts can be analyzed by regarding the Skyrmion crystal as the background field (fcc type) with the inter-distance $L$ between two Skyrmions.
It turns out that the mass difference between $0^{-}$ and $0^{+}$ is still finite for large $L$ (dilute Skyrmion crystal), while it becomes small and the degeneracy is realized for small $L$ (half-Skyrmion).
Therefore, it seems to indicate that the restoration of the broken chiral symmetry would be realized in the latter case.
However, we have to note that the chiral condensate is zero as an averaged value over the Skyrmion crystal, where the non-linear representation of chiral symmetry is assumed, and hence it should be distinguished from the case of the linear sigma model in Refs.~\cite{Sasaki:2014asa,Sasaki:2014wma}.

As a different situation, the masses of $\bar{D}(0^{-})$, $\bar{D}^{\ast}(1^{-})$, $\bar{D}_{0}^{\ast}(0^{+})$ and $\bar{D}_{1}(1^{+})$ mesons are investigated in the dual chiral density wave (DCDW) or the chiral density wave (CDW)~\cite{Suenaga:2015daa}.\footnote{See Ref.~\cite{Buballa:2014tba} for more details about DCDW and CDW.}
The DCDW and the CDW are considered to be realized in high density state of nuclear matter.
In the linear representation, we introduce the light meson field $M=\sigma+i\tau^{a}\pi^{a}$ and the heavy-light meson fields ${\cal H}_{L,R}$ for $(\bar{D}, \bar{D}^{\ast})$ and $(\bar{D}_{0}^{\ast},\bar{D}_{1})$,
 the latter of which is defined in Eq.~(\ref{eq:G_field}).
We consider the effective Lagrangian invariant for the chiral transformation and the heavy quark spin transformation,
\begin{align}
 {\cal L} &= \mathrm{tr} \left( {\cal H}_{L} i v \!\cdot\! \partial \bar{\cal H}_{L} \right) + \mathrm{tr} \left( {\cal H}_{R} i v \!\cdot\! \partial \bar{\cal H}_{R} \right)
 + \frac{\Delta_{m}}{2f_{\pi}} \mathrm{tr} \left({\cal H}_{L} M \bar{\cal H}_{R} + {\cal H}_{R} M^{\dag} \bar{\cal H}_{L} \right) \nonumber \\
& + i \frac{g_{1}}{2f_{\pi}} \mathrm{tr} \left( {\cal H}_{R}\gamma_{5}\gamma^{\mu}\partial_{\mu}M^{\dag} \bar{\cal H}_{L} - {\cal H}_{L}\gamma_{5}\gamma^{\mu}\partial_{\mu}M \bar{\cal H}_{R} \right)
 + i \frac{g_{2}}{2f_{\pi}} \mathrm{tr} \left( {\cal H}_{L}\gamma_{5}/\hspace{-0.5em}\partial M M^{\dag} \bar{\cal H}_{L} - {\cal H}_{R}\gamma_{5}/\hspace{-0.5em}\partial M^{\dag} M \bar{\cal H}_{R} \right) \nonumber \\
& +\frac{g_{3}}{2f_{\pi}} \mathrm{tr} \left( {\cal H}_{R}\gamma^{\mu}\partial_{\mu}M^{\dag} \bar{\cal H}_{L} + {\cal H}_{L}\gamma^{\mu}\partial_{\mu}M \bar{\cal H}_{R} \right)
 +i\frac{g_{4}}{2f_{\pi}} \mathrm{tr} \left( {\cal H}_{L}/\hspace{-0.5em}\partial M M^{\dag} \bar{\cal H}_{L} + {\cal H}_{R}/\hspace{-0.5em}\partial M^{\dag} M \bar{\cal H}_{R} \right) \nonumber \\
& + {\cal O}(\partial^{2}M),
\label{eq:linear_HL_2}
\end{align}
where $\Delta_{m}$ is the mass difference between the multiplets $H$ and $G$, and $g_{1}$, $g_{2}$, $g_{3}$ and $g_{4}$ are the coupling constants.
We notice that this is slightly different from Eq.~(\ref{eq:linear_HL}).
We consider the DCDW configuration
\begin{align}
 M = \phi \cos (2f x) + i\tau^{3} \phi \sin(2fx),
\end{align}
where $\tau^{3}$ is the third component of the Pauli matrices for isospin, $x$ is the distance along the DCDW, $f$ the wave number, $\phi$ the amplitude of the wave.
The energy-momentum dispersion relation is obtained for $\bar{D}(0^{-})$, $\bar{D}^{\ast}(1^{-})$, $\bar{D}_{0}^{\ast}(0^{+})$ and $\bar{D}_{1}(1^{+})$ mesons in the DCDW.
The dispersion relation becomes different from that in vacuum due the $x$-dependence of the $\sigma$ and $\pi$ fields,
and has the minimum at the finite momentum corresponding to the wave number of the DCDW.
When the amplitude $\phi$ is changed, the restoration of chiral symmetry leads to the degeneracy among the dispersion relations.
As far as the value of $\phi$ in vacuum, however, we notice that again the absolute value of the chiral condensate is not changed for this $x$-dependence, hence the change of the dispersion relations in the DCDW would not be relevant to the partial restoration of the broken chiral symmetry.

The in-medium mass of a $D_{0}^{\ast}(\bar{q}c)$ and $\bar{D}_{0}^{\ast}(q\bar{c})$ meson is analyzed in the QCD sum rules~\cite{Hilger:2010zf}.
It was obtained that the mean-value of mass $(m_{D_{0}^{\ast}}+m_{\bar{D}_{0}^{\ast}})/2$ for $D_{0}^{\ast}(\bar{q}c)$ and $\bar{D}_{0}^{\ast}(q\bar{c})$ decreases in nuclear matter, while the mass difference $m_{D_{0}^{\ast}}-m_{\bar{D}_{0}^{\ast}}$ becomes negative.

To reach the conclusion about chiral symmetry, we need to investigate the mass difference between a $D$ meson and a $D_{0}^{\ast}$ meson as well as that between a $D^{\ast}$ meson and a $D_{1}$ meson.
As numerical values, the mass shifts $\Delta m_{D_{0}^{\ast}}=69$ MeV and $\Delta m_{B_{0}^{\ast}}=217$ MeV are obtained~\cite{Wang:2011mj}.
As for the vector and axial-vector mesons, $\Delta m_{D^{\ast}}=-71$ MeV and $\Delta m_{D_{1}}=72$ MeV for charm and $\Delta m_{B^{\ast}}=-380$ MeV and $\Delta m_{B_{1}}=264$ MeV for bottom are obtained~\cite{Wang:2011fv}.
We refer also Ref.~\cite{Wang:2015uya} for updated values including the higher order terms of $\alpha_{s}$ in the quark condensates.
The result shows that the masses of the scalar heavy-light mesons become massive.
We have to note the result 
 that the mass difference between positive parity state and negative parity state becomes larger in nuclear matter.
This is in contrast to the argument about the partial restoration of the chiral symmetry, namely that the mass difference should become smaller (cf.~Ref.~\cite{Bardeen:2003kt}).\footnote{Notice that the technique used in Refs.~\cite{Wang:2011mj,Wang:2011fv} is the same used in Ref.~\cite{Hayashigaki:2000es}. This approach is different from the Weinberg sum rules in Ref.~\cite{Hilger:2011cq}.}

\paragraph{Kondo effect}

One of the most important properties of charm hadrons is simply the heavy mass which is much larger than other light hadrons ($\pi$ mesons, $\rho$ mesons, nucleons and so on).
The heavy mass induces interesting impurity phenomena.
Here we consider the Kondo effect.
The Kondo effect is the phenomena that the electric resistance in metals becomes enhanced when impurity atoms with finite spin is contained as impurity particles~\cite{Kondo:1964}.\footnote{See for example Refs.~\cite{Hewson,Yosida,Yamada} as text books.}
The electric resistance is related to the scattering amplitude between the conducting electron and the impurity particle.
When the interaction between the conducting electron and the impurity particle has the spin-dependence (attraction in spin-singlet channel),
the scattering amplitude suffers from the infrared instability of the Fermi surface,\footnote{The Fermi surface is unstable against attraction (e.g. the Cooper instability in superconductivity), see Ref.~\cite{abrikosov1975methods}.} and it becomes logarithmically divergent as $\sim \ln T/T_{\mathrm{K}}$ at low temperature $T < T_{\mathrm{K}}$.
The temperature $T_{\mathrm{K}}$ characterizing the energy scale of the Kondo effect is called the Kondo temperature.
In perturbative approach, there are four conditions for which the Kondo effect occurs:
(i) heavy impurity particle, (ii) Fermi surface (degenerate state), (iii) quantum fluctuation (loop effect) and (iv) non-Abelian (e.g. spin-dependent) interaction.
As for the non-Abelian interaction, the sign of the coupling is important as discussed below.
As long as those conditions are satisfied, the Kondo effect can occur not only electron systems but also in nuclear matter~\cite{Yasui:2013xr,Yasui:2016ngy} as well as in quark matter~\cite{Yasui:2013xr,Hattori:2015hka,Ozaki:2015sya,Yasui:2016svc}.
Let us pickup two subjects discussed in Refs.~\cite{Yasui:2013xr,Yasui:2016ngy}.

\begin{figure}[t]
\begin{center}
\includegraphics[scale=0.25,bb=0 0 681 163]{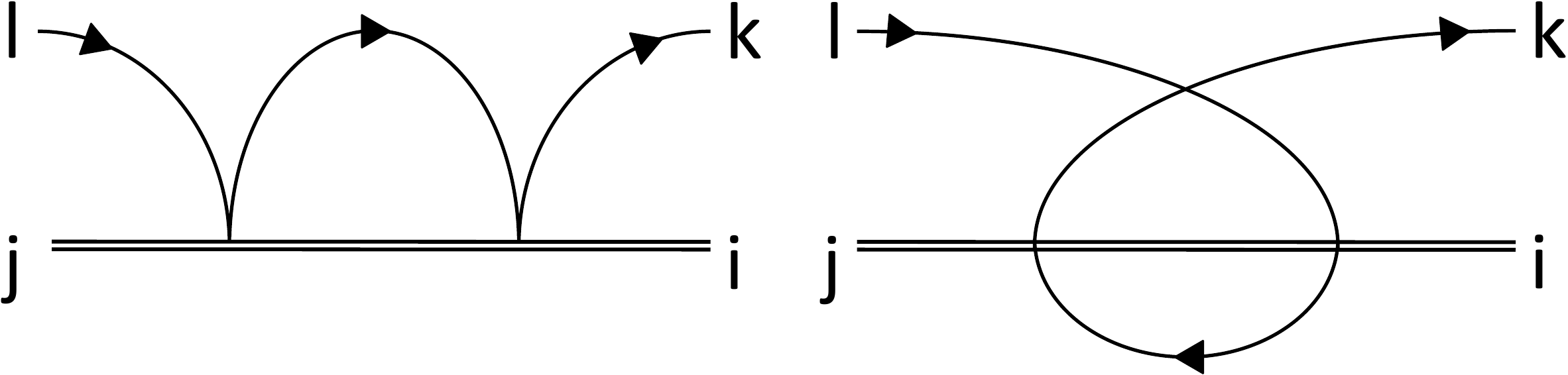} 
\caption{Diagrams for the Kondo effect at one-loop level. The particle loop (left) and the hole loop (right) are shown. The nucleon propagation is represented by the single lines, and the heavy impurity propagation is represented by the double lines, where $i$, $j$, $k$ and $l$ are the indices for fundamental representation of SU($n$) symmetry ($n=2$ for isospin).}
\label{fig:Kondo_pert}
\end{center}
\end{figure}

First of all, we consider a simple example for understanding the Kondo effect.
We assume the interaction Hamiltonian
\begin{align}
 H_{\mathrm{int}} = G \sum_{c=1}^{n^{2}-1} \sum_{kl,ij=1}^{n} \psi_{k}^{\dag} \left( \lambda^{c} \right)_{kl} \psi_{l} \Psi_{i}^{\dag} \left( \lambda^{c} \right)_{ij} \Psi_{j},
 \label{eq:Kondo_simple}
\end{align}
with $G>0$ the coupling constant, $\psi_{k}$ the fermion field composing the Fermi surface, and $\Psi_{i}$ the heavy impurity field, and $\lambda^{c}$ the Gell-Mann matrices ($c=1,\dots,n^{2}-1$) of SU($n$) group.
At the tree level, the scattering amplitude is apparently given by
\begin{align}
 M_{kl,ij}^{(0)} = G \sum_{c=1}^{n^{2}-1} \left( \lambda^{c} \right)_{kl} \left( \lambda^{c} \right)_{ij}.
 \label{eq:M_0}
\end{align}
At the one-loop level, the scattering amplitude is given by
\begin{align}
 M_{kl,ij}^{(1)} = G^{2} \rho_{0} \frac{n}{2} \sum_{c=1}^{n^{2}-1} \left( \lambda^{c} \right)_{kl} \left( \lambda^{c} \right)_{ij} \int_{0} \frac{\mathrm{d}E}{E-i\varepsilon},
 \label{eq:M_1}
\end{align}
with $\rho_{0}$ the state number density at the Fermi surface and $E$ the intermediate energy in the particle and hole loops, which is measured from the Fermi surface, as shown in the diagrams in Fig.~\ref{fig:Kondo_pert}.
Notice that the non-Abelian property of $\lambda^{c}$ leaves the $(\lambda^{c})_{kl}(\lambda^{c})_{ij}$ type-dependence in Eq.~(\ref{eq:M_1}).
First we find that the $M_{kl,ij}^{(1)}$ has a logarithmic divergence in the infrared energy region ($E \simeq 0$).\footnote{It should be noted that the existence of the infrared divergence is generated through the dynamics. This is qualitatively different from the intrinsic ultraviolet divergence due to the four point interaction in Eq.~(\ref{eq:Kondo_simple}).}
Second, by comparing Eqs.~(\ref{eq:M_0}) and (\ref{eq:M_1}), we also find that the $M_{kl,ij}^{(1)}$ can be larger than $M_{kl,ij}^{(0)}$ for any small coupling $G$ due to the logarithmic divergence.
The enhancement of the loop contributions indicate inevitably that the system should be governed by the non-perturbative dynamics for any small coupling.
This is called the Kondo effect.\footnote{The typical scale of the infrared energy relevant to the Kondo effect is called the Kondo scale, which is regarded to be the same as the Kondo temperature $T_{\mathrm{K}}$. This quantity is evaluated by the renormalization group analysis~\cite{Hattori:2015hka}.}
In the above analysis, we can confirm that the four conditions (i)-(iv) play the essential role (see e.g. Ref.~\cite{Hattori:2015hka}).

In Ref.~\cite{Yasui:2013xr}, the $\bar{D}$ ($B$) meson is regarded as the heavy impurity particle in nuclear matter, and the scattering amplitude between the impurity ($\Phi$) and the nucleon ($\psi$) is investigated at one-loop level.
As for the condition (iv), the non-Abelian property of the interaction is played not by spin-exchange but by isospin-exchange, because the $\bar{D}$ ($B$) meson is spin zero and isospin one half.
The interaction Lagrangian is given by the vector-current type
\begin{align}
 {\cal L}_{\mathrm{int}} = - \frac{G_{B}}{2}
 \sum_{a=1}^{n^{2}-1}
 (\bar{\psi}\gamma_{\mu}\lambda^{a}\psi)
 \left( -(i\partial^{\mu} \Phi^{\dag})\lambda^{a}\Phi + \Phi^{\dag} \lambda^{a} i\partial^{\mu} \Phi \right),
\end{align}
with $\lambda^{a}$ ($a=1,\dots,n^{2}-1$) being the Gell-Mann matrices for SU($n$) symmetry in general ($n=2$ for isospin).
Here we assume that the coupling constant $G_{\mathrm{B}}$ is positive to impose the condition that the isospin-singlet channel is attractive.
This assumption would be reasonable when we consider the $\pi$, $\rho$ meson-exchange interaction between a $\bar{D}$ meson and a nucleon.\footnote{See Eq~(\ref{eq:DbarNcontact}) or Eq.~(\ref{eq:DbarNcontactSU8}) for contact interaction with SU(4) or SU(8) symmetry and  Eq.~(\ref{eq:OPEP_DbarN_1/2}) for OPEP with heavy quark symmetry.}
Then, the scattering amplitude of the $\bar{D}$ meson and the nucleon was calculated at one-loop level, and it was obtained that there is a logarithmic enhancement in the infrared momentum scale from the loop ($\bar{D}$-nucleon and $\bar{D}$-hole) effect.
This is called the isospin Kondo effect.
The enhanced scattering amplitude affects transportation properties of a nucleon.

The result that the one-loop contribution becomes divergent indicates that the second order perturbation becomes larger than the first order, hence the perturbative expansion should break down at the infrared momentum scale.
Hence, to obtain the ground state, we need the non-perturbative approach beyond the perturbative one.
There are several theoretical methods to analyze such non-perturbative dynamics~\cite{Hewson,Yosida,Yamada}.
Among them, the mean-field approach is useful as an intuitive understanding.
This method is applied to analyze a $\bar{D}$ ($B$) meson bound in the discrete energy levels in an atomic nucleus, which exhibits the isospin Kondo effect~\cite{Yasui:2016ngy}.
Here, to simplify the model as far as possible, it is supposed that there is only the single valence orbital with energy $\epsilon$ for nucleons, whose degenerate states are labeled by $k=1,\cdots,N$, and the Hamiltonian 
\begin{align}
 H 
 =
 \sum_{k=1}^{N}\sum_{\sigma=\uparrow,\downarrow} \epsilon {c_{k \sigma}}^{\!\dag} c_{k \sigma} +
 g  \sum_{k,k'=1}^{N}  \left( {c_{k' \downarrow}}^{\!\dag} c_{k \uparrow} \, T_{+} + {c_{k' \uparrow}}^{\!\dag} c_{k \downarrow} \, T_{-} + ({c_{k' \uparrow}}^{\!\dag} c_{k \uparrow} \!-\! {c_{k' \downarrow}}^{\!\dag} c_{k \downarrow}) \, T_{3}  \right),
 \label{eq:HK}
\end{align}
is introduced.
Here $c_{k\sigma}^{(\dag)}$ is the annihilation (creation) operator for a nucleon with the orbital state $k$ and the isospin $\sigma=\uparrow,\downarrow$,
and 
$T_{+}$, $T_{-}$ and $T_{3}$ are raising, lowering operators and the third component of the Pauli matrices for the isospin of the bound $\bar{D}$ ($B$) meson.
The sign of the coupling constant is supposed to be positive, $g>0$, to give an attraction in the isospin singlet channel.
Though this is a very simple model, it is essentially important to satisfy the conditions (i)-(iv) for the Kondo effect.

As for Eq.~(\ref{eq:HK}), we notice that the simple mean-field approximation would not be useful because it may be smoothed out by the isospin-fluctuation due to the isospin-exchange interaction.
As a trick, we extend the Fock space spanned by the impurity isospin.
We introduce the quasi-fermion fields $f_{\uparrow,\downarrow}$ corresponding to the isospin up and down of the $\bar{D}$ ($B$) meson, and rewrite the isospin operator for the impurity particle as
\begin{align}
T_{+}={f_{\uparrow}}^{\dag} f_{\downarrow}, \hspace{0.5em}
T_{-}={f_{\downarrow}}^{\dag} f_{\uparrow}, \hspace{0.5em}
T_{3}=\frac{1}{2} \left( {f_{\uparrow}}^{\dag} f_{\uparrow} - {f_{\downarrow}}^{\dag} f_{\downarrow} \right),
\end{align}
provided that the constraint condition,
\begin{align}
 \sum_{\sigma = \uparrow, \downarrow} {f_{\sigma}}^{\dag} f_{\sigma} =1,
\end{align}
is imposed because the number of the impurity particle should be one in average.
The constraint condition can be considered in the modified Hamiltonian
\begin{align}
 \tilde{H} = H + \lambda \left( \sum_{\sigma = \uparrow, \downarrow} {f_{\sigma}}^{\dag} f_{\sigma} -1 \right),
\end{align}
with the Lagrange multiplier $\lambda$ for Eq.~ (\ref{eq:HK}).
With the help of the new operators $f_{\uparrow,\downarrow}$, the three-point interaction in Eq.~(\ref{eq:HK}) is changed to the four-point interaction, and it becomes possible to apply the mean-field approximation by picking up two operators (ex. ${f_{\sigma}}^{\dag} c_{k\sigma}$) from the four operators.
We define the mean-field (energy gap),
\begin{align}
\Delta = -g\sum_{k,\sigma} \langle {f_{\sigma}}^{\dag} c_{k\sigma} \rangle,
\end{align}
as the mixing of the nucleon and the impurity particle.
The price for the simplification is that the Fock space is extended by the introduction of $f_{\uparrow,\downarrow}$.
As a result, it turns out that the energy of the system obtained by the mean-field approximation can be comparable with the exact solution.
The approximate solution becomes much close to the exact one, when the quantum fluctuation around the mean-field is also taken into account by the random-phase approximation.
The mean-field approximation as well as the random-phase approximation would be applicable to realistic model setting for a $\bar{D}$ ($B$) meson in an atomic nucleus.\footnote{Recently, the mean-field approximation in a field-theoretic way is applied to the QCD Kondo effect in light-flavor quark matter with heavy quark distributed as impurity particles, where the non-Abelian interaction is given by the color exchange~\cite{Yasui:2016svc}.}

\paragraph{Spin-isospin correlated nuclear matter}
Concerning $\bar{D}^{(\ast)}$ meson, we have discussed the generation of attractive force induced by the $\bar{D}N$-$\bar{D}^{\ast}N$ mixing~\cite{Yasui:2009bz,GarciaRecio:2008dp,Yamaguchi:2011xb,Yamaguchi:2011qw}. 
This is the mixing effect by the two-body scattering supported by the accompanying nucleon~\cite{Yasui:2012rw,GarciaRecio:2011xt}.
We note that there is apparently no $\bar{D}$-$\bar{D}^{\ast}$ mixing in vacuum.
However, when finite spin-isospin correlation exists in nuclear matter, the $\bar{D}$-$\bar{D}^{\ast}$ mixing as a single-body scattering can happen in nuclear matter~\cite{Suenaga:2014dia}.
It is known that such spin-isospin correlated nuclear matter is realized in the pion condensate~\cite{Kunihiro:PTEP112_123,Kunihiro:PTEP112_197}.
Let us summarize the results about the $\bar{D}^{(\ast)}$ meson mass spectrum in the spin-isospin correlated nuclear matter in Ref.~\cite{Suenaga:2014dia}.

The state with the spin-isospin correlation can be expressed by
\begin{align}
\langle A^{i a} \rangle = \alpha \, \delta^{ia} \hspace{1em} (\mathrm{Pattern}\hspace{0.2em}\mathrm{I}),
\hspace{1em} \mathrm{or} \hspace{1em}
\langle A^{i a} \rangle = \alpha \, \delta^{i3} \delta^{a3} \hspace{1em} (\mathrm{Pattern}\hspace{0.2em}\mathrm{II}),
\end{align}
with the pion axial-vector current $A^{\mu}$, $i=x,y,z$ the space direction and $a=1,2,3$ the isospin direction.
$\alpha$ measures the magnitude of the pion condensate.
Inserting those configurations into Eq.~(\ref{eq:L_heavy_hadron_pion}),
we obtain the mass spectrum of $\bar{D}^{(\ast)}$ meson in the spin-isospin correlated nuclear matter.
In Pattern I, the SU(2)$_{l}$ spin symmetry and the SU(2)$_{I}$ isospin symmetry are coupled each other in the spin-isospin correlated nuclear matter.
As a result, the symmetry is broken to the diagonal SU(2)$_{\mathrm{diag}}$ symmetry.
In Pattern II, the spin-isospin correlated nuclear matter is invariant under $\mathrm{U}(1)_{l} \times \mathrm{U}(1)_{I} \times Z_{2}$ symmetry, where $\mathrm{U}(1)_{l}$ and $\mathrm{U}(1)_{I}$ are spin and isospin symmetries, 
and $Z_{2}$ is the simultaneous transformation by $\mathrm{U}(1)_{l}$ and $\mathrm{U}(1)_{I}$.

First of all, we notice that the $\bar{D}$ and $\bar{D}^{\ast}$ mesons in normal nuclear matter have totally eight degree of freedom due to spin 1/2 and isospin 1/2 of the light quark component ($q$) and spin 1/2 of the heavy antiquark component ($\bar{Q}$).
Those eight states are degenerate in the heavy quark limit.
Because the heavy quark spin is irrelevant to the dynamics in this limit, 
the number of the relevant degrees of freedoms are four from the $q$.
However, the mass degeneracy of those four states is resolved in the spin-isospin correlated nuclear matter.

In Pattern I, the light component $q$ becomes a triplet representation (${\bf 3}_{q}$) or a singlet representation (${\bf 1}_{q}$) of SU(2)$_{\mathrm{diag}}$.
When the heavy antiquark is added, the former becomes ${\bf 3}_{q} \times {\bf 2}_{\bar{Q}} = {\bf 2}_{q\bar{Q}}+{\bf 4}_{q\bar{Q}}$, and the latter becomes ${\bf 1}_{q} \times {\bf 2}_{\bar{Q}} = {\bf 2}_{q\bar{Q}}^{\prime}$.
Thus, in the heavy quark limit, there are six degenerate states from ${\bf 2}_{q\bar{Q}}$ and ${\bf 4}_{q\bar{Q}}$, and there are two degenerate states from ${\bf 2}_{q\bar{Q}}^{\prime}$.

In Pattern II, there are two different charges, $(+,+)$ and $(+,-)$, in correspondence to $\mathrm{U}(1)_{l} \times \mathrm{U}(1)_{I}$ symmetry.
Notice that $(-,-)$ is equivalent to $(+,+)$ and that $(-,+)$ is equivalent to $(+,-)$ because there is $Z_{2}$ symmetry.
Therefore, as for the light quark component, there is a doublet state (${\bf 2}_{q}$) as well as another doublet state (${\bf 2}'_{q}$).
When the heavy antiquark is added, the former becomes ${\bf 2}_{q} \times {\bf 2}_{\bar{Q}} = {\bf 1}_{q\bar{Q}}+{\bf 3}_{q\bar{Q}}$ and the latter becomes ${\bf 2}_{q}^{\prime} \times {\bf 2}_{\bar{Q}} = {\bf 1}_{q\bar{Q}}^{\prime} + {\bf 3}_{q\bar{Q}}^{\prime}$.
Thus, in the heavy quark limit, there are four degenerate states from ${\bf 1}_{q\bar{Q}}$ and ${\bf 3}_{q\bar{Q}}$, and there are another four degenerate states from ${\bf 1}_{q\bar{Q}}^{\prime}$ and ${\bf 3}_{q\bar{Q}}^{\prime}$.

\begin{figure}[tbp]
 \begin{minipage}{0.5\hsize}
   \centering
   \includegraphics[scale=0.7,bb=0 0 260 164]{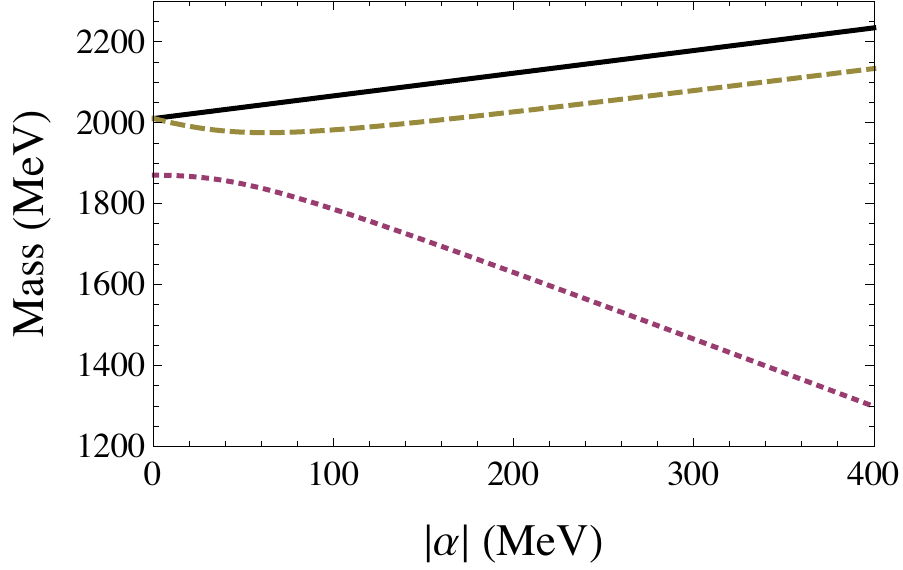}
 \end{minipage}
 \begin{minipage}{0.5\hsize}
    \centering
   \includegraphics[scale=0.7,bb=0 0 260 170]{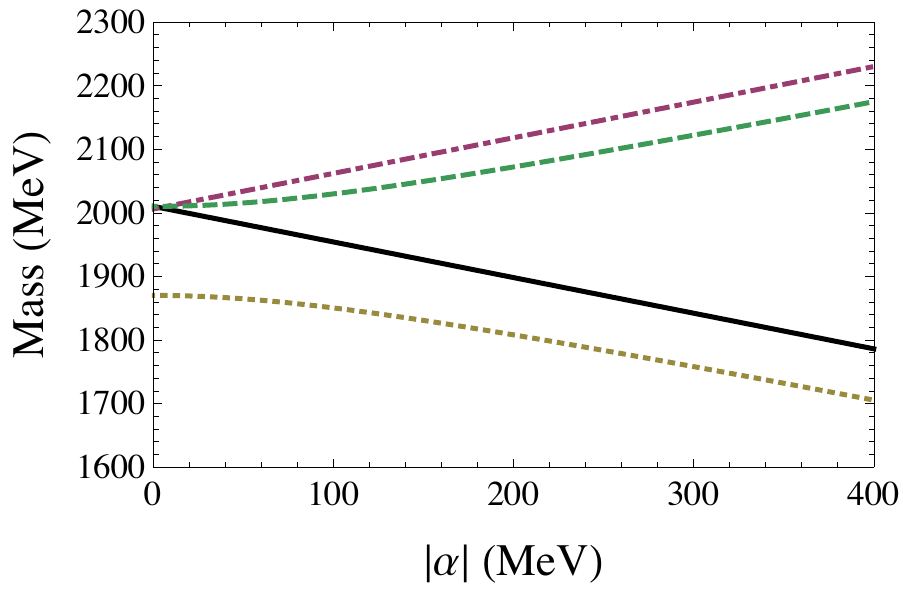}
 \end{minipage}
  \caption{The mass spectrum of $\bar{D}^{(\ast)}$ meson in pion condensate with pattern I (left) and II (right) \cite{Suenaga:2014dia}. In the left, the solid curve indicates ${\bf 4}_{q\bar{Q}}$ state, and the dashed two curves indicate the mixed state of ${\bf 2}_{q\bar{Q}}$ and ${\bf 2}_{q\bar{Q}}^{\prime}$. In the right, four curves indicate the mixing of ${\bf 1}_{q\bar{Q}}$ and ${\bf 1}_{q\bar{Q}}^{\prime}$ for two curves, and the mixing of ${\bf 3}_{q\bar{Q}}$ and ${\bf 3}_{q\bar{Q}}^{\prime}$ for two curves.}
  \label{fig:Suenaga:2014dia}
\end{figure}

The obtained mass spectrum of the $\bar{D}^{(\ast)}$ in the spin-isospin correlated nuclear matter is shown in Fig.~\ref{fig:Suenaga:2014dia}.
In reality, we have to consider the breaking of the heavy quark symmetry at finite charm quark mass, namely the mass difference between a $\bar{D}$ meson and a $\bar{D}^{\ast}$ meson.
In Pattern I, accordingly, the degeneracy of ${\bf 2}_{q\bar{Q}}$ and ${\bf 4}_{q\bar{Q}}$ becomes resolved, and the masses of ${\bf 2}_{q\bar{Q}}$ and ${\bf 4}_{q\bar{Q}}$ becomes different.
Among them, ${\bf 2}_{q\bar{Q}}$ and ${\bf 2}_{q\bar{Q}}^{\prime}$ are mixed each other.
In Pattern II, the degeneracy of ${\bf 1}_{q\bar{Q}}$ and ${\bf 3}_{q\bar{Q}}$ becomes resolved, and the degeneracy of ${\bf 1}_{q\bar{Q}}^{\prime}$ and ${\bf 3}_{q\bar{Q}}^{\prime}$ becomes resolved also.
Then, ${\bf 1}_{q\bar{Q}}$ and ${\bf 1}_{q\bar{Q}}^{\prime}$ are mixed, and ${\bf 3}_{q\bar{Q}}$ are ${\bf 3}_{q\bar{Q}}^{\prime}$ are mixed also.
In this case, due to $\mathrm{U}(1)_{J} \times \mathrm{U}(1)_{I}$ symmetry and $Z_{2}$ symmetry,
the $\Bigl\{(0,+), (0,-)\Bigr\}$ state exists as the degenerate one in correspondence to the $\bar{D}$ meson,
and the $\Bigl\{ (0,+), (0,-)\Bigr\}$, $\Bigl\{(+,+), (-,-)\Bigr\}$ and $\Bigl\{(+,-), (-,+)\Bigr\}$ states exist as the degenerate ones in correspondence to the $\bar{D}^{\ast}$ meson.
Among them, two $\Bigl\{ (0,+), (0,-)\Bigr\}$ states are mixed each other.

As more general form, we may consider the following spin-isospin correlation
\begin{align}
\langle A^{i a} \rangle =  \sum_{j=1,2,3} \alpha_{j} \delta^{ij} \delta^{aj} \hspace{1em} (\mathrm{Pattern}\hspace{0.2em}\mathrm{III}).
\end{align}
In this case, there are four doublets whose masses are all different.

As for other exotic nuclear matter, the $\bar{D}^{(\ast)}$ meson mass spectrum in the Skyrmion matter and the chiral density waves were studied~\cite{Suenaga:2014sga,Suenaga:2015daa}.

\section{Heavy baryons}
\label{sec:charm_baryons}

\subsection{Early studies on $\Lambda_c N$ interaction and $\Lambda_{c}$ nuclei}

Historically, a $\Lambda_c$ nucleus is the first charm nucleus which was studied for the first time as the nucleus with charm flavor in 1970s.
At that time, SU(4) flavor symmetry was applied as a straightforward extension from SU(3) flavor symmetry in hypernuclei with strangeness, and many bound states including excited states in $\Lambda_{c}$ nuclei were discussed.

An early idea can be found in Refs.~\cite{Tyapkin1975,Tyapkin1976,Iwao:1976yi}.
In Ref.~\cite{Iwao:1976yi}, it was already pointed out that the pion exchange potential accompanying the $\Lambda_{c}N$-$\Sigma_{c}N$ mixing gives the attraction for $\Lambda_{c}$ in nucleus.
This is an analogue from the $\Lambda N$-$\Sigma N$ mixing in hypernuclei~\cite{Afnan:1989wb,Afnan:1990vs}.
The binding energy of the most stable state of $\Lambda_{c}$ in nuclear matter is about 28 MeV by assuming the mean-field potential of $\Lambda$ hypernuclear matter with free Fermi gas approximation~\cite{Gatto:1978ka}.
As a small system, the binding energy of a $\Lambda_{c}$ in He nucleus was obtained as about 8 MeV by supposing a mean-field given from $\Lambda$ hypernuclei study.
A $\Lambda_{c}\bar{\Lambda}_{c}$ bound state with binding energy 4.1 MeV was also discussed~\cite{Dover:1977hs}.

Based on SU(4) flavor symmetry, 
$\Lambda_c$ and $\Sigma_c$ bound states were discussed for He, C, O, Ni and Pb nuclei~\cite{Dover:1977jw}.
The interaction between a $\Lambda_{c}$ and a nucleon was provided by a sum of the meson exchange potentials with scalar and vector mesons at long-middle distance ($r>r_{c}$) and the hard core potential at short distance ($r<r_{c}$) with the core radius $r_{c}=0.552$ fm.
This interaction is the same as that in Ref.~\cite{Dover:1977hs}.
The shell model calculation was performed by using the Hartree type potential without spin-orbital potential, and the bound states of $\Lambda_{c}$, $\Sigma_{c}$, $\Xi_{c}$, $\Xi_{c}^{\ast}$ baryons were investigated.
It turned out that the binding energy in He nucleus is about 15 MeV for S-wave and 1 MeV for P-wave, and the binding energy in Pb nucleus is about 60 MeV for S-wave.

More detailed structures were analyzed based on the modern nuclear potentials in Refs.~\cite{Bando:1981ti,Bando:1983yt,Bando:1985up}.
In Ref.~\cite{Bando:1981ti}, based on SU(4) flavor symmetry, the authors used the $\Lambda_{c}N$ potential of a Gaussian type used in $\Lambda N$ potential.
They found that the binding energy was about 3.1 MeV for the bound $\Lambda_{c}N$ system.
Cluster structures of $\Lambda_{c}$ nuclei were also investigated by regarding $^{8}\mathrm{Be}$ nucleus as $\alpha \alpha$ cluster.
They used the folding potential, and obtained the binding energy from 1.47 MeV to 12.28 MeV.
They also showed that the rotation band appears in the energy spectrum.
Furthermore, in Refs.~\cite{Bando:1983yt,Bando:1985up},
the authors made an extension of the Nijmegen one-boson-exchange potential from SU(3) flavor symmetry to SU(4) flavor symmetry for charm, as well as to SU(5) flavor symmetry for bottom, along the line in Ref.~\cite{Dover:1977jw}.
It has the competition between the attraction by $\sigma$ meson exchange and the repulsion by $\omega$ meson exchange.
In comparison to the $\Lambda N$ potential, the $\Lambda_{c}$/$\Lambda_{b}$ potential has the property that the P-wave attraction is enhanced rather than the S-wave attraction because the Majorana type force from $K$, $K^{\ast}$ meson exchange is removed in the $\Lambda_{c}$/$\Lambda_{b}$ case.\footnote{It is known that the $K$ ($K^{\ast}$) meson exchange in the $\Lambda N$ interaction gives a sizable attraction (repulsion) in even (odd) angular momentum due to the Majorana type character.}
The $D$, $D^{\ast}$ meson exchange was considered to be unimportant because of their large masses.
This property leads to the non-locality (momentum-dependence) in the single particle potential of the $\Lambda_{c}$/$\Lambda_{b}$ baryon in nucleus.
The extended Nijmegen potential in this way was applied to calculate the binding energy of $\Lambda_{c}$ in $\alpha$, and it was found that $\Lambda_{c}$ is barely bound, while there is a bound state with the binding energy $3.1$ MeV for $\Lambda_{b}$.
The authors investigated the binding energy of $\Lambda_{c}$ in nuclear matter ($\simeq 20$ MeV) by the Brueckner (G-matrix) approach.

Some comments are in order for the results in Refs.~\cite{Bando:1981ti,Bando:1983yt,Bando:1985up}.
Although the attraction of $\Lambda_{c}$ ($\Lambda_{b}$) is weaker than $\Lambda$,
the $\Lambda_{c}$ ($\Lambda_{b}$) bound state can appear by means of the suppressed kinematic energy due to the heavy mass.
As for the nuclear matter calculation, though the depth of the effective $\Lambda_{c}$ ($\Lambda_{b}$) potential in the G-matrix calculation is about 2/3 of that of the effective $\Lambda$ potential, the number of the bound states increases than that in $\Lambda$ hypernuclei.
As the partial wave components, the P-wave components of $\Lambda_{c}$ ($\Lambda_{b}$) in nuclear matter are increased due to the P-wave attraction rather than $\Lambda$ in nuclear matter.
For example, the most attractive channel in $\Lambda$ hypernuclei is given by $^{3}\mathrm{S}_{1}$+$^{3}\mathrm{D}_{1}$.
This is also the case for $\Lambda_{c}$ nuclei.
In addition to this channel, however, $^{3}\mathrm{P}_{2}$+$^{3}\mathrm{F}_{2}$ shows also a comparable magnitude in $\Lambda_{c}$ nuclei.
This property is interesting as the nuclear many-body dynamics, because it may enable us to probe higher partial wave contributions by injecting $\Lambda_{c}$ ($\Lambda_{b}$) as an impurity.
The application of the mean-field to the finite nuclei was considered in Ref.~\cite{Bando:1985up}.
For example, the authors obtained the binding energy of $\Lambda_{c}$ ($\Lambda_{b}$) about 23 MeV (32 MeV) in Pb nucleus.

The few-body calculation was performed for $\Lambda_{c}$ in $^{3}$He and $^{4}$He nuclei as well as in $^{5}$Li nucleus by assuming a separable type potential~\cite{Gibson:1983zw}.
The binding energy of about 10 MeV was obtained as most.

Afterwards, the study of $\Lambda_c$ in nuclear matter was developed by considering the effect of the quark degrees of freedom~\cite{Froemel:2004ea,Huang:2013zva}.
On the reference of the phenomenological nuclear potential (Nijmegen, AV18)~\cite{Hayashigaki:1998ey,Wiringa:1994wb},
the type of the $\Lambda_{c}N$ potential was classified by the spin and isospin structures, $1$, $\vec{\sigma}_{i} \!\cdot\! \vec{\sigma}_{j}$, $\vec{\tau}_{i} \!\cdot\! \vec{\tau}_{j}$ and $(\vec{\sigma}_{i} \!\cdot\! \vec{\sigma}_{i}) (\vec{\tau}_{i} \!\cdot\! \vec{\tau}_{j})$.
The coupling constants in those terms were changed according to the scaling by a quark wave function in a quark model, and the $\Lambda_{c}N$ potential was obtained.
Intuitively, the vertex strength becomes smaller, because the number of light quarks becomes fewer than light baryons.
The charm dibaryon $\Xi_{c}'N$, $\Xi_{cc}N$ states were investigated, and it was found that the binding energy of $\Xi_{c}'N$ is about 10 MeV at most, and the binding energy of $\Xi_{cc}N$ is a few or a few hundred MeV.
We note, however, that values of the binding energies are different for individual model settings.
In fact, no bound state was found in some cases.
The authors investigated $\Sigma_{c}N$ ($I=3/2,1/2$) also.
Furthermore, multi-charm dibaryons, $\Xi \Xi_{cc}$, $\Xi_{c}'\Xi_{c}'$, $\Xi_{c}'\Xi_{cc}$ and $\Xi_{cc}\Xi_{cc}$, were investigated, and bound states with a few hundred MeV were obtained for some potential models.
One of the reasons for those deep bound states is of heavy masses of the constituent baryons, and another is a weakness of the core repulsion.

\subsection{Recent works on $\Lambda_{c}N$ interaction}

So far, a $\Lambda_{c} N$ potential have been constructed starting from the phenomenological nucleon-nucleon or hyperon-nucleon potentials.
However, because a nucleon as well as a hyperon are regarded as the ``light" baryons,
its naive extension to heavy baryons does not reflect the heavy quark symmetry which is crucially important (Sect.~\ref{sec:heavy_quark_symmetry}).
In Refs.~\cite{Liu:2011xc,Oka:2013iua}, the $\Lambda_{c} N$ potential is developed based on the effective Lagrangian respecting the heavy quark symmetry.
Remember that, in hyperon-nucleon potential, the mixing between $\Lambda N$ and $\Sigma N$ channels provides the strong attraction~\cite{Afnan:1989wb,Afnan:1990vs}.
In analogy, we may think that the mixing between $\Lambda_{c} N$ and $\Sigma_{c} N$ may give the strong attraction as well.
However, the situation is more complicated in a charm sector.
According to the heavy quark symmetry, a heavy-quark spin partner of $\Sigma_{c}$ is $\Sigma_{c}^{\ast}$ with spin-parity $J^{P}=3/2^{+}$ and isospin $I=1$, and the mass splitting between $\Sigma_{c}$ and $\Sigma_{c}^{\ast}$ is quite small as 65 MeV.
Hence, in addition to the $\Lambda_{c}N$-$\Sigma_{c}N$ mixing, the mixing between $\Sigma_{c}$ and $\Sigma_{c}^{\ast}$ provides a strongly attractive contribution.\footnote{This is analogous situation to the $\bar{D}^{(\ast)}N$ potential , for which the mixing between $\bar{D}N$ and $\bar{D}^{\ast}N$ plays the important role~\cite{Yasui:2009bz,Gamermann:2010zz} (cf.~Sect.~\ref{sec:D_mesons}).}
Therefore, it becomes important to investigate the $\Lambda_{c}N$ interaction through the three-channel coupling of $\Lambda_{c}N$-$\Sigma_{c}N$-$\Sigma_{c}^{\ast}N$.

Let us remember that, based on the approximate degeneracy of $\Sigma_{c}$ and $\Sigma_{c}^{\ast}$, we introduced the effective fields $B_{{\bm 6}}$ and $B^{\ast}_{{\bm 6}\mu}$ corresponding to
 $B_{{\bm 6}} = (Q\{qq\}_{I=1,j^{P}=1^{+}})_{I(J^{P})=1(1/2^{+})}$ and 
 $B_{{\bm 6}\mu} = (Q\{qq\}_{I=1,j^{P}=1^{+}})_{I(J^{P})=1(3/2^{+})\mu}$,
with the polarization vector $\mu$, and consider their linear combination as a super-field (\ref{eq:superfield_baryon}) in the heavy quark limit (cf.~Sect.~\ref{sec:heavy_hadron_effective_theory}).
Though $\Sigma_{c}$ and $\Sigma_{c}^{\ast}$ approach the HQS doublet as degenerate states in the heavy quark limit,
$\Lambda_{c}$ approaches the HQS singlet with no state to be paired.
Hence we have to define the effective field $B_{\bar{\bf 3}}$ which is independent of $B_{{\bm 6}}$ and $B^{\ast}_{{\bm 6}\mu}$.
The components in the $\Lambda_{c}N$-$\Sigma_{c}N$-$\Sigma_{c}^{\ast}N$ system are classified according to the internal spin and the angular momentum.
Examples for $J^{P}=0^{+}$ and $1^{+}$ are summarized in Table~\ref{table:LambdacN_channel}~\cite{Liu:2011xc,Oka:2013iua}.

In Refs.~\cite{Liu:2011xc,Oka:2013iua}, the potential between a charm baryon and a nucleon with channel coupling $\Lambda_{c}N$-$\Sigma_{c}N$-$\Sigma_{c}^{\ast}N$ is given by $\sigma$, $\pi$, $\omega$ and $\rho$ exchanges.
It is shown that there are bound states with $J^{P}=0^{+}, 1^{+}$ and binding energies are from a few MeV to a hundred MeV, depending on the model parameters such as coupling constants and momentum cutoff parameters.
Interestingly, the difference between the masses of the $J^{P}=0^{+}, 1^{+}$ states is about 10 MeV, which is quite smaller than the other scales.
This property is understood from the heavy quark symmetry.
When  the $\Lambda_{c}N$ is regarded as a $Qqq$-$N$ system, the brown muck $qq$-$N$ should have spin 1/2 because the spin of $qq$ in $\Lambda_{c}$ must be zero.
When $qq$-$N$ is combined with the heavy quark $Q$ with spin 1/2, the compound states having spin either 0 and 1 become degenerate because the spin direction of the heavy quark is irrelevant to the total energy.
It is also important to note the three-channel coupling, $\Lambda_{c}N$-$\Sigma_{c}N$-$\Sigma_{c}^{\ast}N$, plays the significant role for providing a strong attraction.
For example, when we consider only the $\Lambda_{c}N$ channel, we find that the binding energy becomes smaller or the bound state vanishes.
In fact, the strong attraction is provided by the D-wave mixing given by the pion exchange potential accompanying a tensor force (e.g. $\Sigma_{Q}^{\ast}N(^5\mathrm{D}_{0})$ for $J^{P}=0^{+}$ in Table~\ref{table:LambdacN_channel}), provided that its fraction in the wave function is about a few percents.\footnote{It is shown also that, in $\bar{D}N$-$\bar{D}^{\ast}N$ systems, the pion exchange potential accompanying a tensor force plays the significant role for an attraction~\cite{Yasui:2009bz} (cf.~Sec.~\ref{sec:D_mesons}).}

As another approach, based on the quark model, the $\Lambda_{c}N$ interaction is calculated~\cite{Gal:2014jza,Huang:2013zva}.
In Ref.~\cite{Gal:2014jza}, the possibility of $\Lambda_{c}N$ bound states is discussed by the chiral constituent quark model,\footnote{See Ref.~\cite{Valcarce:2005em} for recent review about baryon-baryon interaction based on the quark model.} and
it is found, however, that there is no bound state both for $^{3}S_{1}$ and $^{1}S_{0}$.
In Ref.~\cite{Huang:2013zva}, the $\Lambda_{c}N$ potential as well as the $\Sigma_{c}N$ are investigated.
It turned out again that there is no bound state in $\Lambda_{c}N$ thought the potential is attractive.
On the other hand, there is a resonance, not a bound state, in $\Sigma_{c} N(^{3}S_{1})$.

\begin{table}[tbp]
\centering
\caption{\label{table_qnumbers} \small Various coupled channels for a 
given quantum number $J^P$~\cite{Liu:2011xc,Oka:2013iua}.} 
\vspace*{0.5cm}
{\small 
\begin{tabular}{ c  | c c c c c c c}
\hline
$J^P$ &  \multicolumn{7}{c}{channels} \\
\hline
$0^+$ &  $\Lambda_{Q}N(^1\mathrm{S}_{0})$ & $\Sigma_{Q}N(^1\mathrm{S}_{0})$ & 
     $\Sigma_{Q}^{\ast}N(^5\mathrm{D}_{0})$ &  & & & \\
$1^+$ &  $\Lambda_{Q}N(^3\mathrm{S}_{1})$ & $\Sigma_{Q}N(^3\mathrm{S}_{1})$ & 
     $\Sigma_{Q}^{\ast}N(^3\mathrm{S}_{1})$ &  $\Lambda_{Q}N(^3\mathrm{D}_{1})$ & $\Sigma_{Q}N(^3\mathrm{D}_{1})$ & $\Sigma_{Q}^{\ast}N(^3\mathrm{D}_{1})$ & $\Sigma_{Q}^{\ast}N(^5\mathrm{D}_{1})$ \\
\hline
\end{tabular}
}
\label{table:LambdacN_channel}
\end{table}

\begin{figure}[tbp]
 \begin{minipage}{0.5\hsize}
   \centering
   \includegraphics[scale=0.3,bb=0 0 604 442]{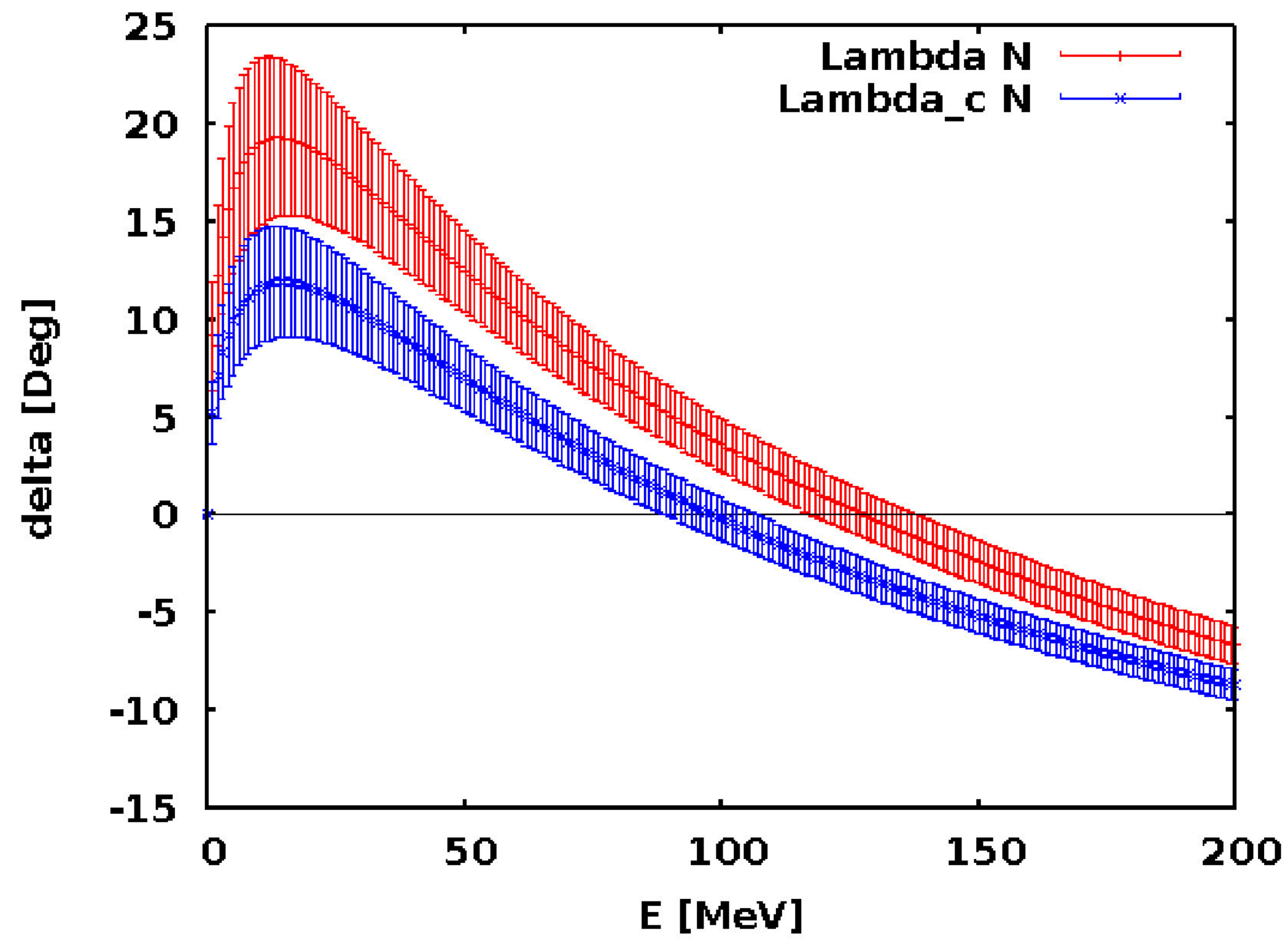}
 \end{minipage}
 \begin{minipage}{0.5\hsize}
    \centering
   \includegraphics[scale=0.3,bb=0 0 608 442]{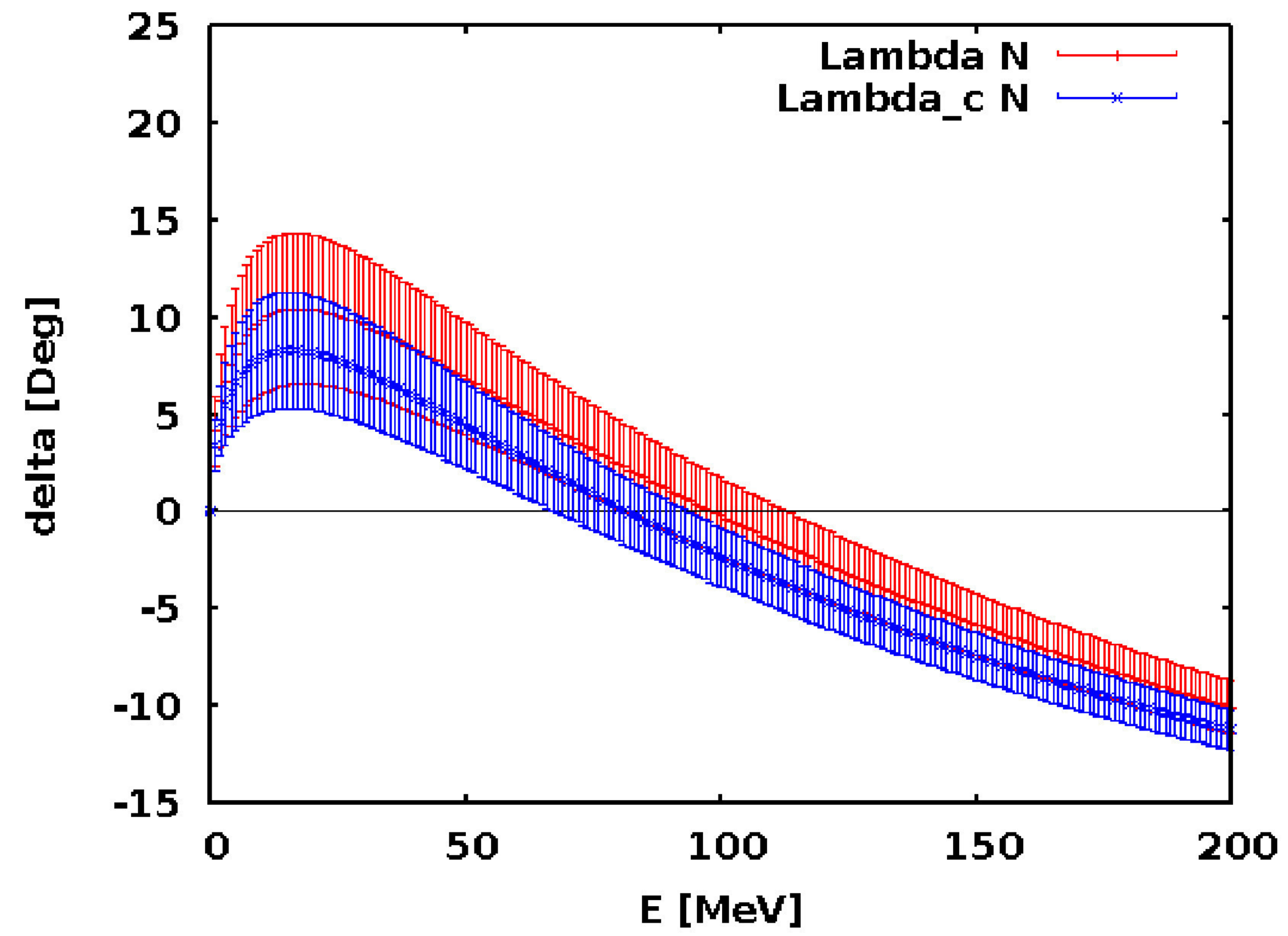}
 \end{minipage}
  \caption{The phase shifts of the scattering of $\Lambda N$ and $\Lambda_c N$ for $m_{\pi}=700$ MeV (left) and $m_{\pi}=570$ MeV (right) from HAL QCD collaboration~\cite{Miyamoto:2016hqo}.}
  \label{fig:Miyamoto:2016hqo}
\end{figure}

Recently, the $\Lambda_{c}N$ potential with $^{1}S_{0}$ channel is investigated by lattice QCD simulation by HAL QCD collaboration~\cite{Miyamoto:2016hqo}.\footnote{See Refs.~\cite{Ishii:2006ec,Aoki:2009ji} for more information about the potential calculation used in HAL QCD collaboration.}
It is found that there is an attraction at low scattering energy and a repulsion at high scattering energy (Fig.~\ref{fig:Miyamoto:2016hqo}).
As a tendency, the $\Lambda_c N$ attraction is slightly weaker than the $\Lambda N$ attraction.
From the behavior of the phase shift, it turned out that the $\Lambda_{c}N$ attraction is not strong enough to form a bound state, though $\Lambda_{c}$ can be bound in nuclear matter due to the (positive) scattering length.\footnote{Note the difference of the definition of the sign of the scattering length in Eq.~(\ref{eq:scattering_massshift}).}
Numerically, the scattering length of $\Lambda_{c}N$ is $a_{\Lambda_c N}=0.43(13)$ fm ($m_{\pi}=700$ MeV) and $a_{\Lambda_c N}=0.29(11)$ fm ($m_{\pi}=570$ MeV).
In those cases, using Eq.~(\ref{eq:scattering_massshift}) by replacing a $D$ meson mass $m_{D}$ to the mass $m_{\Lambda_{c}}$ of $\Lambda_{c}$ baryon, we obtain the $\Lambda_{c}$ mass shifts, 
 which are $\Delta m_{\Lambda_{c}}=-26$ MeV ($m_{\pi}=700$ MeV) and $\Delta m_{\Lambda_{c}}=-18$ MeV ($m_{\pi}=570$ MeV) in normal nuclear matter.
Those numbers are comparable with the binding energy of $\Lambda$ hyperon in nuclear matter.
Its scattering length is $a_{\Lambda N}=0.83(27)$ fm ($m_{\pi}=700$ MeV) and $a_{\Lambda N}=0.39(17)$ fm ($m_{\pi}=570$ MeV). 
Correspondingly, in a similar way to $\Lambda_{c}$, we obtain the mass shift is $\Delta m_{\Lambda}=-67$ MeV ($m_{\pi}=700$ MeV), $\Delta m_{\Lambda}=-32$ MeV ($m_{\pi}=570$ MeV) in normal nuclear matter.\footnote{In literature, it is considered that the binding energy of $\Lambda$ in nuclear matter is about 28 MeV, which is about 2/3 of a nucleon~\cite{Millener:1988hp}.}
At present, it is of course difficult to regard those numbers as the realistic numbers comparable with experimental values due to the large mass of a pion.
For example, the scattering length of $\Lambda N$ is smaller than the empirical values~\cite{Haidenbauer:2013oca}.

\subsection{Few-body systems}

\begin{figure}[tbp]
\begin{center}
\includegraphics[scale=0.4,bb=0 0 720 540]{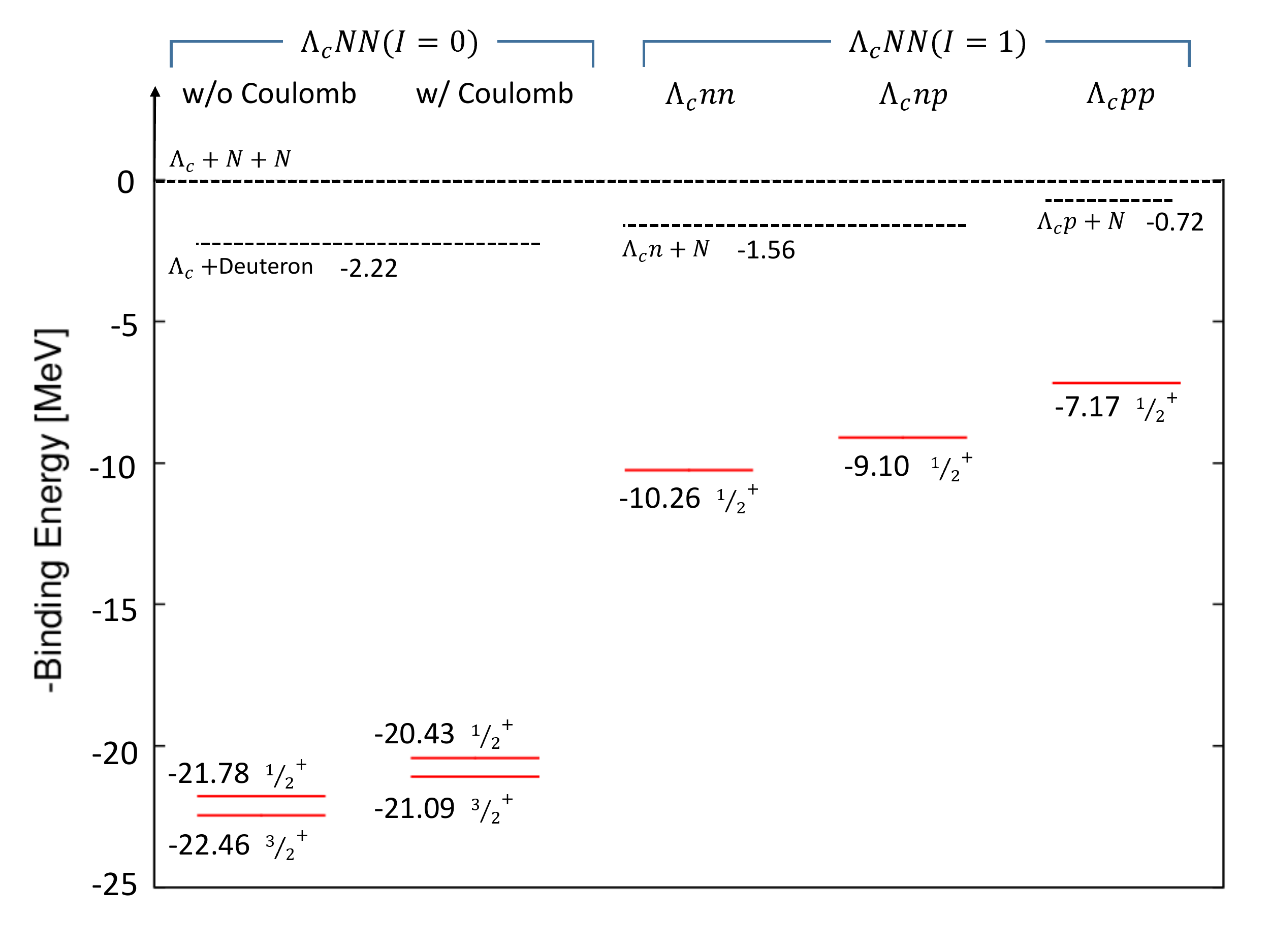}
\caption{Energy levels of $\Lambda_{c}NN$ bound states~\cite{Maeda:2015hxa}.}
\label{fig:Maeda:2015hxa}
\end{center}
\end{figure}

Few-body systems with charm baryons are analogues of few-body hypernuclear systems.
In analogy to $\Lambda NN$-$\Sigma NN$ systems, the possibility of the bound states of $\Lambda_{c} NN$-$\Sigma_{c} NN$ systems are studied based on the baryon-baryon interaction from the chiral quark model~\cite{Garcilazo:2015qha}.
There, the $\Lambda_{c}$-$\Sigma_{c}$ conversions are considered like the $\Lambda$-$\Sigma$ conversions in hypernuclei.
Moreover, in Ref.~\cite{Maeda:2015hxa}, the $\Lambda_{c}N$ potential is constructed including the quark-exchange potential instead of the vector meson exchange at short distance by considering the heavy quark symmetry in analogous way with Ref.~\cite{Liu:2011xc}.
Then, the three-body $\Lambda_{c}NN$ system is investigated by a few-body calculation.
In this work, $\Sigma_{c}$ and $\Sigma_{c}^{\ast}$ degrees of freedom is integrated, and the effective $\Lambda_{c}N$ potential is constructed.
The results of the few-body calculation are shown in Fig.~\ref{fig:Maeda:2015hxa}.
Total isospin $I$ is given by $NN$ because $\Lambda_{c}$ is isospin-singlet.
Here the authors assumed that all partial waves are S-wave.
For the total isospin $I=0$, the $NN$ is isospin-singlet and hence it is spin-triplet due to the Pauli exclusion principle.
As a result, the heavy quark spin structure of $\Lambda_{c}NN$ is assigned to be the HQS doublet.
For the total isospin $I=1$, the $NN$ is isospin-triplet and spin-singlet.
The heavy quark spin structure of $\Lambda_{c}NN$ is assigned to be the HQS singlet.
Such structures of the HQS multiplets are clearly seen for $J^{P}=1/2^{+}$ and $3/2^{+}$ (HQS doublet) for $\Lambda_{c}NN(I=0)$ and $J^{P}=1/2^{+}$ (HQS singlet) in $\Lambda_{c}NN(I=1)$ as shown in Fig.~\ref{fig:Maeda:2015hxa}.

\subsection{Nuclear matter}

The property of a $\Lambda_{c}$ in nuclear matter is discussed by the relativistic mean-field theory~\cite{Tan:2004mt,Tsushima:2002cc,Tsushima:2002ua,Tsushima:2002sm,Tsushima:2003dd}.
As in the conventional mean-field theory~\cite{Serot:1984ey,Serot:1997xg},
we consider that $\sigma$, $\omega$ and $\rho$ mesons make a mean-field, and a nucleon as well as a $\Lambda_{c}$ baryon are coupled to the mean-field.
The couplings are denoted by $g_{\sigma N}$, $g_{\omega N}$, $g_{\rho N}$ for a nucleon and $g_{\sigma \Lambda_c}$, $g_{\omega \Lambda_c}$ for a $\Lambda_{c}$ baryon.
Notice that $\Lambda_c$ does not couple to $\rho$ meson because of the isospin symmetry.
In the mean field theory, the baryon mass in nuclear matter is given by
\begin{align}
M_{N}^{\ast} &= M_{N} - g_{\sigma N} \langle \sigma \rangle, \\
M_{\Lambda_{c}}^{\ast} &= M_{\Lambda_{c}} - g_{\sigma \Lambda_{c}} \langle \sigma \rangle,
\end{align}
with the expectation value of the $\sigma$ field $\langle \sigma \rangle$, hence the partial restoration of the broken chiral symmetry is included in the dynamics~\cite{Tsushima:2002cc,Tsushima:2002ua,Tsushima:2002sm,Tsushima:2003dd}.
The estimate of the coupling strength between the baryon and the meson is analyzed by the quark-meson coupling model.
In this model, the coupling strength is determined by (i) the quark-meson coupling strength ($g_{M}^{q}$) and (ii) the quark wave function inside the baryon from the quark model (the bag model).
The coupling between a $\Lambda_{c}$ baryon and $\omega$
 meson is given as $g_{\omega \Lambda_{c}}=2g_{\omega N}/3$
 for the coupling constant $g_{\omega N}$
 for a nucleon, by assuming $g_{\omega N}=3g_{\omega}^{q}$.
The coupling $g_{\sigma \Lambda_{c}}$ is also obtained as well.
Because the $\sigma$ field has the finite expectation value in the ground state $\langle \sigma \rangle$,
the mass of $\Lambda_{c}$ is sensitive to the partial restoration of the broken chiral symmetry in nuclear matter.
Under the model setting, the authors obtained the binding energy 5.2 MeV for a $\Lambda_{c}$ in Pb nucleus by performing the self-consistent calculation.
The smallness of this number comes from the electric Coulomb repulsion.
In fact, for $\Lambda_{b}$ (electric charge zero), they obtained the binding energy 27 MeV, which is as the same order as the binding energy of a $\Lambda$ hyperon.\footnote{Although the binding energies of the most stable states for $\Lambda_{b}$ and $\Lambda$ in nucleus are almost the same, the number of excited states of the $\Lambda_{b}$ nucleus is larger than that of the $\Lambda$ nucleus.}

Similarly, in Ref.~\cite{Tan:2004mt}, they considered the mean-field $U=g_{\sigma \Lambda_{c}} \langle \sigma \rangle + g_{\omega \Lambda_{c}} \langle \omega_{0} \rangle$ with $g_{\omega \Lambda_{c}}=2g_{\omega N}3$ for a $\Lambda_{c}$ nucleus, and investigated the energy spectrum of the $\Lambda_{c}$ nucleus by changing the $U$ parameter ($U=-10$, $-20$, $-30$, $-40$ MeV).\footnote{Those numbers are comparable with the potential depth of a $\Lambda$ hyperon, namely about 30 MeV.}
As a result, they obtained the $\Lambda_{c}$ bound state in Pb nucleus for $|U|>20$ MeV.

The QCD sum rule is also useful theoretical tools to investigate the $\Lambda_{c}$ bound state in nuclear matter (cf.~Sect.~\ref{sec:QCDSR}).
In Ref.~\cite{Wang:2011hta}, considering the leading order for the change of the chiral condensate and the gluon condensate in nuclear matter,
the authors found that the in-medium masses are $M^{\ast}_{\Lambda_c}=2.335$ GeV and $M^{\ast}_{\Lambda_b}=5.678$ GeV, namely the mass shifts are $\Delta M_{\Lambda_c} = M^{\ast}_{\Lambda_c}-M_{\Lambda_c}=51$ MeV, $\Delta M_{\Lambda_b} = M^{\ast}_{\Lambda_b}-M_{\Lambda_b}=60$ MeV.
On the other hand, the mass shifts of $\Sigma_{c}$ and $\Sigma_{b}$ baryons in nuclear matter are $\Delta M_{\Sigma_{c}}=-123$ MeV and $\Delta m_{\Sigma_{b}}=-375$ MeV, respectively~\cite{Wang:2011yj}. 
Moreover, in Ref.~\cite{Wang:2012xk}, the authors obtained very large mass shifts of $\Xi_{cc}$, $\Omega_{cc}$, $\Xi_{bb}$ and $\Omega_{bb}$ baryons as $\Delta M_{\Xi_{cc}}=-1.11$ GeV, $\Delta M_{\Omega_{cc}}=-0.33$ GeV, $\Delta M_{\Xi_{bb}}=-3.37$ GeV and $\Delta M_{\Omega_{bb}}=-1.05$ GeV, respectively.
Such large value, but with negative sign, for $\Delta M_{\Lambda_{c}}$ is obtained also in Ref.~\cite{Azizi:2016dmr}.

We comment on the possible isospin Kondo effect for charm/bottom baryons in nucleus.
As discussed in Sect.~\ref{sec:D_mesons}, the isospin of a heavy hadron as a heavy impurity particle causes the Kondo effect, namely the logarithmic enhancement of the effective interaction at low energy scattering, which is induced essentially by the non-Abelian (isospin-exchange) type interaction.
Thus it can affect on various properties of the heavy hadrons in nuclear matter.
The Kondo effect can exist, not only for a $\bar{D}$ meson, but also for a $\Sigma_{c}$ baryon as well as for a $\Sigma_{c}^{\ast}$ baryon, because those baryons have the isospin-exchange interaction with a nucleon~\cite{Yasui:2016ngy}.
It can affect the energy spectrum of $\Sigma_{c}$ as well as $\Sigma_{c}^{\ast}$ in a nucleus.
Further studies along this line are an interesting subject.

\section{Summary and future prospects}
\label{sec:summary}

The investigation of 
 hadronic many-body 
systems containing 
different flavors opens a new gate for studying
various aspects of QCD such as 
hadron-hadron interactions, 
modifications of 
the QCD vacuum in the medium
and so forth.

The frontier of nuclear and hadron physics reaches ``heavy-flavor
nuclei'' containing heavy quarks, namely  charm and bottom quarks.
A characteristic 
feature 
of heavy quarks is that 
their masses are much
larger than 
$\Lambda_{\rm QCD}$.
The heavy masses induce the suppression of the kinetic
energy and the emergence of 
the heavy quark symmetry.
The heavy quark spin symmetry 
due to the suppression of the
spin-flip process of the heavy quark
separates
the heavy quark spin and the light degrees of freedom.
These features of heavy quarks produce a new type of nuclear systems, 
e.g. states with the approximate spin degeneracy due to the heavy quark symmetry and the characteristic patterns of decay branching ratios.
However, the different mass thresholds due to the $1/m_Q$ correction may make it difficult to identify the partner in the doublet of the heavy quark spin symmetry.
Nevertheless, such spin structure is one of the new subject not only in view of nuclear and hadron physics but also in view of QCD, which has not 
been observed in the light flavor sectors. 

In this review, we have emphasized the importance of ``symmetries'' and
the ``finite size'' of the heavy-flavor nuclei.
The important symmetries are chiral symmetry and heavy quark symmetry.
Both symmetries are simultaneously manifested in the heavy-flavor nuclei;
the chiral symmetry is realized in the light flavor
sectors of QCD, 
while
the heavy quark symmetry 
emerges 
in the heavy flavor sectors.
The 
interplay
of 
two symmetries is a unique feature of systems with
the heavy and light flavors.
It plays 
a crucial role in the hadron spectra
and the hadron-hadron interactions of the heavy-flavor nuclei.

To make a connection between theories and experiments,
the investigation of ``finite'' systems is important while 
the infinite system 
is 
generally easy to treat.
For light nuclei, we have the Gaussian expansion method to solve few-body systems rigorously. 
For 
heavy nuclei, we have an approach based on the optical potential, which can be constructed from the self-energy of the heavy hadron in nuclear matter. In addition to the strong interaction, the Coulomb interaction is also important in the finite systems.
These studies 
should be helpful to 
understand the hadron many-body
systems theoretically and experimentally.

In this review, 
we have focused on 
the properties of 
heavy hadrons; 
quarkonia ($\bar{Q}Q$),
heavy-light mesons ($\bar{q}Q$ and $q\bar{Q}$) and baryons
($qqQ$), in the nuclear medium.
For each heavy hadron, we discuss the two-body interaction with a nucleon, the few-body systems, and the properties in nuclear matter.

Heavy hadron-nucleon interactions have been investigated by various
approaches.
In particular the quarkonium-nucleon interaction has the unique nature
which is given
by the gluon exchange because the $q\bar{q}$ exchange is suppressed due
to the OZI rule.
The quarkonium-nucleon interaction is described by
the QCD van der Waals potential as a perturbative interaction at short
distance
and the multi-gluon exchange as a nonperturbative interaction at long
distance.
The prediction of the interaction in this picture is quantitatively confirmed by the first principle lattice QCD simulations.
The interaction turns out to be weakly attractive,
while it is not
enough to produce
a two-body quarkonium-nucleon bound state.

On the other hand,
the interactions of $DN$, $\bar{D}N$, $\Lambda_{\rm c}N$  and $\Sigma_{\rm c}N$
have been constructed 
by considering chiral and heavy quark symmetries.
In the light quark sector, the low-energy pion-nucleon interaction is constrained by chiral symmetry. The Yukawa and Weinberg-Tomozawa interactions are the basic ingredients to develop the interaction of heavy hadrons with a suitable generalization.
The heavy quark symmetry requires coupled channels in the heavy hadronic
systems. 
The $DN$, $\bar{D}N$ and
$\Sigma_{\rm c}N$ interactions are considered together with the
$D^\ast N$, $\bar{D}^\ast N$ and $\Sigma^\ast_{\rm c}N$ interactions,
respectively, because $D^\ast$, $\bar{D}^\ast$ and $\Sigma^\ast_{\rm c}$ are the heavy-quark spin partners of $D$, $\bar{D}$ and $\Sigma_{\rm c}$ in the heavy quark limit.
The channel mixing is enhanced 
in the bottom sector where the mass
splitting of the spin partners is 
further reduced in comparison with
the charm sector.

For the interactions of heavy-light mesons, the difference between $D$ 
and $\bar{D}$ 
mesons in nuclear medium 
has been remarked.
The $DN$ 
system is able to couple with the channels of the
meson-heavy baryons and the excited heavy baryons, while the
$\bar{D}N$ 
system is not allowed to couple with other channels at lower
energies.
Hence the $\bar{D}N$ bound state is stable against strong decays.
The magnitude of the interactions is also different.
For the $DN$ potential, many models predict a strong attraction, and
find quasi-bound states of the $DN$ two-body system with
the large binding energy
such as $\Lambda_{\rm c}(2595)$.
On the other hand, the $\bar{D}N$ interactions are obtained as the weak
attraction or repulsion. 
The bound states of the $\bar{D}N$ two-body system
are produced for the models
including $\bar{D}N-\bar{D}^\ast N$ coupled
channels respecting the heavy quark spin symmetry.
For heavy baryon interactions,
the $\Lambda_{\rm c}N-\Sigma_{\rm c}N$ coupling is also non-negligible.
It is the analogous to the $\Lambda N-\Sigma N$ coupling in the
hyperon-nucleon interaction.
The coupled channels of 
$\Lambda_{\rm c}N-\Sigma_{\rm c}N-\Sigma^\ast_{\rm c}N$
give the attractive force which produces the two-body bound states.

The attractive interactions help to 
bind the heavy hadrons both in few-body nuclear systems and in nuclear matter.
For quarkonia,
the weak attraction between a quarkonium and a nucleon 
makes it possible for the quarkonium to be bound by a nucleus.
The properties of the quarkonium in nuclear matter are studied by 
various approaches.
An interesting feature 
is that
it gives us opportunities to investigate the modification of the gluon condensate which is
linked to the QCD vacuum.

In the few-body nuclear systems with 
heavy hadrons,
(quasi-)bound states are 
obtained 
when the attractive two-body interactions are utilized.
It is found 
that the 
properties 
of the few-body systems with heavy hadrons 
are similar to the strangeness nuclear systems, namely $\bar{K}$ nuclei
and hypernuclei.
In nuclear matter,
various phenomena have been discussed as the medium effect for the heavy
mesons and baryons.
The medium effect modifies the mass (or in general the spectral function) of heavy hadrons, as a consequence of the many-body effect.
In addition to them, the coexistent heavy-light flavor
provides interesting aspects,
the chiral symmetry restoration invited by the light
degrees of freedom
and the Kondo effect given by the large mass of a heavy quark.

Heavy quarks in nuclei bring us ideas related to
 the various phenomena and
new approaches to understand QCD.

Based on the above summary of this review, we now remark on future prospects, from the viewpoints of theoretical approaches, and the connections to other systems and experiments.

\vspace{1em}

1.~{\it Theoretical approaches.---}
The theoretical ideas to analyze the nuclear and hadronic systems have
been developed day by day.
Here we list several of new methods and ideas which will help us to understand properties of the
heavy-flavor nuclei in the future.

\begin{itemize}
 \item Lattice QCD \\
       lattice QCD simulation is the powerful method to obtain hadron spectra and
       hadron-hadron interactions from the first-principles, QCD.
       It is useful
       for approaching an interaction
       which is difficult to be determined  by experiments.
       This is of particular importance for providing the basic inputs to the heavy-flavor nuclei.
       There are several well-established methods to investigate the interactions:
       the L\"uscher's finite volume
       method~\cite{Luscher:1990ux,Briceno:2014pka,Prelovsek:2014zga,Yamazaki:2015nka} and
       the HAL QCD
       method~\cite{Ishii:2006ec,Aoki:2009ji,HALQCD:2012aa,Aoki:2012tk,Iritani:2015dhu}.
       The interaction is derived by the 
       energy shift of the two-body system in the L\"uscher's
       method, and by
       the Nambu-Bethe-Salpeter wave functions in the HAL QCD method.
       Recently the lattice calculation is applied  
       not only to two-body interactions but also 
       to few-body interactions being important in the many-body systems
       such as the heavy-flavor nuclei~\cite{Aoki:2012tk,Doi:2011gq}.
       However the many lattice computations have associated with the
       unphysically large quark masses.
       The lattice calculation performing at the physical point is
       one of the important challenges in
       progress~\cite{Aoki:2012oma,Doi:2015oha}.
              
 \item Gauge/gravity correspondence \\
       The gauge/gravity correspondence (holography) provides a
       method to approach strongly coupled gauge theories at large-$N_c$~\cite{Maldacena:1997re}.
       This method is based on the idea of the duality of the strongly
       coupled gauge theory and the weakly coupled gravity theory.
       The original idea of the holography, AdS/CFT correspondence~\cite{Maldacena:1997re},
       is limited to a system possessing the supersymmetry (SUSY).
       However the treatment to break SUSY has been developed as the
       compactification which introduces the breaking
       scale~\cite{Witten:1998zw,Sakai:2004cn,Sakai:2005yt}.
       The holographic approaches have been applied to the investigation of
       the hadron physics 
       such as        
       hadron spectra~\cite{Sakai:2004cn,Sakai:2005yt,Karch:2002sh,Kruczenski:2003be,Kruczenski:2003uq,Erdmenger:2007cm},
       hadron-hadron interactions~\cite{Hashimoto:2009ys}, atomic nuclei~\cite{Hashimoto:2008jq,Hashimoto:2011nm}
       and quark-gluon plasma~\cite{CasalderreySolana:2011us,Natsuume:2014sfa}.
       It has also been attempted to investigate the heavy flavor
       physics~\cite{Erdmenger:2007cm,Jo:2011xq,Hayata:2012rw}.
       In the top-down approaches, however, the presence of the heavy quark symmetry in the
       holographic model is not clear~\cite{Hashimoto:2014jua},
       because the pseudoscalar and vector mesons belong always to the
       same multiplet
       in the gauge theory possessing the SUSY.
       In fact the D3-D7 model predicts the degeneracy of these mesons 
       regardless of flavors.
       Even when the supersymmetry is broken by the
       compactification,
       this symmetry recovers in the UV limit corresponding to the heavy
       quark mass limit~\cite{Kruczenski:2003uq}.

       The phenomenological models of QCD inspired by
       the gauge/gravity correspondence have also been developed.
       The models being the bottom-up approach are called
       AdS/QCD~\cite{Erdmenger:2007cm,Erlich:2005qh,DaRold:2005mxj,Kim:2007rt}.
       This approach is applied to the description of the hadron spectroscopy
       including heavy hadrons~\cite{Branz:2010ub,Gutsche:2011vb}.
       However understanding the connections between QCD and AdS/QCD, and the string
       theory and AdS/QCD remains an open issue.

       The description of QCD including the heavy hadron
       dynamics is a challenging subject in the holographic approach.
             
 \item Methods of many-body systems with finite baryon number\\
       Computing methods for solving many-body systems with finite
       baryon number have been developed in various fields.       
       Those methods are useful to study nuclei composed of
       nucleons.
       In addition, the many-body calculations are also
       utilized to analyze the strangeness nuclear systems 
       containing impurities, which are linked to the heavy-flavor
       nuclei.
       Here a part of the methods applied to the strangeness nuclear
       systems
       is summarized.
       
       Nuclear shell model and Hartree-Fock method are well-known
       approaches based on a mean-field where the many-body system is
       reduced to a single-particle motion in a field given by
       effects of the other particles.
       Nuclear shell model is inspired by the shell structure of
       electrons in atoms.
       The single-particle motion in the mean-field
       potential with the spin-orbit force
       achieves success in explaining the magic number of atomic nuclei.
       The shell model calculation for hypernuclei has been discussed in
       Refs.~\cite{Gal:1971gb,Gal:1972gd,Gal:1978jt,Millener:2008zz,Gal:2011zr}.
       Another widely used method
        to solve a many-body quantum system
       is the Hartree-Fock method.
       The mean-field potential and the wave function of the
       single-particle are obtained by solving the Hartree-Fock equation
       self-consistently~\cite{Zhou:2007zze,Win:2008vw,Schulze:2010zzb,Mei:2015pca}.

       Another important aspect of the nuclear structures is the   clustering phenomena~\cite{PTPS52.89}. 
       Antisymmetrized molecular dynamics (AMD) is the model being able
       to describe properties both of cluster structure wave functions
       and independent-particle motion in a mean
       field~\cite{Ono:1991uz,KanadaEn'yo:2012bj}. 
       Recently AMD is applied to the hypernuclei~\cite{Isaka:2011kz,Isaka:2015iip}.
       
       \item Compositeness\\
       Compositeness characterizes the structure of stable bound states
       close to the thresholds.
       As originally introduced by
	   Weinberg~\cite{Weinberg:1962hj,Weinberg:1965zz}.
       the composite/elementary nature of hadrons is
	      estimated
	      from the field renormalization constant~\cite{Baru:2003qq,Baru:2010ww,Hyodo:2011qc,Aceti:2012dd,Hyodo:2013nka,Sekihara:2014kya}.
       Recently the idea of the compositeness has been extended to the
       quasi-bound states~\cite{Kamiya:2015aea,Guo:2015daa}.
       Since a lot of exotic states have been found as the quasi-bound
       states near the thresholds in the heavy flavor regions,
       the compositeness is helpful to discuss the structure of them.
       For the heavy-flavor nuclei, comparison of the compositeness in
       vacuum and nuclear medium would give us the information on the
      modification of properties of the heavy hadrons in environmental
	      changes.
\end{itemize}

2.~{\it Connections to other systems.---}
The physics of heavy-flavor nuclei is 
 broadly connected to other nuclear and hadronic systems.
The connections encourage us to understand the relations between the high-energy quark-gluon dynamics and the low-energy hadron and nuclear dynamics.
We summarize current status of the neighboring fields which are helpful to study the
heavy-flavor nuclei.

\begin{itemize}
 \item Quarkonia and $XYZ$ resonances\\
       As summarized in this review, 
       the quarkonium-hadron interaction is dominated by the gluonic
       degrees of freedom. 
       While the results of different approaches agree with each other that the quarkonium-nucleon interaction is weakly attractive,
       there are still nontrivial issues to be clarified in the future.
       For instance, a recent lattice QCD calculation of $Z_{\rm c}(3900)$ 
       in Ref.~\cite{Ikeda:2016zwx} indicates 
       the importance of the couplings of $\pi J/\psi-\bar{D}D^\ast$ 
       and $\rho\eta_{\rm c}-\bar{D}D^\ast$. These couplings are considered to be suppressed in phenomenological models, 
       because it is accompanied by the heavy meson exchange interaction.
       The couplings of a quarkonium and a pair of heavy-light mesons
       are also important in the mass shift of
       quarkonia~\cite{Heikkila:1983wd,Eichten:2005ga,Pennington:2007xr}.
       
       The excited charmonia are also interesting objects to be studied. 
       The hadronic loop of the meson-antimeson affects
       the properties of quarkonia, in particular those
       close to thresholds.
       Moreover, there are many candidates of the exotic states, called $XYZ$ resonances,
       which are considered to have non-conventional
       structures such as the compact
       multi-quark states, hadronic molecules and hybrid states with 
       gluonic degrees of freedom~\cite{Brambilla:2010cs}.
       It is an interesting open problem to understand the properties of 
       the quarkonia close to thresholds and
       the $XYZ$ resonances having an exotic structure,
       both in vacuum and in the nuclear medium.
       
 \item Nuclear systems with impurities \\ 
       There have been discussions about nuclei with impurity
       particles.
       The study of the composite systems is a challenging topic to
       analyze both nuclear structures and the modification of the properties of the
       impurities in nuclear medium.
       The impurity particles are not affected by the
       Pauli-blocking
       from 
       the nucleons.
       Therefore the impurity is allowed to enter the deep inside
       of the nucleus,
       and expected to be the probe to study the region which cannot be accessed by nucleons.
       In addition, the impurities would induce the shrinkage 
       effect and change the deformation structure of
       nuclei~\cite{Hashimoto:2006aw,Botta:2012xi,Schulze:2010zzb}.
       On the other hand,
       properties of the impurity itself would also be changed in the nuclear
       medium~\cite{Hayano:2008vn}.
       As an example of mutual change of both nuclear medium and impurities, 
       we have discussed the heavy quark symmetry leading to
       the recombination of the spin correlations inside the nuclear matter.
       For example, the heavy quark symmetry gives the recombination of the spin correlations inside the nuclear matter.

 \item From strangeness to heavy flavor \\
       The quark mass is one of the fundamental parameters in QCD.
       To compare the charm/bottom nuclear physics with the strangeness nuclear physics
       gives us a chance to study the role of flavors in 
       nuclei in terms of QCD. 
       The strangeness nuclear systems have been investigated as
       the nuclear system with impurity particles with strangeness, e.g. kaons and
       hyperons~\cite{Hashimoto:2006aw}.
       The heavy hadron nuclear physics can be situated as an
       extension from strangeness to heavy flavors.
       The properties of the heavy-flavor nuclei should 
        be different from
       the strangeness nuclei due to the properties of the heavy quarks,
       i.e. heavy mass and heavy quark symmetry, as
       discussed in this review.
       The internal excitation pattern and interactions of heavy hadrons
       are 
       different from those in light hadrons. 
       For example, the excitation pattern of charm/bottom baryons may be comparable with the strange baryons,
       as the reduction of the spin-spin interaction is known already
        in the strangeness sector~\cite{Yoshida:2015tia}.

 \item Quark-gluon plasma (QGP) \\
       The investigation of QGP is one of the challenging subjects
       in hot and dense QCD.
       The QGP is the different phase from the hadronic matter,
       and is considered to be realized in the early universe
       and in the relativistic heavy ion collisions (HIC).
       The heavy quarks and quarkonia produced in HIC are used as hard probes to      
       understand the microscopic dynamics of
       QGP~\cite{Matsui:1986dk,Rapp:2009my,Beraudo:2014iva}.
       QGP gives us an opportunity to investigate the heavy quark
       (hadron) dynamics in systems with high
       temperature or high density.
       Recently the Langevin dynamics, a classical description of the
       Brownian motion, of a heavy quark in QGP
       has been investigated in
       Refs.~\cite{Akamatsu:2008ge,Akamatsu:2015kaa}.       
\end{itemize}

3.~{\it Experiments.---}
The interesting physics of heavy-flavor nuclei should eventually be examined in actual experiments. For this purpose, it is necessary to consider how to produce the heavy particles.
We summarize 
discussions of the production of the heavy hadrons and heavy-flavor nuclei in the experiments, together with possible facilities.
 \begin{itemize}
  \item Hadron beams \\
	The charm production by the pion beam on a proton target,
	$\pi^- p \rightarrow D^{\ast -}\Lambda^+_{\rm c}$, are studied
	in Refs.~\cite{Kim:2014qha,Kim:2015ita}.
	The cross section of the process is obtained by using two
	different theoretical frameworks, the effective Lagrangian
	method and the Regge method.
	The coupling constants of the vertices of the charmed meson and
	baryon are assumed to be the same as those for the strangeness
	sector.
	The obtained total cross section of the charm production is
	about $10^{3}-10^{6}$ smaller than that of the strangeness
	production. 
	The hidden-charm baryon production near threshold via the pion
	beam is
	studied in Ref.~\cite{Garzon:2015zva}.
	The enhancement of the total cross section of the 
	$\pi^-p\rightarrow D^-\Sigma^+_{\rm c}$ by the $N^\ast_{\rm cc}$
	resonance is discussed. 	
	The charm production via the pion beam would be performed in
	J-PARC.

	In Ref.~\cite{Yamagata-Sekihara:2015ebw}, the formation of the
	charmed mesic nuclei, $D^- -^{11}$B
	and $D^0 -^{11}$B, in antiproton beam on a $^{12}$C is
	investigated theoretically.
	The optical potential for the charmed meson is given by the
	self-energy where
	the extended Weinberg-Tomozawa interaction is employed as 
	the charmed meson-nucleon interaction in the matter.
	The results suggest the possible observations of the charmed
	mesic nuclei at $\bar{\rm P}$ANDA in FAIR~\cite{Rapp:2011zz} and J-PARC.

  \item Relativistic heavy ion collisions \\
         Heavy ion collisions at Relativistic Heavy Ion
	Collider (RHIC) and Large Hadron Collider (LHC) provide a new stage
	to study the reaction with high multiplicities. Due to the high
	temperature and large volume, the abundance of the yields of the
	produced particles are much larger than those in $e^{+}e^{-}$
	and $pp$ collisions and the fixed target experiments. In addition, the collision source as a rich baryon number
	environment makes it possible to produce,
	not only single hadrons, but also composite nuclei. For example, there are
	reports of observation of anti-hypernuclei in RHIC, whose
	production is difficult to be realized in other
	reactions~\cite{58}. Recently, productions of exotic hadrons,
	containing strangeness, charm and bottom quarks, from
	relativistic heavy ion collisions have been studied in theories
	based on the statistical model and the coalescence
	model~\cite{Cho:2010db,Cho:2011ew}. Importantly, the yields of
	the exotic hadrons depend on the internal configurations of
	composite particles, namely compact mutli-quarks or extended
	hadron molecules. Those approaches are applied to productions of
	charm/bottom nucleus with small baryon numbers.
  \item Photoproduction \\
	The heavy hadron productions in the photon-induced processes have been discussed in the
	literatures.
	The productions of hidden-charm baryon resonances, 
	$N^\ast_{\rm cc}$ and $\Lambda^\ast_{\rm cc}$, including 
	the pentaquarks $P^+_{\rm c}(4380)$ and $P^+_{\rm c}(4450)$ have been
	investigated in
	Refs.~\cite{Huang:2013mua,Wang:2015jsa,Huang:2016tcr}.	
	The effective Lagrangian approach 
	is used in these analyses.
	The photoproductions for the hadronic states with heavy flavor
	are expected to be performed at Jefferson Lab.
  \item Neutrino-nucleus reaction \\
	The baryon production via the neutrino-nucleus reaction has been
	studied for light flavor
	sectors~\cite{Kamano:2012id,Nakamura:2015rta}.	
	The weak interaction also makes it possible to produce the heavy
	hadrons in
	the neutrino-nucleon reactions such as 
	$\nu N\rightarrow l X$ where $l$ and $X$ stand for a lepton and a
	produced heavy baryon resonance, respectively.
	The heavy hadron production via the neutrino-nucleus
	reaction is 
	utilized to produce the heavy-flavor nuclei.
 \end{itemize}

\section*{Acknowledgments}
The authors thank M.~Oka, M.~Harada, A.~Yokota, K.~Suzuki, K.~Ohtani and S.~Maeda for fruitful discussions and useful comments.
This work is supported by JSPS KAKENHI (the Grant-in-Aid for Scientific
Research from Japan Society for the Promotion of Science (JSPS)) with
Grant Nos.~JP24740152 (T.~H.),  JP16K17694 (T.~H.), JP25247036 (S.~Y.), JP15K17641(S.~Y.) and JP26400273 (A.~H.),
by the Yukawa International Program for Quark-Hadron Sciences (YIPQS)
and by the INFN Fellowship Programme.

\appendix
\def\thesection{\Alph{section}}
\def\thesubsection{\thesection.\arabic{subsection}}

\section{Matrix elements of the few-body calculations}
\label{sec:appendix_fewbody_systems}

In this Appendix, we give the matrix elements
which are helpful to solve the eigenvalue problem of
the few-body calculations with the Gaussian expansion method shown in
Sect.~\ref{sec:fewbody_systems}.

For the norm, 
the general form of the $(i,j)$ component of
the spatial part without coefficients
is obtained by
\begin{align}
 N_{ij}&=\int d^3r\int d^3R \,r^{\nu_1}R^{\nu_2}
 \exp{(-\beta_1r^2-\beta_2R^2+\gamma\, \vec{r}\cdot\vec{R})} \notag \\
 &\quad \times\left[Y_{l_1}(\hat{r})\otimes Y_{l_2}(\hat{R})\right]_{LM}
 \left[Y_{l_3}(\hat{r})\otimes Y_{l_4}(\hat{R})\right]_{L^\prime
 M^\prime} 
 \notag \\
 &=\int_{0}^{\infty}dr\int_{0}^{\infty}dR
 r^{\nu_1+2}R^{\nu_2+2}\exp{(-\beta_1r^2-\beta_2R^2)}
 4\pi\sqrt{\frac{\pi}{2\gamma rR}}
\sum_{\lambda=0}^{\infty}I_{\lambda+1/2}(\gamma rR) \notag \\
 &\quad \times
 \left\langle 
 \left[Y_{l_1}(\hat{r})\otimes Y_{l_2}(\hat{R})\right]_{LM}
 \right|
 Y^\ast_{\lambda}(\hat{r})\cdot Y_{\lambda}(\hat{R})
 \left| 
 \left[Y_{l_3}(\hat{r})\otimes Y_{l_4}(\hat{R})\right]_{L^\prime
 M^\prime} 
 \right\rangle
 \notag \\
 &=\sum_{\lambda,n}\frac{\pi}{2}\frac{1}{n!}
 \frac{(\nu_2+\lambda+1)!!(\nu_1+\lambda+2n+1)!!}{(2(\lambda+n)+1)!!}
 \frac{((\nu_2-\lambda)/2)!}{((\nu_2-\lambda)/2-n)!}
 \notag \\
 &\times
 \left(\frac{\gamma}{\sqrt{2\beta^\prime_1 2\beta_2}}\right)^{\lambda+2n}\frac{1}{(2\beta^\prime_1)^{(\nu_1+3)/2}(2\beta_2)^{(\nu_2+3)/2}}
 \notag \\
 &\times
\delta_{LL^\prime}\delta_{MM^\prime}\frac{(-1)^{L^\prime+l_3+l_2}}{4\pi}
 \left\{
 \begin{array}{ccc}
  l_1&l_2 &L \\
  l_4&l_3 &\lambda \\
 \end{array}
 \right\} \notag \\
 &\times
 (2\lambda+1)\sqrt{(2l_3+1)(2l_4+1)}\,
 \clebz{l_3}{\lambda}{l_1} \clebz{l_4}{\lambda}{l_2} \, ,
 \label{eq:normfin}
\end{align}
where $\beta^\prime_1\equiv \beta_1-\gamma^2/4\beta_2$.
The values of $\nu_1$, $\nu_2$, $\beta_1$, $\beta_2$ and $\gamma$
depend on the component $(i,j)$.
The coefficient 
such as
the normalization factor
is omitted.
The summation of $\lambda$ is restricted by the Clebsch-Gordan
coefficients as $|l_1-l_3|\leq\lambda\leq l_1+l_3$ and 
$|l_2-l_4|\leq\lambda\leq l_2+l_4$.
The summation of $n$ is $0\leq n \leq (\nu_2-\lambda)/2$.

The kinetic term is reduced to 
\begin{align}
 \left(\frac{\partial^2}{\partial r^2}-\frac{l(l+1)}{r^2}\right)r^{l+1}\exp{\left(-\frac{r^2}{2b^2_i}\right)}Y_l(\hat{r})
 &=\left(\frac{r^2}{b^4_i}-\frac{2l+3}{b^2_i}\right)\,r^{l+1}\exp{\left(-\frac{r^2}{2b^2_i}\right)}Y_l(\hat{r})
 \, .
\end{align}
Therefore, the matrix element of the kinetic term is obtained in the same
manner as the norm.

The matrix element of a product of angular momentum operators 
$\vec{\cal O}\,^{(L)}_a\cdot\vec{\cal O}\,^{(S)}_a V(r)$
 with the orbital (spin) operator 
$\vec{\cal O}\,^{(L)}_a$ ($\vec{\cal O}\,^{(S)}_a$) and the potential 
function $V(r)$
is given by
\begin{align}
 &\left\langle \Phi^i_{J(LS)}\right|
 \vec{\cal O}\,^{(L)}_a\cdot\vec{\cal O}\,^{(S)}_a V(r)
 \left| \Phi^j_{J(L^\prime S^\prime)}\right\rangle \notag\\
 &=\int_{0}^{\infty}dr\int_{0}^{\infty}dR
 r^{\nu_1+2}R^{\nu_2+2}\exp{(-\beta_1r^2-\beta_2R^2)}
 4\pi\sqrt{\frac{\pi}{2\gamma rR}}
\sum_{\lambda=0}^{\infty}I_{\lambda+1/2}(\gamma rR) \notag \\
 &\quad \times
 (-1)^{L^\prime+J+S}\left\{
 \begin{array}{ccc}
  L^\prime &a &L \\
  S&J&S^\prime\\
 \end{array}
 \right\}
 \left\langle 
 \left[Y_{l_1}(\hat{r})\otimes Y_{l_2}(\hat{R})\right]_{LM}
 \left\| 
 Y^\ast_{\lambda}(\hat{r})\cdot Y_{\lambda}(\hat{R}) 
 \,V(r) {\cal O}\,^{(L)}_a
 \right\| 
 \left[Y_{l_3}(\hat{r})\otimes Y_{l_4}(\hat{R})\right]_{L^\prime
 M^\prime} 
 \right\rangle
\notag \\
&\quad \times
 \left\langle 
 \left[\left[S_1,S_2\right]_{S_{12}},S_3\right]_S
 \left\|
 {\cal O}\,^{(S)}_a
 \right\|
 \left[\left[S^\prime_1,S^\prime_2\right]_{S^\prime_{12}},S^\prime_3\right]_{S^\prime}
 \right\rangle.
\end{align}


%

\end{document}